\documentclass[monophys,vecphys]{svmonou}%
\usepackage{makeidx}
\usepackage{graphicx}
\usepackage{multicol}
\usepackage[bottom]{footmisc}
\usepackage{amsmath}
\usepackage{amsfonts}
\usepackage{amssymb}
\usepackage{graphicx}
\usepackage{lineno}
\usepackage[T1]{fontenc}
\usepackage{blindtext}
\usepackage[latin1]{inputenc}
\usepackage[svgnames]{xcolor}
\usepackage{framed}%
\definecolor{shadecolor}{gray}{0.86}
\makeindex

\newenvironment{myexample}{\begin{shaded}\begin{example}}{\end{example}\end{shaded}}
\begin{document}

\author{Günter Zech}
\title{Analysis of distorted measurements - parameter estimation and unfolding}
\maketitle
\tableofcontents

\newpage

\chapter{Introduction}

\begin{figure}
[ptb]
\begin{center}
\includegraphics[
trim=0.000000in 0.175071in 0.000000in 0.092112in,
height=2.841in,
width=4.0747in
]%
{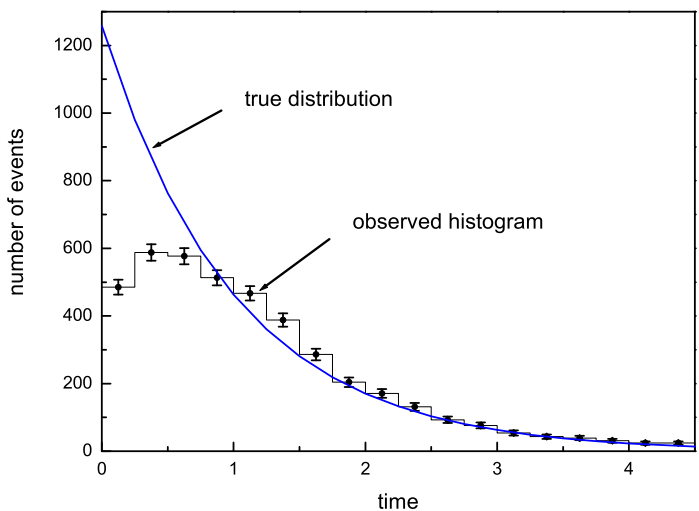}%
\caption{Smeared lifetime distribution.}%
\label{lifetime0}%
\end{center}
\end{figure}

The detectors of physics experiments are never perfect. They suffer from
acceptance losses and from their finite resolution. As a consequence the
distributions that we want to measure are distorted. A typical example is
shown in Fig. (\ref{lifetime0}) where an observed lifetime distribution is
displayed.%

The original distribution, we call it the \emph{true distribution}, is smeared
and events at the low end of the scale are lost. Our goal is to recover as
much as possible the true distribution of the data and in this specific
example, we want to extract the interesting parameter, the true lifetime.

In the past, until about 1990, in most experiments it was adequate to perform
simple corrections to distorted distributions, but in the 90ties when
structure functions were measured in electron proton collisions at the
electron proton collider HERA, more sophisticated unfolding methods had to be
applied. With new data from the Large Hadron Collider (LHC) at CERN the
analysis of distorted data gained new interest. There exist by now a
considerable number of publications on unfolding and parameter reconstruction.
Some early publications \cite{rich72,lucy74,shepp82,vardi85,muelthei86,nara86}
after the pioneering work of Tikhonov \cite{tikhonov} remained essentially
unnoticed by particle physicists. Blobel was the first to promote professional
unfolding in our field \cite{blobel85}. Since then, a large number of
different approaches and studies were published.
\cite{any91,schmelling,lindemann,dagostini95,zech95,hoecker,na38,maga98,dagostini2010,truee,zech,dembinski,blobelbook,kuusela,volobouev,kuuselastark}%
A workshop was held at CERN \cite{cernworkshop} with interesting contributions
but diverging proposals of the participants and an introduction by Lyons
pointing to the essential problems of unfolding which we try to address in
this report. An elementary introduction to unfolding is given in a note by
Cowan \cite{cowan} and a professional introduction to conventional unfolding
with a the focus on singular value decomposition is presented in the book by
Hansen \cite{hansen}.

Most authors from the particle physics community demonstrate the quality of
their proposed methods with Monte Carlo studies, based on a single or very few
data sets and selected distributions. It is not clear how well the obtained
results and conclusions can be generalized and sometimes the choice of the
free parameters of the models is somewhat arbitrary because they cannot be
derived from first principles. The aim of this report is to summarize, test,
extend and compare existing methods, but clearly not all aspects of unfolding
can be covered. The author's hope is that further systematic studies are
stimulated by this report and that in this way a consensus among particle
physicists can be reached on how distorted data should be analyzed.

The experimental situation is usually the following: We are given a sample of
observations which we call events, drawn from a statistical distribution. Each
event is characterized by a set of measured variables like energy, momentum,
time etc.. To simplify the following discussion, we limit ourselves to a
single variable which we denote by $x$. The extension of the results to
several variables is straight forward. To extract useful information from the
sample, we have to understand the detector properties sufficiently well, such
that we can simulate the expected distribution of observed events for a given
undistorted true distribution.

Let us distinguish three different situations and issues:

\begin{enumerate}
\item We have of a \emph{parametric model}. The true distribution from which
the observed events are generated is given up to some unknown parameters. We
want to infer the parameters.

\item We completely ignore the true distribution, but we want to prepare the
data in such a way that they can be quantitatively compared to theoretical
predictions and to results from other experiments. Furthermore, we require
that results from different experiments can be combined. To this end, we
parametrize the true distribution, usually in form of a histogram or by a
spline approximation and we determine the corresponding parameters, i.e. the
content of the histogram bins or the spline coefficients.

\item As in issue 2, we intend to estimate the completely unknown true
distribution, but have the prior information that the distribution is smooth
of which we want to take advantage of. The true distribution has to be
parametrized and the parameters have to be estimated under constraints which
correspond to our smoothness prejudices. Most unfolding procedures refer to
this issue.
\end{enumerate}

As we will see, \emph{issue 1} has a relative simple solution. If we have a
parametric model where $f(x|\vec{\theta})$ is known up to unknown parameters
$\vec{\theta}$, we can estimate the parameters with least square (LS) or
maximum likelihood (ML) methods. We compare the folded true distribution to
the observed distribution and vary the parameter of the true distribution
until the LS statistic is minimum or the likelihood is maximum. In some cases
we can fit the parameters of $f(x|\vec{\theta})$ to the observed sample
ignoring the experimental distortion and correct for the bias of the results
by a Monte Carlo simulation.

The situation is similar for \emph{issue 2}. We have to decide for the kind of
parametrization and choose the number of parameters, i.e. the number of bins
in case the result is presented in form of a histogram. This number has to be
relatively small in order to avoid strong fluctuations, excessive correlations
and huge diagonal errors. Another, but not very realistic possibility which
permits a larger number of bins, would be to restrict the fluctuations by a
smoothing algorithm which than has to be published together with the result.
The same smoothing could then be applied to arbitrary predictions before they
are compared to the experimental result. A quantitative comparison or a
combination of data from different experiments would not be possible.

\emph{Issue 3} is what usually is meant with the notion of unfolding. It
belongs to the field of \emph{parameter density estimation} (PDE)
\cite{scott,porter}. It is less well defined than the other two issues because
there is no precise definition of smoothness. Smoothness is not invariant
under transformations of the random variable and usually our prejudices of
what is smooth depend on the problem that we have to solve. The result of
unfolding a mass distribution will usually be incompatible with that obtained
from unfolding a mass squared distribution. Also the smoothness criteria may
be different when we investigate a line spectrum superposed to a background or
when a transverse momentum distribution in a particle experiment has to be
unfolded. The smoothness constraints improve formally the precision of the
results but introduce a bias. A compromise between precision and bias has to
be found. Since the result and the size of the errors obtained from the
adjustment depend crucially on the smoothness assumptions, the unfolding
result is not suited for the inference of parameters of distributions. However
it provides a semi-quantitative illustration of the true distribution and is
normally closer to the true distribution than a simple parametrization. It can
be used to discard theoretical concepts if the discrepancy between the
unfolding solution and the prediction is large and occasionally the results
may be useful as input for simulation, for instance of structure functions.
Due to the complexity of the unfolding problematic, the corresponding chapter
in this report is much longer than the others.

Some colleagues believe that the issues 2 and 3 can be solved with a single
method. However, it is not clear how the smoothing parameters can be fixed and
how the systematic uncertainties in the parameter estimation can be handled.

Unfolding in higher dimensions suffers from the \emph{curse of dimensionality}%
: The bins of multi-dimensional histograms are often sparsely populated. This
problem can partially be solved with binning-free unfolding methods. A modest,
explorative study is presented in \cite{bohm}.

A technical remark: In many of the figures of this report the titles of the
axes and even scales have been omitted to save space where they are obvious or
not necessary for the understanding of the intended message.

Some parts of this report have been copied with minor modifications from the
book by Bohm and myself \cite{bohm}.


\chapter{\protect\nolinebreak Parameter inference}

In particle physics we are in the lucky situation that in most cases we have a
theoretical description of the data that we collect. The reason is that the
experiments are usually designed with the goal to test a prediction or to
measure parameters of it. We may want, for instance, to determine the lifetime
of a particle from an exponential distribution or the mass and width of a
particle from a Breit-Wigner distribution.

One might think, that first of all, one should unfold the observed
distribution to get rid of the experimental defects and than pursue with the
analysis. However, this is not a good idea, because unfolding is not straight
forward and it is accompanied by a loss of information. It is much better to
fold the prediction and to compare the folded prediction to the observed data.
In this way we can avoid some of the approximations that are necessary in
unfolding methods and we do not have to care about the oscillations that occur
in unfolding procedures due to the statistical fluctuations of the data.

In \cite{blobelbook} a so-called parametric unfolding method is proposed. This
approach has the disadvantage that a response matrix has to be constructed
which depends on the distribution chosen to simulate the smearing. This
dependence can be avoided by iteration but this is not necessary in the much
simpler weighting procedure described in Sect. \ref{weighted}.

\section{Parameter correction method}

Often the true distribution is only slightly distorted by the measurement.
Then we can initially neglect the experimental effects and fit the parameter
we are interested in. We obtain a biased maximum likelihood or least square
estimate $\hat{\theta}^{\prime}$ and an uncertainty $\delta\theta^{\prime}$.
The bias is then estimated by a Monte Carlo simulation based on a value
$\theta$ close to the true value or on $\hat{\theta}^{\prime}$. The bias is
usually within the uncertainties independent of the value chosen in the
simulation. In more than $99\%$ of all measurements in particle physics this
simple method is applied or unnecessary if the bias is negligible.

If the distortions are large, we have to generate the true distribution for a
few values of $\theta$ and simulate the estimation of $\theta^{\prime}$ to
obtain the relation $\theta(\theta^{\prime})$ between the parameter of
interest $\theta$ and the observed quantity $\theta^{\prime}$. The relation
$\theta(\theta^{\prime})$ can usually be taken to be linear in the vicinity of
$\hat{\theta}^{\prime}$ and then the choice of two values of $\theta$ is
enough. The method should become clear in the following examples.%

\begin{figure}
[ptb]
\begin{center}
\includegraphics[
trim=0.200370in 0.156938in 0.194402in 0.100778in,
height=2.2889in,
width=5.98in
]%
{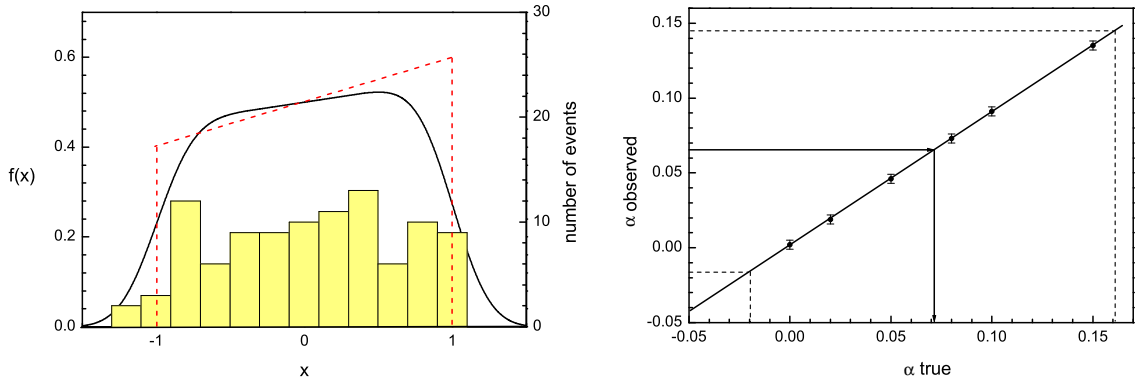}%
\caption{Fit of a linear distribution. Left hand: Histogram of the data and
shape of the true and the distorted distributions. Right hand: Transition from
the observed parameter to its true value.}%
\label{fitlin}%
\end{center}
\end{figure}

\begin{myexample}
Polarization measurements of hyperons often lead to fits of linear cosine
distributions. In Fig. \ref{fitlin} such a distribution is displayed. The
dashed line shows a p.d.f. $f(x)=0.5+1.1x$, where $x$ is confined to the
interval $-1\leq x\leq1$. The folded version of this distribution, with a
kernel following a normal distribution with standard deviation $\sigma
_{s}=0.2$,%
\begin{align*}
g(x)  &  =\int_{-\infty}^{\infty}\mathcal{N}(x^{\prime}|x,\sigma
_{s})f(x^{\prime})dx^{\prime}\;,\\
\mathcal{N}(x^{\prime}|x,\sigma_{s})  &  =\frac{1}{\sqrt{2\pi}\sigma_{s}}%
\exp[-\frac{(x^{\prime}-x)^{2}}{2\sigma_{s}^{2}}]\;,
\end{align*}
extends to regions outside the interval borders of the true distribution. A
histogram from a sample of $100$ events, $x_{1},x_{2},...,x_{100}$ generated
according to the folded distribution is displayed in the same figure. A
maximum likelihood fit of the parameter $\alpha$ to the observed data with the
log-likelihood function%
\[
\ln L(\alpha)=%
{\displaystyle\sum\limits_{i=1}^{100}}
(0.5+\alpha x_{i})
\]
yields the value $\hat{\alpha}^{\prime}=0.065_{-0.082}^{+0.080}$. To correct
this number for the smearing effect, several samples of $10^{5}$ events each,
are generated with different values of $\alpha$. From the corresponding fitted
values a within statistics linear relation $\alpha^{\prime}=0.90\alpha$ is
derived as shown at the lright-hand side of Fig. \ref{fitlin}. This relation
is then used to correct for the bias of $\alpha^{\prime}$. The result is
$\hat{\alpha}=\hat{\alpha}^{\prime}/0.90=0.072_{-0.091}^{+0.089}$. The factor
$0.9$ which relates $\alpha$ and $\alpha^{\prime}$ indicates that the
distortion of the measurement entails a loss in precision of $10\%$.
\end{myexample}

%

\begin{figure}
[ptb]
\begin{center}
\includegraphics[
trim=0.000000in 0.112775in 0.000000in 0.099439in,
height=3.2461in,
width=4.4284in
]%
{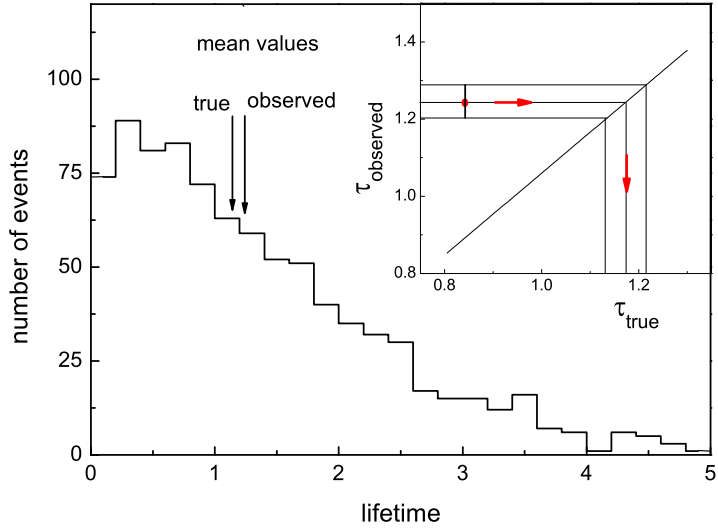}%
\caption{Experimental lifetime distribution. The insert indicates the
transition from the observed to the corrected lifetime.}%
\label{lifemean}%
\end{center}
\end{figure}

\begin{myexample}
In Fig. \ref{lifemean} an observed distorted lifetime distribution is
depicted. The sample mean $\overline{t}$ of a sample of $N$ undistorted
exponentially distributed lifetimes $t_{i}$, is a sufficient estimator of the
mean lifetime \thinspace$\tau$. It contains the full information related to
the parameter $\tau$, the mean lifetime. In case the distribution is distorted
by resolution and acceptance effects, the mean value
\[
\overline{t^{\prime}}=\sum t_{i}^{\prime}/N
\]
of the distorted sample $t_{i}^{\prime}$ will usually still contain almost the
full information relative to the mean life $\tau$. The relation $\tau
(\overline{t^{\prime}})$ between $\tau$ and its approximation $\overline
{t^{\prime}}$ (see insert of Fig. \ref{lifemean}) is generated by a Monte
Carlo simulation. The uncertainty $\delta\tau$ is obtained by error
propagation from the uncertainty $\delta\overline{t^{\prime}}$ of
$\overline{t^{\prime}}$,
\begin{align*}
(\delta\overline{t^{\prime}})^{2}  &  =\frac{(\overline{t^{\prime2}}%
-\overline{t^{\prime}}^{2})}{N-1}\;,\\
\;\;\;\overline{t^{\prime2}}  &  =\frac{1}{N}\sum t_{i}^{\prime2}%
\end{align*}
using the Monte Carlo relation $\tau(\overline{t^{\prime}})$.
\end{myexample}

The correction approach has several advantages:

\begin{itemize}
\item It is not necessary to histogram the observations. A likelihood fit with
individual observations can be performed.

\item Problems due to small event numbers for bins in a multivariate space are avoided.

\item It is robust, simple and requires little computing time if a sufficient
statistic exists. It is ideal for online applications.

\item All approximations are automatically corrected for by the simulation.
\end{itemize}

As we have to perform a likelihood or a $\chi^{2}$ fit, in the cases where we
do not have a sufficient statistic, we occasionally may run into the following
problem: The data can lay outside the range covered by the undistorted p.d.f..
In fact this also happens in the first example of this section. It did not
cause any trouble, we just had to extrapolate the linear p.d.f. $f(x)$ to
values below $-1$ and and above $+1$. However, if the distribution is steep,
some observed data values could correspond to negative function values where
the log-likelihood is not defined. The problem can usually be solved by a
linear transformation of the observed variable in such a way that it is
covered by the range of the variable in which the p.d.f. is defined. In the
example depicted in Fig. \ref{fitlin} this is $[-1,1]$. Here dividing $x$ by
$2$, values outside the interval $[-1,1]$ are excluded. The result obtained
above and its error remain unchanged. The Monte Carlo simulation of the
analysis procedure automatically corrects for the scaling.

Our examples concerned simple, smooth distributions. Complex, multi-modal
distributions with large distortions cannot always be handled with the
correction method without a sizable loss in precision.

\section{Weighting approach}

\label{weighted}Now we turn to the standard method \cite{bohmweighted} that
should be used in all cases where the resolution effects are so large that the
simple correction method does not produce precise results. Now, the
experimental and the simulated data are compared in form of histograms.

\subsection{Negligible statistical error of the Monte Carlo simulation}

If the data sample is not extremely large, we can generate enough Monte Carlo
events such that their statistical error can be neglected.

We compare a data histogram with $B$ bins and bin contents $d_{i}$ to a Monte
Carlo generated histogram with bin contents $t_{i}$. (We use the letters
\emph{d} and \emph{t} to denote \emph{data} and \emph{theory}.)\emph{ }To
produce the Monte Carlo histogram, events are generated according to the
p.d.f. $f(x|\vec{\theta})$ and the detector response is simulated. The
corresponding observed variables $x^{\prime}$ are then histogrammed. We get
$d_{i}$ and a prediction $ct_{i}(\vec{\theta})$ where $c$ is the normalization
constant. Usually the parametric model does not predict the number of observed
events but only the shape of the distribution. Then the normalization $c$ is a
free parameter. We assume that the number of events $d_{i}$ in a bin $i$ is
Poisson distributed with the expected value equal to $ct_{i}$, $d_{i}%
\sim\mathcal{P}(d_{i}|ct_{i})$ with the abbreviation $\mathcal{P}%
(k|\lambda)=e^{-\lambda}\lambda^{k}/k!$.

\subsubsection{The likelihood function}

From the Poisson distribution for bin $i$%
\[
\mathcal{P}(d_{i}|ct_{i})=\frac{e^{-ct_{i}}(ct_{i})^{d_{i}}}{d_{i}!}%
\]
we derive the likelihood and its logarithm. The probabilities for the
different bins are independent. We get
\begin{align}
L(c,\theta)  &  =%
{\displaystyle\prod\limits_{i=1}^{B}}
\frac{e^{-ct_{i}}(ct_{i})^{d_{i}}}{d_{i}!}\;,\label{likelihood}\\
\ln L(c,\theta)  &  =\sum_{i=1}^{B}\left[  -ct_{i}+d_{i}\ln(ct_{i})\right]
+const.\;. \label{loglike}%
\end{align}
The parameter dependence is hidden in the numbers $t_{i}(\theta)$ and the
estimate $\hat{\theta}$ is obtained by maximizing the log-likelihood with
respect to $\theta$. The errors $\delta_{-}$, $\delta_{+}$ are derived in the
usual way from the change of the log-likelihood by half a unit: $\ln
L(\hat{\theta})-1/2=\ln L(\hat{\theta}-\delta_{-})=\ln L(\hat{\theta}%
+\delta_{+})$, or, if the statistics is high enough from the second derivative
of the log-likelihood function at its maximum: $\delta^{2}=\left[  \frac
{d^{2}\ln\theta}{d\theta^{2}}\right]  _{\hat{\theta}}^{-1}$.

An obvious estimate of the parameter $c$ is the ratio of the total number
$N=\Sigma d_{i}$ of observed events and the total number $M=\Sigma t_{i}$ of
simulated events, $\hat{c}=N/M$. This is also the maximum likelihood estimate:
Deriving $\ln L$ with respect to $c$ and setting the derivative equal to
zero$,$%
\[
\frac{d\ln L}{dc}=\sum_{i=1}^{B}(-t_{i}+d_{i}/c)=0\;,
\]
we reproduce the consistent result $\hat{c}=\Sigma d_{i}/\Sigma t_{i}=N/M$.

There is an alternative formulation of the problem: We can calculate the
probabilities $\varepsilon_{i}=t_{i}/\Sigma_{i}t_{i}$ for an event to fall
into bin $i$ and describe the data histogram by a multinomial distribution
\[
\mathcal{M}_{\varepsilon_{1},...,\varepsilon_{B}}^{N}(d_{1},...,d_{B}%
)=\frac{N!%
{\displaystyle\prod\limits_{i=1}^{B}}
\varepsilon_{i}^{d_{i}}}{%
{\displaystyle\prod\limits_{i=1}^{B}}
d_{i}!}%
\]
with the constraint $\Sigma_{i}d_{i}=N$. Here a normalization is obsolete. The
multinomial formulation is equivalent to the multi-Poisson formulation with
normalization fixed to $c=N/M$. It is not recommended to follow the
multinomial way because the errors of the different numbers $d_{i}$ become
correlated and then the calculations are quite clumsy or unnecessary
approximations have to be made.

\subsubsection{Variation of the parameter by weighting the events}

To estimate the parameter of interest $\theta$ we have to maximize the
log-likelihood (\ref{loglike}), varying $\theta$. At first sight one might
think that for each new choice of $\theta$, the complete Monte Carlo
simulation has to be repeated. To proceed in this way does not work because it
would require a huge amount of computer time\footnote{The reason is related to
the fact that a repetition of the simulation induces a large statistical
modification of the likelihood or of $\chi^{2}$ independent of a parameter
change.}. The parameter change is implemented by re-weighting the Monte Carlo
events in the observed histogram. To each event $j$ in bin $i$ with the
observed variable $x_{ij}^{\prime}$, generated with the p.d.f. $f(x_{ij}%
|\theta_{0})$ we associate the weight%
\begin{equation}
w_{ij}(\theta)=f(x_{ij}|\theta)/f(x_{ij}|\theta_{0})\;. \label{weightdef}%
\end{equation}
The weighted event variables $x$ follow the p.d.f. $f(x|\theta)$ and the
weighted observed variables follow the smeared distribution which is compared
to the observed data distribution. To compute the weight, we have to remember
for each Monte Carlo event the true variable value $x_{ij}$. The sum
$\Sigma_{j=1}^{m_{i}}w_{ij}$ of the weights of the $m_{i}$ entries in bin $i$
corresponds to the prediction $t_{i}(\theta)$:
\[
t_{i}(\theta)=\sum_{j=1}^{m_{i}}w_{ij}=\sum_{j=1}^{m_{i}}f(x_{ij}%
|\theta)/f(x_{ij}|\theta_{0})\;.\;
\]

The values $t_{i}(\theta)$ have to be inserted into (\ref{loglike}). The
normalization $c$ can be set, $c=N/\sum_{ij}w_{ij}$ or left free in the fit.
The second possibility is simpler. To simplify the formulas we introduce the
mean value $\bar{w}_{i}$ of the $m_{i}$ weights of the bin $i$:
\begin{equation}
\bar{w}_{i}=\frac{1}{m_{i}}\sum_{j=1}^{m_{i}}w_{ij}\;. \label{wmean}%
\end{equation}

The corresponding log-likelihood is%
\begin{equation}
\ln L(\theta)=\sum_{i=1}^{B}\left[  -cm_{i}\bar{w}_{i}(\theta)+d_{i}\ln
(cm_{i}\bar{w}_{i}(\theta))\right]  +const.\;. \label{loglw}%
\end{equation}
%

\begin{figure}
[ptb]
\begin{center}
\includegraphics[
trim=0.000000in 0.166544in 0.000000in 0.092589in,
height=2.3279in,
width=3.3682in
]%
{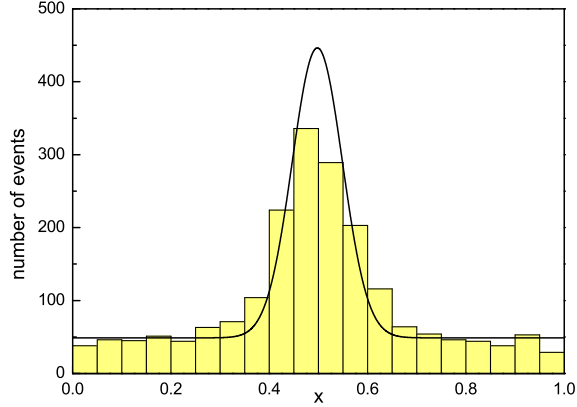}%
\caption{Experimental distribution (histogram) and true distribution used to
generate the data.}%
\label{unigaus}%
\end{center}
\end{figure}

\begin{myexample}
We consider a superposition of a narrow Gaussian with a uniform background
distribution in the interval $[0,1]$. The free parameters are the mean $\mu$,
the standard deviation $\sigma$ of the normal distribution and the background
fraction $\rho$.
\begin{equation}
f(x)=\rho+(1-\rho)\frac{1}{\sqrt{2\pi}\sigma}\exp\left[  -\frac{(x-\mu)^{2}%
}{2\sigma^{2}}\right]  \label{unipeak}%
\end{equation}
A \textquotedblleft data\textquotedblright\ sample is generated with the
parameters $\mu_{d}=0.5$, $\sigma_{d}=0.05$ and $\rho_{d}=0.5$, smeared
according to a normal distribution with standard deviation $\sigma_{s}$ and
histogrammed into $20$ bins. A sample of $2000$ events is displayed in Fig.
\ref{unigaus}. To estimate the parameters a large Monte Carlo event sample
with parameters close to the nominal parameters is generated. The smearing
parameter $\sigma_{s}$ and the number of events are varied. The following
table contains results for different values of the smearing parameter
$\sigma_{s}$ and two different event numbers. The errors are given in
parenthesis and refer to the last few digits of the parameter values.

%

\begin{tabular}
[c]{|r|r|l|l|l|l|}\hline
\# events & \# events MC & $\sigma_{s}$ & $\mu$ & $\sigma$ & $\rho$\\\hline
$2000$ & $100000$ & $0.00$ & $0.5003(21)$ & $0.0523(20)$ & $0.480(15)$\\
$2000$ & $100000$ & $0.02$ & $0.4988(23)$ & $0.0527(23)$ & $0.483(16)$\\
$2000$ & $100000$ & $0.05$ & $0.4977(31)$ & $0.0511(44)$ & $0.491(18)$\\
$2000$ & $20000$ & $0.05$ & $0.4993(31)$ & $0.0512(44)$ & $0.511(18)$\\
$200$ & $10000$ & $0.05$ & $0.5172(103)$ & $0.0640(113)$ & $0.457(56)$\\\hline
\end{tabular}

The results are independent of the parameter values chosen to generate the
Monte Carlo sample. However, if the location of the peak in the simulation
differs strongly from that of the data sample, the weights become large. Then
the number of Monte Carlo events has to be increased to justify the neglect of
the errors. If the simulation parameters are close to the true values, a Monte
Carlo sample which is ten times larger than the data sample is about
sufficient. The normalization is a free parameter in the fit. Fixing it to the
event ratios does not change the results.
\end{myexample}

The following example is more involved.%

\begin{figure}
[ptb]
\begin{center}
\includegraphics[
trim=0.000000in 0.169122in 0.000000in 0.094786in,
height=3.4213in,
width=5.8281in
]%
{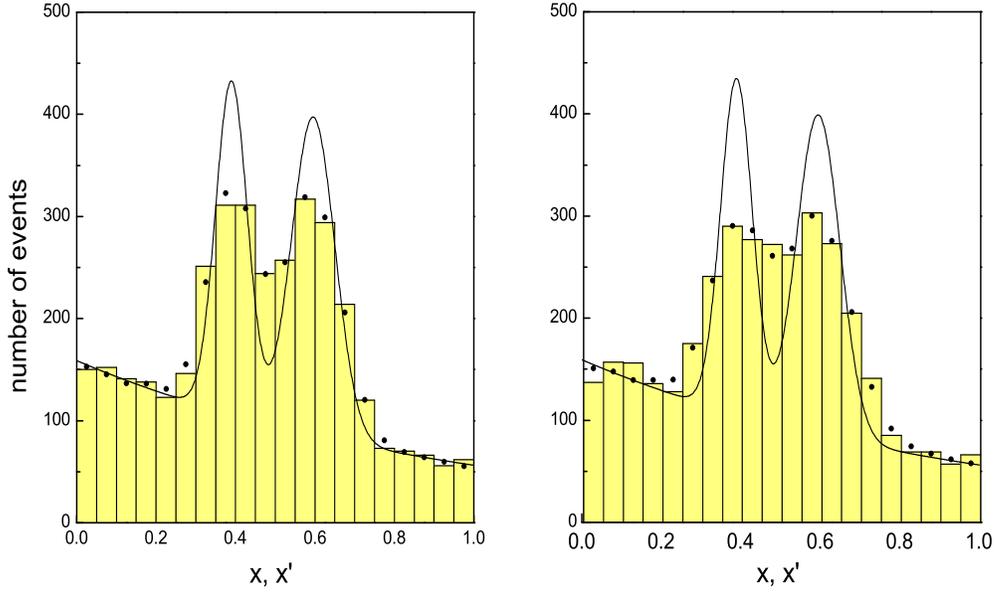}%
\caption{Observed events (histogram), fitted distribution (curve) and
corresponding smeared and normalized histogram (dots). The two plots
correspond to response functions with standard deviations of $0,04$ and
$0.06$, respectively.}%
\label{exp2gaus}%
\end{center}
\end{figure}

\begin{myexample}
The function to be fitted is a superposition of an exponential and two normal
distributions,%
\begin{align*}
f(x|\nu_{ex}  &  ,\nu_{1},\gamma,\mu_{1},\sigma_{1},\mu_{2},\sigma_{2}%
)=\nu_{ex}\gamma\exp(-\gamma x)+\nu_{1}\frac{1}{\sqrt{2\pi}\sigma_{1}}%
\exp\left[  -\frac{(x-\mu_{1})^{2}}{2\sigma_{1}}\right] \\
&  +(1-\nu_{ex}-\nu_{1})\frac{1}{\sqrt{2\pi}\sigma_{2}}\exp\left[
-\frac{(x-\mu_{2})^{2}}{2\sigma_{2}}\right]  \;,
\end{align*}
with seven parameters, $\gamma$, $\mu_{1}$, $\sigma_{1}$, $\mu_{2}$,
$\sigma_{2}$, $\nu_{ex}$, $\nu_{1}$. The data sample consists of $5000$ events
that are generated with the parameters quoted in the second column of the
following table. The Monte Carlo sample contains $50$ times more events.
Events are generated in the interval $[-0.1,1.1]$ but in the fit only events
observed in the range $[0,1]$ are considered. Two different normally
distributed smearing functions with standard deviations $\sigma_{s}=0.04$ and
$\sigma_{s}=0.06$ were applied. The results of the fits are summarized in the
following table and shown in Fig. \ref{exp2gaus}. While the uncertainty of the
slope parameter of the exponential contribution changes by a negligible amount
with the increased smearing, the resolution of the parameters of the two
normal contributions change by about a factor two. Despite the large
distortion of the original distribution in the case $\sigma_{s}=0.06$, the fit
is able to reproduce the true shape quite well.

%

\begin{tabular}
[c]{|c|c|c|c|}\hline
parameter & nominal & $\sigma_{s}=0.04$ & $\sigma_{s}=0.06$\\\hline
$\gamma$ & $1.00$ & $1.072\pm0.093$ & $1.041\pm0.096$\\
$\mu_{1}$ & $0.40$ & $0.392\pm0.007$ & $0.390\pm0.015$\\
$\sigma_{1}$ & $0.05$ & $0.045\pm0.008$ & $0.040\pm0.019$\\
$\mu_{2}$ & $0.60$ & $0.597\pm0.007$ & $0.596\pm0.014$\\
$\sigma_{2}$ & $0.05$ & $0.058\pm0.009$ & $0.055\pm0.015$\\
$\nu_{ex}$ & $0.70$ & $0.677\pm0.020$ & $0.691\pm0.022$\\
$\nu_{1}$ & $0.15$ & $0.143\pm0.015$ & $0.134\pm0.025$\\\hline
\end{tabular}

\end{myexample}

\subsubsection{The least square formulation}

If the observed distribution is not described by Poisson statistics, we cannot
apply the maximum likelihood method and have to turn to the least square
formalism which is the second best choice.

We assume again that the number of simulated events $M$ is much larger than
the number of observed events $N$. The number of events in the histogram bins
are assumed to be Poisson distributed, with the expected value $ct_{i}$. We
form a test quantity $\chi^{2}$,
\begin{equation}
\chi^{2}=\sum_{i=1}^{B}\frac{(d_{i}-ct_{i})^{2}}{\delta_{i}^{2}}\;,
\label{chi2}%
\end{equation}
where $\delta_{i}$ is the expected uncertainty of $d_{i}-ct_{i}$ under the
null hypothesis that the data are described by the prediction, i.e. that the
expected value of the bracket equals zero, $\mathrm{E}(d_{i}-ct_{i})=0$. With
$t_{i}\gg d_{i}$ we can neglect the statistical error of the simulation and in
the denominator remains the expected uncertainty squared of $d_{i}$ which for
the Poisson distribution is the square root of the expected value $ct_{i}$ of
$d_{i}$. We obtain:%
\begin{equation}
\chi^{2}=\sum_{i=1}^{B}\frac{(d_{i}-cm_{i}\bar{w}_{i})^{2}}{cm_{i}\bar{w}_{i}%
}\;. \label{chi2w}%
\end{equation}
Remark that approximating $\delta_{i}^{2}$ simply by $d_{i}$ is inconsistent
and biases the results.

If the measurements are not Poisson distributed, the corresponding estimates
of the errors $\delta_{i}$ have to be inserted in (\ref{chi2}).

Setting $d\chi^{2}/dc=0$, we get the estimate $\hat{c}$:%

\[
\hat{c}=\left[  \frac{1}{B}\sum_{i=1}^{B}\frac{d_{i}^{2}}{t_{i}^{2}}\right]
^{1/2}\;.
\]

Remark that this estimate differs from the MLE and is in disagreement with the
corresponding estimate from multinomial formulation of the problem, however
the difference is negligible in most cases. It does not really matter whether
we fix the normalization to $c=N/\Sigma_{i}m_{i}\bar{w}_{i}$ or leave it as a
free parameter in the fit.

For large event numbers $d_{i}$ the Poisson distribution can be approximated
by a normal distribution and consequently $(d_{i}-ct_{i})/\sqrt{ct_{i}}$ is
normally distributed with variance equal one. Then our test quantity $\chi
^{2}$ follows a $\chi^{2}$ distribution with $B-P$ degrees of freedom, where
$P$ is the number of free parameters in the fit.

Bins with small event numbers are problematic, because the Poisson errors are
strongly asymmetric and LSFs are optimal if the distributions can be
approximated by normal distributions. Therefore for low statistics experiments
one should select the bin width as wide as allowed by the band width of the
distribution. The variation of the resolution with the bin width is estimated
in Appendix 2 for a Gaussian peak.

\paragraph{Parameter estimation in experiments with a large number of events}

Statistical problems decrease with increasing event numbers, but computational
requirements increase. The numerical minimum search that is required to
estimate the wanted parameters can become quite slow. It may happen that we
need of the order of $10^{6}$ or more simulated events. This means that, for
say $10^{4}$ changes of a parameter value during the extremum search that
$10^{10}$ weights have to be computed. This is feasible, but we may want to
speed up the fitting procedure.%
\begin{figure}
[ptb]
\begin{center}
\includegraphics[
trim=0.000000in 0.111821in 0.000000in 0.000000in,
height=3.4769in,
width=2.9107in
]%
{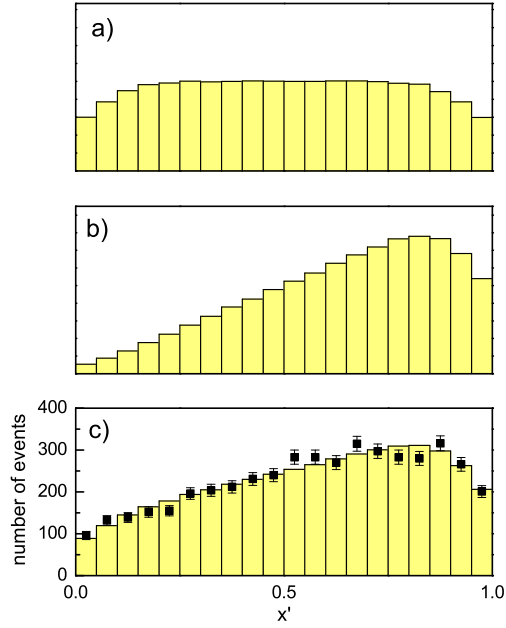}%
\caption{Fit of a linear distorted distribution (squares in c) by the
superposition of two Monte Carlo distributions, a), b).}%
\label{linearfit}%
\end{center}
\end{figure}

The individual weighting of events can be avoided if the parameters appear in
factors depending on the parameters only:%
\begin{equation}
f(x|\theta)=h_{1}(\theta)f_{1}(x)+h_{2}(\theta)f_{2}(x)+...+h_{n}(\theta
)f_{n}(x) \label{superposition}%
\end{equation}
Then we can simulate the smeared versions $f_{1}^{\prime}(x^{\prime})$ to
$f_{n}^{\prime}(x^{\prime})$ of $f_{1}(x)$ to $f_{n}(x)$ and compare the
observed histogram to the superposition of the histograms $t^{(1)}$ to
$t^{(n)}$ of the smeared functions. We replace in (\ref{chi2}) $t_{i}$ by the
sum $t_{i}=h_{1}(\theta)t_{i}^{(1)}+h_{2}(\theta)t_{i}^{(2)}+...+h_{n}%
(\theta)t_{i}^{(n)}$. Since we do not need to compute weights for individual
events, the minimum search is accelerated drastically.

To illustrate the method we choose a simple example.

\begin{myexample}
We fit a straight line. The p.d.f. be $f(x)=1-\theta/2+\theta x$, $0\leq
x\leq1$. With $f_{1}=1$ and $f_{2}=x$ we have $f(x)=(1-\theta/2)f_{1}%
(x)+\theta f_{2}(x)$. We generate Monte Carlo events uniformly distributed in
$x$ and collect the smeared observed values $x^{\prime}$ in a histogram
$\vec{t}^{(1)}$. There may be acceptance losses. We proceed in the same way
with $f_{2}(x)$ and create a histogram $\vec{t}^{(2)}$. Our prediction for
$d_{i}$ is $c[(1-\theta/2)t_{i}^{(1)}+\theta t_{i}^{(2)}]$ with the
normalization factor $c$. The $\chi^{2}$ expression becomes:%
\[
\chi^{2}=\sum_{i=1}^{B}\frac{\left[  d_{i}-c\left(  (1-\theta/2)t_{i}%
^{(1)}+\theta t_{i}^{(2)}\right)  \right]  ^{2}}{c\left(  (1-\theta
/2)t_{i}^{(1)}+\theta t_{i}^{(2)}\right)  }\;.
\]
Fig. \ref{linearfit} shows the results from a numerical example. The
experimental data are simulated with $5000$ events and a slope of $\theta=1$.
The histograms $\vec{t}^{(1)}$ and $\vec{t}^{(2)}$ are derived from $10^{6}$
Monte Carlo events. The Gaussian resolution is $0.1.$ Fig \ref{linearfit} a)
and b) are simulations of the observed uniform and the observed linear Monte
Carlo histograms. A superposition of the two histograms is fitted to the
observed distribution in Fig. \ref{linearfit}c. The fit result is
$\theta=0.955\pm0.033$.
\end{myexample}

In the superposition (\ref{superposition}) the functions $f_{i}(x)$ have to be
positive integrable. For example, a linear distribution of the cosine $z$ of
the polar angle is described by $f(z|\theta)=(1+\theta z)/2$, $-1\leq z\leq1$.
To avoid negative probabilities, we express the p.d.f. by $f(x|\theta
)=(1+\theta)f_{1}+(1-\theta)f_{2}$ with $f_{1}=(1+z)/4$, $f_{2}=(1-z)/4$.

Usually the p.d.f.s are not of the simple form (\ref{superposition}). Then, if
the parameter is known to be close to $\theta_{0}$, we can use a Taylor
expansion of $f(x|\theta)$ at $\theta_{0}$ in powers of the difference
$\Delta=\theta-\theta_{0}$ with respect to the parameter at some preliminary
estimate $\theta_{0}$:%

\begin{align}
f(x|\theta)  &  =f(x|\theta_{0})+\Delta\frac{df(x|\theta)}{d\theta}\left\vert
_{\theta_{0}}\right.  +\frac{\Delta^{2}}{2!}\frac{d^{2}f(x|\theta)}%
{d\theta^{2}}\left\vert _{\theta_{0}}\right.  +\cdots\\
&  =f(x|\theta_{0})\left\{  1+\Delta\frac{1}{f(x|\theta_{0})}\frac
{df(x|\theta)}{d\theta}\left\vert _{\theta_{0}}\right.  +\frac{\Delta^{2}}%
{2!}\frac{1}{f(x|\theta_{0})}\frac{d^{2}f(x|\theta)}{d\theta^{2}}\left\vert
_{\theta_{0}}\right.  +\cdots\right\}  \;.
\end{align}
The powers of the new parameter $\Delta$ factorize. We could now proceed as
above and generate separately Monte Carlo events for each summand, but it is
more economic to generate a single sample following $f(x|\theta_{0})$ and to
apply weights. Let us assume that $\Delta$ is small and that the expansion can
be cut after the third term. The Monte Carlo events are collected into
histograms of the observed values of the variable $x^{\prime}$ are filled into
the histogram $\vec{t}^{(0)}$. Weighting each entry by $w_{1}(x)$ and
$w_{2}(x)$,%
\begin{align}
w_{1}(x)  &  =\frac{1}{f(x|\theta_{0})}\frac{df(x|\theta)}{d\theta}\left\vert
_{\theta_{0}}\right.  \;,\\
w_{2}(x)  &  =\frac{1}{2f(x|\theta_{0})}\frac{d^{2}f(x|\theta)}{d\theta^{2}%
}\left\vert _{\theta_{0}}\right.  \;.
\end{align}
we generate the histograms $\vec{t}^{(1)}$ and $\vec{t}^{(2)}$.

The parameter inference of $\Delta$ is performed by comparing the experimental
histogram bins contents $d_{i}$ to $t_{i}=c(t_{i}^{(0)}+\Delta t_{i}%
^{(1)}+\Delta^{2}t_{i}^{(2)})$:%

\begin{equation}
\chi^{2}=%
{\displaystyle\sum\limits_{i=1}^{B}}
\frac{(d_{i}-ct_{i})^{2}}{ct_{i}}\;. \label{chi20}%
\end{equation}

In many cases the quadratic term can be omitted. In other situations it might
be necessary to iterate the procedure.

This method works only if the number of Monte Carlo events is high enough to
neglect its uncertainty with respect to that of the experimental data.%

\begin{figure}
[ptb]
\begin{center}
\includegraphics[
trim=0.156179in 0.175515in 0.157341in 0.092867in,
height=3.0311in,
width=3.9369in
]%
{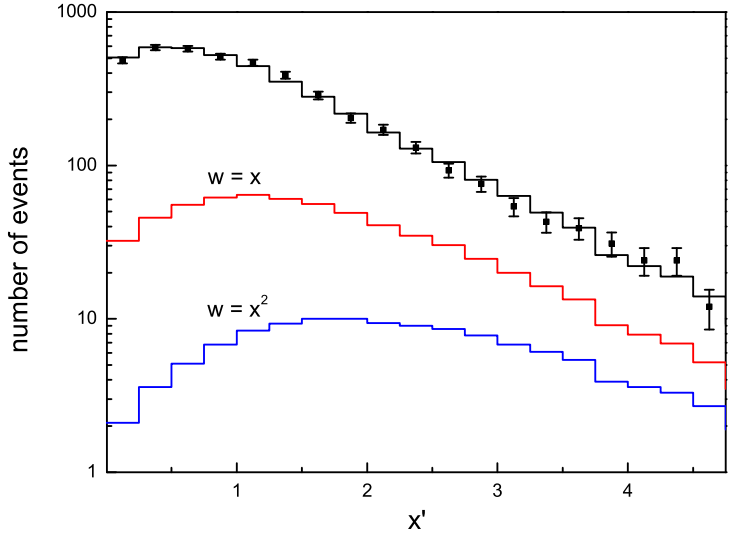}%
\caption{Fit of the slope of a smeared exponential distribution. The fit
(black histogram) is compared to the experimental data (squares). The
contribution of the weighted histograms are displayed as red and blue
histograms.}%
\label{exponfit}%
\end{center}
\end{figure}
To illustrate the method, we consider a lifetime measurement:

\begin{myexample}
We expand the p.d.f.%

\begin{equation}
f(x|\gamma)=\gamma e^{-\gamma x}%
\end{equation}

into a Taylor expansion at $\gamma_{0}$ which is a first guess of the decay
rate $\gamma$:%

\begin{equation}
f(t|\gamma)=\gamma_{0}e^{-\gamma_{0}x}\left\{  1+\frac{\Delta\gamma}%
{\gamma_{0}}(1-\gamma_{0}x)+(\frac{\Delta\gamma}{\gamma_{0}})^{2}(-\gamma
_{0}x+\frac{\gamma_{0}^{2}x^{2}}{2})+\cdots\right\}  \;.
\end{equation}

The Monte Carlo simulation follows the distribution $f_{0}=\gamma
_{0}e^{-\gamma_{0}x}$. Weighting the events by $(1/\gamma_{0}-x)$ and
$(-x/\gamma_{0}+x^{2}/2)$, we obtain the distributions $f_{1}=(1-\gamma
_{0}x)e^{-\gamma_{0}x}$, $f_{2}=(-x+\gamma_{0}x^{2}/2)e^{-\gamma_{0}x}$ and%

\begin{equation}
f(x|\gamma)=f_{0}(x)+\Delta\gamma f_{1}(x)+(\Delta\gamma)^{2}f_{2}%
(x)+\cdots\;.
\end{equation}

If it is justified to neglect the higher powers of $\Delta\gamma/\gamma_{0}$,
we can again describe our experimental distribution this time by a
superposition of three distributions $f_{0}^{\prime}(x^{\prime})$,
$f_{1}^{\prime}(x^{\prime})$, $f_{2}^{\prime}(x^{\prime})$ which are the
distorted versions of $f_{0}(x),f_{1}(x),f_{2}(x)$. The parameter
$\Delta\gamma$ is determined by a $\chi^{2}$ or likelihood fit. In our special
case it is even simpler to weight $f_{0}$ by $x$, and $x^{2}$, respectively,
and to superpose the corresponding distributions $f_{0}$, $g_{1}=xf_{0}$,
$g_{2}=x^{2}f_{0}$ with the factors given in the following expression:%

\begin{equation}
f(x|\gamma)\approx f_{0}(x)\left(  1+\frac{\Delta\gamma}{\gamma_{0}}\right)
-\gamma_{0}g_{1}(x)\left(  \frac{\Delta\gamma}{\gamma_{0}}+(\frac{\Delta
\gamma}{\gamma_{0}})^{2}\right)  +\frac{1}{2}g_{2}(x)\gamma_{0}^{2}\left(
\frac{\Delta\gamma}{\gamma_{0}}\right)  ^{2}\;.
\end{equation}

The parameter $\Delta\gamma$ is then modified until the correspondingly
weighted sum of the distorted histograms agrees optimally with the data.

Fig. \ref{exponfit} illustrates the method with an numerical example. The
decays of $5000$ events are simulated with a decay rate $\gamma=1$ and a large
Gaussian smearing with a standard deviation $\sigma=0.5$. The histogram formed
by these events is indicated by the square dots. $100000$ Monte Carlo events
are generated with a decay rate $\gamma_{0}=1.2$ and histogrammed with the
weights $1$, $x$ and $x^{2}$. The result of the fit is $\Delta\gamma
=-0.196\pm0.020$ and $\gamma=1.004\pm0.020$. The adjusted Monte Carlo
simulation is displayed as the black histogram and the contributions by the
weighted histograms to the fit are indicated as the red and blue histograms.
Due to the large difference between the true value of $\gamma$ and the one
used in the simulation, the quadratic term in $\Delta\gamma$ has to be
included in the expansion. If the Monte Carlo events are generated with
$\gamma_{0}=1.1$, the linear approximation is sufficient.
\end{myexample}

\subsection{Data contaminated by background and correlated errors}

Often the prediction refers to data that are contaminated by background. If we
know the source of the background and its distribution in the true variable
$x$, we can simply reformulate the p.d.f. in such a way that it includes the
background. The common case, however, is that the background distribution is
known only as a function of the observed variable $x^{\prime}$ or has to be
estimated from the observed histogram.

We then subtract the estimated background in each bin and include its
uncertainty $\delta(b_{i})$ in quadrature in the denominator of (\ref{chi2}):
\begin{equation}
\chi^{2}=\sum_{i=1}^{B}\frac{(d_{i}-b_{i}-ct_{i})^{2}}{ct_{i}+\delta^{2}%
(b_{i})}\;. \label{chi2b}%
\end{equation}

The normalization constant becomes $\hat{c}=\Sigma_{i}(d_{i}-b_{i})/\Sigma
_{i}t_{i}$. If we have an absolute prediction $b_{i}$, for instance from an
ancillary measurement with high statistics, the error of Poisson distributed
background is $\delta^{2}(b_{i})=b_{i}$. If an external background measurement
$b_{0i}$ is Poisson distributed and normalized, $b_{i}=\nu b_{0i}$ with given
uncertainty $\delta_{\nu}$ of the normalization, we get instead of the simple
diagonal errors in first order the covariance matrix $C$:%
\begin{equation}
C\equiv\left(
\begin{array}
[c]{cccccc}%
ct_{1}+b_{1}^{2}\delta_{\nu}^{2}/\nu^{2}+(1+\nu)b_{1} & \delta_{v}^{2}%
b_{1}b_{2}/\nu^{2} & . & . & . & .\\
\delta_{v}^{2}b_{2}b_{1}/\nu^{2} & ct_{2}+b_{2}^{2}\delta_{\nu}^{2}/\nu
^{2}+(1+\nu)b_{2} & . & . & . & .\\
. & . & . & . & . & .\\
. & . & . & . & . & .\\
. & . & . & . & . & .\\
. & . & . & . & . & .
\end{array}
\right)  \;. \label{covarianceb}%
\end{equation}
For a simple numerical example with $b_{i}=20$, $\nu=1$, $\delta_{\nu}=0.2$ we
get $b_{i}^{2}\delta_{\nu}^{2}/\nu^{2}+(1+\nu)b_{i}=16+40$ and $\delta_{\nu
}^{2}b_{i}b_{j}/\nu^{2}=16$.

The modified $\chi^{2}$ expression is:%
\begin{equation}
\chi^{2}=\sum_{i=1}^{B}(\vec{d}-\vec{b}-c\vec{t})C^{-1}(\vec{d}-\vec{b}%
-c\vec{t})^{T}\;. \label{chi2corr}%
\end{equation}
The inverse $C^{-1}$ of the covariance or error matrix is the weight matrix.

We have assumed that the errors correspond to Poisson distributions. If this
is not true, then (\ref{covarianceb}) has to be modified accordingly.

\subsection{Including the statistical uncertainty of the simulation}

One should always attempt to generate as many events as necessary to justify
neglecting the statistical uncertainties of the simulation. Usually it is much
cheaper the generate a simulated event than an experimental one. However,
there are exceptions which we will discuss now. We cannot apply anymore the
likelihood method and have to come back to the LS formalism where we have to
minimize $\chi^{2}$,%

\[
\chi^{2}=\sum_{i=1}^{B}\frac{(d_{i}-ct_{i})^{2}}{\delta_{i}^{2}}\;.
\]

The denominator $\delta_{i}$, the error of $d_{i}-ct_{i}$ now has to include
the error of $t_{i}$. To evaluate it, we first have to estimate the expected
number $E(d_{i})=E(ct_{i})=\tau_{i}$.

\subsubsection{Poisson errors}

For a few lines we suppress the index $i$. If not only $d$ but also $t$ is a
Poisson number, $t\sim P(t|\tau/c)$ then the result is%
\begin{equation}
\hat{\tau}=\frac{d+t}{1+1/c} \label{taudt}%
\end{equation}
We derive this relation in the Appendix 2. An intuitive justification of
(\ref{taudt}) is the following: If in an experiment $d$ decays are observed in
the unit time interval and $t$ decays in the interval $(1/c)$, then the rate
is $(d+t)/(1+1/c)$. We have to add the numbers $d$ and $t$ and divide by the
total time.

However, the prediction $t$ is a sum of $m$ weights, $t=\Sigma_{j}mw_{j}$,
where $m$ is a Poisson number and also $w_{j}$ are i.i.d. random numbers. Thus
$t$ follows a so-called \emph{compound Poisson distribution }(CPD). The
expected value of $t$ is $\mathrm{E}(t)=\mathrm{E}(m)\mathrm{E}(w)$ and its
variance $\sigma_{CPD}^{2}$ is equal to $\mathrm{E}(m)\mathrm{E}(w^{2})$. An
obvious estimate of it is the sum of the observed weights squared,
$\widehat{\sigma_{CPD}^{2}}=\Sigma_{k=1}^{m}w_{k}^{2}=m\overline{w^{2}}$. The
estimate of the relative error of $t$ is then%
\[
\frac{\delta_{t}}{t}=\frac{\sqrt{m\overline{w^{2}}}}{m\overline{w}}%
=\frac{\sqrt{\overline{w^{2}}}}{\sqrt{m}\overline{w}}\;.
\]
The error is larger than the usual Poisson value $1/\sqrt{m}$ by the factor
$\sqrt{\overline{w^{2}}}/\overline{w}$. It is convenient to introduce a
fictive number of events $\tilde{m}$, the so-called \emph{equivalent number of
unweighted events} or \emph{effective number of events}. The number $\tilde
{m}$,%
\[
\tilde{m}=m\frac{\overline{w}^{2}}{\overline{w^{2}}}\;,
\]
has the same error $\sqrt{\tilde{m}}$ as a standard Poisson number $\tilde{m}%
$. Sloppily speaking, the weighted sum $t$ has the same statistical
significance as $\tilde{m}$ unweighted events. The effective number of events
is of course smaller than the number $m$ of Monte Carlo events. It comes close
to it, if the dispersion of the weights is small and it agrees with it, if all
weights are equal.

In the large number limit $\tilde{m}\rightarrow\infty$ the distribution of
$\tilde{m}$ and then of course also that of $t$ follows a normal distribution.
In Appendix 1 we show that $\tilde{m}$ can even better be described by a
Poisson distribution, and $t$ which is proportional to $\tilde{m}$,
$t=s\tilde{m}$, $s=\sqrt{\overline{w^{2}}}$. by a scaled Poisson distribution
(SPD). We can assume that we can always afford a sufficiently large number of
Monte Carlo events to justify the asymptotic treatment of the CPD with a
scaled Poisson distribution.

With the approximation of the CPD by the SPD we can use the result
(\ref{taudt}) and get for $\tau=E(d)=E(ct)=E(cs\tilde{m})$%
\[
\tau=\frac{d+\tilde{m}}{1+1/(cs)}\;.
\]
Using $\mathrm{E}(d)=\tau$, and $\mathrm{E}(\tilde{m})=\tau/(cs)$ we obtain
(see Appendix 1)%
\[
\delta^{2}=c(\frac{\sqrt{\overline{w^{2}}}}{\overline{w}}d+t)\;.
\]
Finally we sum over all bins and get the simple result
\begin{equation}
\chi^{2}=\sum_{i=1}^{B}\frac{(d_{i}-ct_{i})^{2}}{(c\frac{\sqrt{\overline
{w_{i}^{2}}}}{\overline{w_{i}}}d_{i}+t_{i})}\;. \label{chi2weight}%
\end{equation}

\subsubsection{Normal approximation}

For completeness we estimate the uncertainty $\delta^{2}$ also for the case of
normally distributed errors. The normal approximation has to be used if the
observed numbers $d_{i}$ contain correction terms. We omit again the bin index
and assume that the expected value $\mathrm{E}(d)$ is equal to the expected
value $\mathrm{E}(ct)=\tau$. For the uncertainty of $d$ we assume that it is
proportional to the expected number of $d$, $\delta d^{2}=c^{\prime}\tau$
while for the simulation we stick to the Poisson error $\delta_{ct}^{2}%
=c^{2}t$. In Appendix 2 we derive%
\[
\hat{\tau}=\left[  \frac{cd^{2}+c^{\prime}c^{2}t^{2}}{c+c^{\prime}}\right]
^{1/2}%
\]
and%
\[
\delta^{2}=\delta_{d}^{2}+\delta_{ct}^{2}=\hat{\tau}(c^{\prime}+c)\;.
\]
In the Poisson limit we get%
\[
\hat{\tau}=\left[  \frac{d^{2}+ct^{2}}{1+1/c}\right]  ^{1/2}%
\]
and%
\[
\delta^{2}=\delta_{d}^{2}+\delta_{ct}^{2}=\hat{\tau}(1+c)\;.
\]

\emph{Remark}: The expected bin content $\tau_{i}$ in bin $i$ is a nuisance
parameter. We have estimated it out, which in principle is a doubtful method.
It is justified in our case because the correlation of the parameters
$\tau_{i}$ with $c$ is negligible.

\subsubsection{Summary of the procedure}

Let us summarize the whole procedure:

1. Simulate the experiment with parameter $\theta_{0}$ and obtain events
$(x_{ik},x_{ik}^{\prime})$ were $x_{ik}^{\prime}$ is the smeared variable of
the k-th event of the $m_{i}$ events in bin $i$ of the histogram of
$x^{\prime}$.

2. Select a starting value for the normalization $c_{0}$.

3. Associate a weight $w_{ik}=1$ to each event.

4. Compute the mean values $\bar{w}_{i}$, $\overline{w_{i}^{2}}$ of each bin
$i$ and $t_{i}=\Sigma w_{i}=m_{i}\bar{w}_{i}$.

5. Compute $\chi^{2}$ according to (\ref{chi2weight}).

6. Let Simplex modify $c$ and $\theta$, recompute weights: $w_{ik}%
=f(x_{ik}|\theta)/f(x_{ik}|\theta_{0})$

7. Go to 4. until the minimum of $\chi^{2}$ is reached.

In most cases we can generate enough Monte Carlo events and apply the simpler
error calculation of the previous section.

As an example we choose a superposition of a normal distribution and a uniform
background :

\begin{myexample}
The $4$ free parameters of the following superposition of a normal and a
uniform distribution%
\begin{equation}
f(x|\mu,\sigma,\phi)=\phi\frac{1}{\sqrt{2\pi}\sigma}\exp[-\frac{(x-\mu)^{2}%
}{2\sigma^{2}}]+(1-\phi);\;0\leq x\leq1 \label{function2}%
\end{equation}
are adjusted. They are the normalization of the observed data to the
simulation $c$, the mean value $\mu$ and the standard deviation $\sigma$ of
the normal distribution and the fraction $\phi$ of the normally distributed
events. With the parameter settings $\mu=0.5$, $\sigma=0.05$, $\phi=0.7$ the
Gaussian is narrow enough to neglect the tails of the distribution outside the
interval $[0,1]$. The Monte Carlo events are generated with the settings
$\mu_{MC}=1.025\mu$, $\sigma_{MC}=1.05\sigma$, $\phi_{MC}=1.05\phi$. The
parameter $c$ is not interesting and not sizably correlated with the other
parameters. Strongly correlated are the estimates of the width $\hat{\sigma}$
and the fraction $\hat{\phi}$. For the selected parameter values the
correlation coefficient is of the order of $0.3$.

The following table summarizes the results of fits averaged over $100$
simulated experiments. The resolutions and the biases of $\mu$, $\sigma$ and
$\phi$ are reported for different combinations of event numbers, $N$ the
number of observed events and $M$ the number of Monte Carlo events. The
standard deviation of the Gaussian smearing $\sigma_{s}$, and the number of
bins $B$ is given.

%

\begin{tabular}
[c]{|l|l|l|l|l|l|l|l|l|l|l|}\hline
$N$ & fit & $M$ & $\sigma_{s}$ & $B$ & $\delta_{\mu}$ & $\delta_{\sigma}$ &
$\delta_{\phi}$ & $b_{\mu}$ & $b_{\sigma}$ & $b_{\phi}$\\\hline
100 & ML & 10000 & 0.05 & 10 & 0.0142 & 0.0187 & 0.0578 & -0.0001 & 0.0025 &
0.0049\\
100 & ML & 20000 & 0.05 & 20 & 0.0109 & 0.0149 & 0.0590 & -0.0010 & -0.0008 &
-0.0020\\
200 & ML & 10000 & 0.05 & 10 & 0.0087 & 0.0126 & 0.0436 & 0.0001 & 0.0003 &
-0.0008\\
200 & LS & 10000 & 0.05 & 10 & 0.0090 & 0.0138 & 0.0924 & 0.0003 & 0.0049 &
\emph{-0.0767}\\
500 & ML & 20000 & 0.05 & 10 & 0.0044 & 0.0066 & 0.0211 & 0.0009 & -0.0006 &
0.0006\\
500 & LS & 20000 & 0.05 & 10 & 0.0043 & 0.0067 & 0.0491 & 0.0008 & -0.0009 &
-0.0064\\
1000 & ML & 50000 & 0.05 & 10 & 0.0037 & 0.0049 & 0.0165 & -0.0002 & 0.0002 &
0.0011\\
1000 & LS & 50000 & 0.05 & 10 & 0.0037 & 0.0049 & 0.0169 & -0.0002 & 0.0000 &
-0.0027\\
1000 & LS & 1000 & 0.05 & 10 & 0.0049 & 0.0073 & 0.0243 & -0.0002 & 0.0004 &
0.0029\\
10000 & LS & 10000 & 0.05 & 10 & 0.0014 & 0.0020 & 0.0070 & 0.0000 & 0.0002 &
0.0001\\
10000 & LS & 10000 & 0.0 & 20 & 0.0013 & 0.0018 & 0.0068 & 0.0001 & 0.0002 &
0.0003\\\hline
\end{tabular}

For a small number of observed events, enough Monte Carlo events can be
generated such that their statistical error can be neglected. This has been
done in the examples of the first eight rows. With only $100$ events a LS fit
fails, because the number of events per bin is too low. In the LS fit bins
with less than $5$ events are excluded, which may introduce a bias. The ML fit
is always successful. With $200$ events the MLE of $\mu$ and $\sigma$ is
slightly more precise than the LS fit. The error $\delta_{\phi}$ is larger in
the LSF than in the MLF. The last three rows correspond to situations where
the statistical fluctuations of the Monte Carlo events have to be taken into
account. Increasing the number of histogram bins slightly reduces the
parameter errors if the number of events is large. The numbers of the last row
are computed for the limiting case were smearing is absent. As expected the
error of the width of the bump is slightly reduced.

As we have $100$ simulations per example, the uncertainties of the reported
biases are a tenth of the corresponding parameter errors, for example
$\delta(b_{\mu})=\delta_{\mu}/10$. Only the result for the fraction parameter
$\phi$ in the case of $200$ observed events is significantly biased.
\end{myexample}

Whenever the number of Monte Carlo events is sufficiently high, the ML fit
should be preferred to a LS fit. The factor that we need depends on the weight
distribution and on how well the parameters used in the simulation agree with
the parameter estimates. To check the validity of the approximation, the Monte
Carlo sample should be subdivided or even better, bootstrap samples should be
fitted \cite{efron}.

\section{Summary}

We have compared event samples suffering from a limited resolution and from
acceptance losses to predictions containing unknown parameters. We distinguish
different situations:

1. If the distortions are moderate, a standard $\chi^{2}$ or likelihood fit
can be performed, comparing the prediction to the observed data. Corrections
to the result can be derived from a Monte Carlo simulations of the measurement process.

2. If the distribution can be written in the form that the parameter functions
factorize,%
\begin{equation}
f(x|\vec{\theta})=%
{\displaystyle\sum}
g_{i}(\vec{\theta})f_{i}(x)\;, \label{parameterfact}%
\end{equation}
the observed distribution can be compared to the superposition $\Sigma
_{i}g_{i}(\vec{\theta})f_{i}^{\prime}(x^{\prime})$ where~$f_{i}^{\prime
}(x^{\prime})$ are the folded versions of the functions $f_{i}$. Histograms
$t_{i}$ corresponding to the functions $f^{\prime}(x^{\prime})$ are generated
in a Monte Carlo simulation. The parameter $\vec{\theta}$ is fitted in a least
square or maximum likelihood fit where the observed data $d_{i}$ are compared
to $\Sigma g_{i}(\vec{\theta})t_{i}$. If the p.d.f. $f(x|\vec{\theta})$ is not
of the form (\ref{parameterfact}), it can be expanded in a Taylor series of
$\mathbf{\Delta\theta}$ around an estimate $\vec{\theta}_{0}$ and then
$\mathbf{\Delta\theta}$ is fitted. The normalization is a free parameter in
the fit.

3. In the standard method, individual Monte Carlo events are weighted with
$f(x|\vec{\theta})/f(x|\vec{\theta}_{0})$ where $x$ is the undistorted
variable and $\theta_{0}$ the parameter value used in the Monte Carlo
simulation. The parameter is adjusted such that the histogram of the weighted
events has the same shape as the histogram of the experimental data.

3.1. If the statistical fluctuation of the simulated number of events can be
neglected, the parameter is estimated in a Poisson MLF.

3.2. In the rare cases where the statistical uncertainties of the simulation
have to be taken into account, the parameter is adjusted in a LSF. The sum of
weights in a bin is described by a compound Poisson distribution (CPD) and can
be approximated by a scaled Poisson distribution (SPD). The calculation of the
denominators (error estimates) of the LS summands based upon a SPD is simpler
than with the normal approximation and the result is more precise.

4. Background can be taken into account is a LS fit. The normalization of the
background contributions leads to correlations which are included in a weight matrix.

\chapter{\protect\nolinebreak Discrete inverse problems and the response
matrix}

\section{Introduction and definition}

We know turn to the problem of unfolding a distorted distribution for which no
parametric prediction for the true distribution is available.

\subsection{An inverse problem}

Folding is described by the integral
\begin{equation}
g(x^{\prime})=\int_{-\infty}^{\infty}h(x^{\prime},x)f(x)dx\;. \label{fredholm}%
\end{equation}
The function $f(x)$ is folded with a response function $h(x^{\prime},x)$,
resulting in the smeared function $g(x^{\prime})$. We call $f(x)$ the
\emph{true distribution} and $g(x^{\prime})$ the \emph{smeared distribution}
or the \emph{observed distribution}. The three functions $g,h,f$ can have
discontinuities but of course the integral has to exist. The integral equation
(\ref{fredholm}) is called Fredholm equation of the first kind with the kernel
$h(x^{\prime},x)$. If the function $h(x^{\prime},x)$ is a function of the
difference $x^{\prime}-x$ only (\ref{fredholm}) is denoted as convolution
integral, but often the terms convolution and folding are not distinguished.
The relation (\ref{fredholm}) describes the \emph{direct process} of folding.
We are interested in the \emph{inverse problem}: Knowing $g$ and $h$ we want
to infer $f(x)$. This inverse problem is classified by the mathematicians as
\emph{ill posed} because it has no unique solution. In the direct process high
frequencies are washed out. The damping of strongly oscillating contributions
in turn means that in mapping $g$ to $f$ high frequencies are amplified, and
the higher the frequency, the stronger is the amplification. In fact, in
practical applications we do not really know $g$, the information we have
consists only in a sample of observations with the unavoidable statistical
fluctuations\footnote{In the statistical literature the fluctuations are
called noise.}. The fluctuations of $g$ correspond to large perturbations of
$f$ and consequently to ambiguities.

The response function often, but not always, describes a simple resolution
effect and then it is called \emph{point spread function} (PSF). There exists
also more complex situations like in positron emission tomography (PET) where
the relation between the observed distribution of two photons and the
interesting distribution of their origin is more involved. In PET and many
other applications the variables $x$ and $x^{\prime}$ are multi-dimensional.

\subsection{The histogram representation}%

\begin{figure}
[ptb]
\begin{center}
\includegraphics[
height=2.7771in,
width=4.0415in
]%
{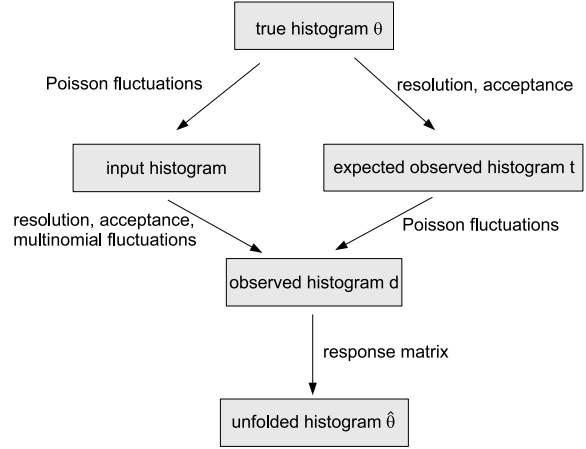}%
\caption{Relations between the histograms involved in the unfolding process.}%
\label{unfoldflow}%
\end{center}
\end{figure}

\subsubsection{Discretization and the response matrix}

The disease of the inverse problem can partially be cured by discretization,
which essentially means that we construct a parametric model. We usually
replace the continuous functions by histograms which can be written as vectors
$\vec{\theta}$ for the true histogram and $\vec{d}$ for the observed histogram
The two histograms are connected by the response function, here by a matrix
$\tens{A}$. We get for the direct process.%

\begin{equation}
\mathrm{E}(\vec{d})=\tens{A}\vec{\theta}\;. \label{folding}%
\end{equation}

\[
\mathrm{E}\left(
\begin{array}
[c]{c}%
d_{1}\\
d_{2}\\
.\\
.\\
.\\
d_{N}%
\end{array}
\right)  =\left(
\begin{array}
[c]{cccc}%
A_{11} & . & . & A_{1M}\\
A_{21} & . & . & A_{2M}.\\
. & . & . & .\\
. & . & . & .\\
. & . & . & .\\
A_{N1} & . & . & A_{NM}%
\end{array}
\right)  \cdot\left(
\begin{array}
[c]{c}%
\theta_{1}\\
.\\
.\\
\theta_{M}%
\end{array}
\right)  \;.
\]

Here $d_{i}$ is the content of bin $i$ of an \emph{observed histogram}.
$\mathrm{E}(\vec{d})$ is the expected value. $A$ is called \emph{response or
folding matrix} and $\theta_{j}$ is the content of bin $j$ of the undistorted
\emph{true histogram} that we want to determine.%
\begin{align}
\theta_{j}  &  =\int_{bin\;j}f(x)dx\nonumber\\
\mathrm{E}(d_{i})  &  =\int_{bin\;i}dx^{\prime}\int_{-\infty}^{\infty
}h(x^{\prime},x)f(x)dx\nonumber\\
A_{ij}  &  =\int_{bin\;i}dx^{\prime}\int_{bin\;j}h(x^{\prime},x)f(x)dx\left/
\theta_{j}\right.  \label{aijdefinition}%
\end{align}
The value $A_{ij}$ represents the probability that the detector registers an
event in bin $i$ that belongs to the true histogram bin $j$. This
interpretation assumes that all elements of $\vec{d}$, $\tens{A}$ and
$\vec{\theta}$ are positive. The number of columns $M$ is the number of bins
in the true histogram and the number of parameters that have to be determined.
The number of rows $N$ is the number of bins in the observed histogram. We do
not want to have more parameters than measurements and require $N\geq M$.
Normally we constrain the unknown true histogram, requiring $N>M$. With $N$
bins of the observed histogram and $M$ bins of the true histogram we have
$N-M$ constraints. The relation between the various histograms is shown in
Fig. \ref{unfoldflow}.

We require that $\tens{A}$ is rank efficient which means that the rank is
equal to the number of columns $M$. Formally, this means that all columns are
linearly independent and at least $M$ rows are linearly independent: No two
bins of the true histogram should produce observed distributions that are
proportional to each other. Unfolding would be ambiguous in this situation but
a simple solution is to combine the bins. More complex cases that lead to a
rank deficiency never occur in practice. A more serious requirement is the
following: By definition of $\tens{A}$, the observed histogram must not
contain events that originate from other sources than the $M$ true bins. In
other words, The range of the true histogram has to cover all observed events.
This requirement often entails that only a small fraction of the events that
contained the border bins of the true histogram are found in the observed
histogram. The correspondingly low efficiency leads to large errors of the
reconstructed number of events in these bins. Most published simulation
studies avoid this complication by restricting the range of the true variable.%

\begin{figure}
[ptb]
\begin{center}
\includegraphics[
trim=0.123424in 0.155488in 0.123785in 0.078123in,
height=2.5371in,
width=5.0892in
]%
{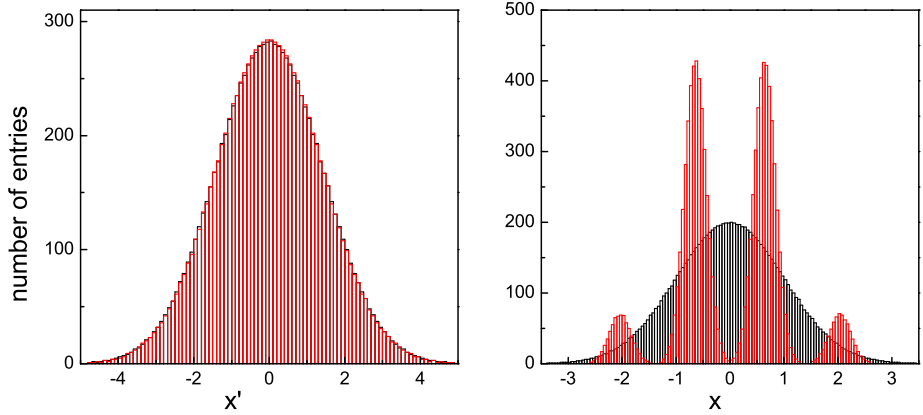}%
\caption{Folded distributions (left) for two different distributions (right).}%
\label{fourgaus}%
\end{center}
\end{figure}
%

\begin{figure}
[ptb]
\begin{center}
\includegraphics[
trim=0.000000in 0.166251in 0.000000in 0.111030in,
height=1.9992in,
width=3.0427in
]%
{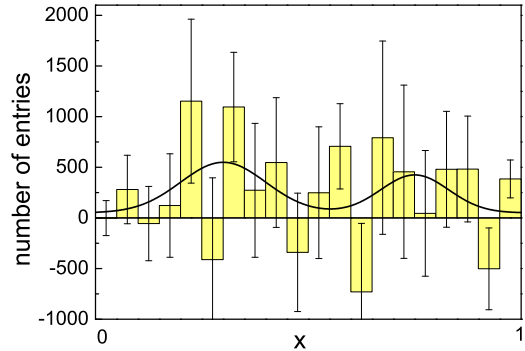}%
\caption{Naive unfolding result obtained by matrix inversion. The curve
corresponds to the true distribution.}%
\label{unfoscillation}%
\end{center}
\end{figure}

Some publications refer to an effective rank and to a null space of the matrix
$\tens{A}$. The null space is spanned by vectors that fulfill $\tens{A}\vec
{\theta}=0$. With our definitions and the restrictions that we have imposed,
the null space is empty and there is no need to introduce an effective rank.

In particle physics the experimental setups are mostly quite complex and for
this reason they are simulated with Monte Carlo programs. To construct the
matrix $\tens{A}$ we generate events following an assumed true distribution
$f(x)$ characterized by the true variable $x$ and a corresponding true bin
$j$. The detector simulation produces the observed variable $x^{\prime}$and
the corresponding observed bin $i$. We will assume for the moment that we can
generate an infinitely large amount of \ "Monte Carlo events" such that we do
not have to care about statistical fluctuations of the elements of $\tens{A}$.
The statistical fluctuations of the observed event numbers be described by the
Poisson distribution.

There is another problem that we neglect but that we have to resume later: The
matrix $\tens{A}$ depends to some extent on the true distribution which is not
known in the Monte Carlo simulation. The dependence is small if the bins of
the true distribution are narrow enough to neglect the fluctuations of $f(x)$
within a bin. This condition cannot always be maintained.

\subsubsection{The need for regularization}

The discrete model avoids the ambiguity of the continuous ill-posed problem
but especially if the response matrix is large, i.e. the bins are narrow
compared to the resolution, the matrix is badly conditioned which means that
the inverse or pseudo-inverse of $\tens{A}$ contains large components. This is
illustrated in Fig. \ref{fourgaus} which shows two different original
distributions and the corresponding distributions smeared with a Gaussian
$\mathcal{N}(x-x^{\prime}|0,1)$. In spite of the extremely different original
distributions, the smeared distributions of the samples are practically
indistinguishable. This demonstrates the sizeable information loss that is
caused by the smearing, especially in the case of the distribution with four
peaks. Sharp structures are washed out and can hardly be reconstructed. Given
the observed histogram with some additional noise, it will be almost
impossible to exclude one of the two candidates for the true distribution even
with a huge amount of data. Since narrow structures in the true distribution
are smeared in the observed distribution and in addition modified by
statistical fluctuations, naive unfolding can produce oscillations as shown in
Fig. \ref{unfoscillation}. Typically, the errors of adjacent bins are strongly
negatively correlated. Combining them would reduce the errors considerably and
produce a histogram that is closer to the true distribution which is shown as
a curve in Fig. \ref{unfoscillation}.

If the matrix $\tens{A}$ is quadratic, we can simply invert (\ref{folding})
and get an estimate $\widehat{\vec{\theta}}$ of the true histogram.
\begin{equation}
\widehat{\vec{\theta}}=A^{-1}\vec{d}\;. \label{inversion}%
\end{equation}
In practice this simple solution usually does not work because, as mentioned,
our observations suffer from statistical fluctuations.%

\begin{figure}
[ptb]
\begin{center}
\includegraphics[
trim=0.133848in 0.193260in 0.135913in 0.126862in,
height=2.2466in,
width=5.5948in
]%
{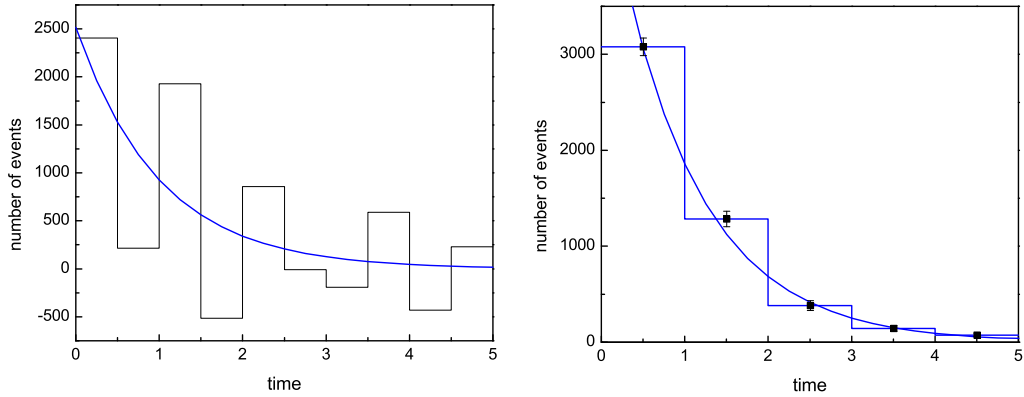}%
\caption{{\small Unfolding by matrix inversion with different binnings.}}%
\label{lifetimebin}%
\end{center}
\end{figure}
In Fig. \ref{lifetimebin} the result of a simple inversion of the data vector
of Fig. \ref{lifetime0} is depicted. The left-hand plot is realized with $10$
bins. It is clear that either fewer bins have to be chosen, see Fig.
\ref{lifetimebin} right-hand plot, or some smoothing has to be applied.

\subsection{Expansion of the true distribution}

\label{splineexpansion}Instead of representing the function $f$ by a
histogram, we can expand it into a sum of functions $B_{i}$. The $B_{i}$ be
normalized, $\int_{-\infty}^{\infty}B_{i}(x)dx=1.$
\begin{equation}
f(x)\approx\sum_{j=1}^{M}{}_{j}\beta_{j}B_{j}(x)
\end{equation}

The response matrix element $A_{ij}$ now is the probability to observe an
event in bin $i$ of the observed histogram that originates from the
distribution $\phi_{j}$:%
\begin{align}
A_{ij}  &  =\int_{bin\;i}dx^{\prime}%
{\displaystyle\sum\limits_{j}}
\int_{-\infty}^{\infty}h(x^{\prime},x)B_{j}(x)\,dx\\
t_{i}  &  \approx A_{ij}\beta_{j}%
\end{align}
In other words, the observed histogram is approximated by a superposition of
the histograms produced by folding the functions $B_{j}$. Unfolding means to
determine the amplitudes $\beta_{j}$ of the basis functions $B_{j}$.

In stead of the expansion into orthogonal functions, $f(x)$ can be
approximated by a superposition of basic spline functions ($b$-splines). For
our applications the $b$-splines of order $2$ (linear), $3$ (quadratic) or $4$
(cubic) are appropriate (see Appendix 3).

Unfolding then produces a smooth function which normally is closer to the true
distribution than a histogram. The disadvantage of spline approximations
compared to the histogram representation is that a quantitative comparison
with predictions or the combination of several results is more difficult.

\emph{Remark:} In probability density estimation (PDE) a histogram is
considered as a first order spline function. The spline function corresponds
to the line that limits the top of the histogram bins. The interpretation of a
histogram in experimental sciences is different from that in PDE. Observations
are collected in bins and then the content of the bin measures the integral of
the function $g$ over the bin and the bin content of the unfolded histogram is
an estimate of the integral of $f$ over that bin. A function can always be
described correctly by a histogram. The description by spline functions is an
approximation. This has to be kept in mind when we compare the unfolding
result to a prediction.

\section{The least square solution and the eigenvalue decomposition}

\subsection{The least square solution}

As mentioned, for a \emph{square matrix} $\tens{A}$, $M=N$ the solution of
$\vec{\theta}$ is simply obtained by matrix inversion, $\widehat{\vec{\theta}%
}=\tens{A}^{-1}\vec{d}$. The error matrix $\tens{C}_{\theta}=\tens{A}^{-1}%
\tens{C}_{d}(\tens{A}^{-1})^{T}$ is derived by error propagation. We omit the
calculation. In the limit where there is no smearing, $\tens{A}$ is diagonal
and describes only acceptance losses.

The choice $M=N$ is not recommended. For $M\leq N$ the least square function
$\chi_{stat}^{2}$ is given by the following relation:
\begin{equation}
\chi_{stat}^{2}=\sum_{i=1}^{N}\frac{(t_{i}-d_{i})^{2}}{t_{i}}=\sum_{i=1}%
^{N}\frac{(%
{\displaystyle\sum\limits_{k=1}^{M}}
A_{ik}\theta_{k}-d_{i})^{2}}{%
{\displaystyle\sum\limits_{k=1}^{M}}
A_{ik}\theta_{k}}\;. \label{chistat}%
\end{equation}

If the numbers $d_{i}$ are not described by a simple Poisson distribution, we
have to insert the weight matrix\footnote{In the literature the error matrix
or covarince matrix is frequently denoted by $\tens{V}$ and the weight matrix
by $\tens{V}^{-1}$.} where $\tens{V}=\tens{C}_{d}^{-1}$ is the inverse of its
error matrix $\tens{C}_{d}$:%
\begin{equation}
\chi_{stat}^{2}=\sum_{i,k=1}^{N}\left[  (t_{i}-d_{i})V_{ik}(t_{k}%
-d_{k})\right]  \;. \label{chistat1}%
\end{equation}

If the data follow a Poisson distribution where the statistics is high enough
to approximate it by a normal distribution and where the denominator of
(\ref{chistat}) can be approximated by $d_{i}$,%
\begin{equation}
\chi_{stat}^{2}=\sum_{i=1}^{N}\frac{(t_{i}-d_{i})^{2}}{d_{i}}=\sum_{i=1}%
^{N}\frac{(%
{\displaystyle\sum\limits_{k=1}^{M}}
A_{ik}\theta_{k}-d_{i})^{2}}{d_{i}}\;, \label{linearLS}%
\end{equation}
the least square minimum can be evaluated by a simple linear matrix calculus.

The linear LS solution is given in standard textbooks. We apply the
transformations%
\begin{align}
\vec{d}\Rightarrow\vec{b}  &  =\tens{A}^{T}\tens{V}\vec{d}\;,\label{unflsq}\\
\tens{A}\Rightarrow\tens{Q}  &  =\tens{A}^{T}\tens{V}\tens{A}\;.
\label{unflsq1}%
\end{align}
We call $Q$ \emph{least square matrix}. We get for the expected value of
$\vec{b}$
\begin{equation}
\mathrm{E}(\vec{b})=\tens{Q}\vec{\theta} \label{btheta}%
\end{equation}
with the LS solution%
\begin{equation}
\widehat{\vec{\theta}}=\tens{Q}^{-1}\vec{b} \label{LSsolution}%
\end{equation}
and the error matrix $\tens{C}_{\theta}$ of the solution%
\[
\tens{C}_{\theta}=\tens{Q}^{-1}\;.
\]

We have simply replaced $\tens{A}$ by $\tens{Q}$ and $\vec{d}$ by $\vec{b}$.
Both quantities are then known. The matrix $\tens{Q}$ is quadratic and can be
inverted if the LS solution exists.

\subsection{Eigenvector decomposition of the least square matrix}%

\begin{figure}
[ptb]
\begin{center}
\includegraphics[
height=4.3013in,
width=3.1449in
]%
{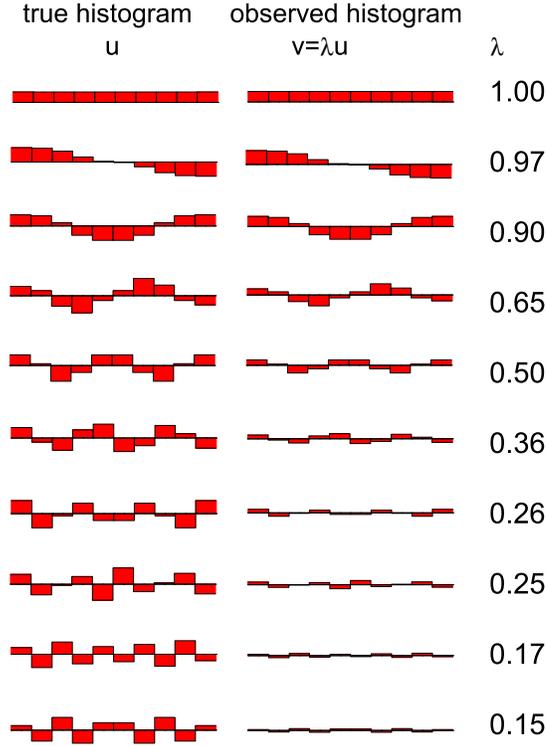}%
\caption{{\small Set of eigenvectors ordered according to decreasing
eigenvalues. A contribution }$\vec{u}_{i}$ {\small in the true histogram
corresponds to a contribution }$\vec{v}_{i}$ {\small to the observed
histogram.}}%
\label{eigenvec2}%
\end{center}
\end{figure}
\index{unfolding!eigendecomposition|see{deconvolution}}%
To understand better the origin of the fluctuations of the LS solution
(\ref{LSsolution}), we factorize the matrix $\tens{Q}$ in the following way:
The matrix\footnote{We require that the square $M\times M$ matrix $\tens{Q}$
has $M$ linearly independent eigenvectors and that all eigenvalues are real
and positive. These conditions are satisfied if a LS solution exists.}
$\tens{Q}=\tens{U}\Lambda\tens{U}^{-1}$ is composed of the diagonal matrix
$\tens{\Lambda}$ which contains the eigenvalues of $\tens{Q}$ and the matrix
$\tens{U}$ whose columns consist of the eigenvectors $\vec{u}_{i}$ of $Q$:%
\[
\tens{Q}=\left(
\begin{array}
[c]{ccccc}%
\vec{u}_{1} & \vec{u}_{2} & . & . & \vec{u}_{M}%
\end{array}
\right)  \left(
\begin{array}
[c]{ccccc}%
\lambda_{1} &  &  &  & \\
& \lambda_{2} &  & 0 & \\
&  & . &  & \\
& 0 &  & . & \\
&  &  &  & \lambda_{M}%
\end{array}
\right)  \left(
\begin{array}
[c]{ccccc}%
\vec{u}_{1} & \vec{u}_{2} & . & . & \vec{u}_{M}%
\end{array}
\right)  ^{-1}\;.
\]%
\begin{equation}
\tens{Q}\vec{u}_{i}=\lambda_{i}\vec{u}_{i}=\vec{v}_{i}\;,\;i=1,...,M\;.
\label{eigen}%
\end{equation}

Software to produce the eigenvector decomposition can be found in most
mathematical computer libraries.

In case of eigenvalues that appear more than once, the eigenvectors are not
uniquely defined. Linear orthogonal combinations can be created by rotations
in the corresponding subspace but they produce the same LS solution.

The solution $\vec{\theta}$ can be expanded into the orthogonal unit
eigenvectors $\vec{u}_{i}$:%
\begin{align*}
\vec{\theta}  &  =%
{\displaystyle\sum\limits_{i=1}^{M}}
a_{i}\vec{u}_{i}\;,\;\;\theta_{k}=%
{\displaystyle\sum\limits_{i=1}^{M}}
a_{i}u_{ik}\;,\\
a_{i}  &  =\vec{\theta}\cdot\vec{u}_{i}\;,\;\;a_{i}=%
{\displaystyle\sum\limits_{k=1}^{M}}
\theta_{k}u_{ik}\;.
\end{align*}
By construction, the amplitudes $a_{i}$ are uncorrelated and the norm
$||\theta||^{2}=\Sigma\theta_{i}^{2}$ of the solution is given by
\[
||\theta||^{2}=\sum_{i=1}^{M}a_{i}^{2}\;.
\]

The transformed observed vector $\vec{b}$ is%
\[
\vec{b}=\sum_{i=1}^{M}a_{i}\lambda_{i}\vec{u}_{i}=\sum_{i=1}^{M}a_{i}\vec
{v}_{i}\;.
\]

In Fig. \ref{eigenvec2} we present an schematic example of a set of
eigenvectors. A contribution $\vec{u}_{i}$ to the true histogram as shown on
the left-hand side will produce a contribution $\vec{v}_{i}$ to the observed
histogram. It is of the same shape but reduced by the factor $\lambda_{i}$ as
shown on the right-hand side. The eigenvalues decrease from top to bottom.
Strongly oscillating components of the true histogram correspond to small
eigenvalues. They are hardly visible in the observed data, and in turn, small
contributions $v_{i}$ to the observed data caused by statistical fluctuations
can lead to rather large oscillating contributions $\vec{u}_{i}=\vec{v}%
_{i}/\lambda_{i}$ to the unfolded histogram if the eigenvalues are small.
Eigenvector contributions with eigenvalues below a certain value cannot be
reconstructed, because they cannot be distinguished from noise in the observed histogram.

The eigenvector decomposition is equivalent to the \emph{singular value
decomposition} (SVD). In the following we will often refer to the term SVD
instead of the eigenvector decomposition, because the former is commonly used
in the unfolding literature.%

\begin{figure}
[ptb]
\begin{center}
\includegraphics[
height=4.078in,
width=5.4869in
]%
{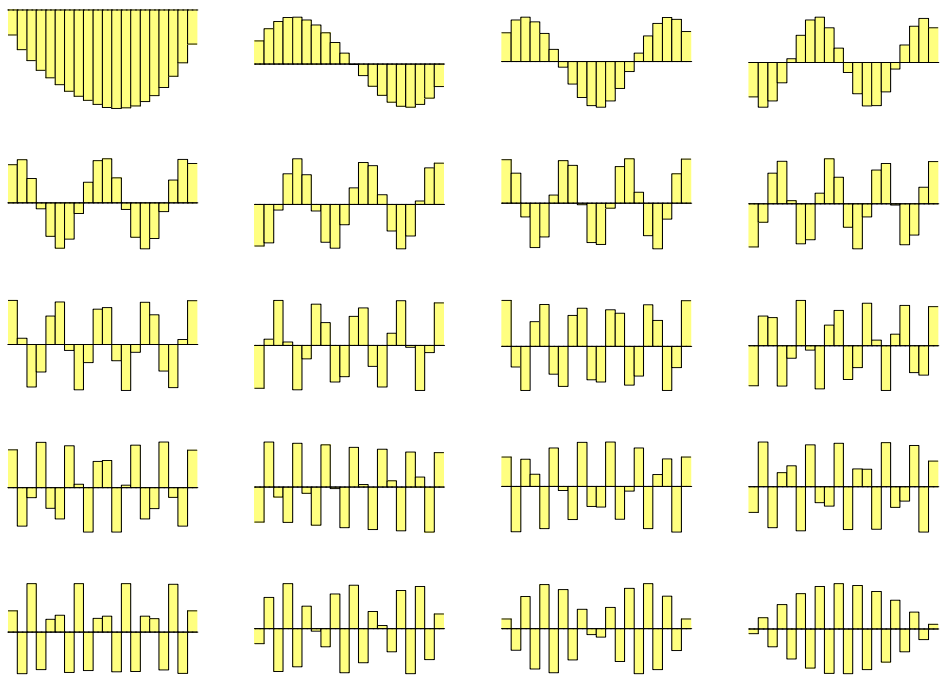}%
\caption{Eigenvectors of the modified LS matrix ordered with decreasing
eigenvalues.}%
\label{einuvec}%
\end{center}
\end{figure}
%

\begin{figure}
[ptb]
\begin{center}
\includegraphics[
trim=0.000000in 0.110916in 0.000000in 0.074052in,
height=3.682in,
width=5.0585in
]%
{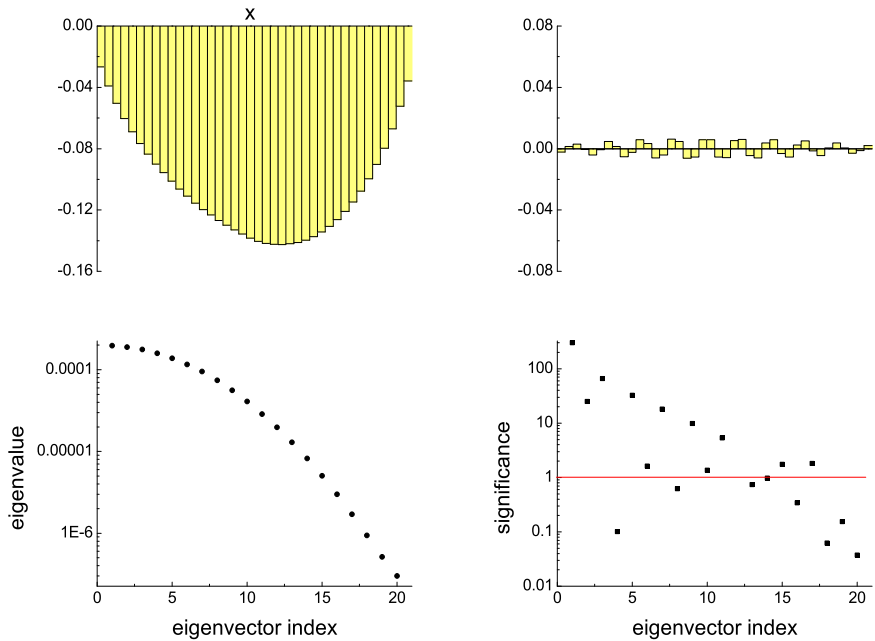}%
\caption{Observed eigenvectors $1$ (top left) and $20$ (top right),
eigenvalues (bottom left) and significance of eigenvector amplitudes (bottom
right).}%
\label{einuu}%
\end{center}
\end{figure}

\begin{myexample}
In Fig. \ref{einuvec} the $20$ eigenvectors of a LS matrix ordered with
decreasing eigenvalue are displayed. The response matrix has $20$ true and
$40$ observed bins. The graph is generated from a sample of $100\,000$
uniformly distributed events in the range of the observed and the true
variables $0<x,x^{\prime}<1$. The response function is a Gaussian with
standard deviation $\sigma_{s}=0.04$. The eigenvectors show an oscillatory
behavior where the number of clusters corresponds roughly to the eigenvector index.

In Fig. \ref{einuu} top the eigenvectors $1$ and $20$ folded with the response
matrix are shown. A contribution of eigenvector $20$ to the observed histogram
is similar to that of noise. The eigenvalues shown at the bottom left graph
vary by about three orders in magnitude. This means that a contribution of the
eigenvector $20$ to the true distribution is suppressed by a factor of $1000$
with respect to a contribution of eigenvector $1$. The bottom lright-hand
graph shows the significance of the amplitudes that are attributed to the
eigenvectors. Significance is defined as the absolute value of the amplitude
divided by its error. As we have indicated above, the significance is expected
to decreases with decreasing eigenvalue. Due to the symmetry of the problem,
the amplitudes with even index should vanish. Statistical fluctuations in the
simulation partially destroy the symmetry. Eigenvector contributions where the
significance is below one, are compatible with being absent within one
standard deviation.
\end{myexample}

\begin{figure}
[ptb]
\begin{center}
\includegraphics[
trim=0.000000in 0.088857in 0.000000in 0.059506in,
height=6.3146in,
width=5.1183in
]%
{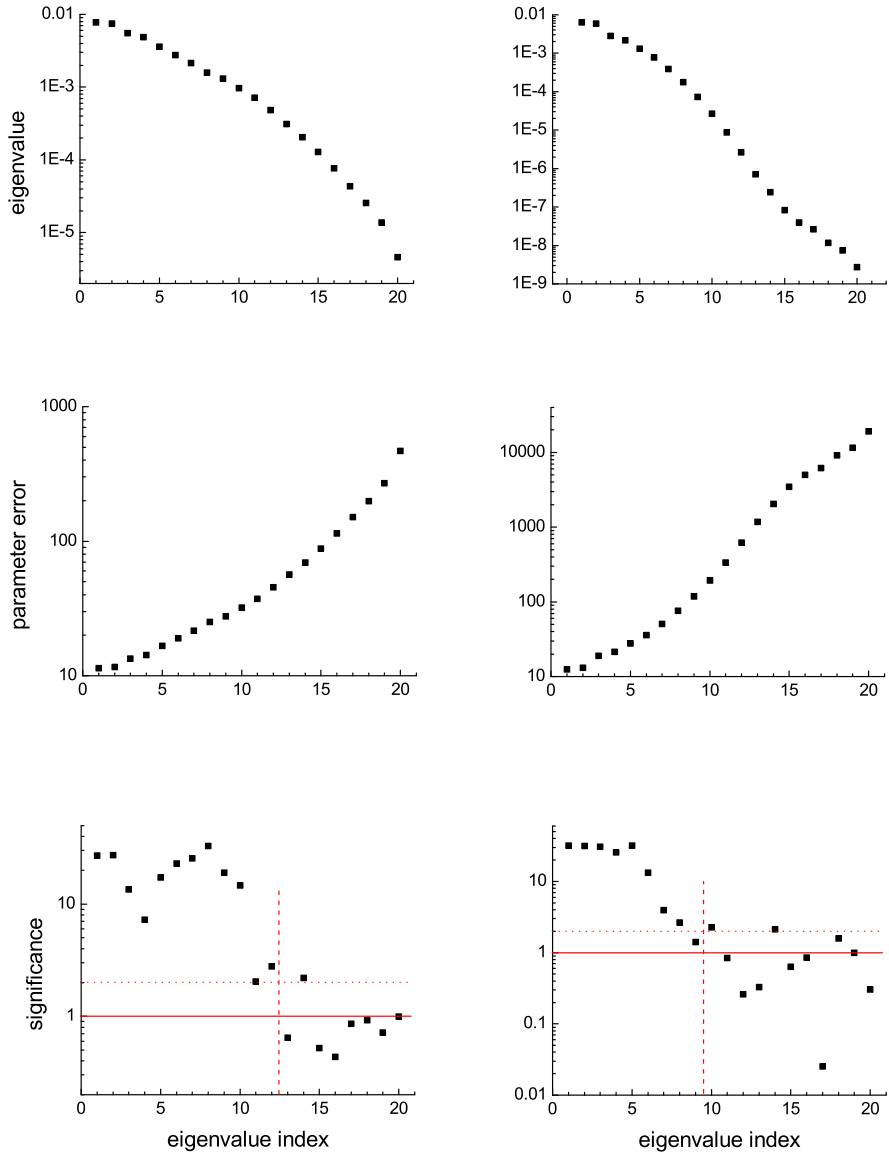}%
\caption{Eigenvector, parameter error and significance as a function of the
eigenvalue index for $5000$ events and resolutions $\sigma_{s}=0.04$ (left
hand) and $\sigma_{s}=0.08$ (right hand).}%
\label{eigen1b5000}%
\end{center}
\end{figure}

\begin{myexample}
In Fig. \ref{eigen1b5000} we compare the eigenvalues, the parameter errors and
the significance of the amplitudes of the eigenvectors for the experimental
resolution $\sigma_{s}=0.04$ (left-hand side) with that of $\sigma_{s}=0.08$
(right-hand side). The values are displayed as a function of the eigenvector
index for a sample consisting of $5000$ events. The eigenvalues decrease by
$4.5$ decades for $\sigma_{s}=0.04$ and by $8.5$ decades for $\sigma_{s}=0.08$
from the first to the last eigenvector. The errors of the amplitudes of the
eigenvectors with large index increase dramatically with $\sigma_{s}$. The
vertical lines in the significance plots indicate the number of eigenvectors
that should be retained to obtain the best agreement of the unfolded histogram
with the true histogram. (As measure of the quality we use the integrated
square error $ISE$ which is explained below.) For $\sigma_{s}=0.04$ this are
$12$ and for $\sigma_{s}=0.08$ only $8$ eigenvectors. For larger numbers the
agreement deteriorates due to spurious oscillations. From the two plots and
the fact that the errors are proportional to the square root of the number of
events we can derive that $200$ times more events with $\sigma_{s}=0.08$ are
necessary to obtain results with the same precision as with $\sigma_{s}=0.04$.
\end{myexample}

%

We conclude that it is difficult to compensate a bad resolution of an
experiment by increasing the statistics! We should always make an effort to
avoid large smearing effects not only because large event numbers are required
but also because the unfolding results then depend strongly on a precise
knowledge of the response function.

\subsection{The effective number of parameters}%

\begin{figure}
[ptb]
\begin{center}
\includegraphics[
trim=0.091374in 0.246089in 0.092522in 0.197173in,
height=2.5463in,
width=5.7517in
]%
{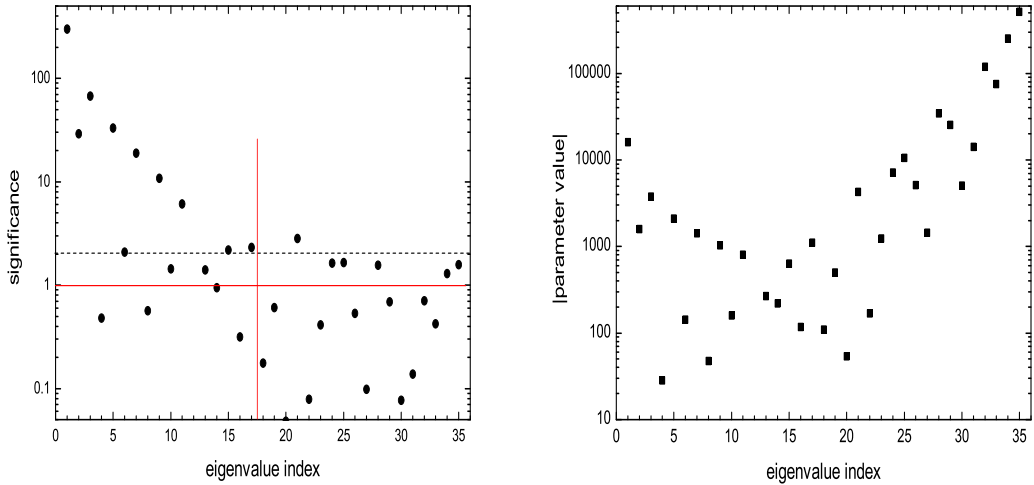}%
\caption{Left hand: Parameter significance as function of the eigenvalue
index. The effective number of parameters is $17$. Right hand: Fitted
parameter values as a function of the eigenvalue index.}%
\label{eigen35}%
\end{center}
\end{figure}

When we unfold a histogram, the number of bins of the unfolded histogram is
the number of free parameters in the fit. The previous example indicates that
the number of parameters that we can determine in a given problem is rather
limited. Below a certain eigenvalue $\lambda_{k}$, all parameters have a
significance close to or below one. We define an effective number of
parameters $N_{eff}=k$ as the number of parameters with eigenvalues above or
equal to this limit but we do not count parameters separated from the dominant
parameters by a gap of two or more parameters with significance below one. In
this way we exclude parameters with small eigenvalues, even if their
significance is above one, because the excess is likely to be caused by
statistical fluctuations. For the example of Fig. \ref{eigen35} with a uniform
distribution the effective number of parameters is $N_{eff}=17$. Because the
significances of parameters $18$ and $19$ are below one, the parameters with
index greater than $19$ are not considered even if their significance is
greater than one. This definition is to a certain extend arbitrary, but it
provides a reasonable estimate of the minimum number parameter that we need to
describe the data. There are also parameters left of index $17$ that are
compatible with being zero. We should not exclude the corresponding
contributions, because the reason for the small values of the significance are
not large errors, but small values of the fitted amplitudes as is indicated in
the lright-hand graph. This graph shows that some amplitudes that correspond
to small eigenvalues become rather large. This is due to the amplification of
high frequency noise in the unfolding. The number of bins in the unfolded
distribution should not be much larger than the effective number of
parameters, because then we keep too much redundant information, but on the
other hand it has to be large enough to represent the highest significant
eigenvector. A reasonable choice for the number of bins is about twice
$N_{eff}$. The optimal number will also depend on the shape of the distribution.

In the following example we study the dependence of $N_{eff}$ and the
significance on the number of true bins.

\begin{figure}
[ptb]
\begin{center}
\includegraphics[
trim=0.127810in 0.137478in 0.128195in 0.092324in,
height=2.8286in,
width=5.6413in
]%
{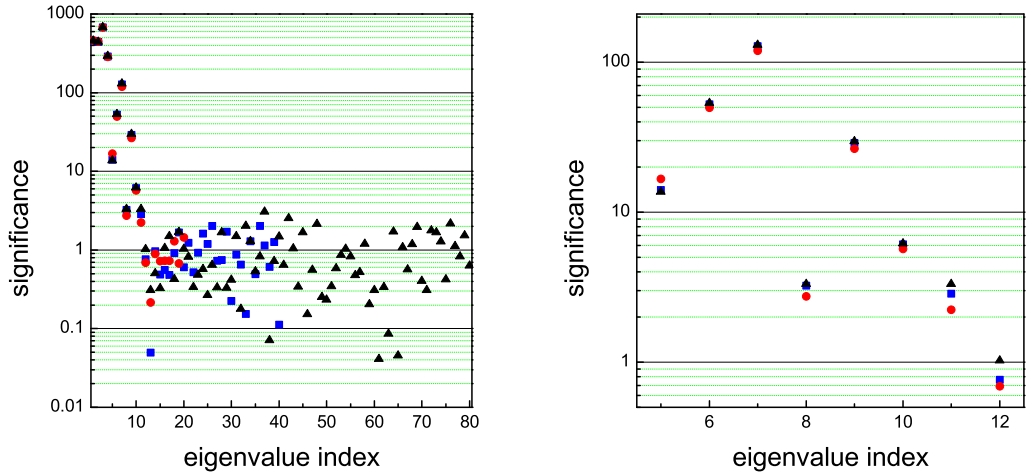}%
\caption{Significance of the eigenvector contributions for three different
binnings ($20$, $40$, $80$) of the true distribution. The expanded graph
(right hand) indicates that the significance slightly increases with the
number of bins.}%
\label{significance1}%
\end{center}
\end{figure}

\begin{figure}
[ptb]
\begin{center}
\includegraphics[
trim=0.000000in 0.136472in 0.000000in 0.091308in,
height=2.308in,
width=3.2138in
]%
{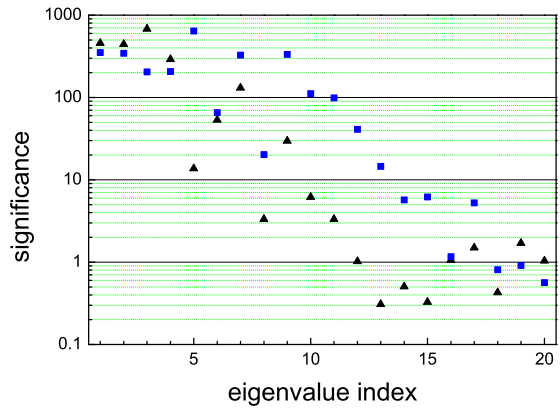}%
\caption{Significance of the eigenvector contributions for two different
resolutions. Reducing the smearing parameter $\sigma_{s}$ by a factor of two
(squares) increases the number of effective parameters from $12$ to $17$.}%
\label{significance2}%
\end{center}
\end{figure}

\begin{myexample}
Fig. \ref{significance1} is derived from a superposition of two Gaussians with
an exponential distribution, see Fig. \ref{exp2gaus}, with $n=1\,000\,000$
events distributed with a Gaussian resolution of $\sigma_{s}=0.08$ into $100$
observed bins. The graph contains results for three different choices of the
number of bins of the unfolded histogram, namely $20$, $40$ and $80$. The
effective number of parameters is $11$ for $20$ and $40$ bins. With $80$ bins
the significance for the dominant eigenvectors is slightly higher than for the
lower bin numbers and we get $N_{eff}=12$. The parameter $N_{eff}$ is mainly
determined by the experimental resolution and less by the number of observed
bins. This is demonstrated in Fig. \ref{significance2}. The $80$ bin result of
the previous figure is compared to the significance obtained with the
resolution increased by a factor of two.
\end{myexample}

%

%

\section{Summary}

We have studied the unfolding problem in form of histograms and a response
matrix. If the number of events is large enough to apply the normal
approximation, it can be solved with a linear least square fit and a simple
matrix formalism. The unfolding solution can be expanded in linearly
independent vectors, the eigenvectors of the least square matrix or
equivalently in those of a singular value decomposition. With decreasing
eigenvalues the eigenvectors present more and more oscillations. Due to the
experimental smearing, the high frequency components are washed out and
difficult to distinguish from noise. As a consequence, only a limited number
of eigenvectors can significantly be reconstructed. This number $N_{eff}$, the
effective number of parameters, can be extracted from the significance of the
eigenvector coefficients. $N_{eff}$ depends strongly on the experimental
resolution and is independent of the number of bins $M$ used in the unfolded
histogram as long as $M$ is larger than about twice $N_{eff}$. For this reason
there is no need to choose the number of bins of the unfolded histogram larger
$2N_{eff}$. The dependence of $N_{eff}$ on the number of bins $N$ in the
observed histogram is negligible if $N$ is greater than about $4N_{eff}$.

\chapter{Unfolding without explicit regularization}

\section{Introduction}

If we want to document the experimental information in such a way that it is
conserved for a comparison with a theory that might be developed in future or
if we want to compare or combine the date with those of another experiment,
the results have to be unbiased. To achieve this condition we could store the
distorted data together with the resolution function. Such a procedure is
optimal in that no information is wasted but it has severe drawbacks. Two
large datasets, the experimental data and the Monte Carlo sample would have to
be published and the whole analysis work would be left to the scientist who
wants to use the data. A less perfect but simple and more practical way is to
unfold the experimental effects and to present the data in form of a histogram
together with an error matrix which than can be used in a future analysis. To
avoid the unpleasant oscillation that we have discussed in the previous
section, we have to choose wide bins. Additional explicit smoothing would bias
the data and has to be omitted. We have to accept that some information will
be lost.

A third possibility which preserves the information that is necessary for a
future quantitative analysis is to unfold the data with a simple explicit
smoothing step and to document the smoothing function. A comparison of the
experimental result with a prediction is then possible but data of different
experiments can not be combined.

In the following we turn to the simple and efficient approach where
oscillations are suppressed by using wide bins in the unfolded histogram. We
call this procedure \emph{implicit regularization}. The bin contents are
fitted either by minimizing the sum of the least squares or by maximizing the likelihood.

We have discussed the LS method already in the previous chapter. The results
of the LS fit are in most cases similar to those of the ML fit. In the
asymptotic limit where the event numbers tend to infinity, and where the
Poisson distribution can be approximated by a normal distribution, the two
methods coincide. Contrary to the LS approach, the MLE usually does not
produce negative entries in the unfolded histogram. With the LS method also
complex situations can be handled, for instance when background has to be
taken care of while the ML method requires Poisson distributed event numbers.

\section{The maximum likelihood approach}

Whenever possible, we should apply a maximum likelihood fit instead of a LS
fit. With Poisson distributed event numbers $d_{i}$ with expected values
$t_{i}=\sum_{j}A_{ij}\theta_{j}$ the probability to obtain $d_{i}$ is%

\[
P(d_{i})=\frac{e^{-t_{i}}t_{i}^{d_{i}}}{d_{i}!}\
\]
and the corresponding log-likelihood is up to an irrelevant constant
\begin{align}
\ln L_{stat}  &  =\sum_{i=1}^{N}\left[  d_{i}\ln t_{i}-t_{i}\right]
\;\nonumber\\
&  =\sum_{i=1}^{N}\left[  d_{i}\ln\sum_{j=1}^{M}A_{ij}\theta_{j}-\sum
_{j=1}^{M}A_{ij}\theta_{j}\right]  \;. \label{likestat}%
\end{align}

Maximizing $\ln L_{stat}$ we obtain an estimate $\widehat{\vec{\theta}}$ of
the true histogram.

Usually we have of the order of $20$ bins and of course the same number of
correlated parameters which have to be adjusted. In this situation the fit
often does not converge very well. Instead of maximizing the log-likelihood
with methods like Simplex, we can compute the solution iteratively.

\section{The Expectation-Maximization algorithm}

The iterations follow the Expectation-Maximization (EM) method \cite{em}. The
EM algorithm is a general iterative method to maximize the likelihood if there
are \emph{missing} (or \emph{latent})\emph{ variables}. It is especially
useful in classification problems in conjunction with p.d.f.s of the
exponential family\footnote{To the exponential family belong among others the
normal, Poisson, exponential, gamma, chi-squared distrribution.}. Applied to
our unfolding problem, the missing information in step $k$ is the fractions of
events $p_{ij}^{(k)}$ in an observed bin $i$ that belong to the true bin $j$.
Hence, there are $M$ missing variables per bin. Its expected value is
$\boldsymbol{E}(p_{ij})=A_{ij}\theta_{j}/\Sigma_{j}A_{ij}\theta_{j}$.The
following alternating steps\footnote{This description of the EM algorithm is
simplified and adapted to our specific problem.} are repeated:

\begin{itemize}
\item Compute the expected log-likelihood given the actual set of parameters
$\vec{\theta}^{(k)}$ and the observed data $\vec{d}$. The expected number of
events that migrates from true bin $j$ to bin $i$ is
\begin{align*}
d_{ij}^{(k)}  &  =d_{i}p_{ij}^{(k)}\\
&  =d_{i}\frac{A_{ij}\theta_{j}^{(k)}}{%
{\displaystyle\sum\limits_{j=1}^{M}}
A_{ij}\theta_{j}^{(k)}}\;.
\end{align*}
For a Poisson distribution the log-likelihood to observe $d_{ij}$ events in
bin $\ i$ that originate from true bin $j$ is up to a constant $\ln
L(\theta)=$ $-A_{ij}\theta_{j}+d_{ij}\ln A_{ij}\theta_{j}$. The expected
log-likelihood of $\vec{\theta}$ is%
\begin{align*}
Q(\vec{\theta}|\vec{\theta}^{(k)})  &  =%
{\displaystyle\sum\limits_{i=1}^{N}}
{\displaystyle\sum\limits_{j=1}^{M}}
[-A_{ij}\theta_{j}+d_{ij}^{(k)}\ln A_{ij}\theta_{j}]\\
&  =%
{\displaystyle\sum\limits_{i=1}^{N}}
{\displaystyle\sum\limits_{j=1}^{M}}
[-A_{ij}\theta_{j}+d_{i}\frac{A_{ij}\theta_{j}^{(k)}}{%
{\displaystyle\sum\limits_{j=1}^{M}}
A_{ij}\theta_{j}^{(k)}}\ln A_{ij}\theta_{j}]\;.
\end{align*}

\item Maximize the expected likelihood $Q(\vec{\theta}|\vec{\theta}^{(k)})$
and obtain $\vec{\theta}^{(k+1)}$. The computation of the maximum of $Q$ is
easy, because the components of the parameter vector $\vec{\theta}$ appears in
independent summands.
\begin{align}
\frac{\partial Q}{\partial\theta_{j}}  &  =%
{\displaystyle\sum\limits_{i=1}^{N}}
[-A_{ij}+d_{i}\frac{A_{ij}\theta_{j}^{(k)}}{%
{\displaystyle\sum\limits_{m=1}^{M}}
A_{ij}\theta_{j}^{(k)}}\frac{1}{\theta_{j}}]=0\;,\nonumber\\
\alpha_{j}\theta_{j}^{(k+1)}  &  =%
{\displaystyle\sum\limits_{i}^{N}}
d_{i}\frac{A_{ij}\theta_{j}^{(k)}}{%
{\displaystyle\sum\limits_{j}}
A_{ij}\theta_{j}^{(k)}}\;,\nonumber\\
\theta_{j}^{(k+1)}  &  =\sum_{i=1}^{N}A_{ij}\theta_{j}^{(k)}\frac{d_{i}}%
{d_{i}^{(k)}}\left/  \alpha_{j}\right.  \;. \label{RLunfolding}%
\end{align}

\end{itemize}

In the second line we have replaced $\Sigma_{i}A_{ij}$ by $\alpha_{j}$ the
average acceptance of the events of true bin $j$.

Before the EM method had been invented, the iterative procedure had been
introduced independently by Richardson and Lucy \cite{rich72,lucy74}
specifically for the solution of unfolding problems. Later it was reinvented
by Shepp and Vardi \cite{shepp82}, Kondor \cite{kondor83}, Mülthei and Schorr
\cite{muelthei86} and D'Agostini \cite{dagostini95} made it popular in
particle physics. That the result of the iteration converges to the maximum
likelihood solution, is a general property of the EM method but was also
proven by Vardi et al. \cite{vardi85} and later independently by Mülthei and
Schorr \cite{muelthei86}. For a discussion of the application to unfolding see
\cite{titterington}.

In the particle physics community, unfolding with the EM method is also called
D'Agostini unfolding, Bayesian unfolding, iterative unfolding or
Richardson-Lucy unfolding\footnote{In most of the the figures of this report
the abreviation R-L is used.}. I propose to agree on the term \emph{EM
unfolding}.%

\begin{figure}
[ptb]
\begin{center}
\includegraphics[
trim=0.000000in 0.098410in 0.000000in 0.034282in,
height=5.7551in,
width=4.5903in
]%
{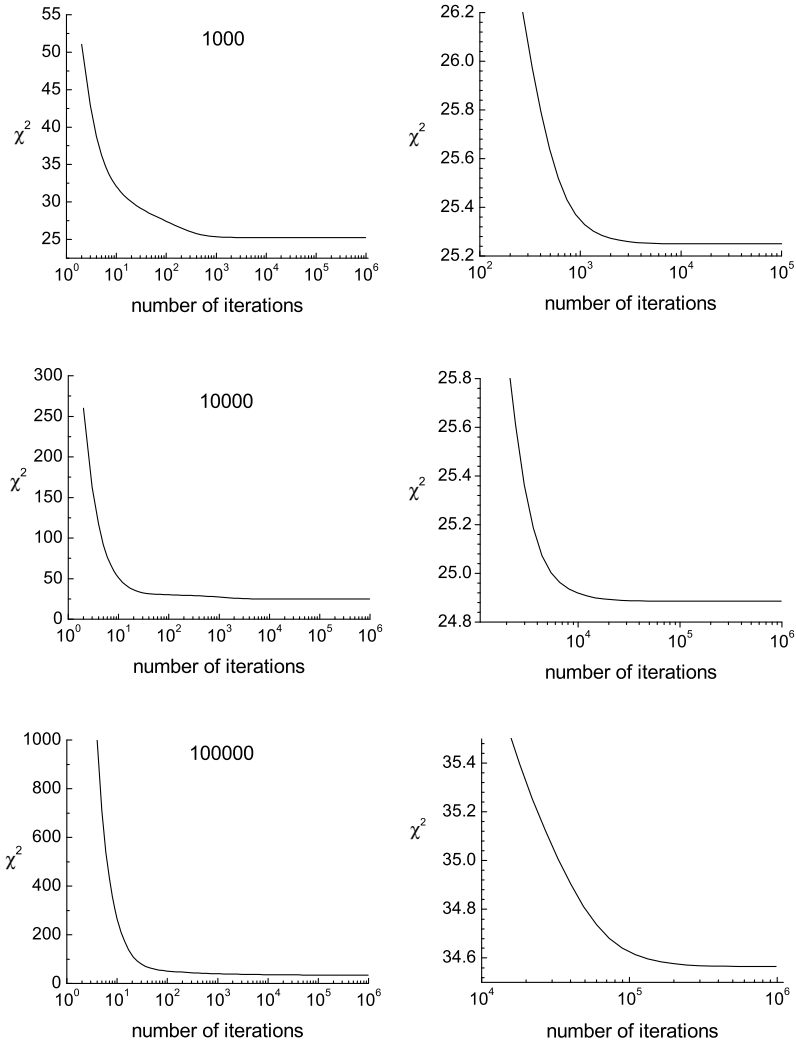}%
\caption{$\chi^{2}$ as a function of the number of iterations for different
event numbers. In the lright-hand plots the tails of the curves are enlarged.
}%
\label{iterconverge}%
\end{center}
\end{figure}
%

\begin{figure}
[ptb]
\begin{center}
\includegraphics[
trim=0.000000in 0.159655in 0.000000in 0.120077in,
height=2.2557in,
width=2.9174in
]%
{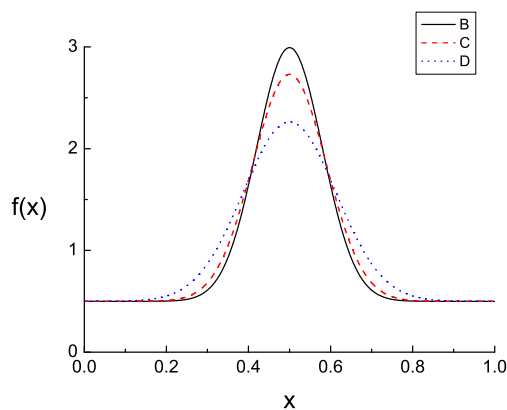}%
\caption{Superposition of a uniform and a normal distribution. The dashed and
dotted curves correspond to the solid curve smeared with Gaussian resolutions
$\sigma_{s}=0.04$ and $\sigma_{s}=0.08$, respectively.}%
\label{curve}%
\end{center}
\end{figure}
%

\begin{figure}
[ptb]
\begin{center}
\includegraphics[
trim=0.127226in 0.993115in 0.130136in 0.126947in,
height=2.1635in,
width=5.9385in
]%
{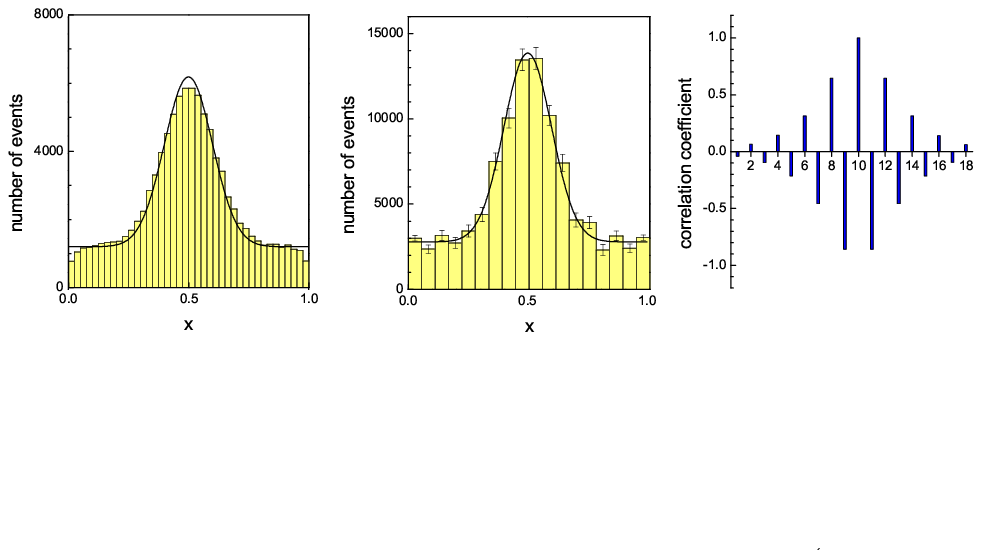}%
\caption{Unfolding without explicit regularization. The left-hand plot shows
the observed distribution, the central plot is the result of the unfolding for
$\sigma_{s}=0.04$ and the lright-hand plot indicates the correlation of bin
$10$ with the other bins of the histogram. The curve represents the true
distribution.}%
\label{rlunfold}%
\end{center}
\end{figure}
\qquad

The iterative method to find the MLE is not only extremely simple, it is also
fast. Approximately $10^{4}$ iterations are executed per second on a simple
laptop computer. The convergence can be accelerated by choosing a starting
distribution that is close to the expected true distribution.%
\begin{figure}
[ptb]
\begin{center}
\includegraphics[
trim=0.132094in 0.158364in 0.132503in 0.132193in,
height=2.4716in,
width=5.7841in
]%
{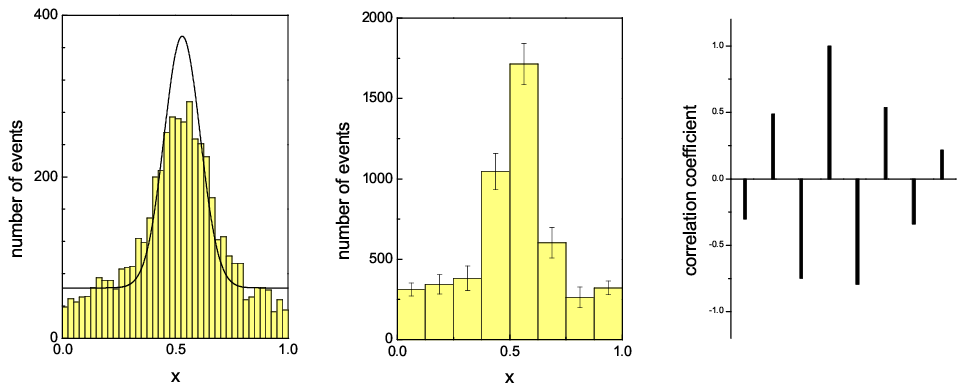}%
\caption{Unfolding without explicit regularization. The left-hand plot shows
the observed histogram for $\sigma_{s}=0.08$, the central plot is the unfolded
histogram and the right hand plot indicates the correlation of bin $4$ with
the other bins of the histogram. The curve represents the true distribution.}%
\label{unf1berr8}%
\end{center}
\end{figure}

Figure \ref{iterconverge} illustrates the convergence of the iteration
sequence for an example generated according to the distribution of Fig.
\ref{curve}. The resolution was set to $\sigma_{s}=0.8$. The observed and the
true distributions had $40$ and $18$ bins, respectively. A uniform starting
distribution is used. The required number of iteration steps increases with
the number of events and the smearing parameter $\sigma_{s}$.%

\begin{figure}
[ptb]
\begin{center}
\includegraphics[
trim=0.000000in 0.132818in 0.000000in 0.101733in,
height=3.6106in,
width=5.0195in
]%
{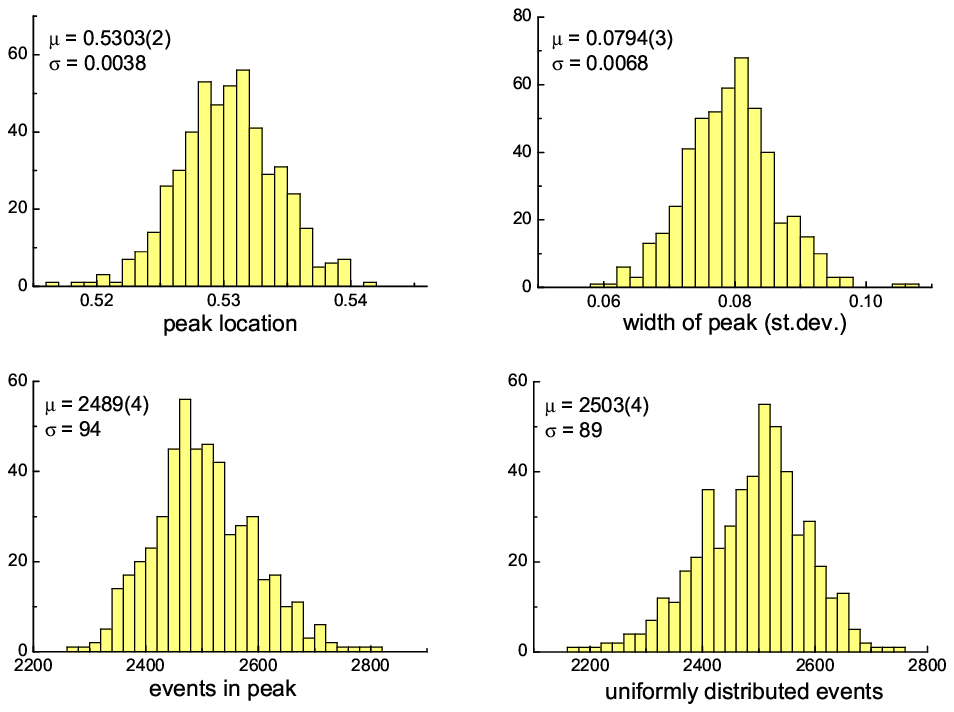}%
\caption{Distribution of the fitted parameters of the one-peak example from
$500$ experiments with $5000$ events, $8$ true bins and resolution $\sigma
_{s}=0.08$. The mean value and the standard deviation of the fitted parameters
are given in the graphs.}%
\label{implicit}%
\end{center}
\end{figure}

\begin{myexample}
We simulate data according to the distribution%
\[
f(x)=0.5\left[  1+\frac{1}{\sqrt{2\pi}0.08}\exp\left(  -\frac{(x-0.5)^{2}%
}{2\cdot0.08^{2}}\right)  \right]
\]
which is a superposition of a uniform and a normal distribution in the
interval $[0,1]$. It is displayed in Fig. \ref{curve}. The tails of the normal
distribution outside the interval are neglected in the normalization. The
response function is also a normal distribution with standard deviation
$\sigma_{s}=0.04$. A total of $100000$ events is generated. The observed
smeared distribution with $40$ bins and the unfolded distribution with $18$
bins are shown in Fig. \ref{rlunfold}. The true distribution, displayed in
Fig. \ref{curve}, is not much modified by the smearing. The height of the peak
is slightly reduced, the peak is a bit wider and at the borders there are
acceptance losses. The central plot shows the unfolded distribution with the
diagonal errors. Due to the strong correlation between neighboring bins, the
errors are about a factor of five larger than $\sqrt{\theta_{i}}$. In the
right-hand plot the correlation coefficients of bin $10$ relative to the other
bins are given. The correlation with the two adjacent bins is negative. It
oscillates with the distance to the considered bins. The correlation
coefficients depend only on the bin width and the smearing function and are
independent of the shape of the distribution.
\end{myexample}

In the following examples we derive the parameters of the true distribution
from the unfolded histogram.%

\begin{figure}
[ptb]
\begin{center}
\includegraphics[
trim=0.000000in 0.183739in 0.000000in 0.100966in,
height=2.1959in,
width=3.1756in
]%
{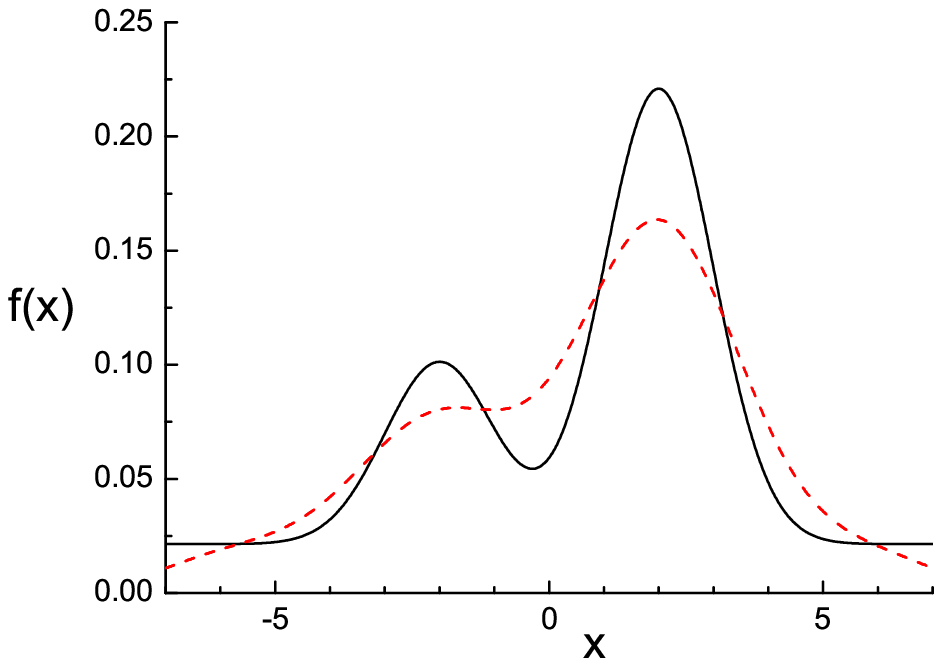}%
\caption{Two-peak distribution. The solid curve corresponds to the true
distribution, the dashed curve contains the experimental smearing. }%
\label{emcurve}%
\end{center}
\end{figure}

\begin{myexample}
We reduce the number of events to $5000$ and increase the smearing to
$\sigma_{s}=0.8$ equal to the standard deviation of the peak. To avoid
oscillations in the unfolded distribution, the number of true bins is reduced
to $8$. The location of the peak is shifted to $\mu=0.53$.
The observed histogram, the unfolded histogram and the correlation with
respect to bin $4$ are presented in Fig. \ref{unf1berr8}. The experiment is
simulated $500$ times and the unfolding result is then used to fit the
parameters of the true distribution. The fit results are summarized in Fig.
\ref{implicit}. The four parameters of the distribution, e.g. the location and
width of the peak and the numbers of events in the uniform and the Gaussian
part of the distribution, are well reproduced. The biases are negligible
compared to the uncertainties. The standard deviations of the distributions of
the fitted parameters agree with the error estimates from the individual fits.
The low number of bins causes some loss in resolution. In the Appendix 4 the
dependence of the resolution of the location and the width of a Gaussian peak
is estimated. In our case the ratio of bin width $w_{b}$ and observed width of
the peak is $w_{b}/\sqrt{\sigma^{2}+\sigma_{s}^{2}}=1.105$. The reductions in
resolution by about $5\%$ and $10\%$ for the peak location and the width
compared to the ideal situation are moderate.
\end{myexample}

%

\begin{figure}
[ptb]
\begin{center}
\includegraphics[
trim=0.141306in 0.122597in 0.142517in 0.104468in,
height=4.4973in,
width=5.9219in
]%
{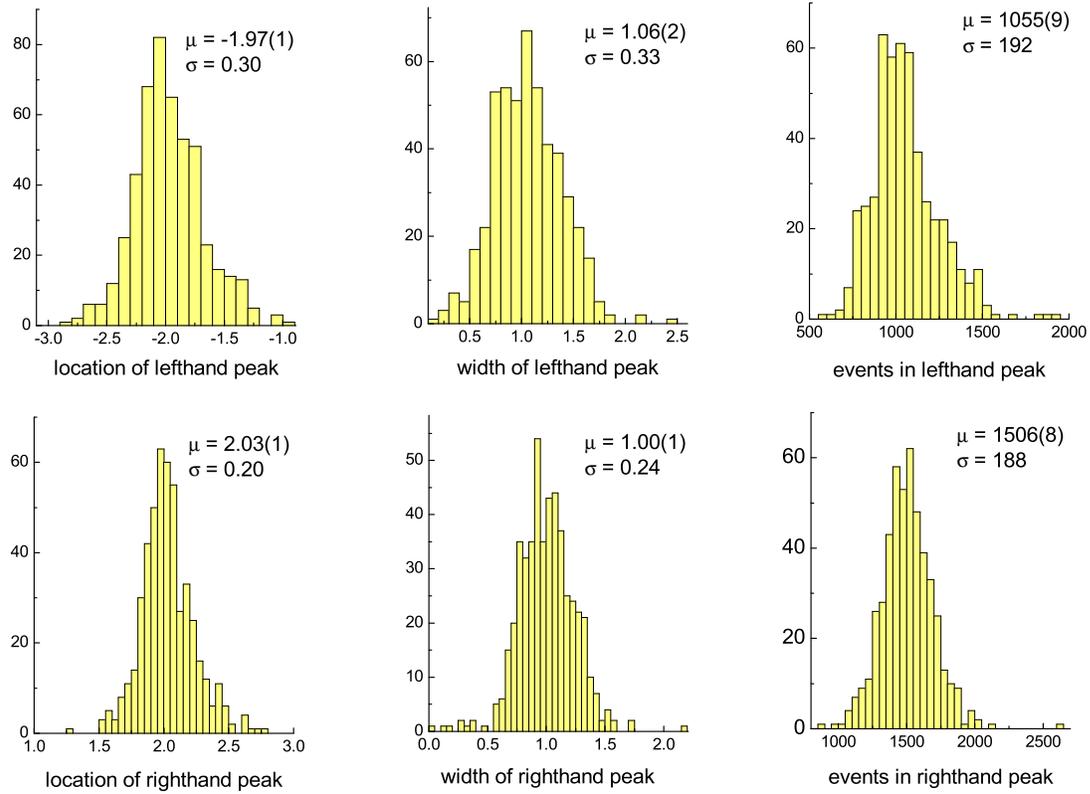}%
\caption{Fit results for the two-peak example.}%
\label{implicit1}%
\end{center}
\end{figure}

\begin{myexample}
We borrow a more involved example from \cite{kuusela}. The distribution now
contains a superposition of a uniform distribution and two normal
distributions,%
\begin{equation}
f(x)=c_{1}\mathcal{N}(x|\mu_{1},\sigma_{1})+c_{2}\mathcal{N}(x|\mu_{2}%
,\sigma_{2})+c_{3}\mathcal{U}(x)\;, \label{kuuseladist}%
\end{equation}
with $7$ parameters. We keep $40$ bins for the observed histogram and choose
$10$ bins for the unfolded histogram. $5000$ events are generated in the range
$-7\leq x\leq7$ according to $f(x)$ with the parameters $c_{1}=0.2$,
$c_{2}=0.5$, $c_{3}=0.3$, $\mu_{1}=-2$, $\sigma_{1}=1$, $\mu_{2}=2$,
$\sigma_{2}=1$. The smearing resolution $\sigma_{s}=1$ is chosen equal to the
standard deviations of the two peaks. The experiment is repeated $500$ times
and each time the parameters are estimated. In $10$ experiments or $2\%$ of
the cases the standard ML fit failed, mainly because the left-hand peak was
not well separated from the larger right-hand peak. The results of the
successful experiments are summarized in Fig. \ref{implicit1}. The observed
bias is again small compared to the statistical error.
\end{myexample}

The examples demonstrate that a relatively small number of bins in the true
histogram is sufficient to infer several parameters of a theoretical
prediction. The precision of the estimates is close to the limit imposed by
the statistical fluctuations of the data. The bias of the results is small
compared to the statistical errors..%

\begin{figure}
[ptb]
\begin{center}
\includegraphics[
trim=0.000000in 0.111551in 0.000000in 0.074685in,
height=3.3997in,
width=4.4118in
]%
{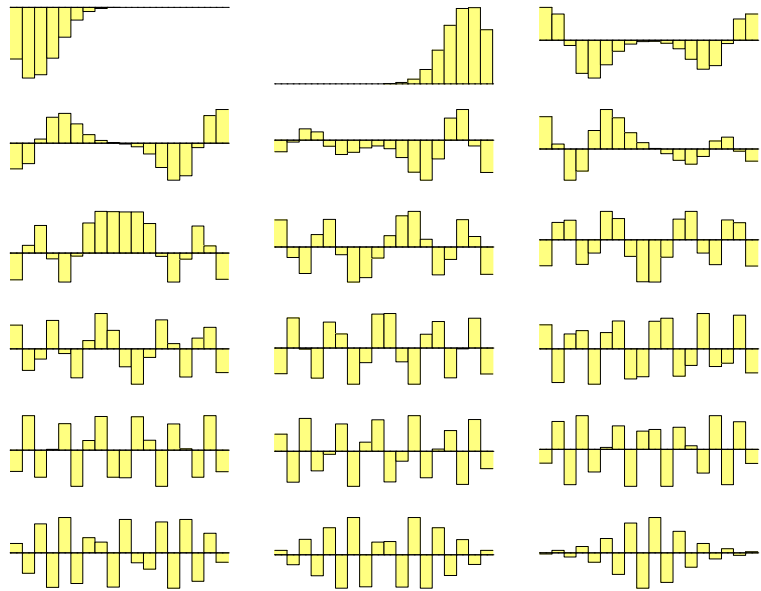}%
\caption{Eigenvctors of the unfolded histogram ordered with increasing
uncertainty of the coefficients.}%
\label{eigenvec}%
\end{center}
\end{figure}

\section{Diagonalizing the error matrix}

We have seen that we can obtain an uncorrelated parameter set in the linear LS
formalism by the eigenvalue decomposition of the LS matrix $\tens{Q}$. The
inverse $\tens{Q}^{-1}$ is the error matrix matrix. Small eigenvalues of the
error matrix correspond to large eigenvalues of $\tens{Q}$.

The diagonalization of the error matrix is not restricted to the linear LS
formalism. It can also be applied to the ML method with the advantage that
small event numbers in the observed histogram can be tolerated. However a
sensible result can only be obtained if the unfolded histogram does not
contain empty or sparsely populated bins. The solution of the unfolding
problem can be formulated as a superposition of the eigenvectors of the error
matrix which satisfy $\vec{v}_{i}\cdot\vec{v}_{j}=\delta_{ij}$. The unfolded
distribution $\widehat{\vec{\theta}}=\sum a_{i}\vec{v}_{i}$ is a superposition
of the eigenvectors with amplitudes given by $a_{i}=\widehat{\vec{\theta}%
}\cdot\vec{v}_{i}$. The amplitudes $a_{i}$ which replace the parameters
$\hat{\theta}_{i}$ are uncorrelated with errors given by the square root of
the eigenvalues of the diagonalized error matrix. Diagonalizing the error
matrix, we can again estimate the effective number of parameters $N_{eff}$.

\begin{figure}
[ptb]
\begin{center}
\includegraphics[
trim=0.142979in 0.182769in 0.160378in 0.171369in,
height=2.5845in,
width=5.8788in
]%
{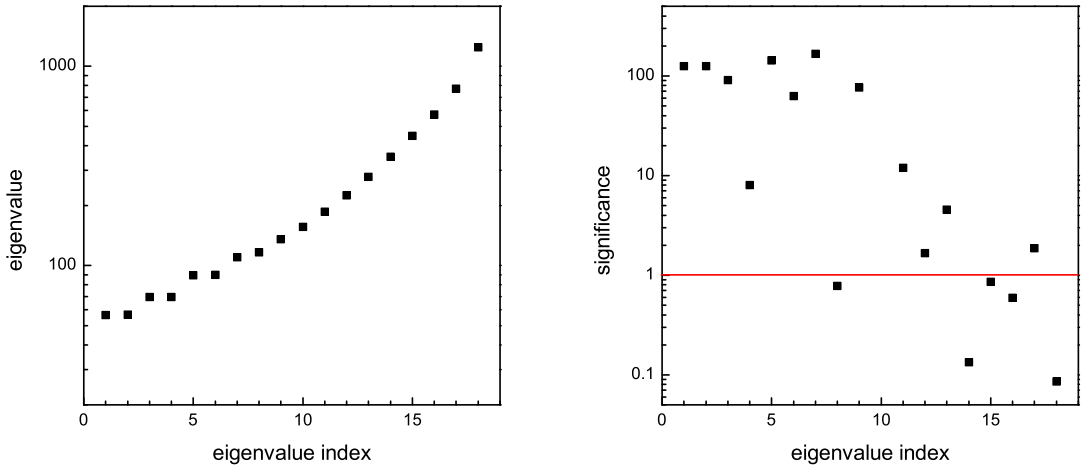}%
\caption{Eigenvalues of the error matrix (left hand) and significance of
eigenvector amplitudes.}%
\label{significance}%
\end{center}
\end{figure}

\begin{myexample}
$100000$ events are generated according to $f(x)=0.5[1+\mathcal{N}%
(x|0.5,\;0.1)]$, smeared with $\sigma_{s}=0.1$ and unfolded. The $18$
eigenvectors $\vec{v}_{i}$, ordered with increasing eigenvalues, which means
with increasing uncertainty of the coefficients, are displayed in Fig.
\ref{eigenvec}. With increasing eigenvalue the components of the eigenvectors
oscillate more and more. The eigenvalues and significances of the eigenvector
amplitudes are displayed in Fig. \ref{significance}. From the plot of the
significance we can estimate that this problem has effectively $13$
independent parameters.
\end{myexample}

%

\section{Choice of the binning}%

\begin{figure}
[ptb]
\begin{center}
\includegraphics[
trim=0.000000in 0.102107in 0.000000in 0.068169in,
height=3.1631in,
width=4.5313in
]%
{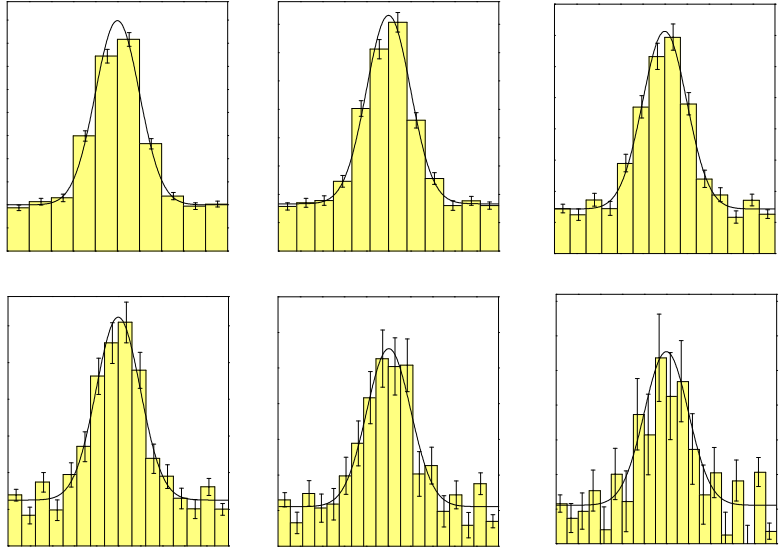}%
\caption{Unfolding results for different binnings of the true histogram.}%
\label{oscil1}%
\end{center}
\end{figure}

The essential parameter that we have to fix, is the bin width of the true
histogram. The optimal binning depends on the available statistics, the band
width of the structure that we want to resolve (which can only be guessed) and
the detector resolution.

Fig. \ref{oscil1} shows how dramatically the result of the MLF depends on the
binning. All plots contain the same data, only the number of bins $M$ increase
from $10$ to $20$ in steps of $2$. The correlations and correspondingly the
diagonal errors increase. The sequence of the histograms helps to choose a
reasonable binning. The diagonal errors should be small enough to justify the
application of error propagation in a comparison with predictions. As a rule
of thumb, the relative error should be less than $0.5$. Sometimes compromises
have to be made and larger errors especially in tails of the distribution can
be accepted. From the visual inspection of Fig. \ref{oscil1} we would probably
choose the binning of the fourth or fifth plot.

Another helpful parameter is the effective number of independent parameters
$N_{eff}$ that we have discussed in the previous section and which can be
derived from the eigenvalue decomposition of the LS matrix and from the error
matrix. The number of bins $M$ has to be larger than $N_{eff}$ but certainly
less than $2N_{eff}$. The $p$-value of the fit has to be acceptable which puts
another lower limit on $M$.

Some physicists prefer to impose a limit on the so-called \emph{purity}.
Loosely speaking, the purity is the fraction of the events that are associated
to a certain true bin which actually originate from the corresponding observed
bin. For a square matrix $\tens{A}$ and a bin $i$ it is $A_{ii}^{-1}%
d_{i}/\theta_{i}$. This quantity depends only on the resolution and does not
take into account the available statistics. It is of some interest because a
low purity signals a strong dependence of the result on a precise knowledge of
the response function.
\begin{figure}
[ptb]
\begin{center}
\includegraphics[
trim=0.131719in 0.187766in 0.134574in 0.156767in,
height=4.0423in,
width=5.8912in
]%
{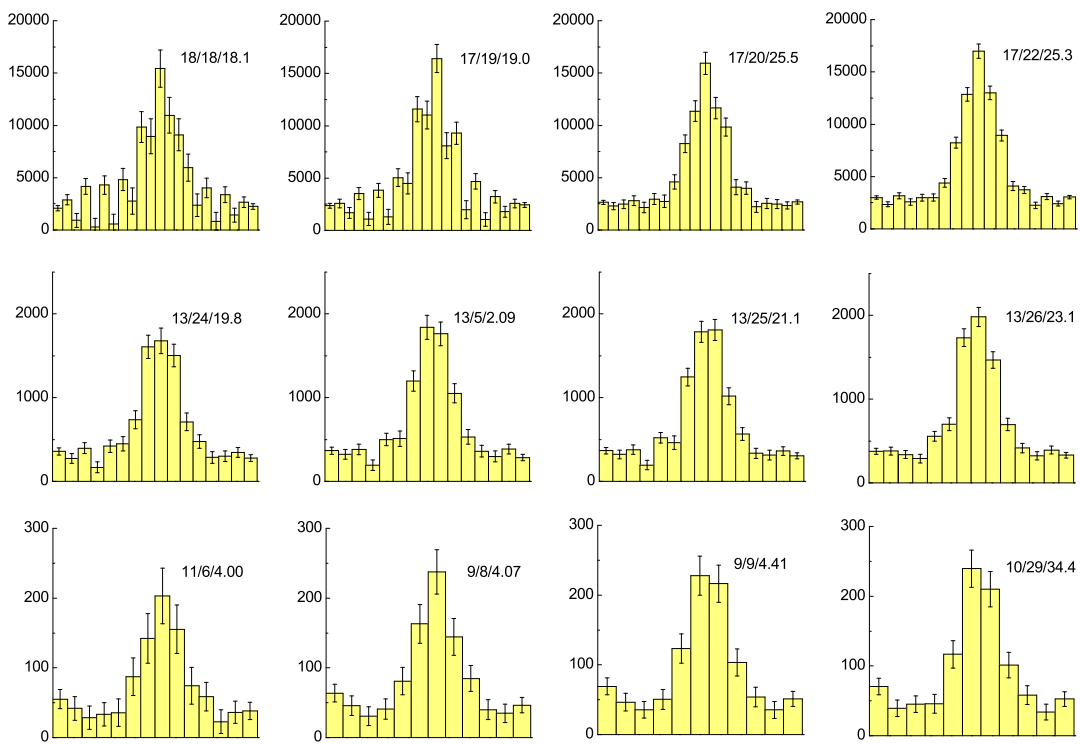}%
\caption{ML fit results for different bin sizes. The number of events
decreases from top to bottom from $100000$ to $10000$ and $1000$. In each plot
$N_{eff}$, the number of constraints in the fit and $\chi^{2}$ are given.}%
\label{fit1b0804}%
\end{center}
\end{figure}
%

\begin{figure}
[ptb]
\begin{center}
\includegraphics[
trim=0.140905in 0.191014in 0.142120in 0.159424in,
height=3.9626in,
width=5.9153in
]%
{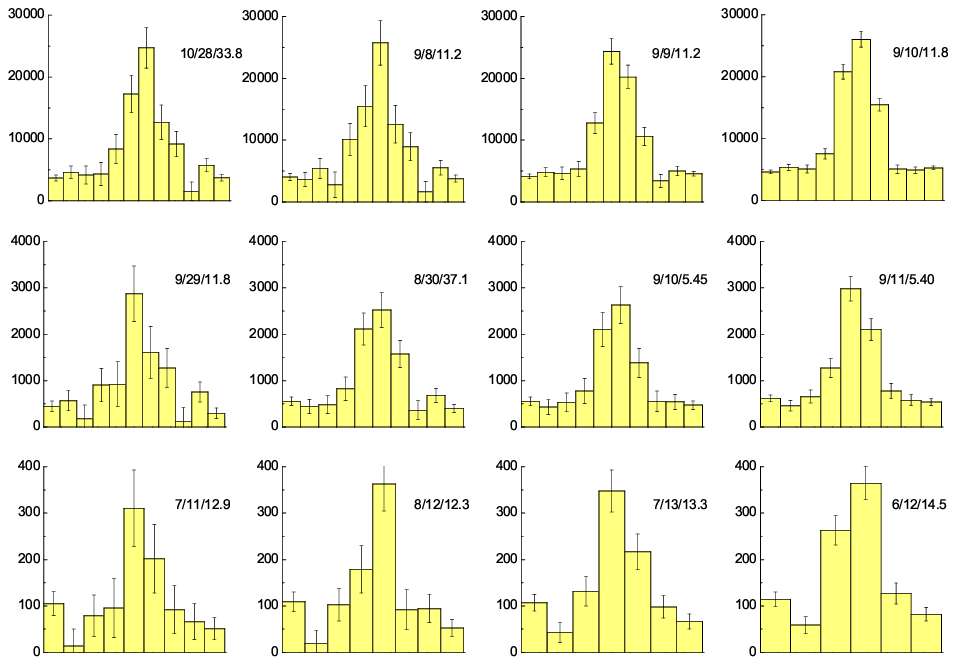}%
\caption{Same as previous figure but for the smearing constant $\sigma_{s}$
increased to $0.08$.}%
\label{fit1b0808}%
\end{center}
\end{figure}

In the following Monte Carlo study the number of bins is varied as a function
of the two essential parameters, i.e. the number of events and the resolution.

\begin{myexample}
We shift the normal distribution $\mathcal{N}(0.53,0.08)$ of the one-peak
distribution in order to avoid specific symmetry preferences. The events are
again equally divided into uniformly and normally distributed ones. The
results for a resolution $\sigma_{r}=0.04$ are shown in Fig. \ref{fit1b0804}.
The top plots contain $100000$ events, the central plots $10000$ events and
the bottom plots only $1000$ events. Within a row, the number of bins
decreases from left to right. For each plot the effective number of parameters
$N_{eff}$, the number of constraints in the fit $N_{c}$ and $\chi^{2}$ of the
fit are given. $N_{c}$ is equal to the difference between the number of
observed bins and the number of true bins. The observed histogram contains
$40$ bins and for comparison in some cases $20$ bins.

The parameter $N_{eff}$ cannot be larger than the number of true bins $M$, but
otherwise it should be independent of the binning. Due to small statistical
fluctuations it varies by one unit. When we look at the first row, we realize
that all $\chi^{2}$ values are acceptable. The chosen number of bins varies
between $22$ and $18$ and is close to the number $N_{eff}=17$ of independent
parameters. The best choice is close to $19$ bins. In the second and the third
row with less statistics one would select the second plot with $15$ and $12$
bins, respectively. The plots $2$ and $3$ in the second row and the plots $3$
and $4$ in the third row differ only in the number of bins $N$ in the observed
histogram. The two histograms are quite similar indicating that the value of
$N$ is of minor importance as long as it is significantly larger than $M$.

In Fig. \ref{fit1b0808} the simulations of the previous figure are repeated
with the experimental resolution reduced by a factor of two, $\sigma_{r}%
=0.08$. This leads to a strong reduction of the number of independent
parameters $N_{eff}$.

In the following table we summarizes the results for $N_{eff}$ for the three
different choices of the number of events and the two simulated resolutions.%

\begin{tabular}
[c]{|c|c|c|}\hline
number of events & $\sigma_{s}=0.04$ & $\sigma_{s}=0.08$\\\hline
$10^{5}$ & $17$ & $9$\\
$10^{4}$ & $13$ & $8$\\
$10^{3}$ & $10$ & $7$\\\hline
\end{tabular}

\end{myexample}

\section{Dependence on the Monte Carlo input distribution}

In the studies presented sofar, the Monte Carlo studies have been performed
with the response matrix generated starting from the true distribution. In
real experiments it has to be based on some guess of the true distribution and
as a consequence the unfolded distribution can be biased.

\subsection{Uniform Monte Carlo distribution}%

\begin{figure}
[ptb]
\begin{center}
\includegraphics[
trim=0.000000in 0.212601in 0.000000in 0.212920in,
height=1.9078in,
width=2.7082in
]%
{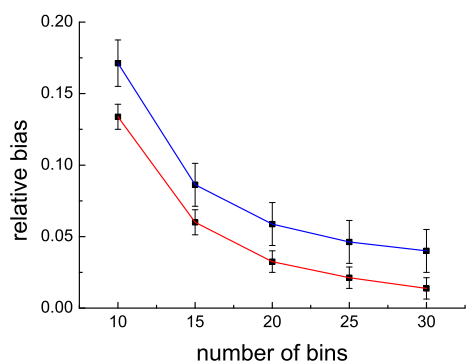}%
\caption{Relative bias of the fitted width $\sigma$ of a normal distribution
due to a biased response matrix. The upper measurements correspond to a
resolution of $\sigma_{s}=0.08$, the lower ones to $\sigma_{s}=0.04$.}%
\label{binbias}%
\end{center}
\end{figure}

We now use a uniform distribution to determine the response matrix to study
the bias caused by this approximation of the true distribution.

\begin{myexample}
We unfold our standard one-peak distribution (\ref{unipeak}) with the
parameters, $\mu=0.53,\sigma=0.08$ and $100000$ events. (The average bias is
independent of the number of events.) We fit the four parameters to the
unfolding result: the position of a Gaussian peak $\mu$, its width $\sigma$,
the number of events $n_{n}$ of the normal distribution and the number of
uniformly distributed events $n_{u}$ and compare them to the nominal values.
Due to the approximative symmetry of the problem the quantities $\mu$, $n_{n}$
and $n_{u}$ are expected to be unbiased, but $\sigma$ is biased toward smaller
values. The bias is the larger the wider the bins are. The relative increase
of the width as a function of the number of bins in the interval $0<x<1$ is
shown in Fig. \ref{binbias} for the two experimental resolutions $\sigma
_{s}=0.04$ and $0.08$. (It is interesting to notice that the fit produces the
best results for $30$ bins where the distributions show extreme oscillations.
Apparently, the error matrix is able to account for the fluctuations even
though the diagonal errors are quite large.) It is also clear that the bias
can be quite sizable for low event numbers and bad experimental resolutions.
For an experiment with $1000$ events and $\sigma_{s}=0.04$ about $10$ bins may
be tolerable which leads to a resulting bias of about $15\%$.
\end{myexample}

%

\begin{figure}
[ptb]
\begin{center}
\includegraphics[
trim=0.000000in 0.195433in 0.000000in 0.163013in,
height=2.8725in,
width=5.3507in
]%
{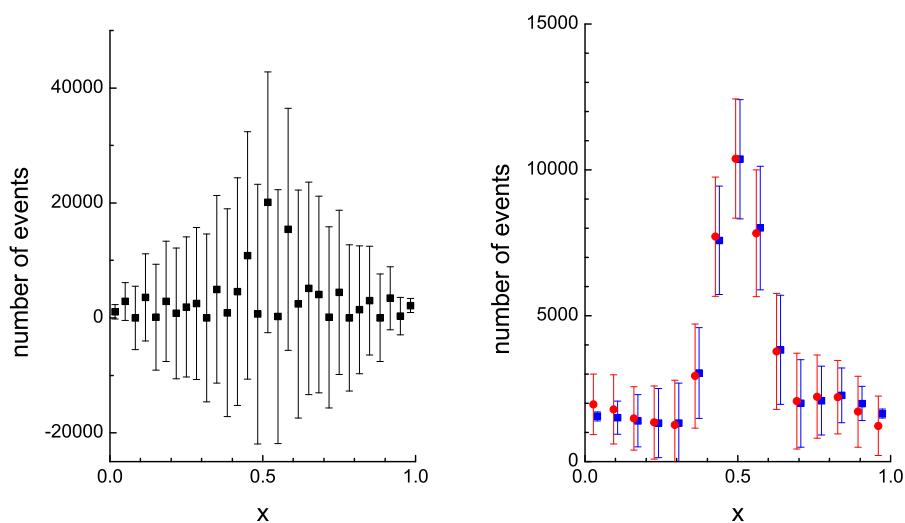}%
\caption{The lright-hand plot is obtained by combining bins of the left-hand
unfolded histogram. The dots correspond to a simple average, the squares to a
weighted average.}%
\label{combine}%
\end{center}
\end{figure}

A simple solution of this bias problem is not known to my knowledge. An
obvious proposal is to unfold with narrow bins and then to combine bins with
the intention to reduce the fluctuations. The content of the bins can either
simply be added or a weighted sum can be computed.

\begin{myexample}
To test this approach, we choose again our standard example with experimental
resolution $\sigma_{s}=0.04$, with $60$ bins in the observed histogram and
$30$ bins in the unfolded histogram where always two adjacent bins are
combined. The result is then a histogram with $15$ bins. Fig. \ref{combine}
shows how extreme the oscillations become with $30$ bins. It is astonishing
that combining bins leads to qualitatively reasonable distributions. However
the computed errors are unrealistically large, probably because simple error
propagation fails. A fit of the parameters to these distributions produces
strongly biased results for all four parameters with large uncertainties.
\end{myexample}

%

\begin{figure}
[ptb]
\begin{center}
\includegraphics[
trim=0.000000in 0.234056in 0.000000in 0.143067in,
height=1.9137in,
width=3.0278in
]%
{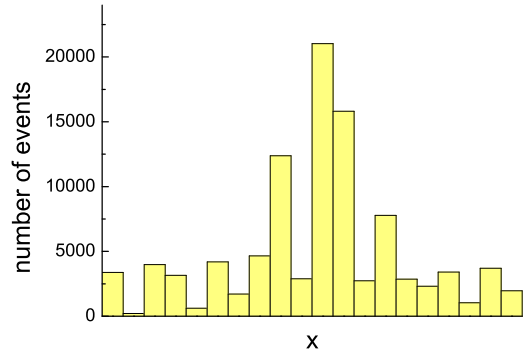}%
\caption{Unfolded histogram obtained by combining bins.}%
\label{combine20}%
\end{center}
\end{figure}

To get more insight into the origin of the failure we compare the result of a
likelihood fit with $20$ bins to a fit with $40$ bins where always two bins
are combined. The histogram obtained by adding the events from two adjacent
bins shown in Fig. \ref{combine20} is much less smooth than the corresponding
unfolding result from a direct fit as presented in Fig. \ref{fit1b0804}.

We have to conclude that our naive method fails. As long as no satisfactory
method is available, we have to model the Monte Carlo input distribution such
that it is in agreement with the observed data and conforms to constraints
hopefully provided by physics.%

\begin{figure}
[ptb]
\begin{center}
\includegraphics[
trim=0.000000in 0.216502in 0.000000in 0.216794in,
height=2.2449in,
width=3.2295in
]%
{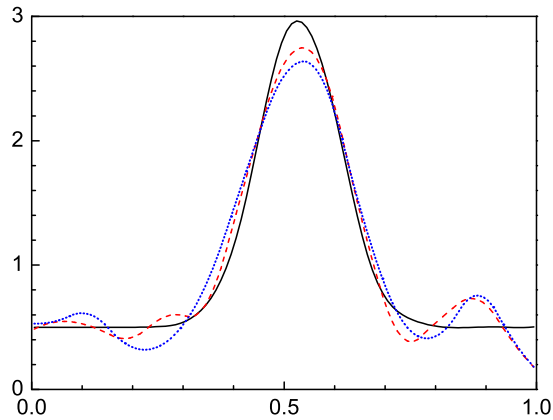}%
\caption{Spline approximation. The black curve is a fit to the true
distribution. The dashed (dotted) curve is the unfolding result of a smeared
distribution with $\sigma_{s}=0.04$ ($0.08$).}%
\label{splinebias}%
\end{center}
\end{figure}

\subsection{Spline approximation of the Monte Carlo input distribution}

A uniform distribution is a very bad approximation of most true distributions.
A possible way to reduce the bias is to approximate the distribution which is
used to determine the response matrix by spline functions. To this end the
observed histogram is unfolded with a smoothing algorithm where the unfolded
distribution is parametrized by $b$-splines, see Sect. \ref{splineexpansion}
and Appendix 3. The result is then used to determine the response matrix with
wide bins. The response matrix will still be slightly biased because of the
applied smoothing, but the bias of the fitted parameters is small compared to
the statistical fluctuations.

\begin{myexample}
$1000$ events following the one-peak distribution (\ref{unipeak}) are
generated with experimental resolutions $\sigma_{s}=0.04$ and $0.08$. The data
are histogrammed into $40$ bins. The unfolded distribution is parametrized
with $15$ quadratic $b$-splines. The spline coefficients are determined with a
MLF with a curvature penalty.
The result, shown in Fig. \ref{splinebias}, is then used to determine the
response matrix. Using this matrix a sample of $100000$ events is generated
and unfolded without explicit regularization with the iterative ML method to
determine the remaining bias. From the resulting $10$ bin histogram together
with the error matrix the $4$ parameters of the one-peak example are
determined in a LSF. The results are shown in the following table in the last
two rows. The width of the peak is still slightly biased, but this bias has to
be compared to the statistical fluctuation for $1000$ events which would be a
factor of $10$ larger than the error given in the table. The bias is about a
factor $6$ smaller than the statistical uncertainty and can be tolerated. Our
study corresponds to a small number of events. Of course the spline
approximation for the distribution used to compute the response matrix would
improve with increased statistics and thus the bias would become smaller.%

\begin{tabular}
[c]{|l|l|l|l|}\hline
& $\mu$ & $\sigma$ & $\rho$\\\hline
nominal & 0.53 & 0.08 & 0.5\\
true spline & 0.5293 (6) & 0.0795 (8) & 0.500 (3)\\
unfold spline $\sigma_{s}=0.04$ & 0.5298 (6) & 0.0786 (8) & 0.502 (3)\\
unfold spline $\sigma_{s}=0.08$ & 0.5295 (8) & 0.0777 (14) & 0.501 (4)\\\hline
\end{tabular}

\end{myexample}

\begin{figure}
[ptb]
\begin{center}
\includegraphics[
trim=0.000000in 0.165256in 0.000000in 0.092220in,
height=3.2835in,
width=4.2142in
]%
{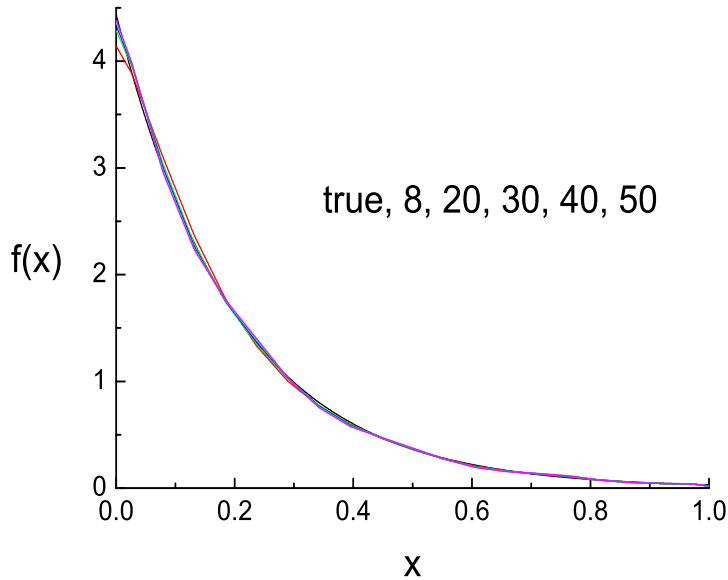}%
\caption{Linear spline approximations, obtained by unfolding a sample of
$5000$ events. The true distribution and the curves obtained for $8$, $20$,
$30$, $40$ and $50$ iterations are displayed.}%
\label{splinept}%
\end{center}
\end{figure}

\begin{myexample}
We repeat the same procedure for an exponential distribution with $40$
observed bins. To determine the Monte Carlo input distribution used to compute
the response matrix, we generate a data sample of $5000$ events drawn from the
p.d.f. $5\cdot\exp(-5x)$, $0<x<1$ with resolution $\sigma_{s}=0.08$ and unfold
it to a superposition of $20$ linear $b$-splines, again with the iterative EM
method with early stopping. As always a uniform starting distribution is
chosen. The results for $8$, $20$, $30$, $40$ and $50$ iterations are
summarized in Fig. \ref{splinept} and compared to the true distribution.
Obviously the response matrix is rather insensitive to the number of applied
iterations. The unfolding result for $30$ iterations is finally chosen to
generate the response matrix. The parameter estimates of the slope parameter
averaged over $1000$ event samples yielded a bias of $0.005\pm0.003$ to be
compared to the $1$ st. dev. statistical uncertainty of $\delta=\pm0.096$ for
the fit from a single sample. (With a perfect detector the uncertainty of the
slope parameter derived from the unbinned sample would be $\pm0.072$.)
\end{myexample}

%

The solution presented here is not very elegant and it is not clear how well
it works for more complex distributions. Eventually, systematic errors have to
be added to the unfolding result. Most distributions in particle physics will
be smoother than the one-peak distribution. Then the proposed method will
certainly provide satisfactory results.

\section{Statistical uncertainties introduced by the simulation}

In the majority of all cases it is possible to generate enough simulated
events to neglect the statistical fluctuation of the elements of the response
matrix. If the number of collected events is very large and comparable to the
number of simulated events, it may be necessary to include it. An analytic
estimate of the error introduced by the uncertainty of the response matrix is
complicated. A simple numeric, but computer time consuming solution of the
problem is provided by the bootstrap method \cite{efron}:

Let us assume that we have $n$ observed events and $m$ Monte Carlo events used
to compute $\tens{A}$. Then we draw form the set of observed events $n$ events
with replacement and from the set of simulation events $m$ events with
replacement. This means that some of the drawn events are identical. With this
bootstrap set we perform the fit of the unfolded distribution. The procedure
is repeated many, say $k=100$ times. The standard deviation of the
distribution of the results for each bin is an estimate of the uncertainty of
the true bin content, e.g. of the diagonal error. From the diagonal errors and
the correlation matrix which is almost identical in all sets, we can compute
the covariance matrix. If the numbers $n$ and $m$ are very large, a faster
method is to subdivide the data and the Monte Carlo sample in many, say
$k=100$ subsamples and perform the fit for all subsamples. The fluctuation of
the fitted values provides an estimate of the corresponding diagonal errors.
These errors then have to be scaled down by a factor $\sqrt{k}$ ($=10$ if
$100$ subsamples were used). The covariance matrix is obtained again by
scaling the correlation matrix to the diagonal error elements. The relative
uncertainty of the error estimate is in both cases $\sqrt{k}$.

\section{Summary and recommendations}

Unfolding with wide bins avoids excessive fluctuation in the unfolded
histogram and produces unbiased results with well defined errors. The
histogram is obtained with a likelihood fit if the data are Poisson
distributed in the observed histogram or otherwise by a LS fit. The MLE is
computed with the iterative the EM procedure. The effective number of
parameters $N_{eff}$ can be estimated with the eigenvalue decomposition of the
LS matrix as was shown in the previous chapter, or from the diagonalization of
the error matrix. The number of bins in the unfolded histogram should be
larger than $N_{eff}$ and small enough to avoid large oscillations and errors
that forbid linear error propagation. If the distribution contains narrow
peaks, the unfolding result depends on the shape of the distribution that is
used to compute the response matrix. This shape can be approximated by spline
functions which are fitted to the data in a regularized unfolding procedure.
Errors of the unfolded histogram due to statistical uncertainties of the
response matrix can be estimated by bootstrap methods.

\chapter{Unfolding with explicit regularization}

\section{General considerations}

\subsection{Regularization methods}

The main field where professional unfolding methods are applied lies in image
reconstruction. In medicine unblurring of tomographic pictures of arterial
stenoses, of tumors or orthopedic objects are important. Other areas of
interest are unblurring of images of astronomical surveys, of geographical
maps, of tomographic pictures of tools or mechanical components like train
wheels to detect damages. Also pattern recognition for example of fingerprints
or the iris is an important field of interest. The goal of most applications
is to dig out hidden or fuzzy structures from blurred images, to remove noise
and to improve the contrast, while in physics applications we are mainly
interested in quantitative results. We want to be able to combine data, to
estimate parameters and to document the results of experiments. This is
achieved with regularization by wide binning as discussed in the previous
chapter. On the other hand, we want also to take advantage of the fact that
physics distributions are rather smooth. Often we can remove the roughness of
an unfolding result without affecting very much the real structures of the
true distribution. We then can represent it with many more bins than without a
smoothing algorithm and obtain a much clearer picture of it. We have a penalty
to pay: We cannot safely quantify the uncertainties of our results. This is a
general problem in probability density estimation.

To obtain a smooth distribution, we have to implement a mechanism that
suppresses oscillations caused by the noise in the observed data.

In the majority of situations in experimental physics we unfold
one-dimensional distributions. Most regularization methods permit to extend
their recipes to multi-dimensional problems.

We restrict the study to three different explicit regularization methods in
their basic form without sophisticated extensions:

\begin{enumerate}
\item \emph{Truncation methods}: In the eigenvalue decomposition of the LS
matrix ( equivalent to the SVD) low eigenvalue contributions to the unfolding
solution are suppressed or eliminated.

\item \emph{Penalty methods}: A penalty term which is sensitive to unwanted
fluctuations is introduced in the LS or ML fit of the unfolding solution.
Typically, deviations from a uniform or a linear distribution are suppressed.
Standard methods penalize curvature, low entropy or a large norm of the
unfolding distribution.

\item \emph{Iterative fitting with early stopping}: A smooth distribution is
iteratively modified and adjusted to the observation. The iteration process
based on the EM method is stopped before \textquotedblleft
unacceptable\textquotedblright oscillations emerge. Alternatively, the
iterative unfolding result is smoothed after each iteration. Then the
iteration sequence converges automatically.
\end{enumerate}

Some commercial plotting programs offer smoothing algorithms that can be
applied to arbitrary distributions. We will not discuss those because the
results are difficult to interpret. A simple bin-by-bin correction method has
been used in the past in some particle physics experiments. The ratio of the
numbers in the observed and the true histogram in the simulation is used to
correct the observed histogram. This approach should be discarded because it
can produces wrong or strongly biased results.

In a recent publication \cite{kuuselastark} external constraints (positivity,
monotonicity, convexity) are introduced into the unfolding process of specific
distributions like $p_{t}$ distributions of jets. The method provides global
coverage\footnote{All predictions that disagree with the error bounds in at
least a single bin are excluded at the given confidence level.} and avoids the
regularization with a necessarily doubtfully defined regularization
parameter.
The user has to be sure that the constraints do not exclude unexpected
interesting phenomena.

In the following, we first discuss some general properties of regularization
approaches and devote the remaining part to the description of the standard
methods and to a comparative study with specific distributions. We will mainly
stick to our simple standard example with Poisson distributed data of a
distribution consisting of a peak over a uniform background and vary the
statistics, the width of the peak and the Gaussian resolution.

\subsection{Acceptance losses}

The correction of acceptance losses is straight forward. The correction can be
applied either after unfolding the smearing effects or it can be included in
the response matrix. The latter method should be preferred because we want to
smooth the true distribution and not the distribution which is affected by the
acceptance losses. Furthermore as is emphasized in \cite{dembinski}, the
acceptance can depend on the observed variable. Then a correction of the true
distribution is not sufficient but the losses are correctly taken into account
in the response matrix.

\subsection{Variable dependence and correlations}

We have already stressed that smoothness criteria cannot be derived from basic
principles. Smoothness is not invariant against variable transformations.

Because of the subjective character of smoothing, we can compare the quality
of different methods only with selected examples. However, there is a
desirable property of unfolding approaches: They should take the specific
properties of the smearing process into account: The events in a bin of the
observed histogram originate from different bins of the true histogram.
Consequently, in the unfolded histogram the corresponding bin contents are
negatively correlated. If the experimental distortions are caused by a simple
resolution effect that is described by a point spread function, the
correlation occurs predominantly between adjacent bins but in more complicated
situations the distortions can lead to correlations between bins that
correspond to quite different variable values. In the absence of correlations
between bins and especially if the resolution is perfect, smoothing should not
be active. This feature is inherent in truncated SVD and the EM unfolding with
early stopping, but is not realized in penalty regularization
methods\footnote{Of course this can always be achieved manually by setting the
regularization parameter to zero.}.

Most unfolding methods have the convenient property that the unfolding result
does not depend on the ordering of the bins in the unfolded distribution. This
means that multi-dimensional distributions can be represented by
one-dimensional histograms. An exception is regularization with a curvature penalty.

The dependence of the smoothness criteria on the chosen variable can be used
to adapt the regularization approaches to specific problems. If, for instance,
penalties favor a uniform distribution, we can choose a variable in which we
expect that the distribution is roughly uniform but in most cases it is better
to adapt the penalty function. On might for instance not penalize the
deviations from uniformity for a nearly exponential distribution but the
deviation from an exponential. The corresponding procedure in the iterative EM
method is to select the starting distribution such that it corresponds to our
expectation of the true distribution.

\subsection{Measures of the unfolding quality in Monte Carlo experiments}

To get a feeling for a reasonable setting of a regularization parameter, we
can compare the unfolded distribution $\widehat{\vec{\theta}}$ to the true
distribution $\vec{\theta}$ in toy experiments. For a quantitative comparison,
we introduce a variable $X^{2}$, defined by%
\[
X^{2}=\sum_{i=1}^{N}\frac{(\hat{\theta}_{i}-\theta_{i})^{2}}{\theta_{i}}\;
\]
which resembles the $\chi^{2}$statistic used in goodness-of-fit tests with
Poisson distributed histogram entries. (The choice of the parameter $X^{2}$ as
a test quantity is somewhat arbitrary, $X^{2}$ is not $\chi^{2}$ distributed.)

In PDE the standard measure of the agreement between the true and the
constructed distributions is the \emph{integrated squared error} ($ISE$). For
a functions $f(x)$ and its PDE $\hat{f}(x)$ it is defined by
\begin{equation}
ISE=\int_{-\infty}^{\infty}\left[  \hat{f}(x)-f(x)\right]  ^{2}dx\;.
\label{ise}%
\end{equation}

(Other common measures are the integrated absolute error $\int|\hat
{f}(x)-f(x)|dx$ and the Kullback-Liebler distance $\int|\hat{f}(x)\ln[\hat
{f}(x)/f(x)]dx$ which is related to the likelihood ratio. Contrary to $ISE$,
these measures are dimensionless.)

The expected value of $ISE$ is called \emph{mean integrated square error
}($MISE$).
\[
MISE=\mathrm{E}\left[  \int_{-\infty}^{\infty}(\hat{f}(x)-f(x)^{2}dx\right]
\;.
\]

$ISE$ is not defined for histograms in the way as physicists interpret them.
To adapt the $ISE$ concept to our needs, we modify the definition such that is
measures the difference of the estimated content of the histogram $\hat
{\theta}_{i}$ and the prediction $\theta_{i}$.
\begin{equation}
ISE=%
{\displaystyle\sum\limits_{i=1}^{M}}
\left(  \hat{\theta}_{i}-\theta_{i}\right)  ^{2} \label{iseintegral}%
\end{equation}
In the following sections the value has been normalized to the event numbers
and the numbers of bins. In the comparisons only the relative values are
important and whether we choose (\ref{ise}) or (\ref{iseintegral}) which is
simpler to compute, does not matter. $ISE$ defined by (\ref{iseintegral})
depends on the binning and, as has been pointed out by Volobouev
\cite{volobouevprivate}, it is zero for a single bin. The $ISE$ attributes
more weight to regions where there are many events than $X^{2}$, but the
application of the $ISE$ and the quantity $X^{2}$ usually lead to similar
conclusions. \ (The denotation of the modified quantities by the names of well
defined parameters $ISE$ and $MISE$ is unfortunate but for technical reasons
it cannot be changed any more in this report. The definition
(\ref{iseintegral}) is sensible if the unfolding result is compared to a
prediction but if it is used for the visualization of the distribution or for
a simulation the PDE definition (\ref{ise}) is relevant.

\subsection{Choice of the regularization strength}

\label{regustrength}A critical parameter in all unfolding procedure is the
regularization strength which regulates the smoothness of the unfolding result
and which determines bias and precision. The optimal value of the
regularization parameter depends on the specific application. To fix it, we
must have an idea about the shape of the true distribution. We might choose it
differently for a structure function, a Drell-Yan distribution with possible
spikes, a transverse momentum distribution and the distribution of the cosmic
background radiation. We need some prior knowledge. Based on purely
statistical arguments, we cannot define smoothness and the optimal
regularization parameter. From the data we can only estimate an upper limit of
the regularization parameter: The unfolded distribution has to be
statistically compatible within the measurements. Most unfolding methods try
to approach this limit and eliminate fluctuations that are compatible with
noise. There is no scientific justification for this pragmatic choice and one
has to be aware of the fact that in this way real structures may be eliminated
that can be resolved with higher statistics.

\subsubsection{Visual inspection}

If we resign to the idea to use the unfolded distribution for parameter fits,
it seems tolerable to apply subjective criteria for the choice of the
regularization strength. By inspection of the unfolding results obtained with
increasing regularization, we are to some extent able to distinguish
fluctuations caused by the procedure from structures in the true distribution
and to select a reasonable value. Probably this method is in most cases as
good as the following approaches which are partially quite involved.

\subsubsection{Goodness-of-fit approaches}

An obvious quantity to look at is certainly the $\chi^{2}$ statistic. In a
standard LSF, without regularization, with normally distributed errors, $N$
observed quantities and $M$ fitted parameters, we expect that $\chi_{0}^{2}$
follows a $\chi^{2}$ distribution with $NDF=$ $N-M$ degrees of freedom. With
regularization the value $\chi^{2}$ will be larger than $\chi_{0}^{2}$ by
$\Delta\chi^{2}$. It is not reasonable to cut on $\chi^{2}$, the relevant
parameter is $\Delta\chi^{2}$. With the usual approximations, we find $1$,
$2$, $3$ standard deviation error limits of the fitted parameters by
increasing $\chi_{0}^{2}$ by $\Delta\chi^{2}=1$, $4$, $9$ with confidence
$cl$,%
\begin{equation}
cl=\int_{0}^{\Delta\chi^{2}}f_{NDF}(u)du \label{cl1}%
\end{equation}
where $f_{NDF}(u)$ is the $\chi^{2}$ distribution with $NDF$ degrees of
freedom. With $NDF$ of the order of $20$, the confidence that the true
solution is contained in the one standard deviation interval is small. To be
independent of the $NDF$, we turn to the $p$-value defined by%

\begin{equation}
p=\int_{\Delta\chi^{2}}^{\infty}f_{NDF}(u)du=1-cl \label{pvalue}%
\end{equation}
Requiring a certain value $cl$ or $p$, we can derive from (\ref{pvalue}) a
corresponding $\chi^{2}$ boundary $\chi_{c}^{2}$ in the $M$-dimensional
parameter space. The parameter $p$ corresponds to the frequency in which the
true parameter point is located outside the boundary in a large number of measurements.

A LSF or MLF without regularization produces the best estimate of the
parameters together with a goodness of fit quantity $\chi_{0}^{2}$ and a
confidence interval limited by the selected values of $\Delta\chi^{2}$. A
small value of $cl$ and a correspondingly large $p$-value of the regularized
parameter indicate that the fit is well compatible with the measurement. We
could fix the regularization parameter by choosing a limiting $p$-value
$p_{reg}$ which corresponds to a value $\Delta\chi_{reg}^{2}$ by which the
regularization is allowed to increase $\chi_{0}^{2}$. For example choosing
$p_{reg}=0.9$, the unfolded distribution should have $\chi^{2}$ smaller than
the true distribution in $90\%$ of all cases.%

\begin{figure}
[ptb]
\begin{center}
\includegraphics[
trim=0.151528in 0.203737in 0.164317in 0.174674in,
height=1.8564in,
width=5.892in
]%
{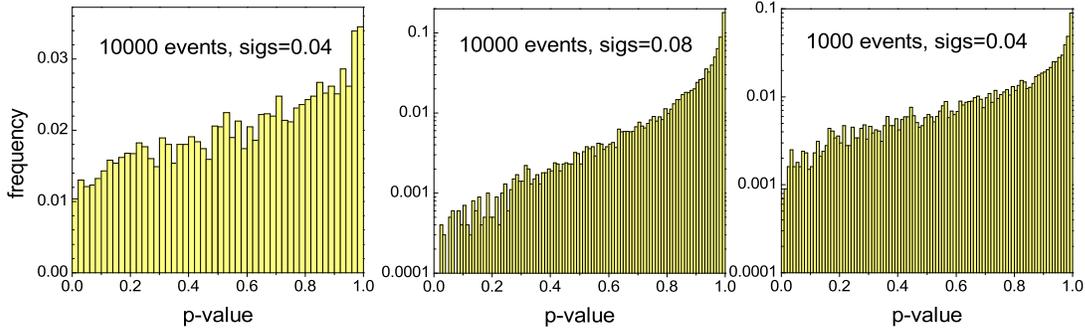}%
\caption{Distribution of the pseudo $p$-value for different experimental
resolutions and event numbers.}%
\label{pvaluedistribution}%
\end{center}
\end{figure}

Unfortunately the parameter estimate $\hat{\theta}$ without regularization is
degenerated. LSFs produce bins with negative entries and the related
unphysical fit results are hard to accept as best estimates. The MLF avoids
negative values, but while the LSF produces negative entries, often empty bins
are obtained. The MLF is to be preferred to the LSF but the constraint to
positive event numbers entails an increase of $\chi^{2}$ relative to the
nominal value for a LSF. We cannot expect that the expected value of our test
quantity $\Delta\chi^{2}$ obeys a $\chi^{2}$ distribution and the distribution
of the pseudo $p$-value $p^{\prime}$ derived from (\ref{pvalue}) will not be
uniform. Its distribution can be generated in simulations where we know the
true distribution. Since the $p^{\prime}$ distribution depends mainly on the
resolution, the number of events and the binning and less on the shape of the
true distribution, it can be estimated also for real experiments.

\begin{myexample}
We generate events, perform a MLF and compute $\chi_{0}^{2}$. Folding the true
distribution and comparing it to the observed distribution we get $\chi
_{true}^{2}$ which is larger than $\chi_{0}^{2}$ by $\Delta\chi^{2}$. The
number $\Delta\chi^{2}$ is converted to $p^{\prime}$ by (\ref{pvalue}). The
results for the pseudo $p$-value of the one-peak distribution with $20$ true
bins, $40$ observed bins, different event numbers and resolutions derived from
$10000$ simulated experiments are displayed in Fig. \ref{pvaluedistribution}.
The distributions are far from being uniform. The lower the event number and
the worse the resolution is, the more the distribution is peaked towards large
pseudo $p$-values.
\end{myexample}

It is obvious that we have to take into account the expected shape of the
distribution when we derive the regularization parameter. The value of the cut
has to be calibrated and it is not guaranteed that the same value is optimal
independent of binning, event numbers and resolution.

Regularization based upon $\chi^{2}$ or a $p$-value have been proposed in
\cite{schmelling, zech}.

\subsubsection{Truncation approaches}

The unfolding result can be expanded into orthogonal components which are
statistically independent.

We have studied above the eigenvector decomposition of the modified least
square matrix (equivalent to SVD) and realized that the small eigenvalue
components $\lambda_{i\text{ }}$cause the unwanted oscillations. A smooth
result is obtained by cutting all contributions with eigenvalues $\lambda
_{i},i=1,...,k)$ below a cut value $\lambda_{k}$. This procedure is called
\emph{truncated singular value decomposition} (TSVD).The value is chosen such
that eigenvectors are excluded with statistically insignificant amplitudes.
The truncation in the framework of the LSF has its equivalence in the ML
method. As shown above, we can order the eigenvectors of the covariance matrix
of a MLF according to decreasing errors and retain only the dominant
components. This method is attractive theoretically, but in concrete
applications technical difficulties may arise due to the fact that the error
matrix of the unregularized fit may not be well conditioned. A possible way
out of the dilemma could be to perform a MLF with a soft regularization by a
penalty term and to base the final regularization on the corresponding well
conditioned error matrix. The amplitudes corresponding to the orthogonalized
covariance matrix which are statistically significant are retained.

The physicist community is still attached to the - for historical reasons -
popular linear matrix calculus. Nowadays computers are fast and truncation
based on the diagonalized covariance matrix derived from a non-linear LS or a
ML fit is probably the better choice than TSVD.

\subsubsection{Empirical Bayes selection}

Kuusela and Paranetos \cite{kuusela} form the product%
\begin{equation}
p=p_{1}(\vec{d}|\vec{\theta})p_{2}(\vec{\theta}|r) \label{kuusela1}%
\end{equation}
where the first factor is simply the probability to observe $\vec{d}$ given
the true distribution $\vec{\theta}$ and corresponds to the usual Poisson
likelihood. The second factor is a kind of smoothness probability. It is
roughly of the form $\exp[-rR(\vec{\theta})]$ with $r$ the regularization
parameter and $R$ a measure of the smoothness, here derived from the
curvature. The log-likelihood of $p$ corresponds to our relation
(\ref{lnlpenalty}). Now the regularization parameter $r$ is chosen such that
the probability marginalized with respect to $\vec{\theta}$ is
maximal\footnote{The marginalization with respect to $\vec{\theta}$ protects
against overfitting $r$. A justification of the method is indicated in
\cite{kuusela} and a relevant reference is given in the publication.}.%
\begin{equation}
\hat{r}=\arg\max_{r}\int p_{1}(\vec{d}|\vec{\theta})p_{2}(\vec{\theta}%
|r)\vec{d}\vec{\theta} \label{kuusela}%
\end{equation}
This means that $r$ is derived from the data. The multi-dimensional integral
is considered as intractable numerically, but is solved with the EM algorithm.
The estimate $\hat{r}$ is plugged into (\ref{kuusela1}) to obtain the
unfolding result $\widehat{\vec{\theta}}$. The method is attractive because
the regularization constant is fixed in a unique way by a simple principle.
Naively, one would optimize $r$ directly from (\ref{kuusela1}) together with
$\vec{\theta}$ but then the result would be $r=0$, i.e. no regularization.

To derive error bands with $95\%$ coverage, an iterative bias correction is
applied to the point estimate. The correction has a very similar effect as a
decrease of the regularization constant. This is shown in the following example.%

\begin{figure}
[ptb]
\begin{center}
\includegraphics[
trim=0.119750in 0.212937in 0.121688in 0.201617in,
height=2.0224in,
width=5.9111in
]%
{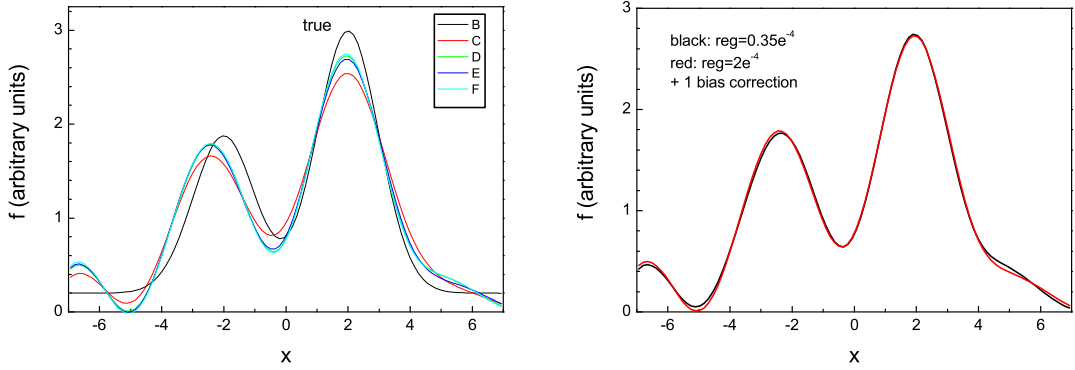}%
\caption{Bias correction. The left-hand graph shows the true distribution (B),
the unfolded distribution (C) and unfolded distributions with sequential bias
corrections (D, E, F). The lright-hand graph shows the equivalence of the bias
correction and a reduced regularization strength. }%
\label{biascorrect}%
\end{center}
\end{figure}

\begin{myexample}
The distribution \ref{kuuseladist} is used to generate $1000$ observed events
with a Gaussian resolution of $\sigma_{s}=1$. The events are histogrammed into
$40$ bins. The unfolded distribution is parametrized by $30$ cubic $b$-splines
with $28$ knots inside the interval $[-7,7]$ and $2$ knots at the borders.
Unfolding is performed with a curvature penalty to suppress strong variations
of the unfolded distribution. Fig. \ref{biascorrect} left shows the true
distribution and the unfolded distribution together with the results obtained
by bias corrections. The bias correction which is computed with $10^{6}$
simulated events, brings the unfolded distribution closer to the true
distribution, but the fake bump at the left-hand side is also enhanced,
because it is corrected as well for an assumed bias. A very similar result is
obtained by decreasing the regularization strength as shown in the right-hand
graph. A bias corrected distribution is compared to one where only the
regularization strength is decreased such that the heights of the larger peaks
agree. The two distributions are hardly distinguishable in this specific example.
\end{myexample}

In this way the final choice of the regularization parameter used to compute
location and width of the error bands is shifted to the choice of the number
of bias iterations that have to be applied. Thus a part of the beauty of the
method is lost. The bias correction is only applied to the error bands, the
original point estimate is retained \cite{kuuselapoint}. Thus, the errors are
not centered at the point estimates.

\subsubsection{Iterative minimization of the integrated square error $ISE$}

The best choice of the regularization strength depends mainly on the
resolution, i.e. the response matrix and less on the shape of the true
distribution. A crude guess of the latter can be used to estimate the
regularization parameter $r$. (Here $r$ is a generic name for the number of
iterations in the EM method, the penalty strength or the cut in truncation
approaches.) The true distribution is approximated by the unfolding result and
then $r$ can be optimized with simulated experiments. A goodness-of-fit
measure $Q$ of the quality of the agreement between the distribution used in
the simulation and the distribution obtained in the unfolding has to be
defined. A natural choice is $Q=ISE$ as defined above. The process can be
iterated, but since the shape of the true distribution is not that critical,
this will not be necessary in the majority of cases. The procedure consists of
the following steps:

\begin{enumerate}
\item Unfold $\vec{d}$ with varying regularization strength $r$ and select the
\textquotedblleft best\textquotedblright\ value $\tilde{r}$ and
$\mathbf{\tilde{\theta}}^{(0)}$ by visual inspection of the unfolded histograms.

\item Use $\mathbf{\tilde{\theta}}^{(0)}$ as input for typically $n=100$
simulations of \textquotedblleft observed\textquotedblright\ distributions
$\tilde{d}_{i}$, $i=1,n$.

\item Unfold each \textquotedblleft observed\textquotedblright distribution
with varying $r$ and select the value $\tilde{r}_{i}$ that corresponds to the
smallest value of the goodness-of-fit measure $Q$. The value of $Q$ is
computed by comparing the unfolded histogram with $\mathbf{\tilde{\theta}%
}^{(0)}$.

\item Take the mean value $\bar{r}$ of the regularization strengths $\tilde
{r}_{i}$, unfold the experimental distribution and obtain $\mathbf{\tilde
{\theta}}^{(1)}$. If necessary, go back to 2., replace $\mathbf{\tilde{\theta
}}^{(0)}$ by $\mathbf{\tilde{\theta}}^{(1)}$ and iterate.
\end{enumerate}

The procedure is independent of the regularization method and the quality
measure. An example of its application is presented in Sect. \ref{emunfolding}.

Dembinski and Roth \cite{dembinski} propose to minimize the $MISE$ in the
observed space. This quantity is expected to be strongly correlated to $MISE$
in the true space, however, their starting relation (19) $MISE=\int
V\,dx^{\prime}+ISE$ (in different notation, the first term of the right-hand
side is the variance integrated over the observed space) is not correct. If
the bias is small compared to the integrated variance, the right-hand side is
on average approximately twice as large as the left-hand side.

\subsubsection{The $L$-curve approach}

Large fluctuation imply a large value of the norm $||\theta^{2}||_{2}$ of the
solution,%
\[
||\vec{\theta}^{2}||=%
{\displaystyle\sum\limits_{i=1}^{M}}
\theta_{i}^{2}\;.
\]
For a given number of events $\Sigma_{i}\theta_{i}$ in the unfolded histogram,
the norm is minimal for a uniform distribution. In most unfolding methods, the
norm decreases with increasing regularization. In turn the residual norm%
\[
||\tens{A}\vec{\theta}-\vec{d}||^{2}%
\]
which measures the discrepancy of the solution with the observed histogram
increases with the regularization strength. The Tikhonov regularization
penalizes solution with a large norm of the solution. A balance has to be
found between small $||\vec{\theta}^{2}||$ and small $||\tens{A}\vec{\theta
}-\vec{d}||$. The log-log relation of these two quantities, with the
regularization strength as parameter, is called $L$-curve \cite{hansen}.
Ideally, for a rectangular $L$-shaped curve one would select the
regularization corresponding to the corner of the $L$-curve where both norms
are small. For other regularization penalties the norm $||\theta^{2}||$ is
replaced by the corresponding expression. Physicists would probably replace
the residual norm by $\chi^{2}$.

\subsection{Error assignment to unfolded distributions}

The regularization introduces a bias and decreases the error $\delta_{s}$
obtained in the fit. The height of peaks is reduced, the width is increased,
valleys are partially filled. The true uncertainties $\delta$ depend on the
nominal error $\delta_{s}$ and the bias $b$, $\delta^{2}=\delta_{s}^{2}+b^{2}%
$. Increasing the regularization strength reduces $\delta_{s}$ but increases
the bias $b$. As the bias cannot be calculated, we have to guess its size when
we choose the regularization strength. The art is to find the optimal balance
between the nominal error and the assumed bias introduced by the regularization.

\subsubsection{Calculation of the nominal error}

A scientific measurement is of limited use if it is not accompanied by an
uncertainty estimate. We have pointed out above that the assignment of errors
to regularized unfolding results is problematic. Putting aside our concerns
and neglecting the bias, we can calculate the errors in different ways:

\begin{enumerate}
\item The usual way is to apply error propagation starting from the observed
data $\vec{d}$. To be consistent\footnote{Error propagation starting from the
observed data insted of the best estimate can lead to inconsistent results. A
striking example is known as Peelle's pertinent puzzle \cite{peelle}.} with
the point estimate, the best estimate of the folded distribution $\hat{d}%
_{i}=\Sigma_{k}A_{ik}\hat{\theta}_{k}$ should be used instead. Error
propagation is quite sensitive to non-linear relations which occur with low
event numbers. To avoid the problem, $\theta$ can be re-fitted starting from
$\hat{d}$ and the errors can be provided by the fit program.

\item The errors can be derived from the curvature matrix at the LS or ML
estimates. This is the standard way in which symmetric error limits are
computed in the common fitting programs. In principle also likelihood ratio or
$\chi^{2}$ contours can be computed. The parameters $\vec{\theta}$ is varied,
the corresponding histogram $\vec{d}$ is computed and compared to
$\widehat{\vec{d}}=\tens{A}\widehat{\vec{\theta}}$. Values of $\vec{\theta}$
that change the difference $\ln L(\widehat{\vec{\theta}})-\ln L(\vec{\theta})$
by $1/2$ fix the standard likelihood ratio error bounds.

\item We can use bootstrap resampling techniques \cite{efron}. In short, the
data sample is considered as representative of the true folded distribution.
From the $N$ observed events, $N$ events are drawn with replacement. They form
a bootstrap sample $\vec{d}^{\ast}$ which is histogrammed and unfolded. This
procedure is repeated many times and in this way a set of unfolded
distributions is generated. from which the fluctuations, confidence intervals
and correlations can be extracted. For example, in a selected bin the standard
$68\,\%$ confidence interval contains $68\,\%$ of the bootstrap results.
Alternatively we can start from the fitted unfolded distribution $\hat
{d}\tens{A}\widehat{\vec{\theta}}$. Bootstrap histograms are then constructed
from a random Poisson process. For bin $i$ we choose $d_{i}^{\ast}%
\sim\mathcal{P}_{\mathbf{\hat{d}}}$. The latter method includes the
fluctuation in the total number of events and is consistent. Bootstrap errors
include flip-flop effects: Small differences in the observed distribution can
lead to large differences in the eigenvector decomposition and in TSVD to
sizable changes in the result. Also a curvature penalty may introduce similar effects.
\end{enumerate}

A more detailed and professional discussion of the error estimation with
bootstrap methods is presented in \cite{kuusela}.

Contrary to claims \cite{blobelbook}, the nominal errors of the unfolded
distribution can be computed for all unfolding approaches with the three
mentioned methods\footnote{In the publications
\cite{dagostini95,dagostini2010} by D'Agostini, the error estimation is not
correct.}. In the EM method, the errors computed with methods 1 and 2 do not
include the regularization constraint due to early stopping. Bootstrap
sampling takes it into account. The error margins are correspondingly smaller.

\subsubsection{Problems related to the errors assignment}

Frequentist coverage in the context of unfolding means: For a given coverage
probability $cl$, an \emph{arbitrary true distribution} $f(x)$ has to be
accepted in the fraction $cl$ of a large number of experiments, and
specifically, parameters of the true distribution estimated from the unfolding
result should within the computed error intervals contain the true parameter
values in the fraction $cl$ of many experiments. Coverage is violated on
purpose by the regularization, as the goal of regularization is to exclude
strongly fluctuating distributions that are compatible with the data.
Regularized histograms are biased and correctly calculated errors of biased
distributions that do not include a possible bias in the error estimate cannot
cover even for smooth distributions, contrary to claims in some publications.
That the specific distribution which is selected in a toy Monte Carlo
experiments is covered by the unfolding solution within the given errors
should be a triviality, but even this is not realized by most approaches.%

\begin{figure}
[ptb]
\begin{center}
\includegraphics[
trim=0.000000in 0.237063in 0.000000in 0.166690in,
height=2.3362in,
width=3.3466in
]%
{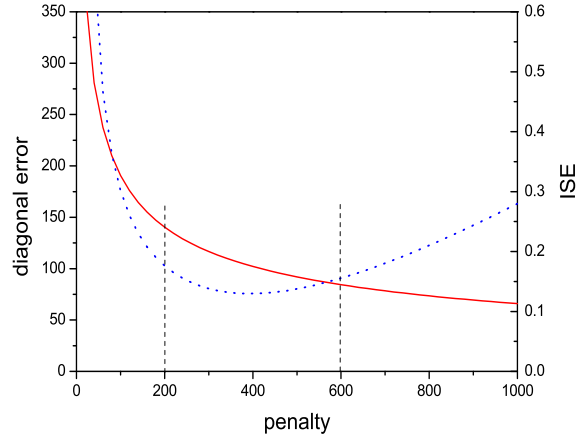}%
\caption{Error dependence on the regularization strength. (solid line). The
dotted curve indicates the integrated square error as a function of the
regularization strength. The units are arbitrary. The vertical lines
qualitatively indicate the uncertainty of the regularization parameter.}%
\label{errorpenalty}%
\end{center}
\end{figure}

Another problem is the dependence of the estimated errors on the
regularization strength. In Fig. \ref{errorpenalty} the diagonal error of the
highest bin of the one-peak example is plotted as a function of the
regularization strength, here for entropy regularization. A qualitative
estimate of the range of reasonable regularizations is indicated in the
figure. The interval corresponds to regularization penalties which lead to a
satisfactory agreement of the unfolded distribution with the true
distribution. Within this range the error varies considerably. Each unfolding
approach has its own error definition,

Since the regularization excludes variations of the unfolded distribution that
are compatible with noise, the smoothing is more effective in low statistics
experiments than in experiments with a large number of events. The assigned
errors do not scale with the square root of the number of events and thus it
is impossible to combine two unfolded data sets, even if they are produced
under exactly the same conditions. Small real structures of a distribution
that are resolved with high statistics may be excluded in experiments with a
small number of events.

The errors that we can assign to the unfolded distribution are at most a rough
indication of the uncertainties and have to be considered with great care. The
presentation of the results in form of histograms indicate only the diagonal
errors but the errors of adjacent bins are correlated, in some schemes
negatively in others positively which makes a big difference if bins are
combined. In publications the covariance matrices or some information about
the correlations should be given.

Due to the mentioned correlation problems, sometimes bin-wise coverage is
attempted or error bands are introduced \cite{dembinski,kuusela}. Predictions
should then remain for instance in the fraction $\alpha$ of the bins inside
the error limits. Error bands with a high confidence limit, say $90\%$ give a
good qualitative indication whether a predictions is incompatible with the
unfolding result. However, quantitative conclusions should not be drawn.

Approximate coverage is realized in \ref{kuusela} by separating the point
estimate from the interval estimate. The regularization for the error estimate
is weaker than for the point estimate. In this way the bias introduced in the
interval estimate is small. In \cite{kuuselastark} constraints which may be
available in some specific cases are used to guaranty (over)coverage.

The restrictions that we have summarized do not mean that explicit unfolding
is obsolete. It permits to discover structures in exploratory experiments and
its results help to visualize the unknown distribution much better than the
histograms produced by implicit unfolding with wide bins.

\section{EM unfolding with early stopping}

In the previous chapter we have seen that the EM algorithm produces the MLE of
the unfolded histogram. To suppress the fluctuations of the MLE we stop the
iteration once the result is compatible with the data and the bin-to-bin
fluctuations are still acceptable. We have to fix the starting distribution
and the stopping condition.

\subsection{A few examples}

We resume our standard one-peak example for $50000$, $5000$ and $500$ events
and two different Gaussian resolutions of $\sigma=0.04$ and $0.08$. The
starting distribution is uniform. The results are summarized in the following
figures. The unfolded histograms are compared to the true histogram indicated
by squares. The number of iterations is given in each plot. For each of the
settings the test quantities as defined above, $X^{2}$ and the $ISE$ are
plotted as a function of the number of applied iterations. The optimal number
of iterations is similar for both parameters.%

\begin{figure}
[ptb]
\begin{center}
\includegraphics[
trim=0.129093in 0.183734in 0.131891in 0.184027in,
height=4.0473in,
width=5.8962in
]%
{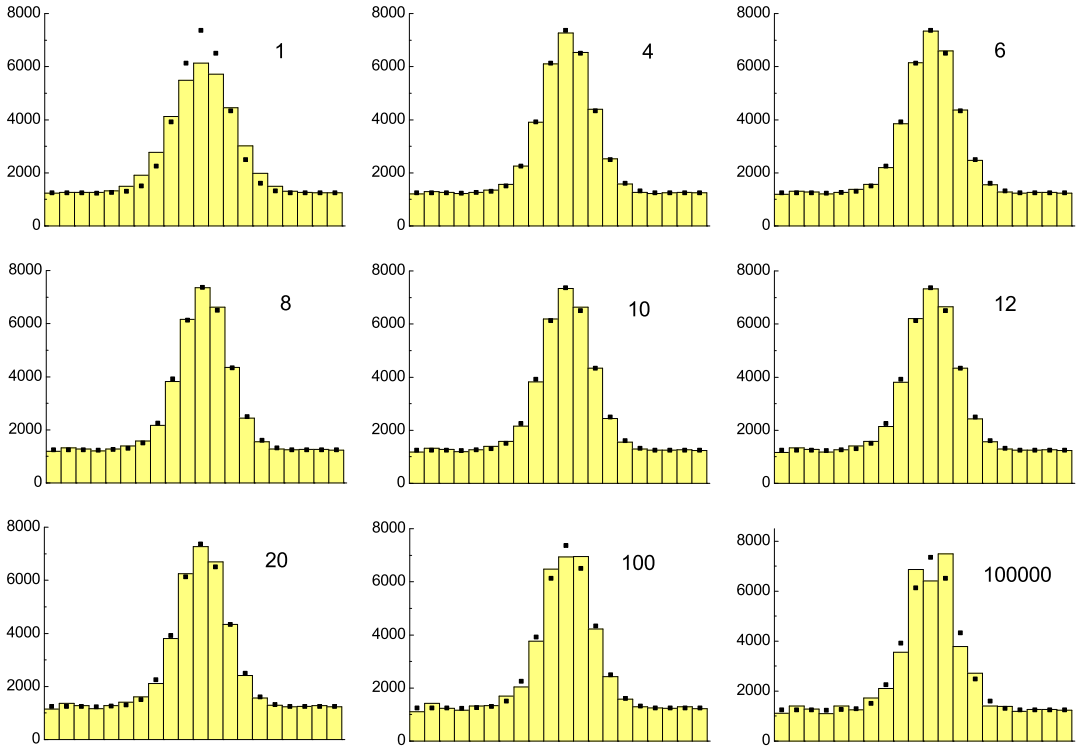}%
\caption{EM iterative unfolding results for $50000$ observations and
resolution $\sigma_{s}=0.04$. The number of iterations is indicated in the
plots. }%
\label{iter1b50000_04}%
\end{center}
\end{figure}
%

\begin{figure}
[ptb]
\begin{center}
\includegraphics[
trim=0.134773in 0.203615in 0.137749in 0.163167in,
height=1.9161in,
width=4.7289in
]%
{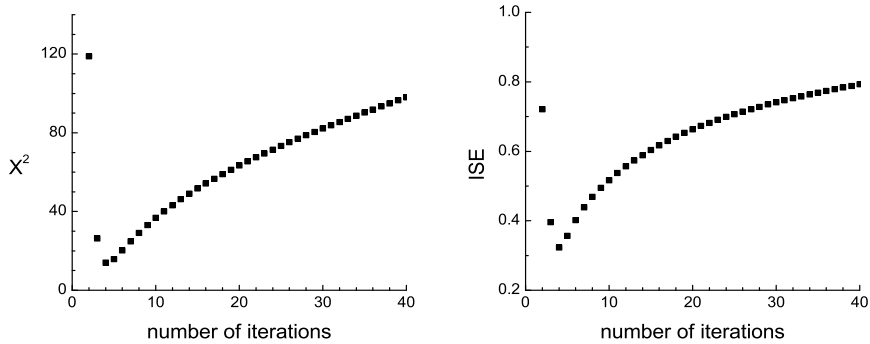}%
\caption{$X^{2}$ and the $ISE$ as functions of the number of iteration for
$50000$ events and resolution $\sigma_{s}=0.04$. }%
\label{iter1b50000_04c}%
\end{center}
\end{figure}
%

\begin{figure}
[ptb]
\begin{center}
\includegraphics[
trim=0.144281in 0.181686in 0.153074in 0.151844in,
height=4.1486in,
width=5.9053in
]%
{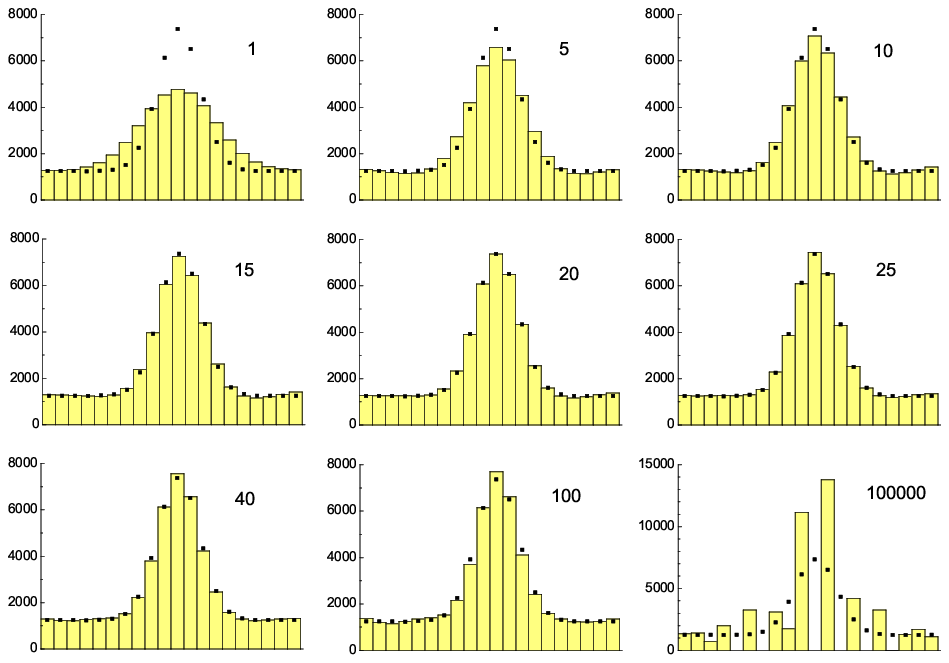}%
\caption{Same as Fig. \ref{iter1b50000_04} but for resolution $\sigma
_{s}=0.08$ and $50000$ events.}%
\label{iter1b50000_08}%
\end{center}
\end{figure}
%

\begin{figure}
[ptb]
\begin{center}
\includegraphics[
trim=0.000000in 0.175341in 0.000000in 0.175537in,
height=1.8846in,
width=4.9223in
]%
{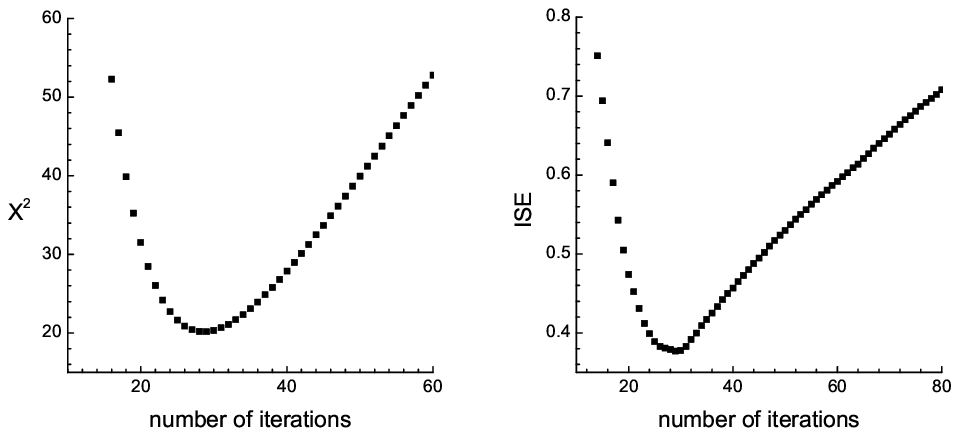}%
\caption{Same as Fig. \ref{iter1b50000_04c} but for resolution $\sigma
_{s}=0.08$ and $50000$ events.}%
\label{iter1b50000_08c}%
\end{center}
\end{figure}
%

\begin{figure}
[ptb]
\begin{center}
\includegraphics[
trim=0.165582in 0.180483in 0.175348in 0.167174in,
height=4.2059in,
width=5.9302in
]%
{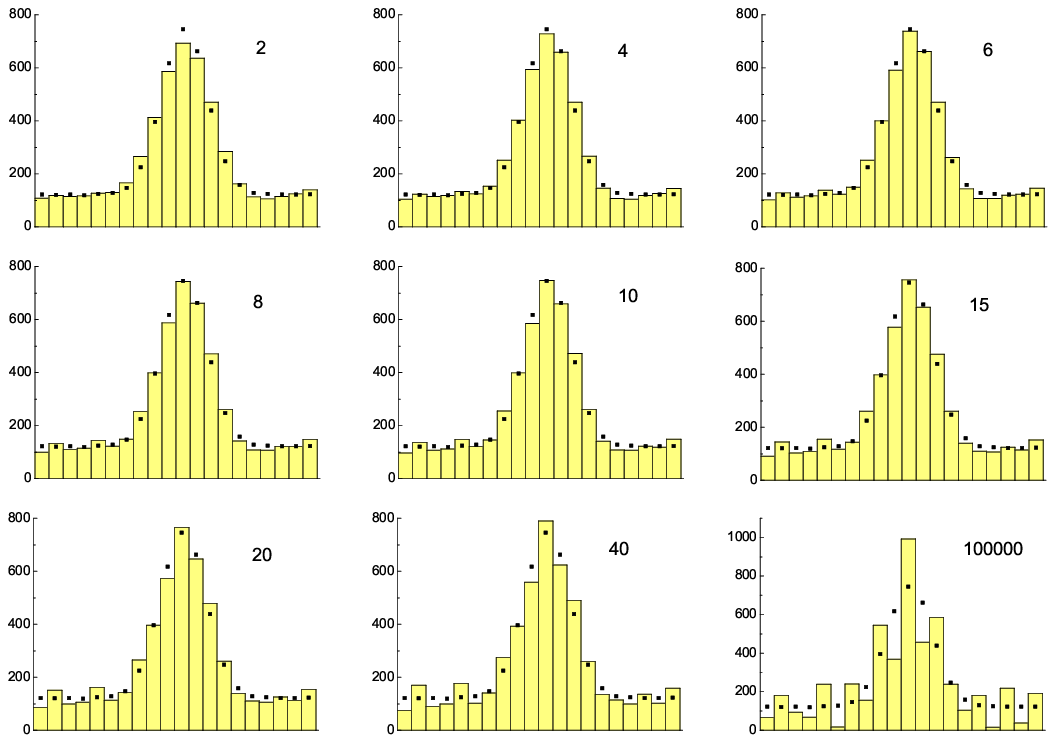}%
\caption{Same as Fig. \ref{iter1b50000_04} but for $5000$ events, $\sigma
_{s}=0.04$.}%
\label{iter1b5000_04}%
\end{center}
\end{figure}
%

\begin{figure}
[ptb]
\begin{center}
\includegraphics[
trim=0.000000in 0.203630in 0.000000in 0.189279in,
height=1.8472in,
width=4.8078in
]%
{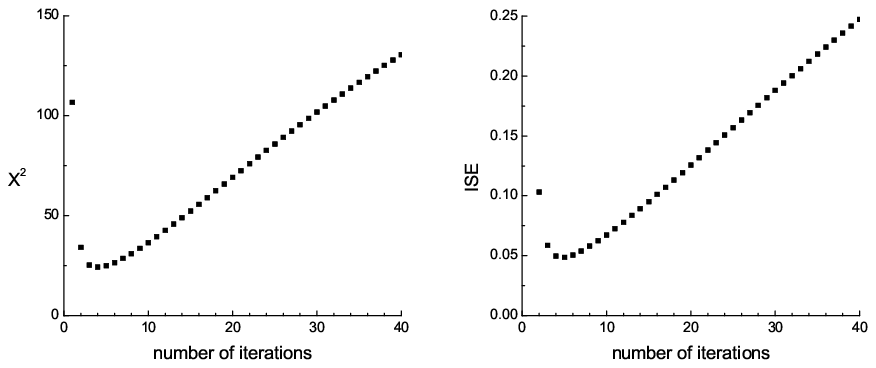}%
\caption{Same as Fig. \ref{iter1b50000_04c} but for $\sigma_{s}=0.04$ and
$5000$ events.}%
\label{iter1b5000_04c}%
\end{center}
\end{figure}
%

\begin{figure}
[ptb]
\begin{center}
\includegraphics[
trim=0.142498in 0.184054in 0.144718in 0.177237in,
height=4.1038in,
width=5.892in
]%
{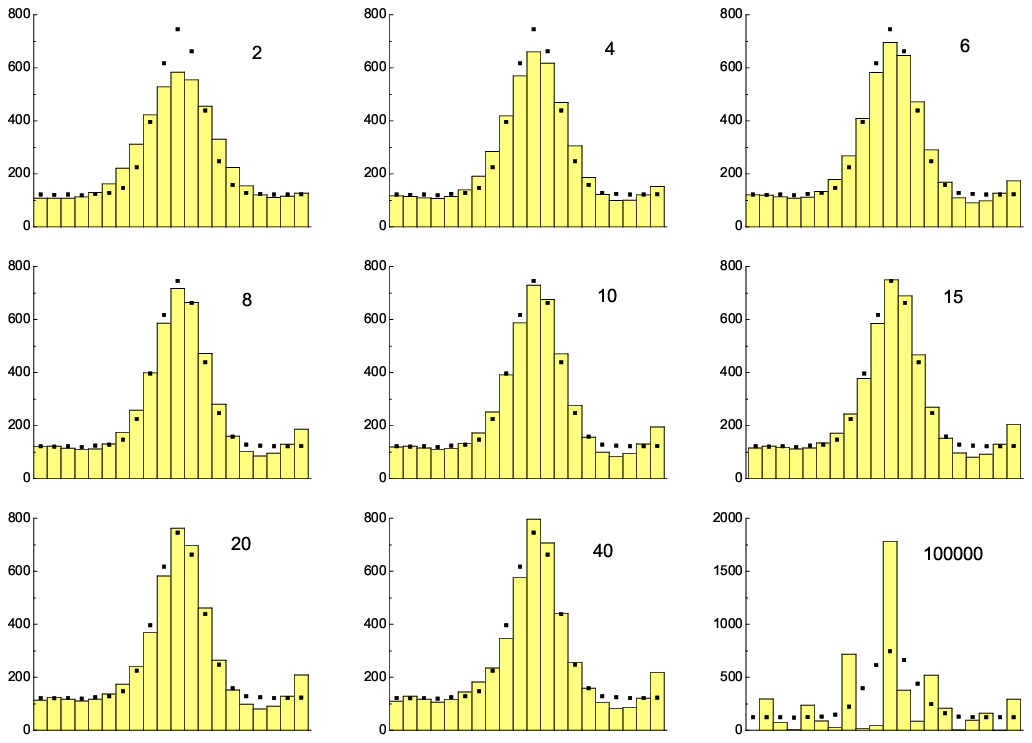}%
\caption{Same as Fig. \ref{iter1b5000_04} but for resolution $\sigma_{s}=0.08$
and $5000$ events.}%
\label{iter1b5000_08}%
\end{center}
\end{figure}
%

\begin{figure}
[ptb]
\begin{center}
\includegraphics[
trim=0.000000in 0.194729in 0.000000in 0.216527in,
height=1.8232in,
width=4.8078in
]%
{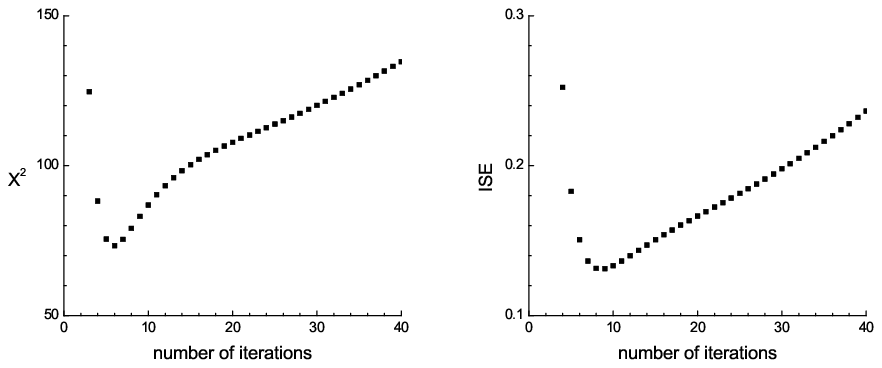}%
\caption{Same as Fig. \ref{iter1b5000_04c} but for resolution $\sigma
_{s}=0.08$ and $5000$ events.}%
\label{iter1b5000_08c}%
\end{center}
\end{figure}
%

\begin{figure}
[ptb]
\begin{center}
\includegraphics[
trim=0.122315in 0.153307in 0.123510in 0.153599in,
height=4.1386in,
width=5.8912in
]%
{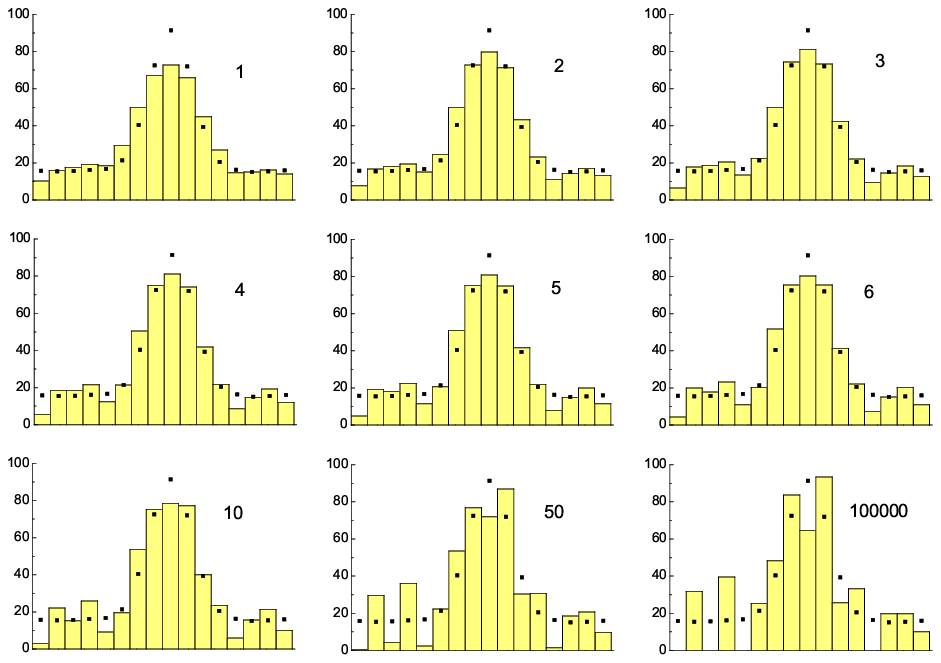}%
\caption{Same as Fig. \ref{iter1b50000_04} but for $500$ events, $\sigma
_{s}=0.04$.}%
\label{iter1b500_04}%
\end{center}
\end{figure}
%

\begin{figure}
[ptb]
\begin{center}
\includegraphics[
trim=0.000000in 0.157001in 0.000000in 0.138406in,
height=2.2648in,
width=4.9655in
]%
{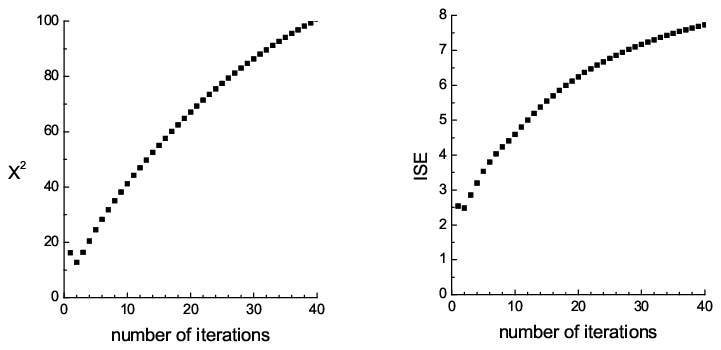}%
\caption{Same as Fig. \ref{iter1b50000_04c} but for $500$ events, $\sigma
_{s}=0.04$.}%
\label{iter1b500_04c}%
\end{center}
\end{figure}
%

\begin{figure}
[ptb]
\begin{center}
\includegraphics[
trim=0.124705in 0.145992in 0.126697in 0.134290in,
height=4.21in,
width=5.9244in
]%
{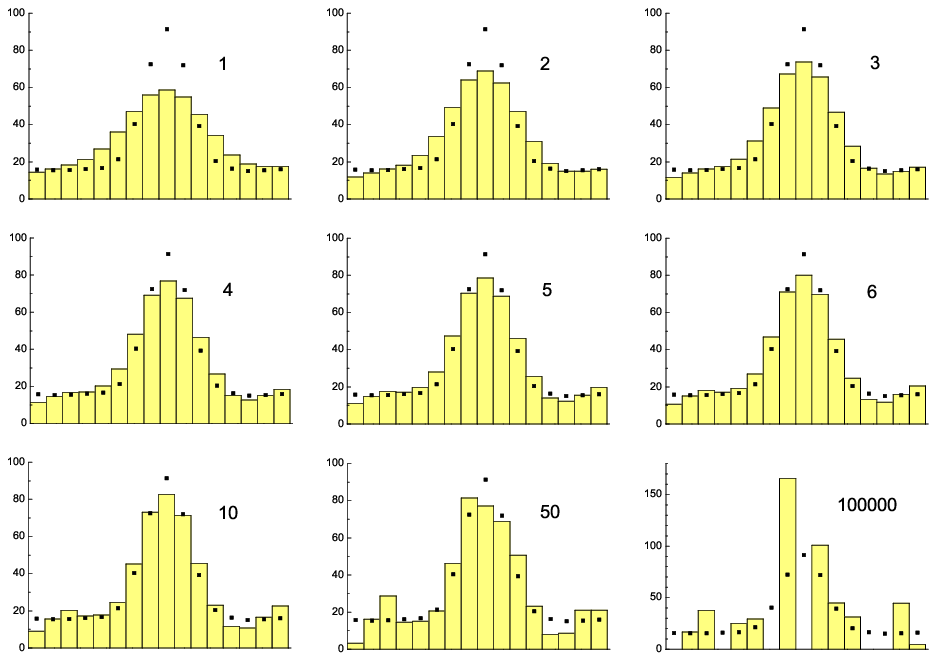}%
\caption{Same as Fig. \ref{iter1b50000_04} but for resolution $\sigma
_{s}=0.08$ and $500$ events.}%
\label{iter1b500_08}%
\end{center}
\end{figure}
%

\begin{figure}
[ptb]
\begin{center}
\includegraphics[
trim=0.000000in 0.155666in 0.000000in 0.124740in,
height=2.2673in,
width=4.9838in
]%
{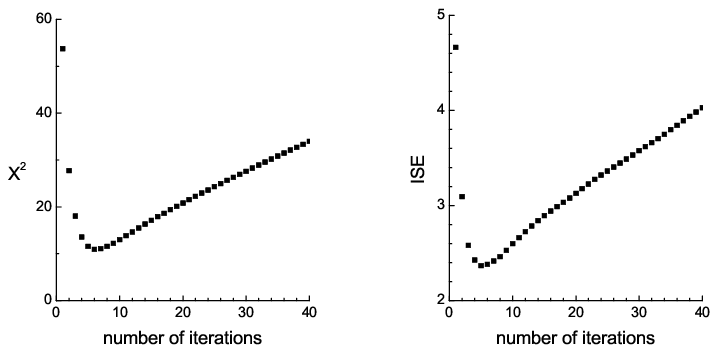}%
\caption{Same as Fig. \ref{iter1b50000_04c} but for resolution $\sigma
_{s}=0.08$ and $500$ events.}%
\label{iter1b500_08c}%
\end{center}
\end{figure}

The optimal result for $50000$ events and $\sigma_{s}=0.04$ is obtained with
$3$ iterations, but qualitatively with $20$ iterations the adjustment is
satisfactory, too. If we increase $\sigma_{s}$ to $0.08$ about $30$ iterations
are required to unfold the data. Generally, the optimal number of iterations
increases with the Gaussian smoothing parameter $\sigma_{s}$ and for
$\sigma_{s}=0$ no iteration would but necessary. The convergence is always
quite fast. With only $500$ events, the method does not succeed anymore to
reproduce height and width of the peak very well.

\subsection{Reduced iteration speed}

Occasionally, already the first or second iteration minimizes the test
quantities. Then it may be desirable to have a finer step size, to slow down
the convergence. This is achieved with a modified unfolding function. We just
have to introduce a parameter $\beta>0$ into (\ref{RLunfolding})%

\[
\hat{\theta}_{j}^{(k+1)}=\left[  \sum_{i=1}^{M}A_{ij}\hat{\theta}_{j}%
^{(k)}\frac{d_{i}}{d_{i}^{(k)}}/\alpha_{j}+\beta\hat{\theta}_{j}^{(k)}\right]
/(1+\beta)\;.
\]

The value $\beta=0$ corresponds to the original sequence (\ref{RLunfolding}).
The value $\beta=1$ slows down the convergence by about a factor of two and
with $\beta=\infty$ the parameter $\theta$ remains unchanged.

Applied to the one-peak example with $500$ events and $\sigma_{s}=0.04$ where
the minimum of $X^{2}$ is reached at the second iteration, $\beta=1$ moves the
minimum to the forth iteration. However, the result is slightly worse, the
$ISE$ is increased from $0.0487$ to $0.0514$. The slowing down of the
convergence does not simply interpolate between the results of the standard iteration.

\subsection{Choice of the regularization strength}

\label{emunfolding}A heuristic method to estimate the best number of
iterations based on the $p^{\prime}$ dependence on the number of iterations
has been proposed in \cite{zech}. However, the method does not work, if the
starting histogram is close to the true histogram and it has been tested only
with very few event samples and distributions. A better way to select the
number of iterations is provided by the method which tries to minimize the
$ISE$ explained in Sect. \ref{regustrength}.%
\begin{figure}
[ptb]
\begin{center}
\includegraphics[
trim=0.130120in 0.183648in 0.131275in 0.183830in,
height=2.308in,
width=5.6588in
]%
{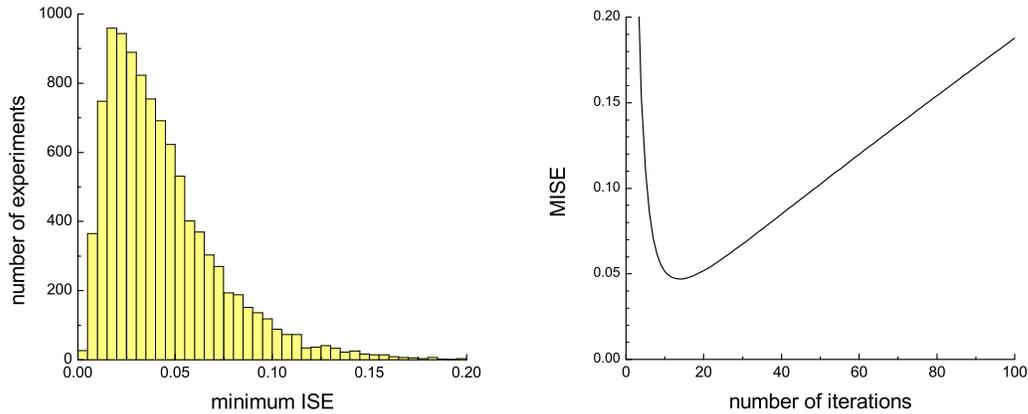}%
\caption{Distribution of the minimal $ISE$ from $10000$ simulations (left
hand) and the estimate of $MISE$ as a function of the number of iterations
(right hand).}%
\label{figmise}%
\end{center}
\end{figure}

\begin{myexample}
To test the method, let us look at the distribution of the lowest $ISE$ that
can be obtained with the standard example with $5000$ events and smearing
resolution $\sigma_{s}=0.08$. The results from $10000$ simulations are given
in Fig. \ref{figmise}. Each time the number of iterations $k$ for which the
$ISE$ is minimum is selected. The distribution of the minimal values of the
$ISEs$ (Fig. \ref{figmise} left hand) has a mean value of $0.043$. The mean
integrated square error ($MISE$) as a function of the number of iterations is
shown at the right-hand side of the same figure. The minimum is obtained at
$14$ iterations. Choosing always $k_{0}=14$ iterations a mean value
$MISE=0.047$ is obtained, not much larger than the mean of the individually
optimized $ISE$ values. Apparently, the $MISE$ depends only weakly on the
number of iterations. In the range from $10$ to $20$ iterations it varies by
less than $10\%$.
\end{myexample}

In real experiments we do not know the true distribution needed to estimate
the optimal number of iterations, but a preliminary unfolded histogram
$\vec{\theta}^{(0)}$ can be used instead. Starting from $\vec{\theta}^{(0)}$ a
large sample of observed histograms can be generated from which the number of
iterations $k_{0}$ can be derived that minimizes the $MISE$.

\begin{figure}
[ptb]
\begin{center}
\includegraphics[
trim=0.000000in 0.262325in 0.000000in 0.197004in,
height=2.2424in,
width=2.983in
]%
{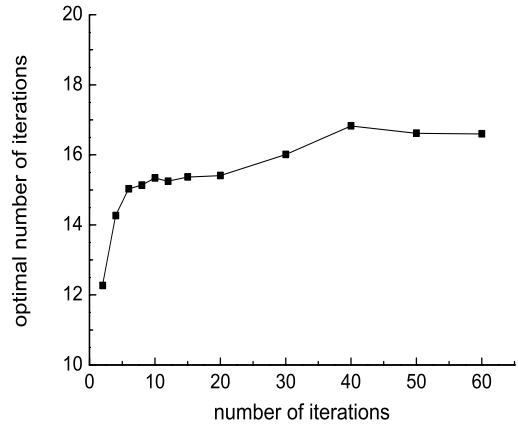}%
\caption{Iterative method to select the optimal number of iterations. The
estimate $k_{0}$ (ordinate) varies little with the number of iterations $k$
(abscissa) used to obtain the unfolded histogram $\vec{\theta}^{(0)}$ used in
the simulation.}%
\label{bestiter}%
\end{center}
\end{figure}

The best estimate of $k_{0}$ depends little on the number $k$ of iterations
used to determine the preliminary histogram $\vec{\theta}^{(0)}$. This is
shown In Fig. \ref{bestiter} where $k_{0}$ is plotted as a function of $k$ for
the previous example. The reason behind this behavior is that the optimal
regularization depends mainly on the experimental resolution and less on the
details of the true distribution. In cases where the dependence is larger, the
procedure can be iterated.%

\subsection{Introducing a final smoothing step}

It has been proposed \cite{dagostini95,dagostini2010,volobouev} to apply after
the iteration sequence a final smoothing step: After iteration $i$ the result
$\vec{\theta}^{(i)}$ is folded with a smoothing matrix $g$, yielding
$\vec{\theta}^{(i)\prime}$, $\theta_{k}^{(i)\prime}=\sum_{l}g_{kl}\theta
_{l}^{(i)}$. If $\vec{\theta}_{k}^{(i)\prime}$ agrees with $\vec{\theta}%
_{k}^{(i-1)\prime}$ within given limits, the iteration sequence is terminated.
In this way, convergence to a smooth result is imposed. In
\cite{dagostini2010} it is proposed to add after the convergence one further
iteration to $\vec{\theta}^{(i+1)\prime}$.

The parameters of the smoothing matrix which define the regularization
strength have to be adjusted to the specific properties of the problem that
has to be solved. The approach may be very successful in problems where prior
knowledge about the shape of the true distribution is available, but in the
general case it is not obvious how to choose the smoothing step. The intention
of the additional iteration is to avoid a too strong influence of the
smoothing step on the final result \cite{dagostini2010}.

\subsection{Dependence on the starting distribution}

So far we have used a uniform starting distribution for the EM iteration. If
there is prior knowledge of the approximative shape of the true distribution,
for instance from previous experiments, then the uniform histogram can be
replaced by a better estimate. Experience shows that the influence of the
starting histogram on the unfolding result is rather weak.%

\begin{figure}
[ptb]
\begin{center}
\includegraphics[
trim=0.000000in 0.199413in 0.000000in 0.159648in,
height=1.9485in,
width=4.6766in
]%
{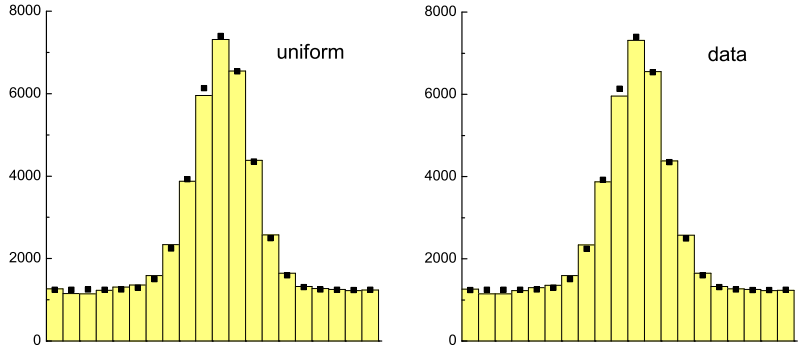}%
\caption{Iterative unfolding with two different starting histograms, left
uniform and right experimental.}%
\label{startfg}%
\end{center}
\end{figure}

\begin{myexample}
We repeat the unfolding of the distribution with $50000$ events and
experimental resolution $\sigma_{s}=0.08$ starting with the histogram of the
observed events. (The choice of the example with a large number of events is
less sensitive to statistical fluctuations than an example with low statistics
and should indicate possible systematic effects.) The two results displayed in
Fig. \ref{startfg} are qualitatively indistinguishable. Starting with the
uniform histogram, the lowest $ISE=0.0964$ is obtained after $40$ iterations
with $\chi^{2}=35.1$. With the observed histogram the values obtained after
$38$ iterations are $ISE=0.0940$ with the same value $\chi^{2}=35.1$. In the
low statistics example with $500$ events and resolution $\sigma_{s}=0.04$ the
minimum is reached already after $2$ iterations with the $ISE=0.0488$ and
$0.0487$, respectively and values $\chi^{2}=36.0$ and $36.3$.
\end{myexample}

The influence of the starting distribution on the unfolding result should be
checked but in the majority of cases is not necessary to deviate from the
uniform histogram.

\section{SVD based methods}

SVD unfolding was first applied in particle physics by Hoecker and Kartvelishvili.

\subsection{Truncated SVD}

The SVD decomposes the unfolded histogram into statistically independent
vectors, $\vec{\theta}_{0}=\Sigma_{i=1}^{M}a_{i}\vec{u}_{i}$, and provides an
ordering of the vectors according to their sensitivity to noise. In this way
it offers the possibility to obtain a stable solution by chopping off
eigenvectors with low eigenvalues. Only contribution with eigenvector indices
less than or equal to the index $m$ are kept:
\[
\vec{\theta}_{reg}=\sum_{i=1}^{m}a_{i}\vec{u}_{i}\;.
\]
The choice of of the cut-off $m$ is based on the significance $S_{i}%
=a_{i}/\delta_{i}$ of the eigenvector contributions $a_{i}$ which is provided
by the LS fit. The amplitudes of the eliminated eigenvectors should be
compatible with zero within one or two standard deviations.

The application of the method, called truncated SVD (TSVD) is simple and
computationally fast. The idea behind TSVD is attractive but it has some limitations:

\begin{itemize}
\item The SVD solution is obtained by a linear LS fit. This implies that low
event numbers in the observed histogram are not treated correctly. Combining
bins with low event numbers can reduce the problem.

\item The eigenvalue decomposition is mainly related to the properties of the
response matrix and does not sufficiently take into account the shape of the
unfolded distribution. Small eigenvalues may correspond to significant
structures in the true distribution and the corresponding eigenvectors may be
eliminated by the truncation. The combination of the vectors belonging to
several \textquotedblleft insignificant\textquotedblright\ amplitudes may
contribute significantly to the true distribution.
\end{itemize}

%

\begin{figure}
[ptb]
\begin{center}
\includegraphics[
trim=0.138527in 0.156387in 0.146201in 0.125425in,
height=2.2889in,
width=5.4869in
]%
{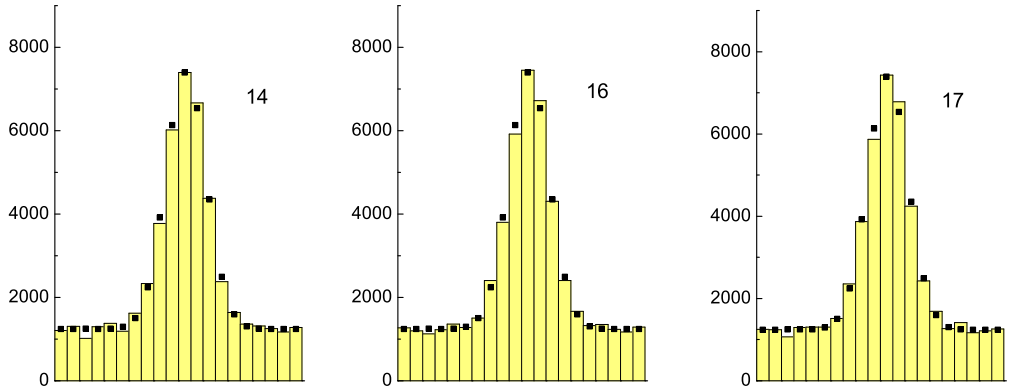}%
\caption{Truncated SVD unfolding results for $50000$ events and resolution
$\sigma_{s}=0.04$. The number of eigenvectors that have been included is
indicated. The central plot corresponds to the minimum of the $ISE$.}%
\label{unfsvd50000r4}%
\end{center}
\end{figure}
%

\begin{figure}
[ptb]
\begin{center}
\includegraphics[
trim=0.000000in 0.149316in 0.000000in 0.112280in,
height=2.2026in,
width=5.0369in
]%
{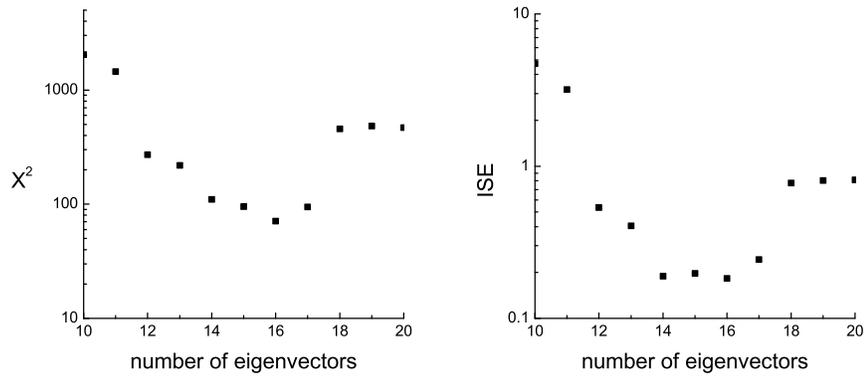}%
\caption{Distribution of $X^{2}$ and $ISE$ as a function of the number of the
included eigenvectors for $50000$ events and resolution $\sigma_{s}=0.04$.}%
\label{missvd50000r4}%
\end{center}
\end{figure}
%

\begin{figure}
[ptb]
\begin{center}
\includegraphics[
trim=0.122410in 0.147466in 0.123561in 0.118078in,
height=2.2881in,
width=5.4877in
]%
{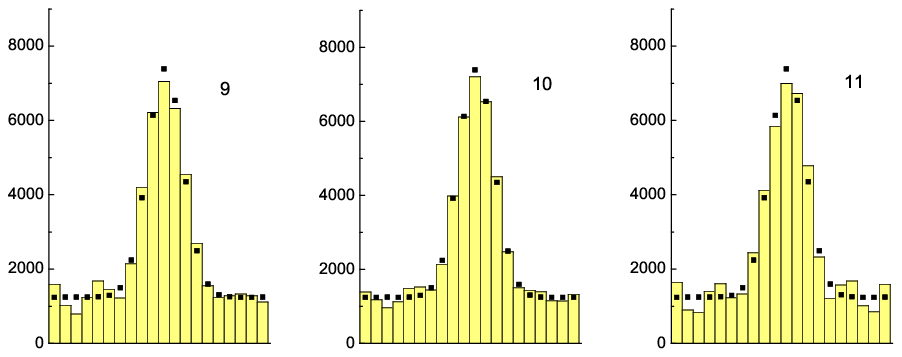}%
\caption{Same as Fig. \ref{unfsvd50000r4} but with resolution $\sigma
_{s}=0.08$ and $50000$ events.}%
\label{unfsvd50000r8}%
\end{center}
\end{figure}
%

\begin{figure}
[ptb]
\begin{center}
\includegraphics[
trim=0.000000in 0.166026in 0.000000in 0.128475in,
height=2.347in,
width=5.0079in
]%
{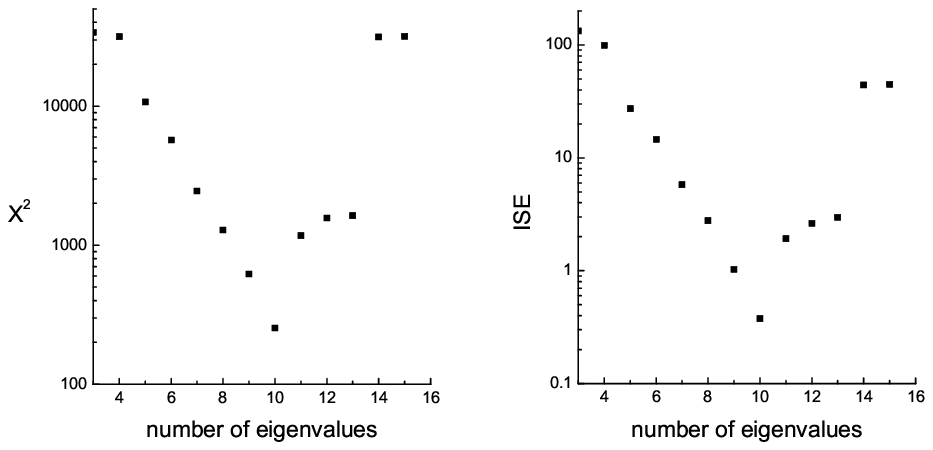}%
\caption{Same as Fig. \ref{missvd50000r4} but with resolution $\sigma
_{s}=0.08$ and $50000$ events.}%
\label{missvd50000r8}%
\end{center}
\end{figure}
%

\begin{figure}
[ptb]
\begin{center}
\includegraphics[
trim=0.142427in 0.155688in 0.143569in 0.124550in,
height=2.3014in,
width=5.4661in
]%
{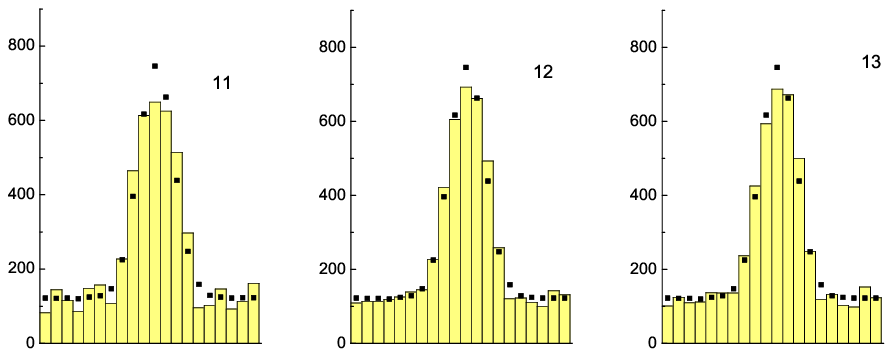}%
\caption{Same as Fig. \ref{unfsvd50000r4} but with $5000$ events and
resolution $\sigma_{s}=0.04$.}%
\label{unfsvd5000r4}%
\end{center}
\end{figure}
\begin{figure}
[ptbptb]
\begin{center}
\includegraphics[
trim=0.000000in 0.153940in 0.000000in 0.115733in,
height=2.0099in,
width=4.8709in
]%
{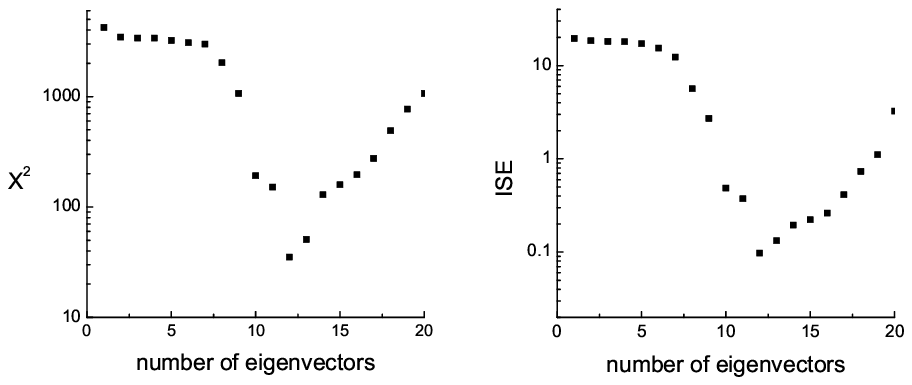}%
\caption{Same as Fig. \ref{missvd50000r4} but with $5000$ events and
resolution $\sigma_{s}=0.04$.}%
\label{missvd5000r4}%
\end{center}
\end{figure}
%

\begin{figure}
[ptb]
\begin{center}
\includegraphics[
trim=0.135953in 0.165877in 0.137857in 0.135413in,
height=2.3171in,
width=5.5575in
]%
{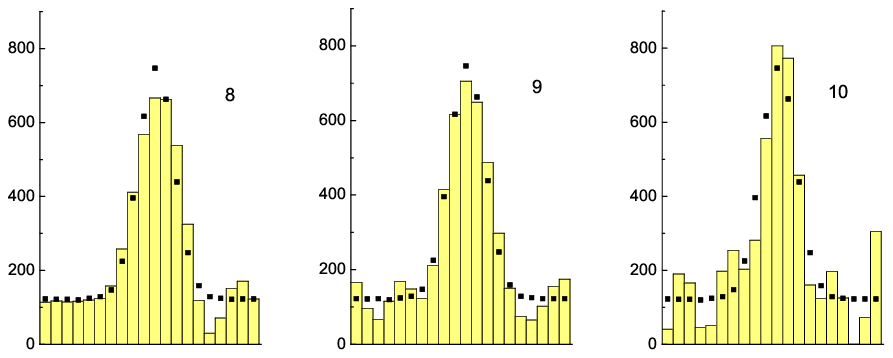}%
\caption{Same as \ref{unfsvd50000r4} but with $5000$ events and resolution
$\sigma_{s}=0.08$.}%
\label{unfsvd5000r8}%
\end{center}
\end{figure}
%

\begin{figure}
[ptb]
\begin{center}
\includegraphics[
trim=0.000000in 0.183907in 0.000000in 0.138022in,
height=1.8945in,
width=4.8211in
]%
{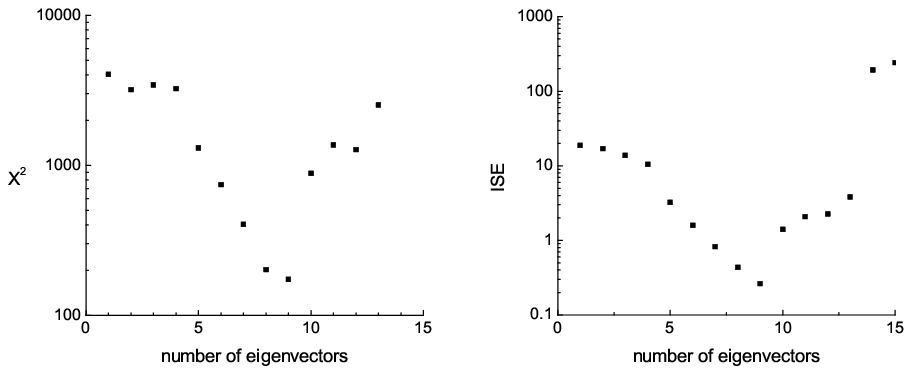}%
\caption{Same as Fig. \ref{missvd50000r4} but with $5000$ and resolution
$\sigma_{s}=0.08$.}%
\label{missvd5000r8}%
\end{center}
\end{figure}
%

\begin{figure}
[ptb]
\begin{center}
\includegraphics[
trim=0.139490in 0.163385in 0.141375in 0.133297in,
height=2.3404in,
width=5.5658in
]%
{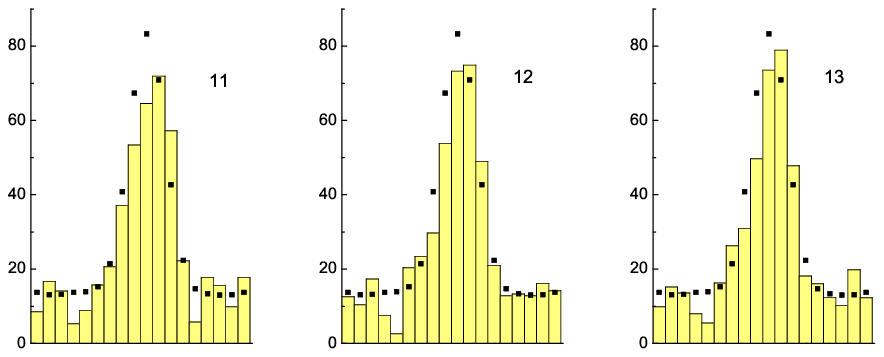}%
\caption{Same as \ref{unfsvd50000r4} but with $500$ events and resolution
$\sigma_{s}=0.04$.}%
\label{unfsvd500r4}%
\end{center}
\end{figure}
\begin{figure}
[ptbptb]
\begin{center}
\includegraphics[
trim=0.000000in 0.170933in 0.000000in 0.126792in,
height=1.995in,
width=4.9506in
]%
{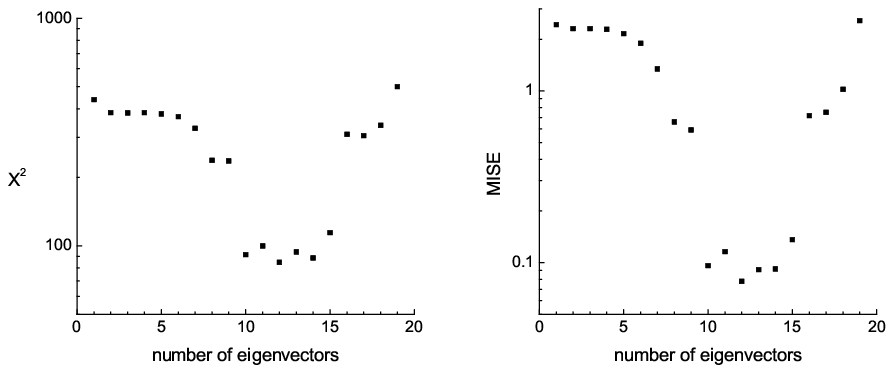}%
\caption{Same as \ref{missvd50000r4} but with $500$ events and resolution
$\sigma_{s}=0.04$.}%
\label{missvd500r4}%
\end{center}
\end{figure}
%

\begin{figure}
[ptb]
\begin{center}
\includegraphics[
trim=0.136851in 0.152318in 0.137982in 0.122030in,
height=2.3769in,
width=5.5359in
]%
{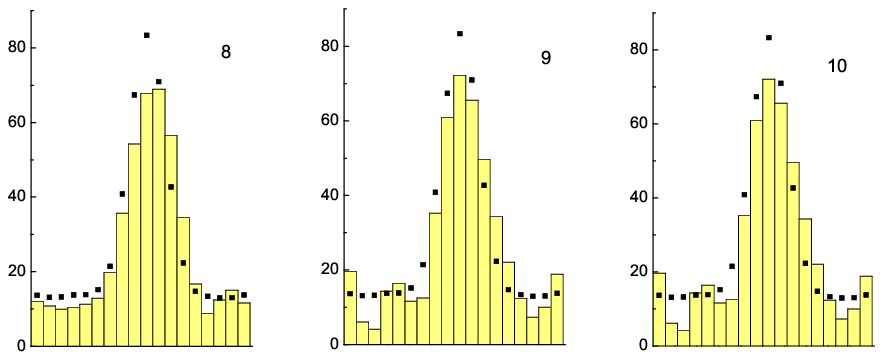}%
\caption{Same as \ref{unfsvd50000r4} but with $500$ events and resolution
$\sigma_{s}=0.08$.}%
\label{unfsvd500r8}%
\end{center}
\end{figure}
\begin{figure}
[ptbptb]
\begin{center}
\includegraphics[
trim=0.000000in 0.146155in 0.000000in 0.109880in,
height=2.0108in,
width=5.1449in
]%
{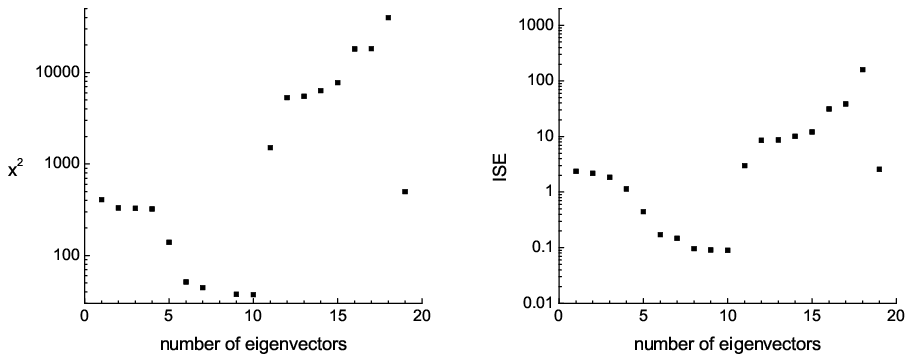}%
\caption{Same as \ref{missvd50000r4} but with $500$ events and resolution
$\sigma_{s}=0.08$.}%
\label{missvd500r8}%
\end{center}
\end{figure}

Figs. \ref{unfsvd50000r4} to \ref{unfsvd500r8} show unfolding results from
TSVD. The same data as in the previous section have been taken. In all figures
the central histogram corresponds to the smallest value of the $ISE$ that is
attainable. The numbers in the plots indicate how many eigenvectors have been
included. The results for the examples with only $500$ events are
unsatisfactory, probably because in the uniform part of the histogram only
$12.5$ events per bin are expected which makes a linear LSF problematic. In
the two examples the reconstructed number of events is significantly higher
than the true number. The increases are $5\%$ and $7\%$ for the resolutions
$\sigma_{s}=0.08$ and $\sigma_{s}=0.04$. The agreement of the unfolded
histogram with the true histogram is worse than in the EM method. The values
of $X^{2}$ and $ISE$ as functions of the number of eigenvectors are not very
smooth and make it difficult to choose the number of retained eigenvectors.

\subsection{Smooth truncation}%

\begin{figure}
[ptb]
\begin{center}
\includegraphics[
trim=0.000000in 0.174216in 0.000000in 0.116451in,
height=1.829in,
width=2.533in
]%
{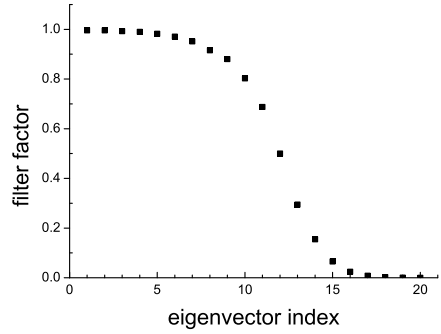}%
\caption{Filter factor as a function of the eigenvector index.}%
\label{filterfactor}%
\end{center}
\end{figure}
\begin{figure}
[ptb]
\begin{center}
\includegraphics[
trim=0.129695in 0.199647in 0.132595in 0.199807in,
height=1.7708in,
width=5.6322in
]%
{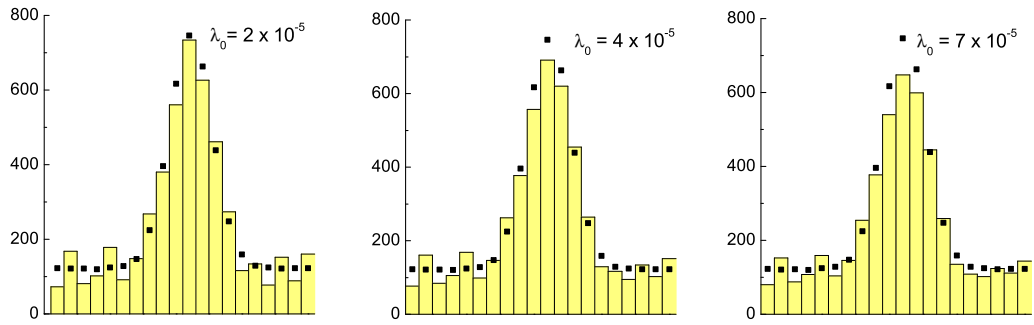}%
\caption{SVD unfolding with smooth truncation. The data sample contains $5000$
events, the resolution is $\sigma_{s}=0.04$.}%
\label{unfsmoothr4}%
\end{center}
\end{figure}
\begin{figure}
[ptbptb]
\begin{center}
\includegraphics[
trim=0.000000in 0.316347in 0.000000in 0.253489in,
height=1.8298in,
width=4.1727in
]%
{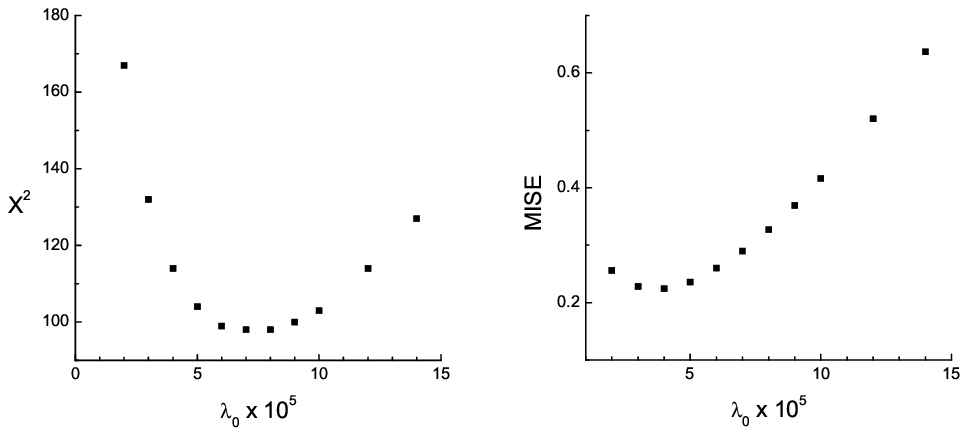}%
\caption{SVD unfolding with smooth truncation. $X^{2}$ and $ISE$ for $5000$
events and resolution $\sigma_{e}=0.04$.}%
\label{mismoothr4}%
\end{center}
\end{figure}
\begin{figure}
[ptb]
\begin{center}
\includegraphics[
trim=0.154212in 0.192708in 0.155480in 0.192866in,
height=1.8124in,
width=5.8754in
]%
{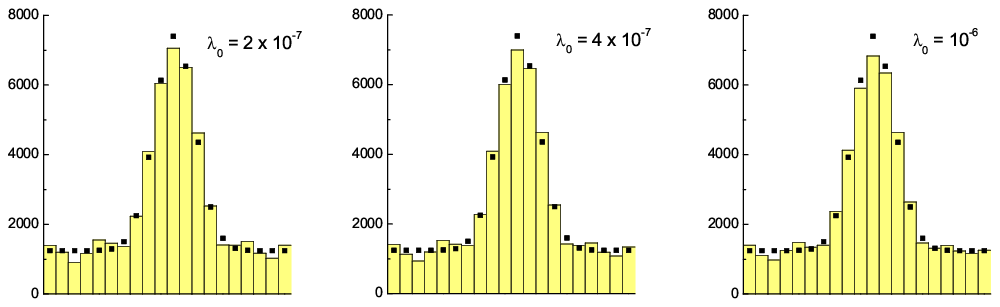}%
\caption{SVD unfolding with smooth truncation. The data sample consists of
$50000$ events, the resolutionis $\sigma_{s}=0.08$.}%
\label{unsmoothr8}%
\end{center}
\end{figure}
\begin{figure}
[ptb]
\begin{center}
\includegraphics[
trim=0.000000in 0.332209in 0.000000in 0.290205in,
height=1.6413in,
width=4.5645in
]%
{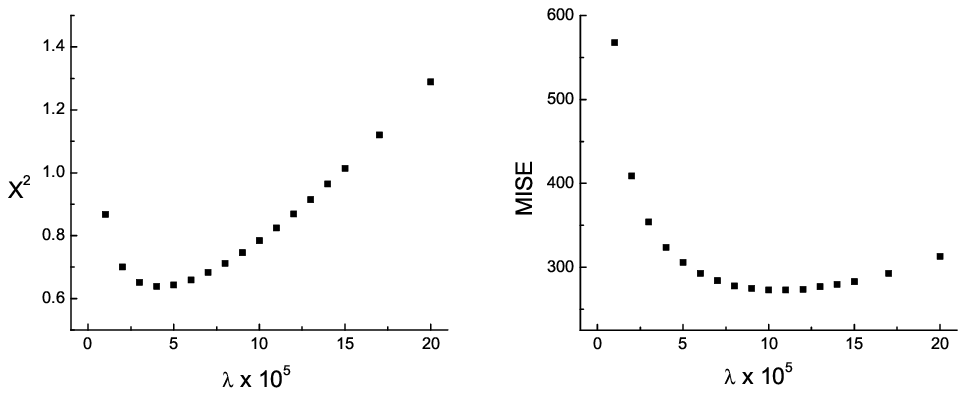}%
\caption{SVD unfolding with smooth truncation. $X^{2}$ and $ISE$ for $50000$
events and resolution $\sigma_{s}=0.08.$}%
\label{mismoothr8}%
\end{center}
\end{figure}

It has been proposed \cite{hansen, blobel} to replace the brut force chopping
off of the noise dominated components by a smooth cut. This is accomplished by
filter factors%
\begin{equation}
\varphi(\lambda)=\frac{\lambda^{2}}{\lambda^{2}+\lambda_{0}^{2}}
\label{filter}%
\end{equation}
where $\lambda_{0}$ is the eigenvalue which fixes the degree of smoothing and
$\lambda$ is the eigenvalue corresponding to the coefficient which is to be
multiplied by $\varphi(\lambda)$. The solution is then%
\[
\vec{\theta}_{reg}=\sum_{i=1}^{M}\varphi(\lambda_{i})a_{i}\vec{u}_{i}\;.
\]
The function \ref{filter} is displayed in Fig. \ref{filterfactor}. The
amplitude of the eigenvector with eigenvalue $\lambda=\lambda_{0}$ is reduced
by a factor $2$. For large eigenvalues $\lambda$ the filter factor is close to
one and for small values it is close to zero. The SVD components with large
eigenvalues are hardly affected while components with small eigenvalues are
strongly damped. As an example we simulate and unfold $5000$ events with
resolution $\sigma_{s}=0.04$ and $50000$ events with resolution $\sigma
_{s}=0.08$. As before, the minimum of the $ISE$ is determined by varying the
parameter $\lambda_{0}$, see Figs. \ref{mismoothr4} and \ref{mismoothr8}. The
central plots of Figs. \ref{unfsmoothr4} and \ref{unsmoothr8} correspond to
the minima of the $ISE$. The locations of the minima of $X^{2}$ and $ISE$
differ considerably. The $ISEs$ in this specific example are $0.224$ and
$0.639$, considerably larger than those of the standard truncated SVD which
are $0.098$ and $0.376$. This strong effect may be accidental.%

%

%

It is a strange compromise to reduce the amplitude of a component $m$ and to
include a fraction of the amplitude of a less significant component $n>m$.

In \cite{hansen} it is shown that the filtered SVD solution is equivalent to
Tikhonov's norm regularization under the condition that the uncertainties of
the observations correspond to white noise (normally distributed fluctuations
with constant variance). We will come back to the norm regularization below.

\subsection{Selective SVD}

Truncated SVD eliminates eigenvector contributions that suffer from large
uncertainties. It happens that modeling the true distribution, high frequency
contributions are required and that the corresponding amplitudes are
significant. Not the absolute error but the significance is relevant for the
decision whether to include a component or not. Instead of a vertical cut in
the bottom plots of Fig. \ref{eigen1b5000} we could apply a horizontal cut and
eliminate for example all components with significance below a certain value,
for instance $2$ standard deviations (SSVD).

At first sight this idea seems attractive. Why should components that are
compatible with noise be kept? Well, typical distributions in particle and
astrophysics have no periodic regularities that can be associated to specific
frequencies and eigenvectors. If an eigenvector $k_{0}$ is required to
describe the distribution than usually all other eigenvectors $k<k_{0}$ with
lower frequency are required as well. If their significance is low, then also
their amplitude is low and they will not strongly influence the result. An
exception are some angular distributions with preferred and forbidden
frequencies. As a consequence it is reasonable to keep all eigenvectors with
eigenvalues larger or equal to the eigenvalue of any significant eigenvector.%

\begin{figure}
[ptb]
\begin{center}
\includegraphics[
trim=0.000000in 0.160135in 0.000000in 0.117432in,
height=2.1229in,
width=2.8975in
]%
{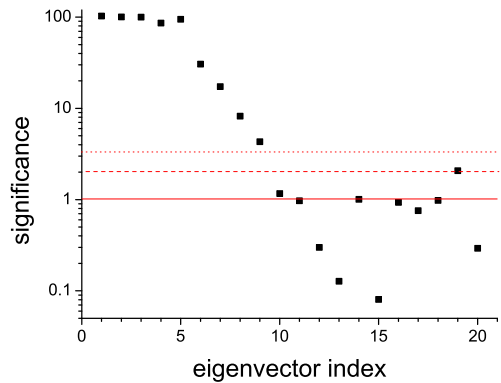}%
\caption{Significance as a function of the eigenvector index for $50000$
events.}%
\label{eigen1b500008}%
\end{center}
\end{figure}

In Fig \ref{eigen1b500008} the plot of Fig. \ref{eigen1b5000} bottom left is
repeated with tenfold statistics. One and two standard deviations are
indicated with solid and dashed lines. The dotted line corresponds to
$\sqrt{10}$ standard deviations which are expected for a signal of one
standard deviation in the corresponding case with only $5000$ events. The
observed signals with more than two standard deviations for the eigenvectors
$10$ and $14$ with $5000$ events are not present in the high statistics case
and therefore obviously due to statistical fluctuations. Independent on
whether we choose truncated or selected SVD the scatter plot
\ref{eigen1b500008} would suggest to retain $10$, $11$ or $12$ eigenvectors
while the minimum of the $ISE$ is obtained with $16$ contributions. This shows
how difficult it is to select the eigenvector threshold from the significance plots.

Sometimes distributions obey symmetry constraints. For example, a distribution
of $x$ may be known to be symmetric with respect to the center $x_{c}$ of the
variable range. Then only odd components should be present and the even
eigenvectors can be ignored \cite{hansen}, but a better way to take advantage
of the symmetry is to replace $x$ by the absolute value $|x-x_{c}|$. We can
proceed in the same way with more complex symmetries. To infer the symmetry
from the observed histogram is dangerous because statistical fluctuations
destroy the exact symmetry and then it is impossible to distinguish between
fluctuations and real effects.

We conclude that truncated SVD should be preferred to selective SVD.

\section{Penalty regularization}

The EM and truncated SVD methods are very intuitive and general. If we have
specific ideas about what we consider as smooth, we can penalize deviations
from the wanted features by introduction of a penalty term $R$ in the
likelihood or LS fit:
\begin{align}
\ln L  &  =\ln L_{stat}-R\;,\label{lnlpenalty}\\
\chi^{2}  &  =\chi_{stat}^{2}+R\;. \label{lnlchi2}%
\end{align}
Here $\ln L_{stat}$ and $\chi_{stat}^{2}$ are the expressions given in
(\ref{likestat}) and (\ref{chistat}).The sign of $R$ is positive such that
with increasing $R$ the unfolded histogram becomes smoother. If we prefer a
uniform distribution, $R$ could be chosen proportional to the norm
$||\theta||^{2}=\sum_{1=1}^{N}\theta_{i}^{2}$. This is the simple Tikhonov
regularization \cite{tikhonov}. Popular are also the entropy regularization
which again favors a uniform solution and the curvature regularization which
prefers a linear distribution. Entropy regularization is frequently applied in
astronomy and was introduced to particle physics by Schmelling. All three
methods have the tendency to reduce the height of peaks and to fill up
valleys, a common feature of all regularization approaches. More sophisticated
penalty functions can be invented if a priori knowledge about the true
distribution is available. In particle physics, distributions often have a
nearly exponential shape. Then one would select a penalty term which is
sensitive to deviations from an exponential distribution.

\subsection{Curvature regularization}

An often applied regularization function $R$ is,
\begin{equation}
R(x)=r_{c}\left(  \frac{\D^{2}f}{\D x^{2}}\right)  ^{2}\;. \label{regukrumm}%
\end{equation}
It increases with the curvature of $f$ and favors a linear unfolded
distribution. The regularization constant $r_{c}$ determines the power of the regularization.

For a histogram of $M$ bins with constant bin width we approximate
(\ref{regukrumm}) by
\begin{equation}
R=r_{c}\sum_{i=2}^{M-1}\frac{(2\theta_{i}-\theta_{i-1}-\theta_{i+1})^{2}%
}{n^{2}}\;. \label{regubin}%
\end{equation}
with $n$ the total number of events and $r_{c}$ the parameter that fixes the
regularization strength.

\begin{figure}
[ptb]
\begin{center}
\includegraphics[
trim=0.140229in 0.186650in 0.142970in 0.155688in,
height=2.2076in,
width=5.6413in
]%
{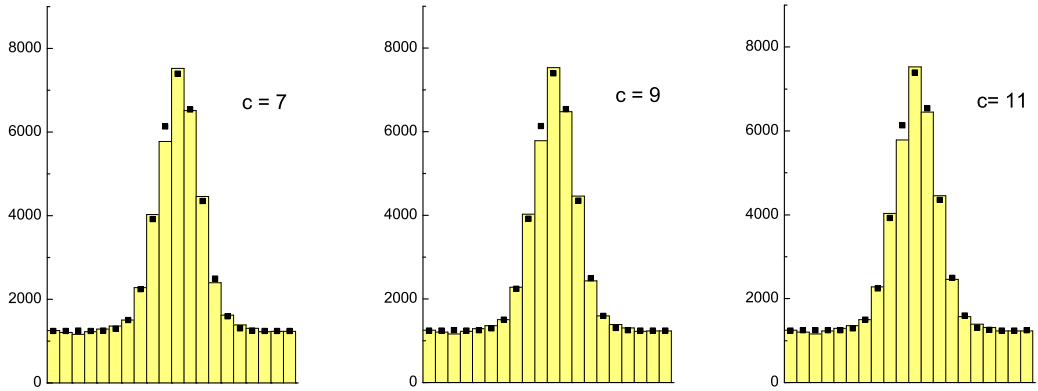}%
\caption{Unfolded histograms with curvature regularization for three different
valus of the regularization constant. The central plot corresponds to the
smallest $ISE$. $50000$ events have been generated with resolution $\sigma
_{s}=0.04.$}%
\label{unfcur50000r4}%
\end{center}
\end{figure}

\begin{figure}
[ptb]
\begin{center}
\includegraphics[
trim=0.000000in 0.203702in 0.000000in 0.203897in,
height=1.8514in,
width=4.7065in
]%
{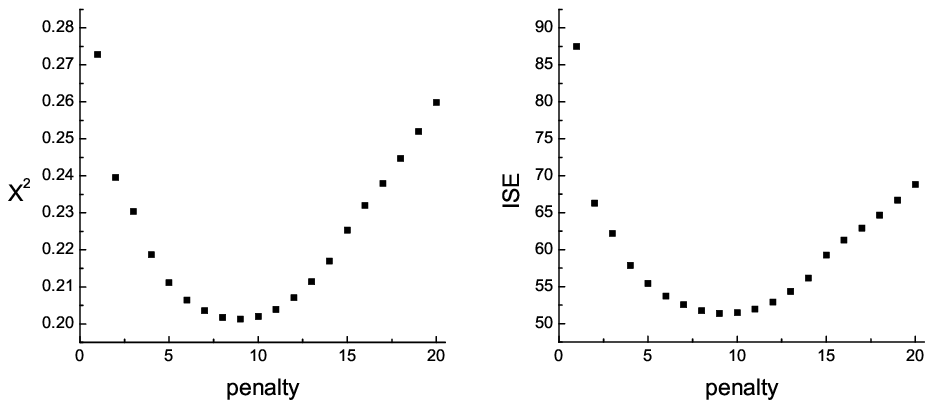}%
\caption{Test quantities $X^{2}$ and the $ISE$ as a function of the
regularization constant for $50000$ generated events and resolution
$\sigma_{s}=0.04$.}%
\label{miscur50000r4}%
\end{center}
\end{figure}

\begin{figure}
[ptbptb]
\begin{center}
\includegraphics[
trim=0.126981in 0.155681in 0.128164in 0.124779in,
height=2.1777in,
width=5.7144in
]%
{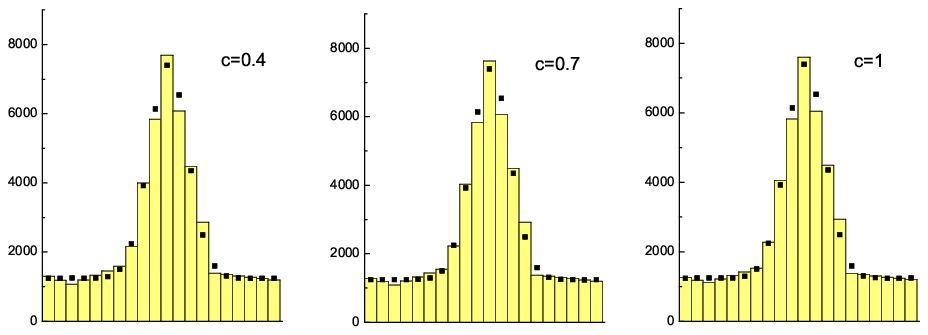}%
\caption{Same as Fig. \ref{unfcur50000r4} but with resolution $\sigma
_{s}=0.08$ and $50000$ events.}%
\label{unfcur50000r8}%
\end{center}
\end{figure}

\begin{figure}
[ptbptbptb]
\begin{center}
\includegraphics[
trim=0.000000in 0.172931in 0.000000in 0.148408in,
height=1.7758in,
width=4.8825in
]%
{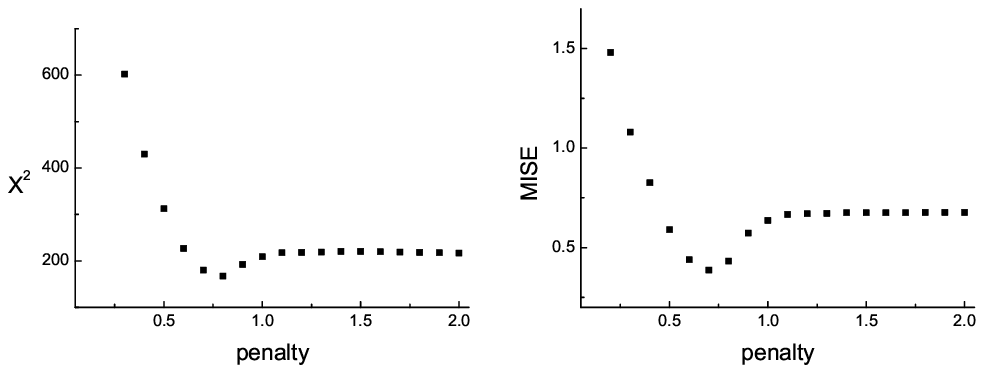}%
\caption{Same as Fig. \ref{miscur50000r4} but for resolution $\sigma_{s}=0.08$
and $50000$ events.}%
\label{miscur50000r8}%
\end{center}
\end{figure}

\begin{figure}
[ptb]
\begin{center}
\includegraphics[
trim=0.136298in 0.163734in 0.138196in 0.121576in,
height=2.3562in,
width=5.6106in
]%
{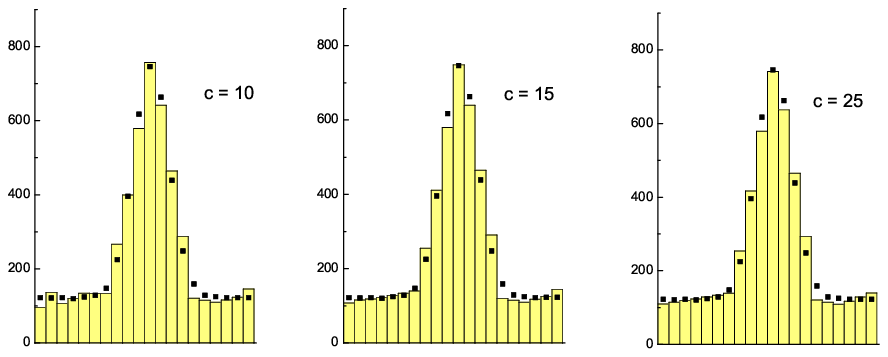}%
\caption{Same as Fig. \ref{unfcur50000r4} but for $5000$ events and
$\sigma_{s}=0.04$.}%
\label{unfcur5000r4}%
\end{center}
\end{figure}

\begin{figure}
[ptbptb]
\begin{center}
\includegraphics[
trim=0.000000in 0.100355in 0.000000in 0.067090in,
height=2.171in,
width=4.9199in
]%
{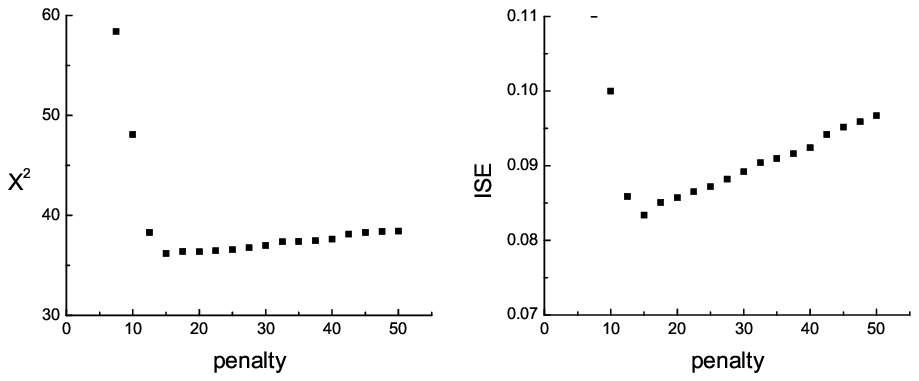}%
\caption{Same as Fig. \ref{miscur50000r4} but for $5000$ events, $\sigma
_{s}=0.04$.}%
\label{miscur5000r4}%
\end{center}
\end{figure}

\begin{figure}
[ptbptb]
\begin{center}
\includegraphics[
trim=0.000000in 0.175354in 0.000000in 0.210502in,
height=2.166in,
width=4.9522in
]%
{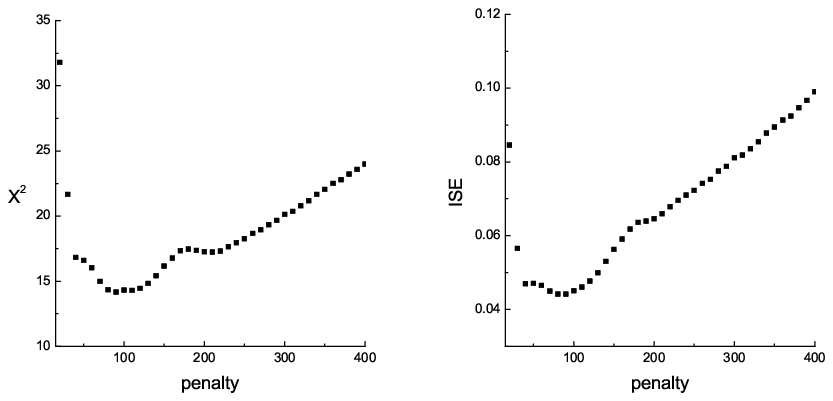}%
\caption{Same as Fig. \ref{miscur50000r4} but for $500$ events, $\sigma
_{s}=0.04$.}%
\label{miscur500r4}%
\end{center}
\end{figure}

\begin{figure}
[ptb]
\begin{center}
\includegraphics[
trim=0.140902in 0.156856in 0.142113in 0.125584in,
height=2.1345in,
width=5.7708in
]%
{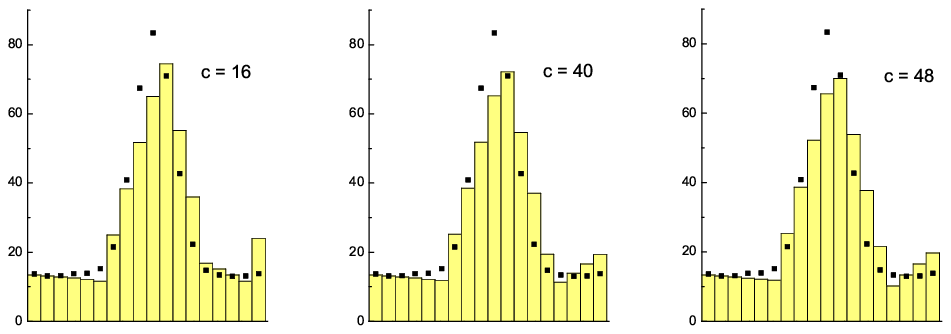}%
\caption{Same as Fig. \ref{unfcur50000r8} but for $500$ events, $\sigma
_{s}=0.08$.}%
\label{unfcur500r8}%
\end{center}
\end{figure}

The Figs. \ref{unfcur50000r4} to \ref{unfcur500r8} show the unfolding results
of the same samples as studied in the previous sections. The central histogram
corresponds to a minimum of the $ISE$. The regularization parameters used to
unfold the histograms are indicated in the figures. (To avoid large numbers,
the numbers are not equal but proportional to $r_{c}$.) Qualitatively the
unfolding results are similar to those obtained with the methods discussed
above. Below each unfolding histogram $X^{2}$ and the $ISE$ are presented for
varying regularization parameters. The dependence of these quantities on the
regularization strength is more complex than in the EM method. The minima of
the test quantities are relatively shallow if the event numbers or the
resolution are low and occasionally secondary minima occur. The agreement of
the unfolded histograms with the true histogram is significantly worse than in
the EM approach.
%

%

%

\begin{figure}
[ptb]
\begin{center}
\includegraphics[
trim=0.140498in 0.161695in 0.142517in 0.116594in,
height=2.0971in,
width=5.5965in
]%
{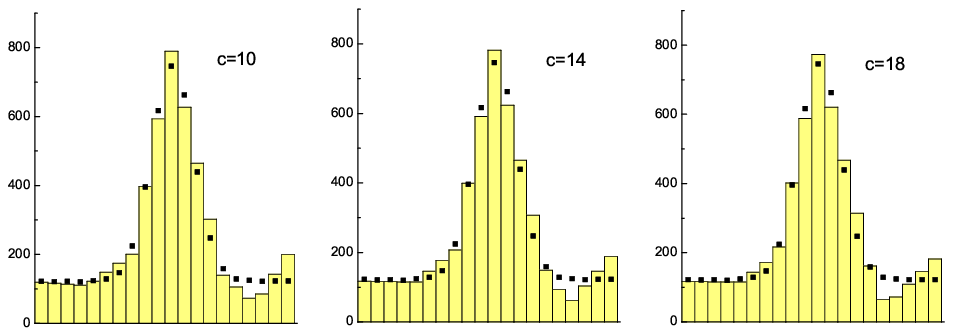}%
\caption{Same as Fig. \ref{unfcur5000r4} but for $\sigma_{s}=0.08$ and $5000$
events.}%
\end{center}
\end{figure}
%

\begin{figure}
[ptb]
\begin{center}
\includegraphics[
trim=0.000000in 0.171202in 0.000000in 0.149161in,
height=1.8223in,
width=4.9298in
]%
{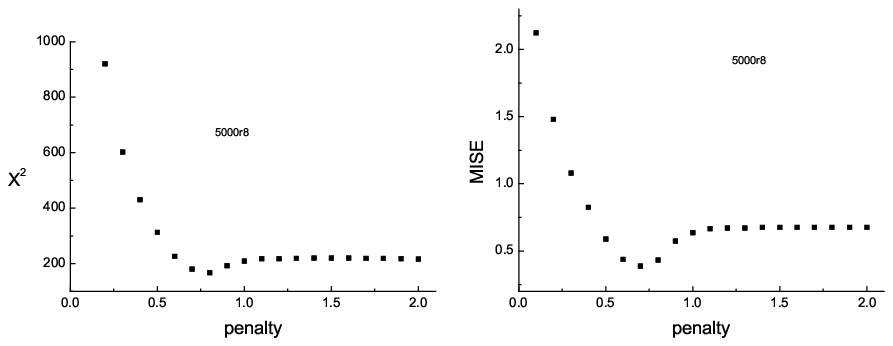}%
\caption{Same as \ref{miscur5000r4} but for $\sigma_{s}=0.08$ and $5000$
events.}%
\end{center}
\end{figure}
%

\begin{figure}
[ptb]
\begin{center}
\includegraphics[
trim=0.135073in 0.182452in 0.136992in 0.152189in,
height=2.269in,
width=5.6604in
]%
{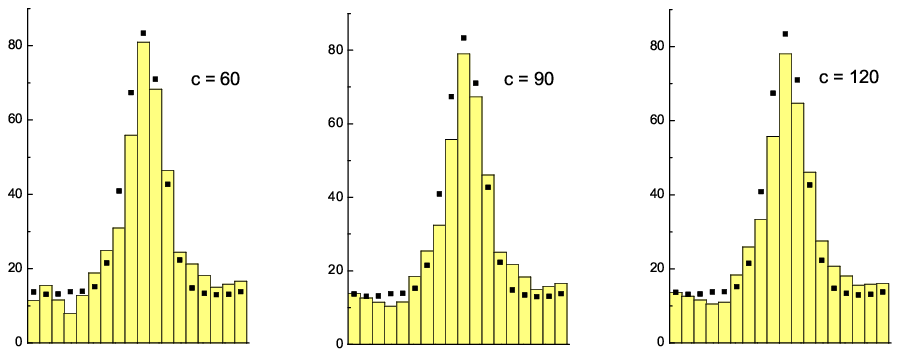}%
\caption{Same as Fig. \ref{unfcur50000r4} but with $500$ events, $\sigma
_{s}=0.04$.}%
\end{center}
\end{figure}
%

%

\begin{figure}
[ptb]
\begin{center}
\includegraphics[
trim=0.000000in 0.207538in 0.000000in 0.207741in,
height=1.9029in,
width=4.8493in
]%
{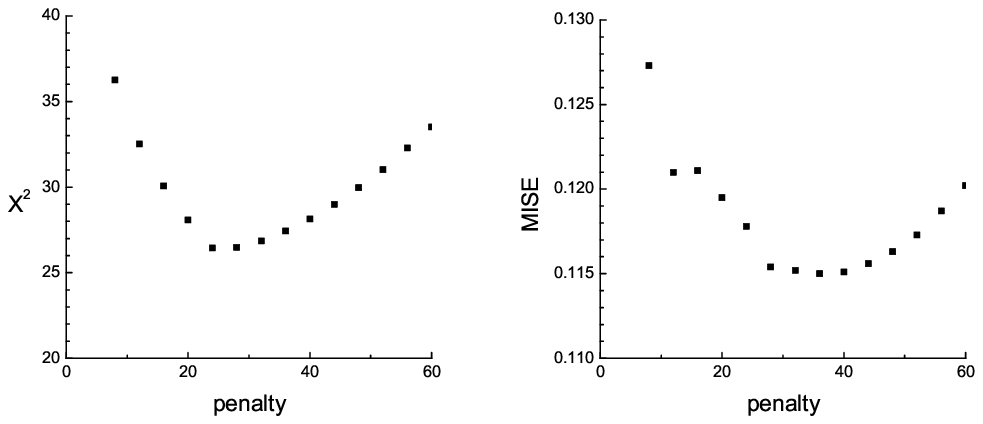}%
\caption{Same as Fig.\ref{miscur50000r8} but for $500$ events, $\sigma
_{s}=0.08$.}%
\label{miscur500r8}%
\end{center}
\end{figure}

\begin{figure}
[ptb]
\begin{center}
\includegraphics[
trim=0.114237in 0.144717in 0.115984in 0.117998in,
height=2.5413in,
width=5.8771in
]%
{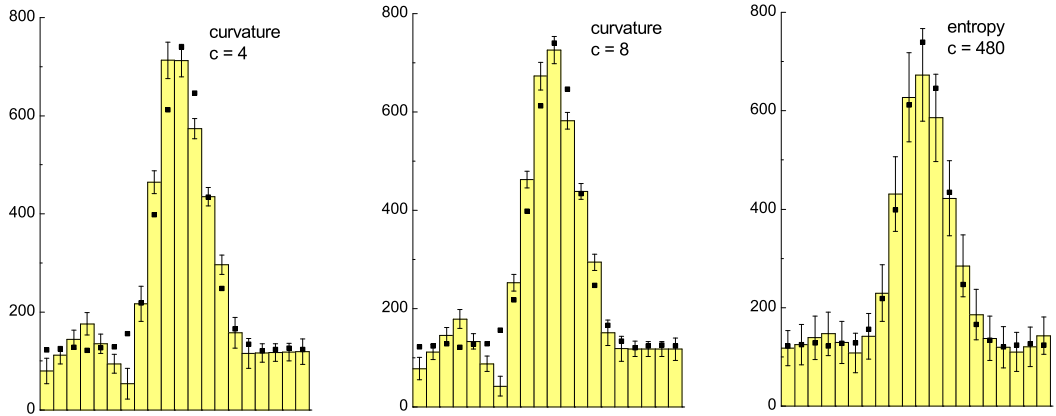}%
\caption{Unfolding with a curvature penalty for two different regularization
strength and comparison with entropy regularization.}%
\label{ucurvspecial}%
\end{center}
\end{figure}

The curvature penalty tries to find a piecewise linear distribution. Sometimes
the unfolding result can be disappointing. An example is presented in Fig.
\ref{ucurvspecial} where a fake triangular peak is generated which is absent
in all other smoothing approaches that have been investigated. To illustrate
the significance of the excess of events, the nominal errors provided by the
fit are included in the graph.%

\subsubsection{Special treatment of the border bins}

The curvature penalty is a function of the content of three adjacent bins. It
is not very efficient at the two border bins of the histogram. In the field of
PDE specific methods have been developed to avoid the problem \cite{scott}.
More smoothing at the edges of the histogram can be achieved by increasing the
bin size of the border bins or by increasing the penalty. The latter solution
is adopted in \cite{kuusela}.

\subsection{Entropy regularization}

We borough the entropy concept from thermodynamics, where the entropy $S$
measures the randomness of a state and the maximum of $S$ corresponds to the
equilibrium state which is the state with the highest probability. It has also
been introduced into information theory and into Bayesian statistics to fix
prior probabilities. However, there is no intuitive argument why the entropy
should be especially suited to cure the fake fluctuations caused by the noise.
It is probably the success of the entropy concept in other fields and its
relation to probability which have been at the origin of its application in
unfolding problems. We penalize a low entropy and thus favor a uniform distribution.

The entropy $S$ of a discrete distribution with probabilities $p_{i}%
\;,i=1,\ldots,M$ is defined through the relation:
\[
S=-\sum_{i=1}^{M}p_{i}\ln p_{i}\;.
\]

For a random distribution the probability for one of the $n=\Sigma\theta_{i}$
events to fall into true bin $i$ is given by $\theta_{i}/n$. The maximum of
the entropy corresponds to an uniform population of the bins, i.e. $\theta
_{i}=const.=n/M$, and equals $S_{max}=\frac{1}{M}\ln M$, while its minimum
$S_{min}=0$ is found for the one-point distribution (all events in the same
bin $j$) $\theta_{i}=n\delta_{i,j}$. We define the entropy regularization
penalty with the regularization strength $r_{e}$ of the distribution by
\begin{equation}
R=r_{e}\sum_{i=1}^{M}\frac{\theta_{i}}{n}\ln\frac{\theta_{i}}{n}\;.
\label{reguent}%
\end{equation}

Adding a term proportional to $R$ to $\chi^{2}$ or subtracting it from $\ln L$
can be used to smoothen a distribution.

A draw-back of a regularization based on the entropy or the norm is that
distant bins are related, while smearing is a local effect. Entropy
regularization is popular in astronomy \cite{nara86,maga98}. It has been
introduced to particle physics applications by Schmelling \cite{schmelling}.

%

\begin{figure}
[ptb]
\begin{center}
\includegraphics[
trim=0.148341in 0.165004in 0.149584in 0.132037in,
height=2.0498in,
width=5.623in
]%
{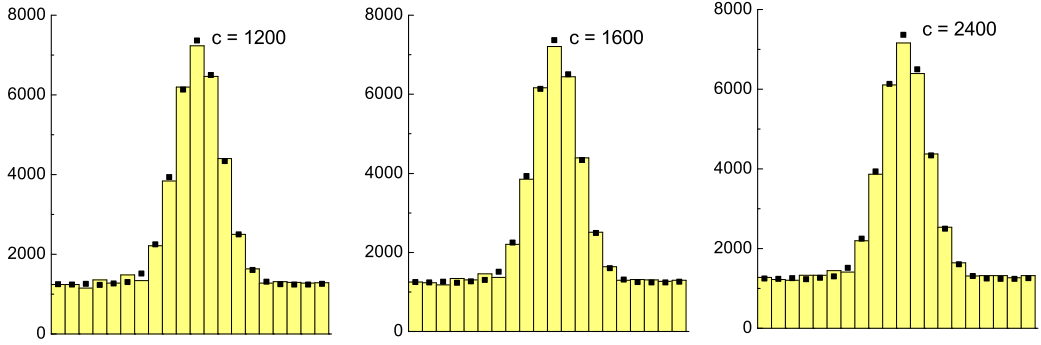}%
\caption{Unfolded histograms with entropy regularization for three different
valus of the regularization constant. The central plot corresponds to the
smallest $ISE$. $50000$ events have been generated with resolution $\sigma
_{s}=0.04.$}%
\label{uent50000r4}%
\end{center}
\end{figure}
%

\begin{figure}
[ptb]
\begin{center}
\includegraphics[
trim=0.000000in 0.168679in 0.000000in 0.126653in,
height=1.8721in,
width=4.802in
]%
{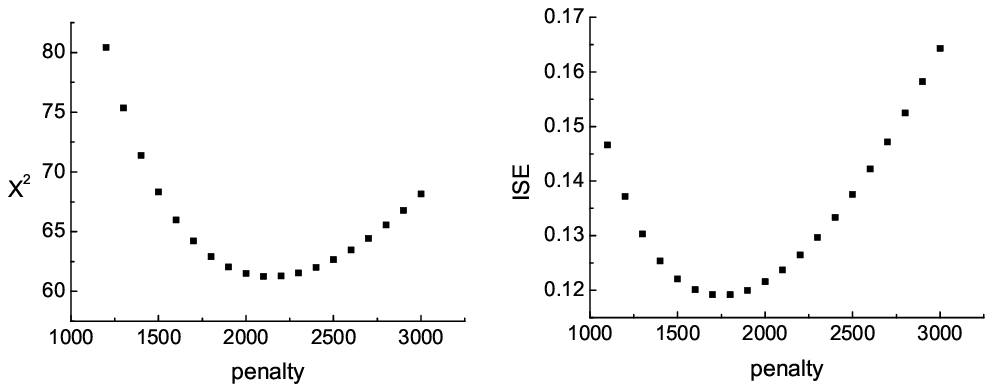}%
\caption{Test quantities $X^{2}$ and $ISE$ as a function of the regularization
constant for $50000$ generated events and resolution $\sigma_{s}=0.04$.}%
\label{misent50000r4}%
\end{center}
\end{figure}
%

\begin{figure}
[ptb]
\begin{center}
\includegraphics[
trim=0.142730in 0.160191in 0.143953in 0.128287in,
height=2.1071in,
width=5.7135in
]%
{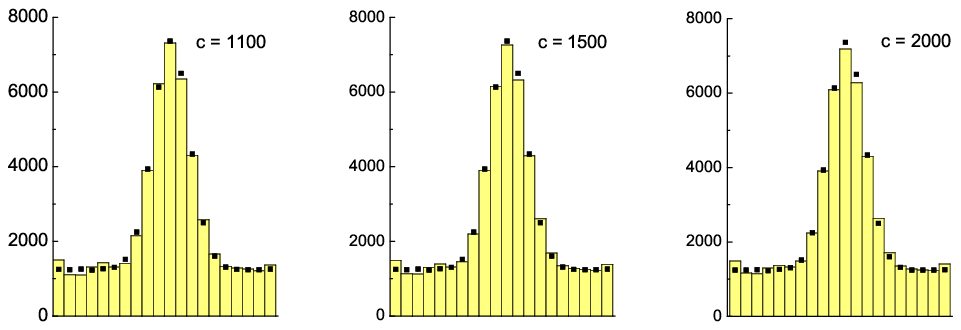}%
\caption{Same as Fig. \ref{uent50000r4} but with resolution $\sigma_{s}=0.08$
and $50000$ events.}%
\end{center}
\end{figure}
%

\begin{figure}
[ptb]
\begin{center}
\includegraphics[
trim=0.000000in 0.170759in 0.000000in 0.128264in,
height=1.8904in,
width=4.7953in
]%
{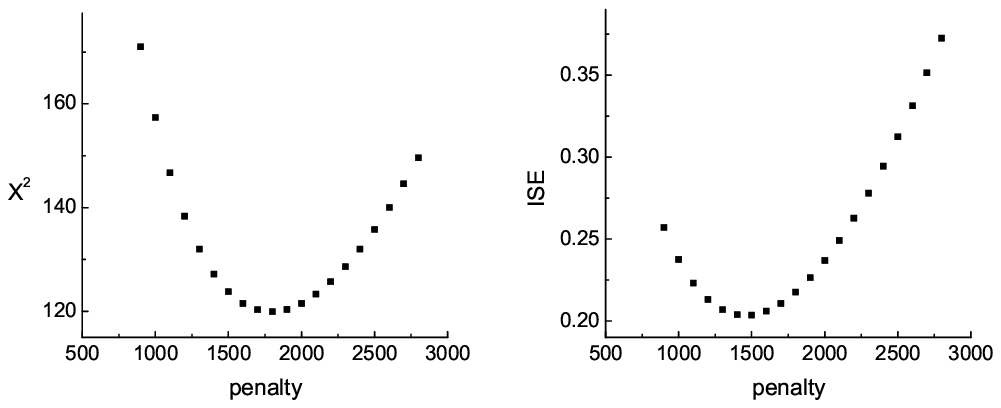}%
\caption{Same as Fig. \ref{misent50000r4} but for resolution $\sigma_{s}=0.08$
and $50000$ events.}%
\end{center}
\end{figure}
%

\begin{figure}
[ptb]
\begin{center}
\includegraphics[
trim=0.139577in 0.161654in 0.140784in 0.129555in,
height=2.0589in,
width=5.5815in
]%
{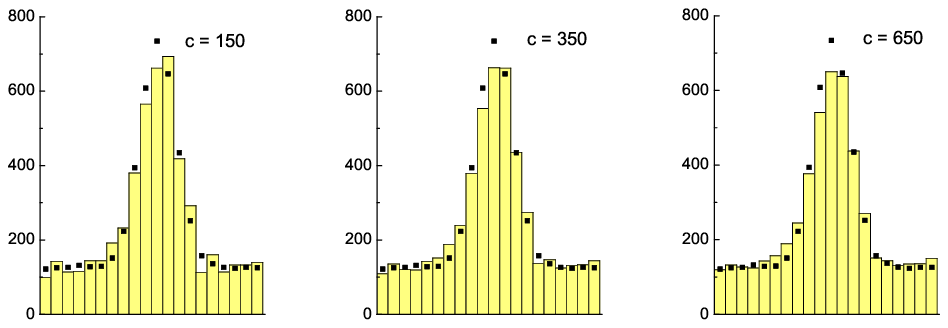}%
\caption{Same as Fig. \ref{uent50000r4} but with $5000$ events, $\sigma
_{s}=0.04$.}%
\end{center}
\end{figure}
\qquad%

\begin{figure}
[ptb]
\begin{center}
\includegraphics[
trim=0.000000in 0.165553in 0.000000in 0.124213in,
height=1.9303in,
width=4.8028in
]%
{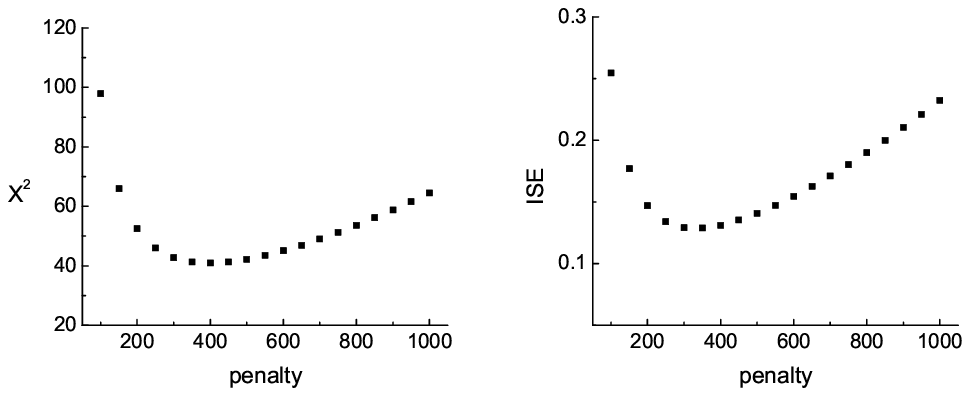}%
\caption{Same as Fig. \ref{misent50000r4} but for $5000$ events, $\sigma
_{s}=0.04$.}%
\end{center}
\end{figure}
%

\begin{figure}
[ptb]
\begin{center}
\includegraphics[
trim=0.143728in 0.159523in 0.144939in 0.127786in,
height=2.1179in,
width=5.6712in
]%
{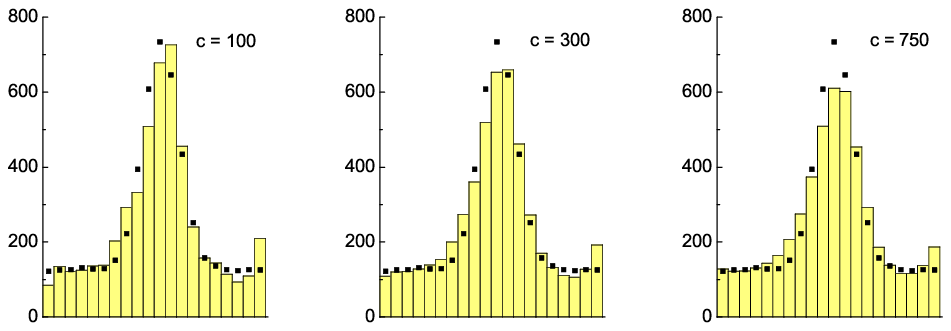}%
\caption{Same as Fig. \ref{uent50000r4} but with $5000$ events and resolution
$\sigma_{s}=0.08$.}%
\end{center}
\end{figure}
%

\begin{figure}
[ptb]
\begin{center}
\includegraphics[
trim=0.183470in 0.157409in 0.184730in 0.125967in,
height=2.1154in,
width=4.7073in
]%
{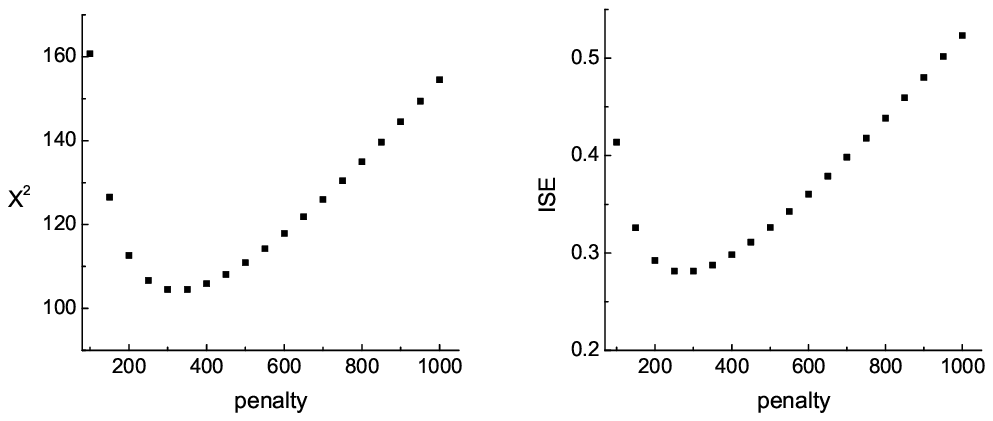}%
\caption{Same as Fig. \ref{misent50000r4} but for $5000$ events and resolution
$\sigma_{s}=0.08$.}%
\end{center}
\end{figure}
%

\begin{figure}
[ptb]
\begin{center}
\includegraphics[
trim=0.138252in 0.158521in 0.139447in 0.126950in,
height=2.132in,
width=5.6405in
]%
{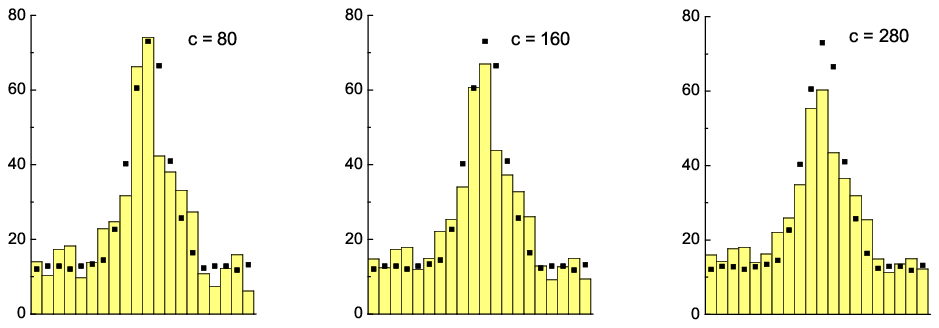}%
\caption{Same as Fig. \ref{uent50000r4} but with $500$ events, $\sigma
_{s}=0.04$.}%
\end{center}
\end{figure}
%

\begin{figure}
[ptb]
\begin{center}
\includegraphics[
trim=0.000000in 0.171010in 0.000000in 0.128305in,
height=1.8107in,
width=4.7073in
]%
{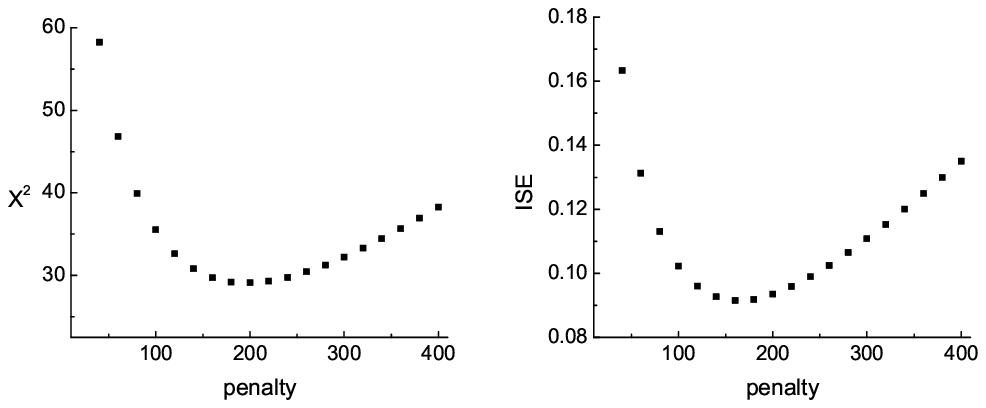}%
\caption{Same as Fig. \ref{misent50000r4} but for $500$ events, $\sigma
_{s}=0.04$.}%
\end{center}
\end{figure}
%

\begin{figure}
[ptb]
\begin{center}
\includegraphics[
trim=0.183273in 0.155681in 0.184468in 0.124779in,
height=2.1768in,
width=5.599in
]%
{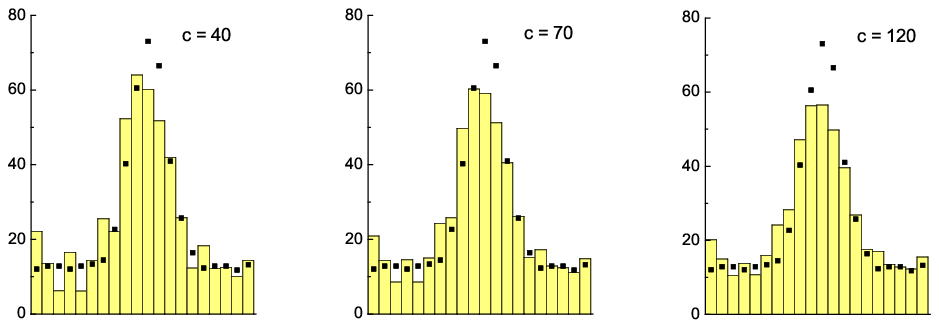}%
\caption{Same as Fig. \ref{uent50000r4} but with $500$ events and resolution
$\sigma_{s}=0.08$.}%
\end{center}
\end{figure}
%

\begin{figure}
[ptb]
\begin{center}
\includegraphics[
trim=0.000000in 0.168188in 0.000000in 0.126423in,
height=1.8447in,
width=4.7837in
]%
{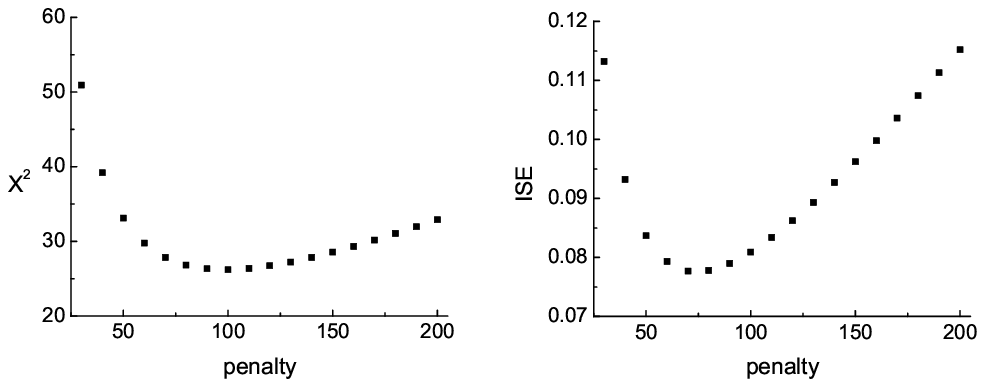}%
\caption{Same as Fig. \ref{misent50000r4} but for $500$ events and resolution
$\sigma_{s}=0.08$.}%
\label{misent500r8}%
\end{center}
\end{figure}

Simulation results are summarized in the same way as for the curvature
regularization in Figs. \ref{uent50000r4} to \ref{misent500r8}.

\subsection{Tikhonov or norm regularization}

The most obvious and simplest way to regularize unfolding results is to
penalize a large value of the norm squared $||\vec{\theta}||^{2}$ of the
solution:%
\begin{equation}
R=\frac{r_{n}}{n^{2}}%
{\displaystyle\sum\limits_{i=1}^{M}}
\theta_{i}^{2}\;. \label{regunorm}%
\end{equation}
The norm regularization has first been proposed by Tikhonov \cite{tikhonov}.
Minimizing the norm implies a bias towards a small number of events in the
unfolded distribution. To avoid this effect, contrary to the originally
proposed penalty, we normalize the norm to the number of events squared
$n^{2}$. The normalized norm can still bias the result, but in the following
examples the bias is negligible.%

\begin{figure}
[ptb]
\begin{center}
\includegraphics[
trim=0.145248in 0.148465in 0.146383in 0.118936in,
height=2.2549in,
width=5.6704in
]%
{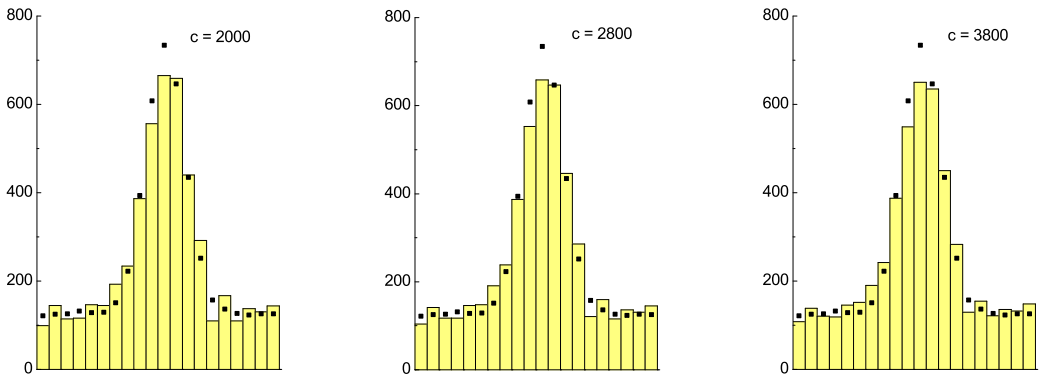}%
\caption{Norm regularization for $5000$ events and resolution $\sigma
_{e}=0.04$. The central plot correspond to the minimum of the $ISE$. The value
of the regularization parameter is indicated in each plot.}%
\label{unor5000r4}%
\end{center}
\end{figure}
%

\begin{figure}
[ptb]
\begin{center}
\includegraphics[
trim=0.000000in 0.137862in 0.000000in 0.092040in,
height=2.1038in,
width=4.6808in
]%
{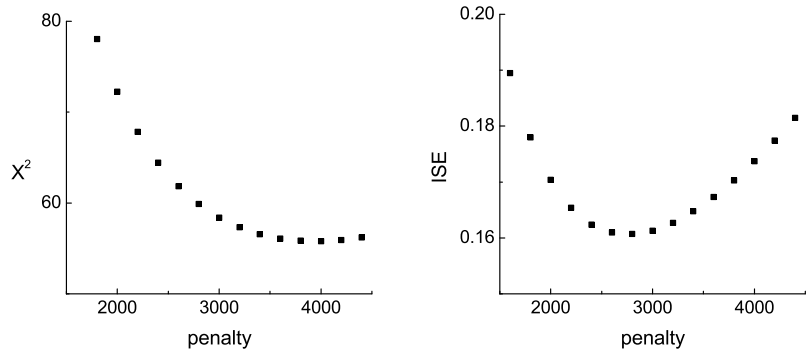}%
\caption{Distribution of $X^{2}$ and $ISE$ as a function of the norm
regularization parameter with $5000$ events and resolution $\sigma_{e}=0.04$.}%
\label{misnor5000r4}%
\end{center}
\end{figure}
%

\begin{figure}
[ptb]
\begin{center}
\includegraphics[
trim=0.139196in 0.149778in 0.140331in 0.119920in,
height=2.2316in,
width=5.6372in
]%
{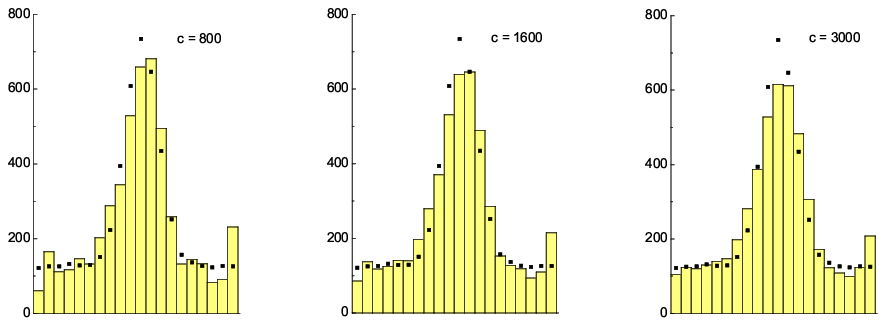}%
\caption{same as Fig. \ref{unor5000r4} but for resolution $\sigma_{s}=0.08$
and $5000$ events.}%
\end{center}
\end{figure}
%

\begin{figure}
[ptb]
\begin{center}
\includegraphics[
trim=0.183633in 0.138034in 0.184029in 0.092088in,
height=2.2972in,
width=4.7214in
]%
{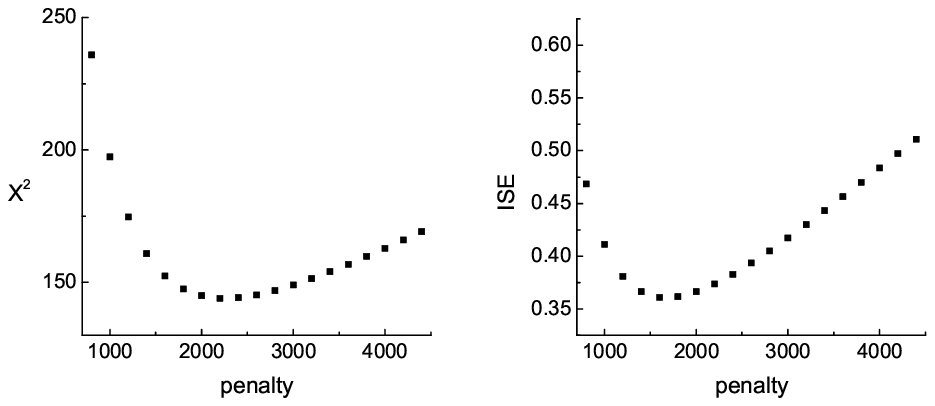}%
\caption{same as Fig. \ref{misnor5000r4} but for resolution $\sigma_{s}=0.08$
and $5000$ events.}%
\end{center}
\end{figure}
%

\begin{figure}
[ptb]
\begin{center}
\includegraphics[
trim=0.000000in 0.159949in 0.000000in 0.128123in,
height=2.0656in,
width=5.6704in
]%
{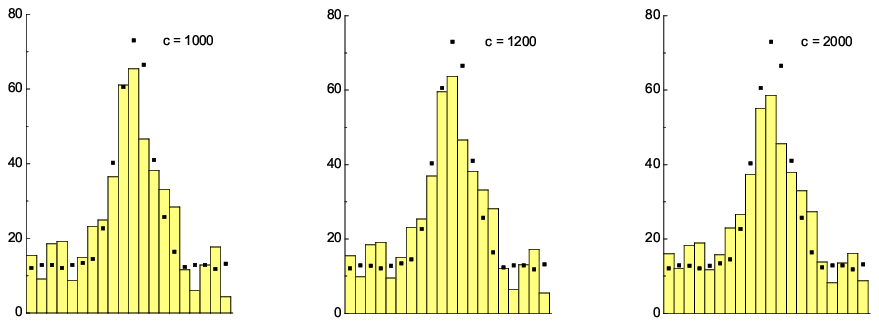}%
\caption{Same as Fig. \ref{unor5000r4} but for $500$ events, $\sigma_{s}%
=0.04$.}%
\end{center}
\end{figure}
%

\begin{figure}
[ptb]
\begin{center}
\includegraphics[
trim=0.000000in 0.119406in 0.000000in 0.079796in,
height=2.1137in,
width=4.7638in
]%
{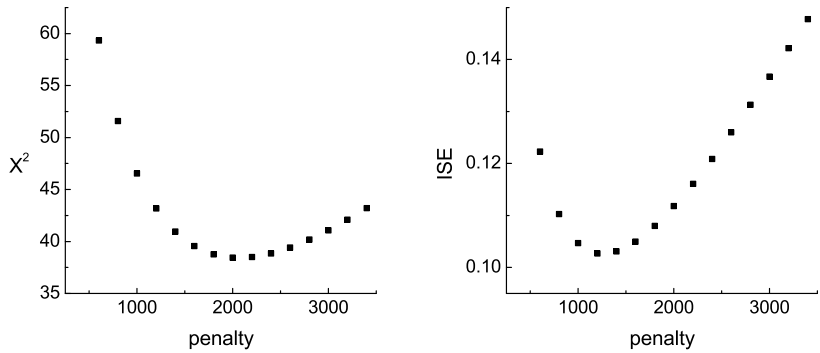}%
\caption{Same as Fig. \ref{misnor5000r4} but for $500$ events, $\sigma
_{s}=0.04$.}%
\label{misnor500r4}%
\end{center}
\end{figure}

The unfolding results, especially in the case of $500$ events, are less
convincing than those of the other penalty methods. We notice a large
difference of the penalty values which minimize $X^{2}$ and $ISE$.

\section{Spline approximations}

Simulations of particle experiments often are based on PDEs. For instance the
proton structure functions are required to predict cross sections in proton
proton collisions at the Large Hadron Collider at CERN. For these kind of
simulations coarse binned histograms are not optimal and smooth unfolding
results are preferred which can be obtained with spline approximations.
Unfolding to spline functions was first introduced by Blobel \cite{blobel85}.
Spline approximations are also proposed in \cite{kuusela} in conjunction with
a curvature penalty and in \cite{dembinski} with entropy regularization. In
all three cases the unfolded distribution is approximated by cubic
$b$-splines. It is not obvious though that cubic splines are better suited
than quadratic splines. The latter are more stable at the border bins of the
unfolded histogram. For steeply falling distributions like transverse momentum
distributions approximations by linear $b$-splines are appropriate.

The representation of the unfolded function by a superposition of spline
functions reduces the dependence of the unfolding result on the function used
in the simulation of the response matrix. In the methods with penalty
regularization, the construction of a response matrix and the dependence of
the unfolding result on the distribution used in the Monte Carlo distribution
can be avoided altogether with the parameter estimation method explained in
Chapter 2. This possibility is also realized in \cite{dembinski} in
conjunction with entropy regularization.

It has to be noted that independently of the regularization a systematic error
is introduced by the fact that the true distribution is approximated by the
spline curve. It can happen that this approximation is poor but normally it is
excellent within the statistical uncertainties.

The relevant formulas are given in Sect. \ref{splineexpansion}.

Once the elements of the response matrix have been computed, the unfolding
proceeds in the same way as with histograms. The unfolding procedure is
completely analogous to that with histograms. The coefficients $\beta_{i}$ are
fitted to the observed data vector $\vec{d}$ in the same way as in the
histogram representation. Alternatively, the coefficients of the $b$-splines
can be directly fitted (see Sect. \ref{weighted}).

\subsection{Curvature penalty}

The curvature penalty of the spline representation is computed following the
analytic method used in \cite{kuusela}. The total squared curvature $R$%
\begin{align*}
R  &  =%
{\displaystyle\sum\limits_{ij}}
\beta_{i}\Omega_{ij}\beta_{j}\;,\\
\Omega_{ij}  &  =\int B_{i}^{\prime\prime}(x)B_{j}^{\prime\prime}(x)dx\;,
\end{align*}
is a simple function of the second derivatives of the cubic $b$-splines.

The fit of the $b$-splines to the data is much less problematic than that of
histograms. No secondary minima of $ISE$ are observed.%

\begin{figure}
[ptb]
\begin{center}
\includegraphics[
trim=0.132337in 0.184519in 0.133544in 0.184814in,
height=4.073in,
width=5.9053in
]%
{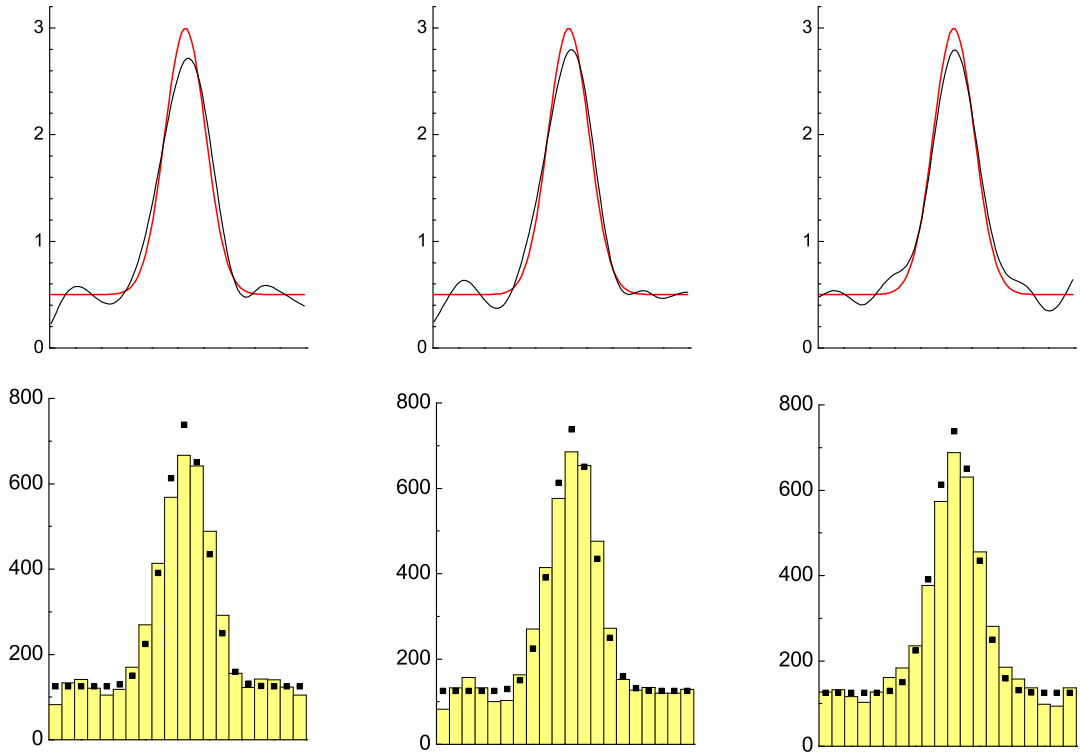}%
\caption{Unfolding of three samples to a cubic spline function with curvature
penalty. The bottom plots are projections of the spline distribution into
histogram bins.}%
\label{curvespline}%
\end{center}
\end{figure}

\begin{myexample}
We apply the method to three data samples of our standard one-peak example
with $5000$ events and smearing resolution $\sigma_{s}=0.08$. Twenty cubic
$B$-splines are fitted to the data. Each time a curvature penalty is applied
such that the $ISE$ is minimal.

In Fig. \ref{curvespline} the unfolding results are compared to the true
distribution. The unfolded distributions are transformed back into the
histogram presentation and the $ISE$ values for the histograms are computed.
In this way we are able to compare the results to the direct histogram fits.
The resulting values are $0.184$, $0.134$ and $0.121$. The $ISE$ values
$0.325$, $0.279$ and $0.445$ from the direct histogram fit are in all cases
significantly worse than those of the spline fits.
\end{myexample}

\subsection{EM unfolding}

The spline approximation can be implemented in all unfolding methods with
minimal changes in the computer programs. Of special interest is the
performance of the iterative method which is especially successful in the
histogram representation. The likelihood of the spline representation converge
to the MLE$.$

\begin{figure}
[ptb]
\begin{center}
\includegraphics[
trim=0.147183in 0.196607in 0.148430in 0.196900in,
height=3.7509in,
width=5.9568in
]%
{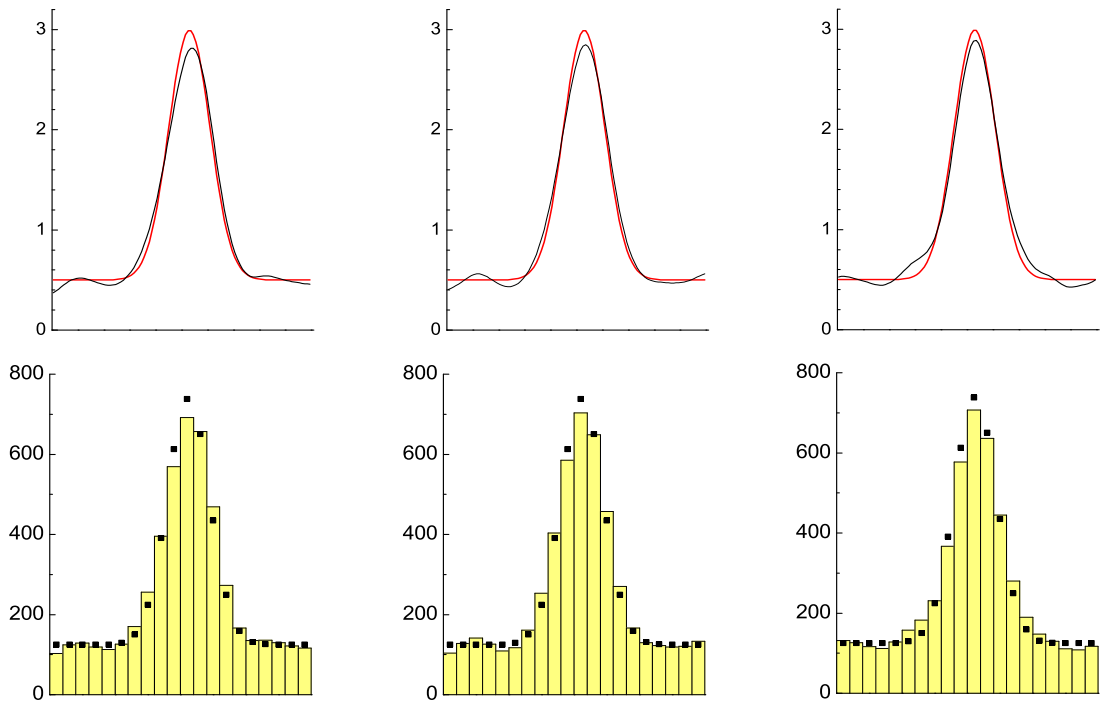}%
\caption{Same as Fig. \ref{curvespline} but with EM unfolding.}%
\label{iterspline}%
\end{center}
\end{figure}

\begin{figure}
[ptb]
\begin{center}
\includegraphics[
trim=0.000000in 0.203172in 0.000000in 0.149014in,
height=2.0548in,
width=2.8792in
]%
{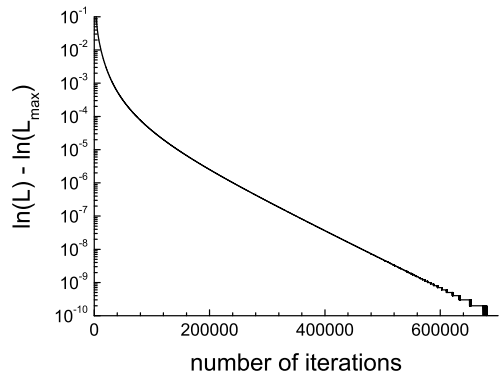}%
\caption{Difference of the log-likelihood from the value at $10^{6}$
iterations as a function of the number of iterations. }%
\label{splineconverge}%
\end{center}
\end{figure}

\begin{myexample}
In Fig. \ref{iterspline} results for the EM unfolding are depicted. The same
data sets are used as in Fig. \ref{curvespline}. The convergence of the
log-likelihood is displayed in Fig. \ref{splineconverge}. The convergence is
initially fast and then the residual value decays exponentially.

Compared to curvature regularization the bias is smaller and the size of the
wiggles at the borders is reduced. The $ISE$ values for the three samples
$0.081$, $0.053$ and $0.080$ are substantially lower than those obtained with
the curvature regularization. This may partially be due to the uniform
starting distribution which suppresses fluctuations in the flat region of the distribution.
\end{myexample}

More detailed studies are necessary to establish the promising performance of
the EM-unfolding into a superposition of $b$-splines.%

%

\section{Quality tests of the unfolding methods and comparison}

A comparison of regularization methods has to be based on selected examples.
Examples can be chosen such that a specific penalty function favors the
corresponding distribution and thus a comparison is always partially biased.
Furthermore, a quantitative comparison of the different unfolding methods is
difficult, because often clear rules how to choose the regularization
strengths are missing. To avoid this problem, we always select the
regularization such that the $ISE$ is minimal. In the following we compare in
the Monte Carlo simulations always the best (relative to the $ISE)$ achievable
unfolding results. To judge the quality, we compare the $ISE$, $X^{2}$ and
$\chi^{2}$ which is computed by folding the solution and comparing it to the
observed histogram. In all cases the reference distribution is assumed to be
uniform. The observed and the true histogram contain always $40$ and $20$
bins, respectively. While the value of the $ISE$ depends on the binning, the
quality comparison of the methods is expected to be rather insensitive to it.

We restrict ourselves to the basic versions of the EM method, the simple TSVD
and to ML fits with curvature, entropy and norm penalty terms. The adjustment
of the regularization parameter to a minimal $ISE$ requires many fits. For
this reason, we limit the comparison to only a few examples.

\subsection{The one-peak distribution}

The results for the one-peak example are summarized in Table \ref{tabx2iser4}
for smearing resolution $\sigma_{s}=0.04$ and in Table \ref{tabx2iser8} for
$\sigma_{s}=0.08$. The last line shows the result when the observed
distribution is taken as true distribution. It is comforting that with the
exception of one case, the considered unfolding approaches produces values of
the two test quantities that are better than those obtained without unfolding.

The winner of the comparison is the EM method. There is no clear tendency
which would allow a classification of the other approaches. A curvature
regularization seems to be more successful than the norm regularization. It is
astonishing that the performance of truncated SVD is relatively good for the
cases with only $500$ events where we would expect that the linear least
square fit is inferior to the methods based on the likelihood function.

Another quantity of interest is the unfolding bias with respect to the total
number of events. Truncated SVD regularization in the $500$ event sample
looses $7\%$ of the events for resolution $\sigma_{s}=0.4$ and $5\%$ for
$\sigma_{s}=0.8$. The bias is below $1\%$ in all other methods.

\begin{table}[ptb]
\caption{One-peak example, $X^{2}$, $ISE$ and $\chi^{2}$, resolution
$\sigma_{s}=0.04$}%
\label{tabx2iser4}
\begin{center}%
\begin{tabular}
[c]{|l|c|c|c|c|c|c|c|c|c|}\hline
events & \multicolumn{3}{c|}{500} & \multicolumn{3}{c|}{5000} &
\multicolumn{3}{c|}{50000}\\\hline
statistic & $X^{2}$ & $ISE$ & $\chi^{2}$ & $X^{2}$ & $MISE$ & $\chi^{2}$ &
$X^{2}$ & $MISE$ & $\chi^{2}$\\\hline
EM & $15.1$ & $0.049$ & $36$ & $24.3$ & $0.049$ & $36$ & $24.6$ & $0.067$ &
$24$\\
TSVD & $25.0$ & $0.078$ & $34$ & $35.2$ & $0.098$ & $37$ & $71.2$ & $0.183$ &
$21$\\
curvature & $14.2$ & $0.044$ & $42$ & $36.2$ & $0.083$ & $35$ & $51.4$ &
$0.202$ & $24$\\
entropy & $34.0$ & $0.122$ & $50$ & $51.5$ & $0.133$ & $41$ & $91.9$ & $0.237$
& $29$\\
norm & $42.4$ & $0.123$ & $49$ & $90.3$ & $0.206$ & $40$ & $116$ & $0.262$ &
$29$\\
observed & $118$ & $0.548$ & $483$ & $70.9$ & $0.191$ & $4820$ & $454$ &
$1.230$ & $-$\\\hline
\end{tabular}
\end{center}
\end{table}

\begin{table}[ptb]
\caption{One-peak example, $X^{2}$, $ISE$ and $\chi^{2}$, resolution
$\sigma_{s}=0.08$}%
\label{tabx2iser8}
\begin{center}%
\begin{tabular}
[c]{|l|c|c|c|c|c|c|c|c|c|}\hline
events & \multicolumn{3}{c|}{500} & \multicolumn{3}{c|}{5000} &
\multicolumn{3}{c|}{50000}\\\hline
statistic & $X^{2}$ & $ISE$ & $\chi^{2}$ & $X^{2}$ & $MISE$ & $\chi^{2}$ &
$X^{2}$ & $MISE$ & $\chi^{2}$\\\hline
EM & $23.9$ & $0.098$ & $29$ & $73.3$ & $0.131$ & $47$ & $41.2$ & $0.096$ &
$35$\\
TSVD & $23.1$ & $0.090$ & $28$ & $174.1$ & $0.264$ & $49$ & $253$ & $0.376$ &
$33$\\
curvature & $26.4$ & $0.115$ & $33$ & $117$ & $0.199$ & $47$ & $167$ & $0.388$
& $33$\\
entropy & $31.3$ & $0.132$ & $35$ & $104$ & $0.173$ & $52$ & $154$ & $0.301$ &
$37$\\
norm & $46.5$ & $0.174$ & $36$ & $153$ & $0.224$ & $48$ & $464$ & $1.060$ &
$36$\\
observed & $138$ & $0.558$ & $476$ & $239$ & $0.796$ & $4762$ & $2629$ &
$9.08$ & $-$\\\hline
\end{tabular}
\end{center}
\end{table}

\subsection{A two-peak distribution}

\begin{figure}
[ptb]
\begin{center}
\includegraphics[
trim=0.000000in 0.137872in 0.000000in 0.092112in,
height=3.7418in,
width=5.6762in
]%
{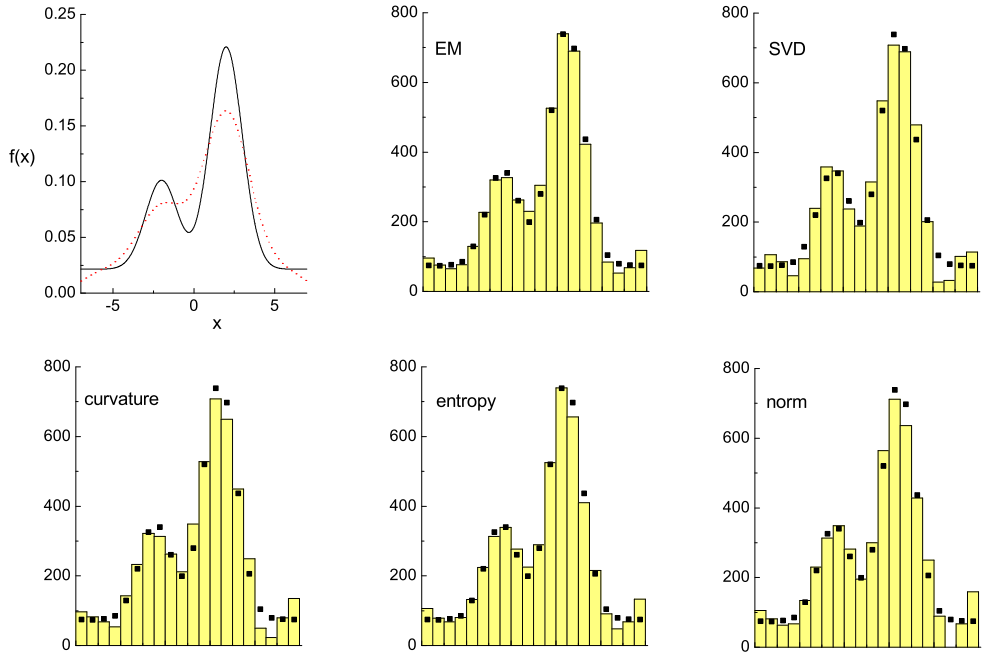}%
\caption{Unfolding results from different methods. The top left-hand plot
shows the true distribution and its smeared version. The squares correspond to
the true distribution.}%
\label{compare2b}%
\end{center}
\end{figure}

\begin{figure}
[ptb]
\begin{center}
\includegraphics[
trim=0.000000in 0.137731in 0.000000in 0.091945in,
height=2.4118in,
width=5.4935in
]%
{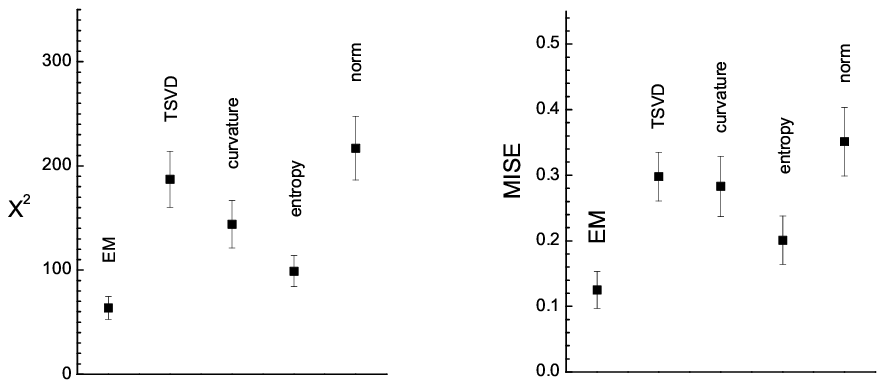}%
\caption{Two-peak distribution: Mean values of $X^{2}$ and $ISE$ from $10$
experiments with $5000$ events each.}%
\label{compare2bmise}%
\end{center}
\end{figure}

\begin{figure}
[ptb]
\begin{center}
\includegraphics[
trim=0.000000in 0.137798in 0.000000in 0.092293in,
height=2.3645in,
width=3.2329in
]%
{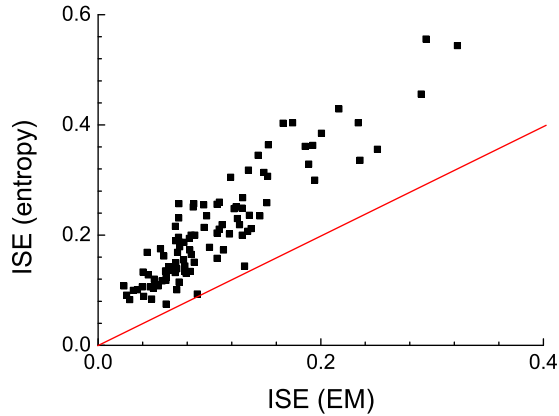}%
\caption{Scatter plot of $ISE$ values obtained from $100$ simulated
experiments for the two-peak example in the EM method and the entropy penalty
fit. }%
\label{compement}%
\end{center}
\end{figure}

We turn to the distribution%
\[
f(x)=0.2\mathcal{N}(-2,1)+0.5\mathcal{N}(2,1)+0.3\mathcal{U}%
\]
defined in the interval $[-7,7]$ with smearing $\sigma_{s}=1$ which has been
used in \cite{kuusela}. The function and its smeared version are displayed in
Fig. \ref{compare2b} top left. The unfolded distributions obtained with the
EM, the truncated SVD and three penalty methods for the first of $10$ samples
with $5000$ events are depicted in the same figure. The optical inspection
does not reveal large differences between the results. The mean values of of
$X^{2}$and $ISE$ from the $10$ samples are presented in Fig.
\ref{compare2bmise}. They indicates that truncated SVD and curvature
regularization perform less well than the other approaches and that the $MISE$
obtained with \ the EM method is significantly smaller than the values of the
competing approaches. In Fig. \ref{compement} the $MISEs$ of the EM iteration
and the entropy penalty fit of $100$ samples are compared. There are large
fluctuations from sample to sample, the results are correlated but the values
of the EM values are always lower than those of the entropy penalty fit
method. The goodness-of-fit statistic $\chi^{2}$ evaluated for the minimum of
the $MISE$ is very similar in all approaches.%

%

%

It has not been attempted to improve also the statistics for the curvature
penalty method to a similar extent. This would be very time consuming, because
to find the minimum of the $ISE$ for this method cannot be automatized.
Instead we repeat the study with the number of events increased to $50000$ and
a different weighting of the contributions to the true distribution:
$f(x)=0.3\mathcal{N}(-2,1)+0.5\mathcal{N}(2,1)+0.2\mathcal{U}$. It is expected
that there the fluctuations of the results from sample to sample are smaller.

\begin{figure}
[ptb]
\begin{center}
\includegraphics[
trim=0.074293in 0.183347in 0.075525in 0.183547in,
height=2.176in,
width=5.2179in
]%
{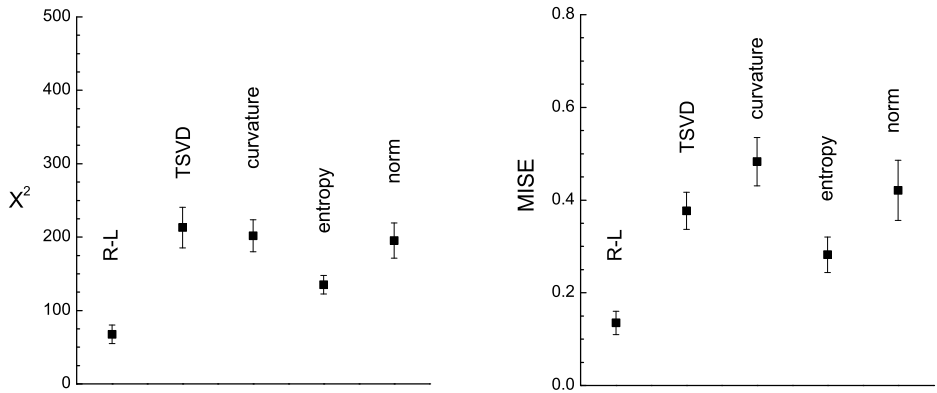}%
\caption{Two-peak distribution: Mean values of $X^{2}$ and $ISE$ from $5$
experiments with $50000$ events each.}%
\label{x2mise50000}%
\end{center}
\end{figure}

In Fig. \ref{x2mise50000} $X^{2}$ and $ISE$ averaged over $5$ samples are
plotted for the considered regularization methods. Again the results of the EM
approach are considerably better than those of all other approaches.%

\subsection{A lifetime distribution}

\begin{figure}
[ptb]
\begin{center}
\includegraphics[
trim=0.126826in 0.188801in 0.128865in 0.189076in,
height=3.6579in,
width=5.8671in
]%
{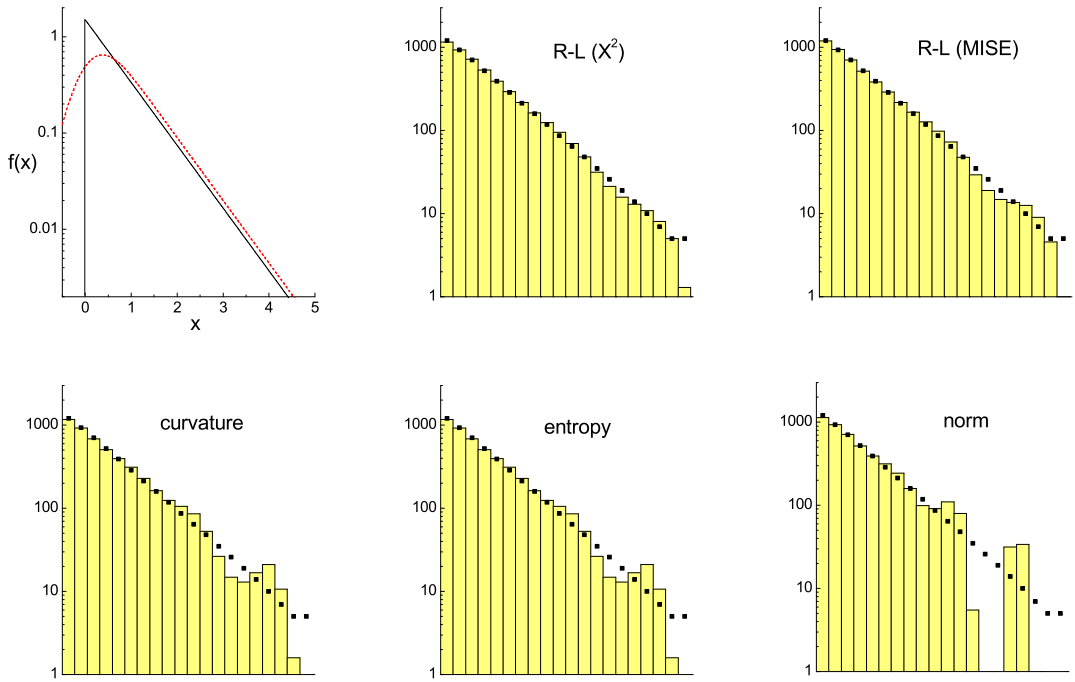}%
\caption{Unfolding an exponential distribution.}%
\label{lifetimeunf}%
\end{center}
\end{figure}

Fig. \ref{lifetimeunf} top left shows an exponential distribution
$f(x)=e^{-x}$, ranging from zero to infinity. It is observed with a resolution
of $\sigma_{s}=1$ in the interval $[-1,5]$ which is subdivided into $40$ bins.
The true histogram contains $20$ bins from which $19$ are $0.25$ units wide.
The $20th$ bin covers all true values from $4.75$ to infinity. In this way it
is guaranteed that all observed values have a true partner. The last wide bin
is excluded from the $ISE$ calculation. $5000$ events have been generated.%

The unfolding results summarized in Fig. \ref{lifetimeunf} demonstrate a
rather good performance of the EM approach. The results for the methods with
curvature and entropy penalties are similar but slightly worse, while norm
regularization is unable to reproduce the true distribution. TSVD fails
technically because the linear LS fit cannot cope with the small event numbers
in some bins of the observed histogram.

Of course the performance of all approaches could be improved: We could fit
the deviation from a first guess of an exponential distribution and penalize
the deviations. In real experiments one would anyway fit the relevant
parameters with the method described in Chapter 2.

\subsection{A pt distribution}

The following example is similar to the previous one but is of interest
because the smearing uncertainty is not constant and because acceptance losses
are introduced explicitly.%

\begin{figure}
[ptb]
\begin{center}
\includegraphics[
trim=0.113937in 0.148679in 0.115099in 0.119270in,
height=2.8991in,
width=5.9153in
]%
{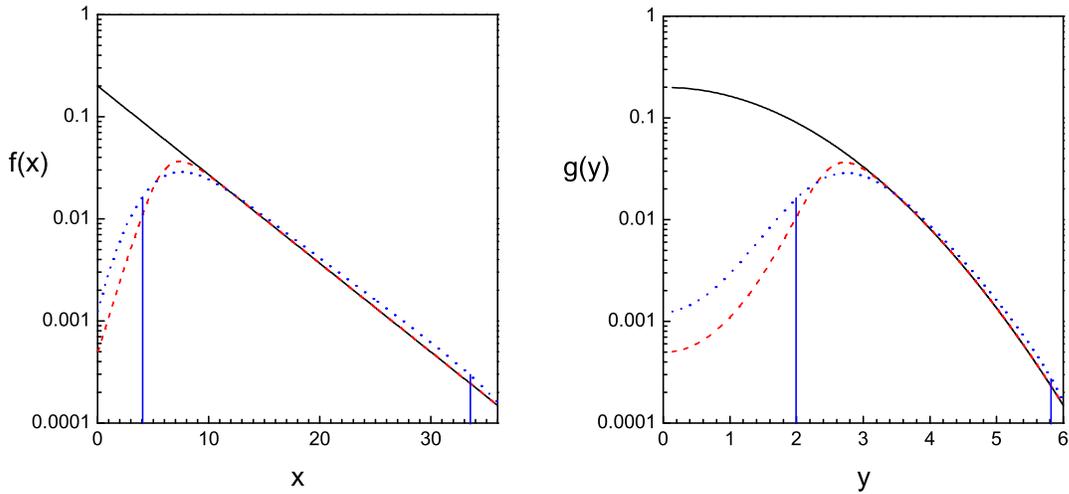}%
\caption{Transverse momentum distribution. The true distribution corresponds
to the full line of the left hand plot. The accepted part is given by the
dashed line. Folding with the resolution function produces the dotted line.
The unfolded histogram is evaluated in the range limited by the vertical
lines. The transformation to the variable $y=\sqrt{x}$ is presented in the
lright-hand plot.}%
\label{ptdistribution}%
\end{center}
\end{figure}
%

\begin{figure}
[ptb]
\begin{center}
\includegraphics[
trim=0.129163in 0.187507in 0.131188in 0.187798in,
height=3.9352in,
width=5.858in
]%
{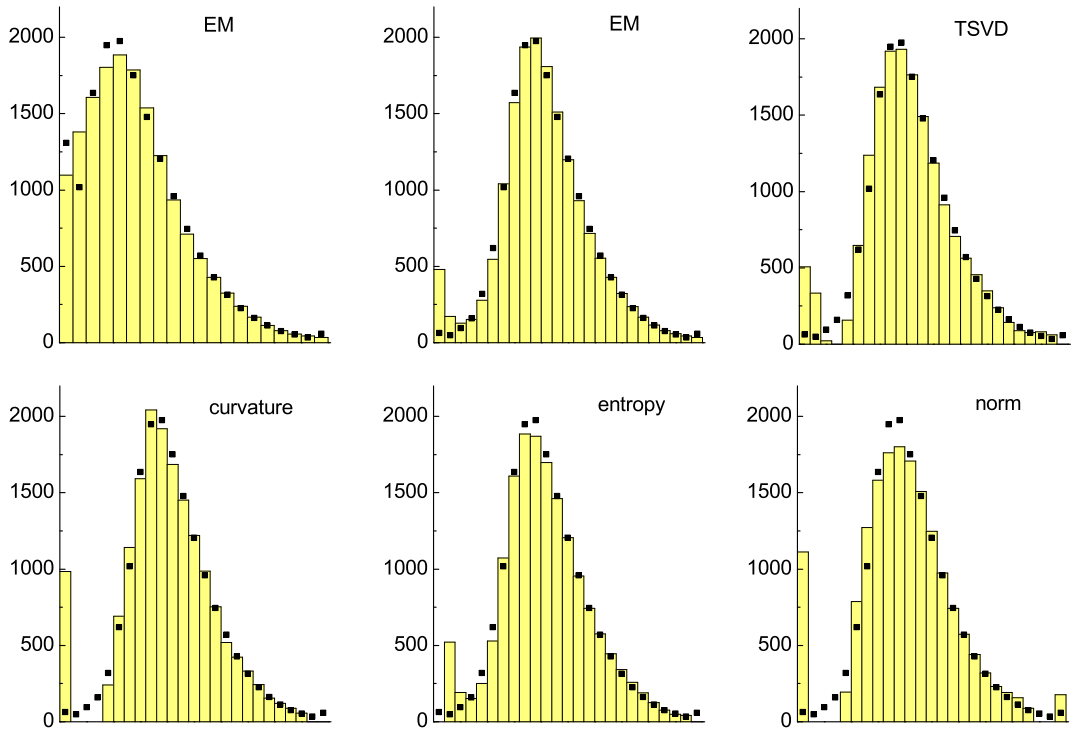}%
\caption{Unfolded pt distributions. The drop at low pt values id due to the
limited acceptance. Top left: EM method, original binning, Top center: EM
method with modified binning. The following plots correspond to TSVD,
curvature, entropy and norm regularization.}%
\label{unfpt}%
\end{center}
\end{figure}
%

\begin{figure}
[ptb]
\begin{center}
\includegraphics[
trim=0.000000in 0.178566in 0.000000in 0.214479in,
height=1.7501in,
width=5.6363in
]%
{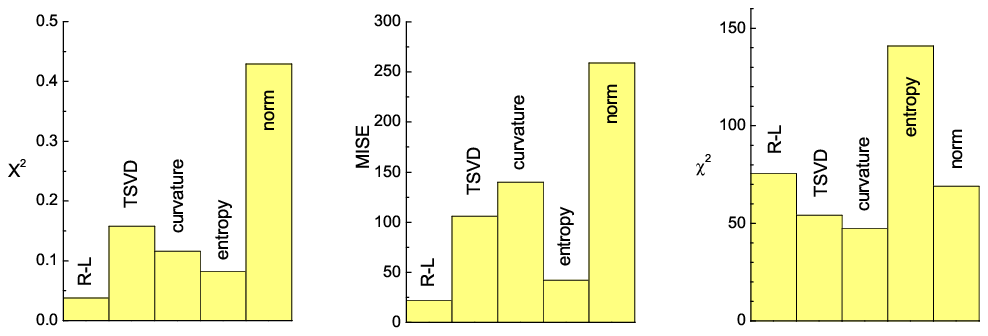}%
\caption{$X^{2}$, $MISE$ and $\chi^{2}$ of different unfolding results of a
$pt$-distribution.}%
\label{misept}%
\end{center}
\end{figure}

The distribution of the transverse momentum squared in particle experiments
follows in many cases approximately an exponential distribution. In our toy
experiment we simulate again a simple exponential $f(x)=\gamma e^{-\gamma x}$
in the interval $[0,\,\infty]$ with $\gamma=0.2$. This time we assume that the
acceptance $\varepsilon$ is low for small values of $x$, $\varepsilon
=1/[1+e^{-(x-2)}]$. The sigmoid function produces $\varepsilon(2)=0.5$,
$\varepsilon(-\infty)=0$ and $\varepsilon(\infty)=1$. The true distribution
$f(x)$, the accepted part and the smeared version are displayed in Fig.
\ref{ptdistribution}$.$

Momentum measurements have the tendency that the error increases with the
momentum. Here we assume $\sigma_{s}(x=1)=0.8$ and $\sigma_{s}\sim\sqrt{x}$.
For the graphical representation and to simplify the programming, it is useful
to have uncertainties which are proportional to the bin size. As in
\cite{blobel} we apply the transformation $y=\sqrt{x}$ to the distribution
$f(x)$ and get $g(y)=2y\gamma e^{-\gamma y^{2}}$ with smearing uncertainty
$\sigma_{sy}=0.4$ independent of $y$. The unfolding is then performed in $y$.
The result can be transformed back to the original variable $x$ where we
obtain a histogram with bin sizes that increase with $x$. We choose the
interval $[2,\,6]$ for the observed variable $y^{\prime}$ with $40$ bins $0.1$
wide and discuss the results in the $y$ system and do not apply the acceptance
correction. This is not the way physicists like the presentation, but it
better suited for the unfolding study. The pure Poisson fluctuations can be
estimated from the event numbers in the linear ordinate and the resolution is
constant across the abscissa. The central $18$ bins of the $20$ bins of the
true histogram are $0.2$ wide and contained in the range $2.2<y<5.8$. The two
border bins cover the intervals $[0,\,2.2]$ and $[5.8,\,\infty]$ and are
discarded after the fit of the unfolded distribution.

The result of the unfolded $y$ histogram for the EM method are shown in the
top left plot of Fig. \ref{unfpt}. The reconstruction of the first two bins is
not satisfactory. The reason is the low efficiency for the number of events in
the first bin. The performance can be improved by subdividing the bin
containing the underflow. The procedure has been repeated with $25$ true bins
and $50$ observed bins where now the underflow bin covers $[0,\,1.2]$. The
remaining bins, starting from $1.2$ are $0.2$ wide except the last bin which
is kept as before. The first $5$ bins and the last bin are not considered in
the unfolding test.

The central plot of the top row in Fig. \ref{unfpt} shows the improved
unfolding result of the EM method and the following plots those of TSVD and
the three penalty methods (curvature, entropy and norm). The results for the
test quantities are displayed in Fig. \ref{misept}. Again EM performs best.
Norm regularization is not competitive.

\begin{figure}
[ptb]
\begin{center}
\includegraphics[
trim=0.149959in 0.124951in 0.144589in 0.125241in,
height=4.1029in,
width=5.9269in
]%
{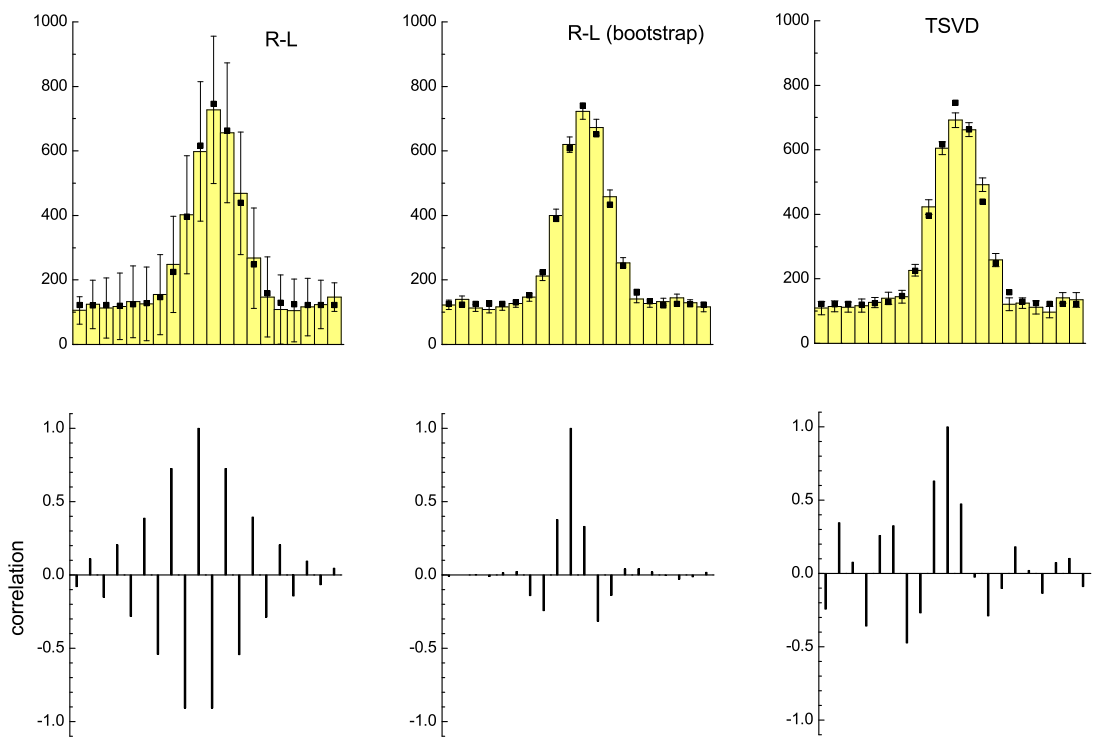}%
\caption{Unfolded histograms with assigned errors. The squares indicate the
true distribution. Below each histogram the correlation of the content of bin
10 with the other bins is plotted.}%
\label{unferr1}%
\end{center}
\end{figure}
%

\begin{figure}
[ptb]
\begin{center}
\includegraphics[
trim=0.148624in 0.156655in 0.137920in 0.163003in,
height=3.9676in,
width=5.9086in
]%
{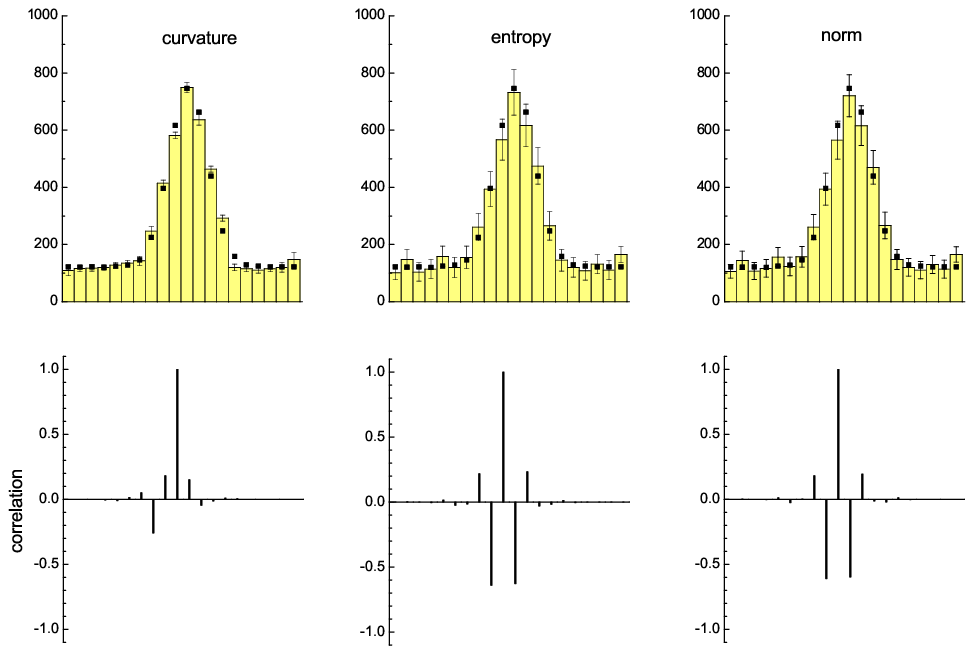}%
\caption{Same as previous figure but for penalty regularization.}%
\label{unferr2}%
\end{center}
\end{figure}

\section{Parameter estimation from unfolded histograms}

Cousins has proposed the following \emph{bottom-line test }\cite{cousins}:
\textquotedblleft If the unfolded spectrum and supplied uncertainties are to
be useful for evaluating which of two models is favored by the data (and by
how much), then the answer should be materially the same as that which is
obtained by smearing the two models and comparing directly to data without
unfolding \textquotedblright.

In this report a similar test is applied which maintains the idea behind
Cousins' proposal: In parametric models parameter estimates obtained from the
unfolded histogram should be as precise as those obtained directly from the
data. While Cousins emphasizes goodness-of-fit we concentrate on parameter
estimation. The two approaches should lead to the same conclusions.

We use the unfolding results of the five methods as applied to four simulated
samples to estimate the $4$ parameters of the one-peak distribution. This
example is chosen because the smearing and unfolding effects are best studied
in examples that have sharp structures.

Simulation results for the single peak example and moderate smearing
$\sigma_{s}=0.04$ are shown in the Figs. \ref{unferr1} and \ref{unferr2}. The
errors are derived from the curvature matrix of the log-likelihood function at
its maximum, except for the bootstrap errors of the EM approach which are
obtained from $10000$ bootstrap samples derived from the observed sample after
each time $4$ EM iterations\footnote{Due to an initial programming mistake,
the random data used for the boostrap approach differ from those used in the
other methods.}. The correlation of all bins relative to bin $10$ indicate how
the regularization reduces the negative correlation. As curvature
regularization acts locally, the correlation between adjacent bins is
positive. The size of the errors in the EM method are not affected by the
regularization. The diagonal errors are large and there is a strong negative
correlation between adjacent bins.

\begin{figure}
[ptb]
\begin{center}
\includegraphics[
trim=0.000000in 0.173892in 0.000000in 0.174171in,
height=1.9294in,
width=3.1in
]%
{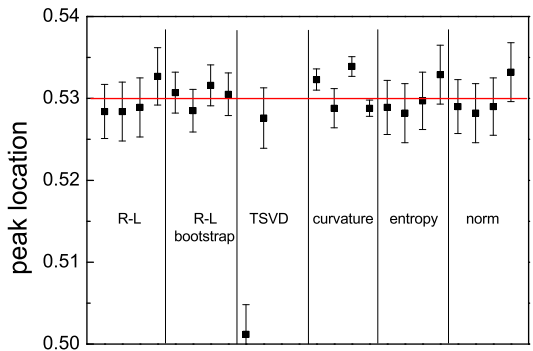}%
\caption{Fit results for the location of the peak.}%
\label{cousins1}%
\end{center}
\end{figure}

\begin{figure}
[ptb]
\begin{center}
\includegraphics[
trim=0.000000in 0.182265in 0.000000in 0.194267in,
height=1.7783in,
width=2.8916in
]%
{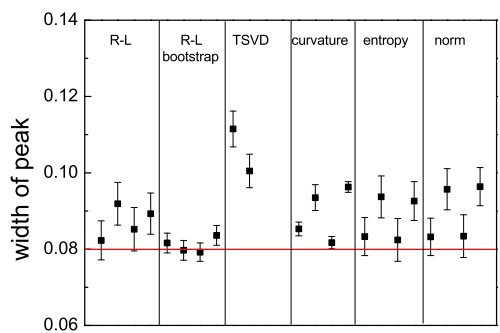}%
\caption{Fit results for the width (st. dev.) of the peak.}%
\label{cousins2}%
\end{center}
\end{figure}

\begin{figure}
[ptb]
\begin{center}
\includegraphics[
trim=0.000000in 0.173536in 0.000000in 0.173807in,
height=1.8464in,
width=3.0859in
]%
{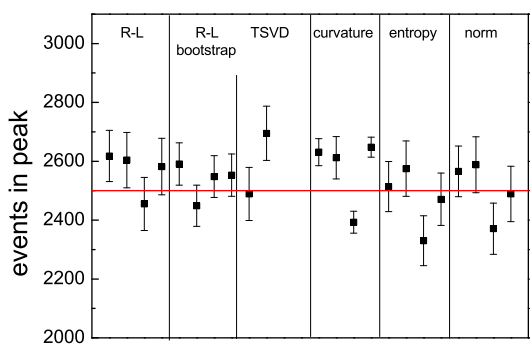}%
\caption{Fit results for the number of events that are associated to the
peak.}%
\label{cousins3}%
\end{center}
\end{figure}

\begin{figure}
[ptb]
\begin{center}
\includegraphics[
trim=0.000000in 0.172723in 0.000000in 0.184111in,
height=1.8796in,
width=3.1523in
]%
{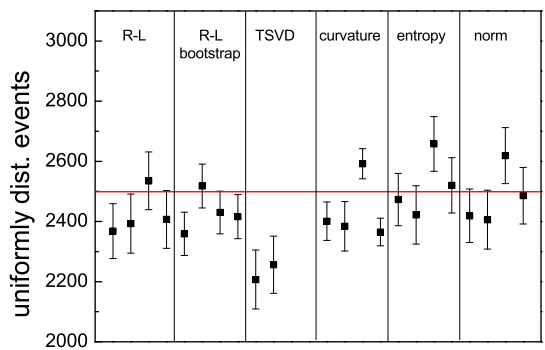}%
\caption{Fit results for the number of uniformly distributed events.}%
\label{cousins4}%
\end{center}
\end{figure}

The results of the parameter fits are displayed in Figs. \ref{cousins1},
\ref{cousins2}, \ref{cousins3} and \ref{cousins4}. The fit failed in two cases
with the TSVD approach which we do not consider further.

The peak location, Fig. \ref{cousins1}, is found with similar precision in the
EM method and the three penalty methods. The error estimates of the curvature
regularization fail to cover the true value which is indicated by the
horizontal line. This is no surprise, as the penalty term in the fit reduces
the errors. The error assignments obtained of the entropy and norm
regularization seem to be adequate.

As expected, the width of the peak, Fig. \ref{cousins2}, is reconstructed too
wide in all methods except in the EM method with bootstrap errors. The small
bias in the latter case is hard to explain.

The event numbers are reasonably well estimated in all methods except in TSVD
and EM with bootstrap errors.

We conclude that TSVD and EM iterative unfolding with bootstrap errors should
be discarded and that the error estimates of the unfolding procedure with a
curvature penalty are doubtful. The performance of the remaining methods is
similar. There is a sizable and unavoidable bias in the reconstructed width of
the peak. The bias is approximately as large as the statistical error. The
number of simulated experiments is too small for more detailed conclusions.%

%

%

%

\begin{figure}
[ptb]
\begin{center}
\includegraphics[
trim=0.141465in 0.092294in 0.141844in 0.062839in,
height=4.5214in,
width=5.9551in
]%
{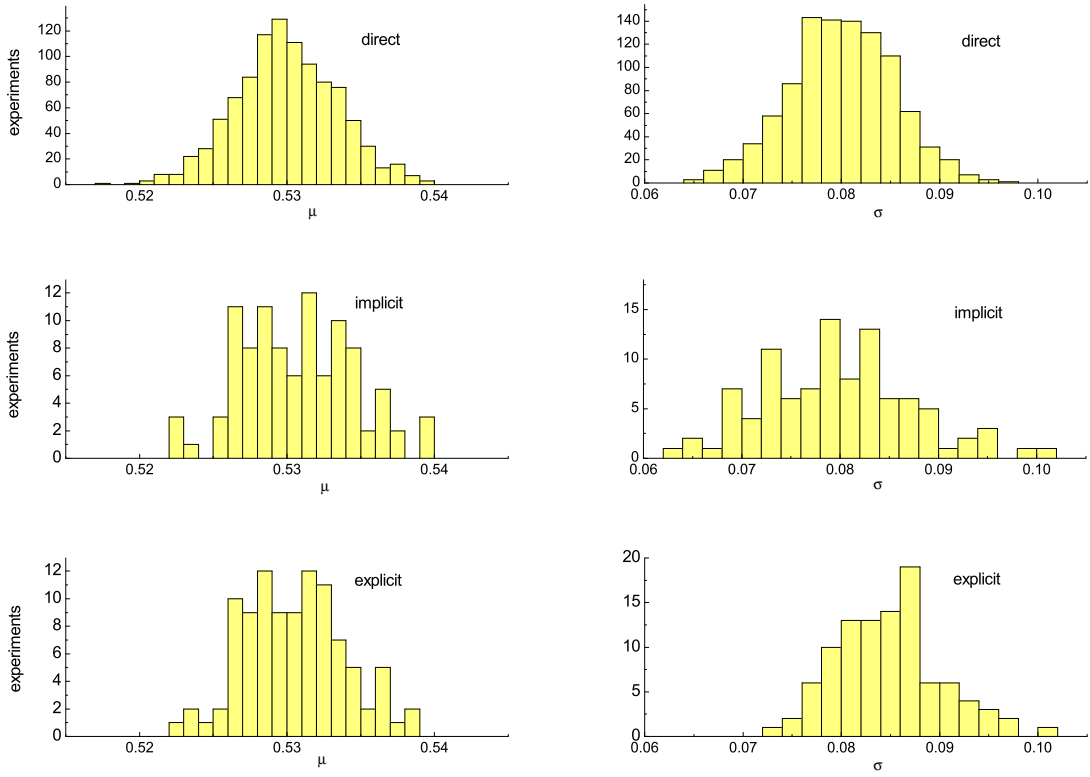}%
\caption{Comparison of fits of the peak location (left) and its width (right)
of the one-peak example with $\sigma_{s}=0.08$. From top to bottom: direct fit
to the data, implicit regularization by binning, explicit entropy
regularization. }%
\label{compregu}%
\end{center}
\end{figure}

In the following we compare the parameter estimates for the one-peak example
with a widths of the peak, $\sigma=0.08$ a) directly from the observed sample
with the weighting method, b) from the unfolded histogram with $8$ bins
without explicit regularization, c) from the unfolded histogram with entropy
regularization, including the full error matrix, d) from the unfolded
histogram with entropy regularization with the diagonal errors only. $100$
experiments with $5000$ events are simulated. The results are summarized in
Fig. \ref{compregu} and Tables \ref{tabfit4}. The mean values of the fitted
location of the peak and its width are given together with the errors. The
corresponding errors for a single experiment are larger by a factor of $10$.

The direct fit exhausts all information and is more precise than all other
methods. The results obtained for the width with entropy resolution are
strongly biased especially if only the diagonal errors are considered. The
bias is about as large as the statistical error. We conclude that Cousins'
test fails.

The errors of the fits with implicit regularization by wide binning are
moderately larger than those with explicit regularization but are unbiased.
They could probably be reduced by optimizing the bin width.

The situation will be different for rather smooth distributions. It can even
happen that the regularization penalty reduces the errors to below the
statistical limits of an undistorted distribution. This is acceptable for a
probability density estimate but not for a standard measurement. Smooth
distributions can be unfolded with wide bins without much loss in information
and then an explicit regularization is obsolete.%

\begin{table}[ptb]
\caption{Estimation of the parameters of a peak with width $\sigma=0.08$.}%
\label{tabfit4}%
\centering
\begin{tabular}
[c]{|l|l|l|l|l|l|l|}\hline
& \# experiments & $\overline{\mu}$ & $\delta\bar{\mu}$ & $\overline{\sigma}$
& $\delta\bar{\sigma}$ & bias$_{\sigma}$\\\hline
nominal &  & 0.5300 &  & 0.08 &  & \\
direct & 100 & 0.52997 & 0.00033 & 0.08059 & 0.00052 & 0.00059\\
implicit & 100 & 0.53070 & 0.00039 & 0.07969 & 0.00077 & -0.00031\\
explicit & 100 & 0.53052 & 0.00033 & 0.08483 & 0.00052 & 0.00483\\
explicit diagonal & 100 & 0.53097 & 0.00040 & 0.08527 & 0.00063 &
0.00527\\\hline
\end{tabular}
\end{table}

\section{Summary and recommendations}

We have considered five different unfolding methods:

\begin{itemize}
\item \emph{Methods based on Singular Value Decomposition:} The SVD offers the
possibility to discard the disturbing high frequency components of the
unfolded histogram. Truncated SVD is to be preferred to selective SVD. A
smooth cut-off is not recommended. The application of TSVD is technically
simple, but from the considered examples we have to conclude that the
performance is not competitive with that of the other considered methods. TSVD
fails in some cases with low event numbers.

\item \emph{EM unfolding:} The EM method is extremely simple, effective and
fast. It can be applied also to spline representations of the unfolded
distribution. In the considered examples it performs better than the competing methods.

\item \emph{Unfolding with curvature regularization: }This popular method
favors linear distributions. A positive feature is the local smoothing which
is adapted to point spread distortions. It is more difficult to fix the
regularization strength than in other methods. Due to the strong bin to bin
correlations the fit of the unfolded histogram has to be assisted to find the
absolute minimum. Its application to multi-dimensional histograms is
problematic. The assigned errors occasionally exclude the true distribution
even in toy experiments and in some low statistics examples the method fails
completely to reproduce the true distribution. Some of these difficulties are
avoided if the distribution is approximated by a superposition of $b$-splines.

\item \emph{Unfolding with entropy regularization}: The method, like norm
regularization, favors uniform distributions. In most of the considered
examples, the unfolding results obtained with entropy regularization are more
precise than those with curvature or norm regularization.

\item Tikhonov's\emph{ norm regularization:} The performance in the considered
examples is inferior to entropy regularization from which it differs only by
the weighting of the of the fluctuations.
\end{itemize}

TSVD and the EM method are both based on plausible general concepts while the
penalties used to regularize the results, favor specific features of the distributions.

All methods lead to biased results. The nominal errors that can be attributed
to the unfolded histogram are to be considered as rough estimates. The size of
the error bars depends on the regularization parameter, the errors do not
cover and sometimes exclude even rather smooth distributions. The assigned
errors are partially subjective and should be considered as an educated guess
of the author. An exception are the errors derived in the very special method
\cite{kuuselastark}. In the EM method they cover approximately, but due to the
strong negative correlation of adjacent bins, the diagonal errors are large
and exaggerate the uncertainties in the graphical representation where only
the diagonal errors can be shown. The unfolding results are usually closer to
the true distribution than those obtained with the wide bin method. They help
to visualize the true distribution, are useful to exclude qualitatively
predictions and can be applied to simulate the distribution. The point
estimates derived from the unfolded distributions are biased and less precise
than the values derived directly from the observed data. The bottom-line test
fails. Because of the bias of the point estimates and the partially arbitrary
size of the errors, they cannot be used in quantitative goodness-of-fit tests
and parameter estimation.

Further requirements and remarks:

\begin{itemize}
\item It is recommended to fix the regularization parameter by visual
inspection or with the iterative minimization of the $ISE$.

\item The unfolding results depend on the distribution used to determine the
response matrix. The dependence can be kept small, if in a first step the data
are unfolded into a superposition of $b$-splines. The result is then used to
determine the response matrix.

\item There are indications that a representation of the unfolded distribution
in form of a superposition of $b$-splines is closer to the true distribution
than that with a histogram. Furthermore in the methods with penalty
regularization it is not necessary to construct a response matrix and the
problem related to its Monte Carlo construction is avoided.

\item The considered range of the true variable has to cover all observed
variable values. In most Monte Carlo studies this problem is avoided by
artificially restricting the true space.

\item It is important to study possible systematic errors due to the limited
knowledge of the response function. The statistical errors introduced by the
response matrix can be estimated with resampling techniques.

\item Programs off the shelf should be used with caution. The user and not the
author of the program has the responsibility of the correctness of the results
and therefore must understand the assumptions and approximations that are made.

\item It would be interesting to compare the more sophisticated methods
\cite{truee,dembinski,kuusela} to the simple iterative EM unfolding.

\item Publications should always include unbiased results like those provided
by histograms with wide bins without explicit regularization. In this way the
requirements for the presentation of the unfolding results obtained with
explicit regularization can be relaxed.
\end{itemize}

\chapter{Summary of the summaries and a personal recommendation}

\begin{itemize}
\item If a theoretical prediction exists, parameters can be extracted directly
from the distorted data. It is not necessary to unfold and to construct a
response matrix.

\item The probability density estimates (PDEs) obtained by unfolding smeared
data with explicit regularization provide a good illustration of the
underlying true distribution but cannot be used to derive quantitative
conclusions. The iterative EM unfolding with early stopping performs
significantly better than the other considered unfolding methods.

\item If no prediction exists, it is recommended to use the EM method with
early stopping to obtain a preliminary spline approximation of the true
distribution. This result is then used to determine the optimal number of the
EM iterations by the iterative optimization of the $ISE$ and to generate an
improved response matrix. The data are unfolded a second time, again to a
spline approximation. Qualitative error bands derived with bootstrap
resampling methods can be associated to the result. They should be accompanied
by an explanation of their relevance. They can be obtained with bootstrap
resampling. From the regularized unfolding result a final response matrix is
constructed which then is used to unfold the data into an effectively unbiased
histogram with wide bins and to compute an error matrix. In this way the
probability density estimate (PDE) is accompanied by the information that is
needed to estimate parameters and to perform goodness-of-fit tests.
\end{itemize}

It is hoped that this report will stimulate further systematic studies.

\chapter*{Acknowledgement}

I am very grateful to Mikael Kuusela and Victor Panaretos for their patience
in explaining to me their unfolding methods and for related discussions. I
thank Hans Dembinski for an interesting exchange of ideas concerning the
optimization of the regularization, Bob Cousins for helpful comments and his
interest in my work, Gerhard Bohm and Vato Kartvelishvily for useful comments.
I am especially grateful to Igor Volobouev for pointing out to me some
mistakes in formulas of this report and I appreciate his critical comments.

\chapter*{Appendices}

\section*{Appendix 1: The Compound Poisson Distribution}

\label{appendixcpd}If a parameter of a distribution is itself randomly
distributed, then we have a compound distribution. The \emph{compound Poisson}
\emph{distribution }(CPD) describes the sum of a Poisson distributed number of
very independent and identical distributed weights. In the case we are
interested in, we have a continuous weight distribution. A number $n$ of
weights $w_{i}$, $i=1,..,n$, are randomly chosen from a p.d.f. $g(w)$ where
$n$ follows a Poisson distribution $\mathcal{P}(n|\lambda)=e^{-\lambda}%
\lambda^{n}/n!$. The distribution $f(x)$ where $x=\Sigma_{i=1}^{n}w_{i}$ has a
complicate analytic form, but its moments are easy to calculate.

The first two moments of $f(x)$ follow directly from the definition:%
\begin{align*}
\mu &  =\lambda\boldsymbol{E}(w)\\
\sigma^{2}  &  =\lambda\boldsymbol{E}(w^{2})
\end{align*}
Skewness and kurtosis are also simple functions of the expected values of
powers of the weight \cite{CPD}:%
\begin{align}
\gamma_{1}  &  =\frac{\boldsymbol{E(}w^{3})}{\lambda^{1/2}\boldsymbol{E}%
(w^{2})^{3/2}}\;,\label{gamma1}\\
\gamma_{2}  &  =\frac{\boldsymbol{E}(w^{4})}{\lambda\boldsymbol{E(}w^{2})^{2}%
}\;. \label{gamma2}%
\end{align}

In the special case that all weights are equal, skewness $\gamma_{1p}%
=\lambda^{-1/2}$ and kurtosis $\gamma_{2p}=\lambda^{-1}$ of the Poisson
distribution are reproduced. For a narrow weight distribution, more precisely
if $\boldsymbol{E(}w^{2})^{1/2}\approx\boldsymbol{E(}w^{3})^{1/3}%
\approx\boldsymbol{E(}w^{4})^{1/4}\approx\boldsymbol{E(}w)$, the two shape
parameters are close to those of a Poisson distribution.%
\begin{figure}
[ptb]
\begin{center}
\includegraphics[
trim=0.000000in 0.183939in 0.000000in 0.184196in,
height=2.3271in,
width=4.5205in
]%
{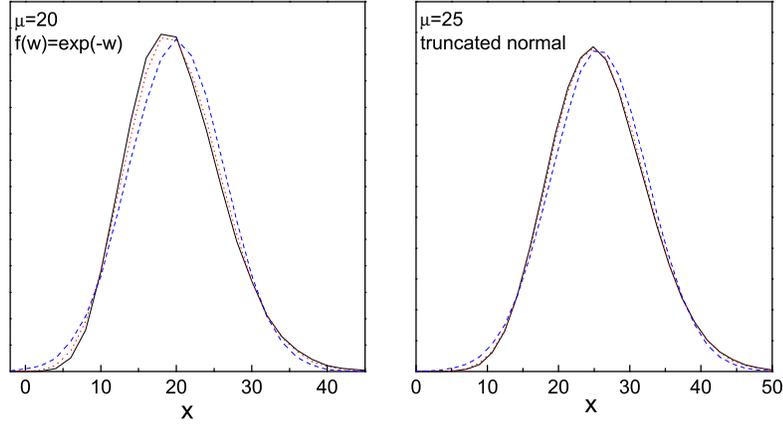}%
\caption{Comparison of the approximations of a compound Poisson distribution
CPD) by a scaled Poisson distribution (dotted) and a normal distribution
(dashed) to the CPD.}%
\label{spd}%
\end{center}
\end{figure}

This observation suggests to approximate the CPD by a scaled Poisson
distribution (SPD), $f(x)\approx\tilde{f}$ $(\tilde{x})$ where $\tilde
{x}=s\tilde{n}$ and $\tilde{n}\sim\mathcal{P}_{\tilde{\lambda}}$. The Poisson
distributed number $\tilde{n}$ is multiplied by a scaling factor $s$ which
depends on the weights such that the result $\tilde{x}$ has a similar
distribution as $x$. To reproduce the first two moments of the CPD,
$\mathrm{E}(\tilde{x})=\mathrm{E}(x)=\mu$, $\mathrm{Var}(\tilde{x}%
)=\mathrm{Var}(x)=\sigma^{2}$, we select $\tilde{\lambda}$ and the scaling
factor $s$ in the following way:%

\begin{align}
\tilde{\lambda}  &  =\lambda\frac{\mathrm{E}(w)^{2}}{\mathrm{E}(w^{2}%
)}\;,\label{enoe}\\
s  &  =\frac{\mathrm{E}(w^{2})}{\mathrm{E}(w)}\;.
\end{align}

In applications, the expected values of the powers of the weight are not
known. We have only a set of observed weights. We replace the expected values
by the empirical mean values $\bar{w}$, $\overline{w^{2}}$, set $\tilde
{n}=n\bar{w}^{2}/\overline{w^{2}}$ and use the observation that $\tilde{n}$ is
approximately Poisson distributed. The number $\tilde{n}$ is called\emph{
equivalent number of unweighted events} or \emph{effective number of events}
because the $n$ observed events provide the same statistical significance as
$\tilde{n}$ Poisson distributed events.

It can be shown that the SPD is a much better approximation of the CPD than a
normal distribution \cite{CPD}. This conclusion relies on simulations and on
the analytic result that the SPD shape parameters are closer to the CPD values
than those of the normal distribution where we have $\gamma_{1}=\gamma_{2}=0$:%

\begin{align}
\gamma_{1}(N)  &  =0<\gamma_{1}(SPD)\leq\gamma_{1}(CPD)\;,\label{gamma1spd}\\
\gamma_{2}(N)  &  =0<\gamma_{2}(SPD)\leq\gamma_{2}(CPD)\;. \label{gamma2spd}%
\end{align}

Two examples are shown in Fig. \ref{spd}. In the left-hand graph the sum of
$n$ exponentially distributed weights is displayed. The number $n$ is Poisson
distributed with mean $20$. Even though the weight distribution is very wide,
the SPD approximates the CPD (the discrete distribution has been approximated
by a polygon) very well, much better than the normal approximation. In the
right-hand graph the weight distribution is a truncated normal distribution
$\mathcal{N}(x>0|1,1)$. Here the agreement is so good that the SPD and the CPD
are hardly distinguishable.

\section*{Appendix 2: Error calculation of $(d-ct)$}

\subsection*{Poisson evaluation}

\label{weighterror}Given be two observed Poisson distributed numbers $N_{1}$,
$N_{2}$, distributed according to $N_{1}\sim\mathcal{P}_{\lambda}$ and
$N_{2}\sim\mathcal{P}_{a\lambda}$ with given $a$ and unknown $\lambda$. The
log-likelihood of $\lambda$ is%
\[
\ln L(\lambda)=-\lambda-a\lambda+N_{1}\ln\lambda+N_{2}\ln(a\lambda)\;.
\]
The value $\hat{\lambda}$ that maximizes $L$ obeys
\[
\frac{d\ln L}{d\lambda}=0=-(1+a)+N_{1}/\hat{\lambda}+N_{2}/\hat{\lambda}\;.
\]
The result is%
\[
\hat{\lambda}=\frac{N_{1}+N_{2}}{1+a}\;.
\]

With $N_{1}=d$, $N_{2}=\tilde{m}$, we get for the expected value
$\lambda=\mathrm{E}(d)=\mathrm{E}(ct)=\mathrm{E}(cs\tilde{m})$ and with
\textrm{E}$(\tilde{m})=\lambda/(cs)$we find%

\[
\hat{\lambda}=\frac{d+\tilde{m}}{1+1/(cs)}\;.
\]
The error $\delta$ of the difference $d-ct=d-cs\tilde{m}$ is%
\begin{align*}
\delta^{2}  &  =\delta^{2}(d)+\delta^{2}(cs\tilde{m})\\
\hat{\lambda}  &  =\delta^{2}(d)+c^{2}s^{2}\delta^{2}(\tilde{m})\;.
\end{align*}
The errors have to be calculated from the expected values $\mathrm{E}%
(d)=\lambda$, $E(\tilde{m})=\lambda/(cs)$:%

\begin{align*}
\delta^{2}  &  =\lambda(1+cs)\\
&  \approx\frac{d+\tilde{m}}{1+1/(cs)}(1+cs)\\
&  \approx cs(d+\tilde{m})\\
&  \approx c(\frac{\sqrt{\overline{w^{2}}}}{\overline{w}}d+t)\;.
\end{align*}

\subsection*{Normal approximation}

Under the assumption that the expectation of $d$ and $ct$ is $\lambda$, the
errors are $\delta_{d}^{2}=\lambda$ and $\delta_{t}^{2}=\lambda/c$. We form a
LS expression%
\[
\chi^{2}=\frac{(d-\lambda)^{2}}{\lambda}+\frac{(ct-\lambda)^{2}}{c\lambda}%
\]
which we minimize with respect to $\lambda:$%
\[
\frac{d\chi^{2}}{d\lambda}=\frac{-2\lambda}{c\lambda^{2}}\left[  c\left(
d-\lambda)-(ct-\lambda\right)  \right]  -\frac{1}{c\lambda^{2}}\left[
c(d-\lambda)^{2}+(ct-\lambda)^{2}\right]  =0\;.
\]
We find%
\[
\hat{\lambda}=\left[  \frac{d^{2}+ct^{2}}{1+1/c}\right]  ^{1/2}%
\]
and%
\[
\delta^{2}=\delta_{d}^{2}+\delta_{ct}^{2}=\hat{\lambda}(1+c)
\]

Now we consider the more general case$\delta_{d}^{2}=c^{\prime}\lambda$.%

\[
\chi^{2}=\frac{(d-\lambda)^{2}}{c^{\prime}\lambda}+\frac{(ct-\lambda)^{2}%
}{c\lambda}%
\]
and get%
\[
\hat{\lambda}=\left[  \frac{cd^{2}+c^{\prime}c^{2}t^{2}}{c+c^{\prime}}\right]
^{1/2}%
\]
and%
\[
\delta^{2}=\delta_{d}^{2}+\delta_{ct}^{2}=\hat{\lambda}(c^{\prime}+c)
\]

\section*{Appendix 3: Spline approximation}

\subsection{Quadratic $b$-splines}

\label{appendixspline}%
\begin{figure}
[ptb]
\begin{center}
\includegraphics[
trim=0.000000in 0.229965in 0.000000in 0.184233in,
height=1.8755in,
width=4.0872in
]%
{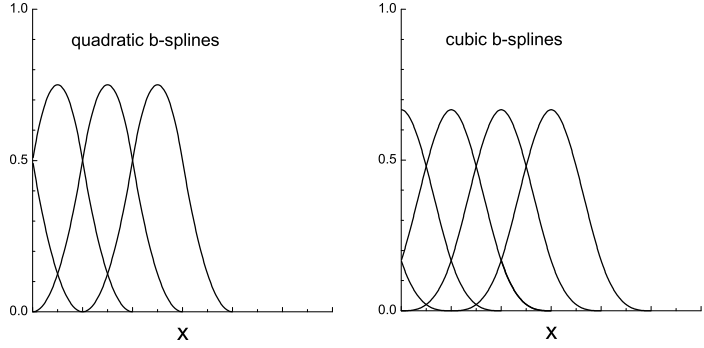}%
\caption{Quadratic and cubic b-spline functions.}%
\label{bspline1}%
\end{center}
\end{figure}
A quadratic $b$-spline (see Fig. \ref{bspline1}) centered at $x_{0}$ with a
bin width $b$ is defined by
\begin{align*}
B(x|x_{0},b)  &  =\frac{1}{2b}\left(  \frac{x-x_{0}+3/2b}{b}\right)
^{2}\mbox{
for }\;x_{0}-3b/2\leq x\leq x_{0}-b/2\;,\\
&  =\frac{1}{2b}\left[  \frac{3}{2}-2\left(  \frac{x-x_{0}}{b}\right)
^{2}\right]  \mbox{ for }x_{0}-b/2\leq x\leq x_{0}+b/2\;,\\
&  =\frac{1}{2b}\left(  \frac{x-x_{0}-3/2b}{b}\right)  ^{2}\mbox{ for
}x_{0}+b/2\leq x\leq x_{0}+3b/2\;,\\
&  =0\;\;\mbox{ else}\;.
\end{align*}

For a given interval [$x_{\min},x_{\max}$] subdivided into $n$ bins of width
$b=(x_{\max}-x_{\min})/n$ the number of quadratic $b$-spline functions is
$n+2$. Two $b$-splines are centered outside the interval at $x_{0}=x_{\min
}-b/2$ and $x_{\max}+b/2$, respectively.

\subsection*{Cubic $b$-splines}

Cubic $b$-splines are defined as follows:
\begin{align*}
B(x|x_{0},b)  &  =\frac{1}{6b}\left(  2+\frac{x-x_{0}}{b}\right)  ^{3}\mbox{
for }x_{0}-2b\leq x\leq x_{0}-b\;,\\
&  =\frac{1}{6b}\left[  -3\left(  \frac{x-x_{0}}{b}\right)  ^{3}-6\left(
\frac{x-x_{0}}{b}\right)  ^{2}+4\right]  \mbox{ for }x_{0}-b\leq x\leq
x_{0}\;,\\
&  =\frac{1}{6b}\left[  3\left(  \frac{x-x_{0}}{b}\right)  ^{3}-6\left(
\frac{x-x_{0}}{b}\right)  ^{2}+4\right]  \mbox{ for }x_{0}\leq x\leq
x_{0}+b\;,\\
&  =\frac{1}{6b}\left(  2-\frac{x-x_{0}}{b}\right)  ^{3}\mbox{ for }x_{0}%
+b\leq x\leq x_{0}+2b\;,\\
&  =0\;\;\mbox{ else}\;.
\end{align*}

For a given interval [$x_{\min},x_{\max}$] subdivided into $n$ bins of width
$b=(x_{\max}-x_{\min})/n$ the number of cubic $b$-spline functions is $n+3$.
Two $b$-splines are centered outside the interval at $x_{0}=x_{\min}-b$, ,
$x_{\max}+b$ and two at the borders.

\section*{Appendix 4: Choice of the bin width for parametric models}

\label{appendixbin}When we determine a parameter $\theta$ of a distribution
$f(x|\theta)$ from a data sample of independent and identically distributed
events ($x_{1,}x_{2},...,x_{N}$), the first choice of the method is a
binning-free likelihood fit. Often this is not possible or too computer time
consuming and we have to group the data in bins.

Wide bins have the advantage that the fluctuations of the number of entries
can be approximated by a normal distribution, and that correlations due to the
limited experimental resolution are small, but the disadvantage is that the
parameter resolution can suffer. For instance the precision with which we can
infer the location and width of a narrow peak over a smooth background
decreases with increasing bin width. The exact relation between bin width and
resolution depends on the shape of the distribution and the parameters of
interest but we can derive a rule of thumb based on the Nyquist-Shannon theorem:

\textquotedblleft If a function $y(t)$ contains no frequency higher than
$\nu_{\max}$, it is completely determined by giving its values at a series of
points spaced $1/(2\nu_{\max})$ apart\textquotedblright

The Nyquist-Shannon theorem is of fundamental importance in the field of
digital signal processing. It cannot be applied directly to our problem but it
provides a hint for reasonable choices of the bin width. We can associate a
maximum spatial frequency $\nu_{\max}$to the the function $f(x)$ that we allow
and apply the Nyquist-Shannon theorem to infer the number of points, here the
number of histogram bins that we need to fix the function parameters. Instead
of considering the frequency we turn to its inverse, the
bandwidth\footnote{Unfortunately this definition is opposite to the definition
of bandwidth in technical applications where is denotes the range of
frequencies.} and get $w<h_{f}/2$ with $w$ the bin width of the histogram and
$h_{f}$ the band width of the narrowest structure of the function $f(x)$. Let
us assume that the lowest bandwidth $h_{f}$ is due to a narrow Gaussian peak
with standard deviation $\sigma$. The bandwidth of normal distribution can be
approximated by $h_{f}=\sqrt{2}\sigma$. (This is used in kernel density
estimation.) With our crude estimate (a factor of two can hardly be argued),
we find:%
\[
w<\sigma/\sqrt{2}%
\]
%

\begin{figure}
[ptb]
\begin{center}
\includegraphics[
trim=0.000000in 0.236542in 0.000000in 0.185148in,
height=2.6227in,
width=3.8065in
]%
{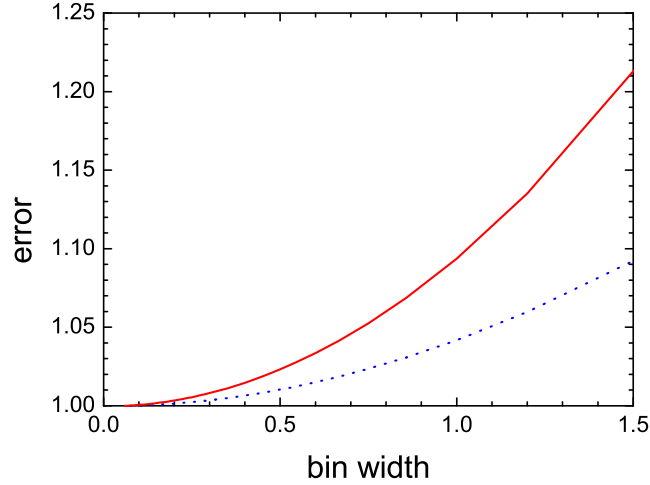}%
\caption{Error estimates for the mean (dotted curve) and standard deviation
(solid curve) of a normal distribution obtained from a data histogram. The
errors are given in units of the error for infinitely small bins and the bin
width is in units of the nominal standard deviation.}%
\label{resolbin}%
\end{center}
\end{figure}

The bin width should not be larger than the standard deviation of the peak.

A quantitative result can be obtained in the following simple example:

A large data sample is used to determine the mean $\mu$ and the standard
deviation $\sigma$ of a normal distribution. The data are histogrammed into
bins of equal width $b$. The nominal values of the distribution be $\mu_{0}=0$
and $\sigma_{0}=1$ and the data are contained in the interval $-3<x<3$. The
LSF test statistic $\chi^{2}$ with $B=6/b$ bins is%
\[
\chi^{2}=\sum_{i=1}^{B}\frac{(cn_{i}-t_{i})^{2}}{t_{i}}\;.
\]

Here $n_{i}$ is the number of events in bin $i$ which, multiplied by a
normalization constant $c$, is compared to the integral of the normal
distribution over the interval $[x_{1i},x_{2i}]$ given by the bin boundaries:%

\begin{align*}
t_{i}  &  =\int_{x_{1i}}^{x_{2i}}\frac{1}{\sqrt{2\pi}\sigma}\exp\left[
-\frac{(x-\mu)^{2}}{2\sigma^{2}}\right]  dx\\
&  =\operatorname{erf}\left[  \frac{x_{2i}-\mu}{\sqrt{2}\sigma}\right]
-\operatorname{erf}\left[  \frac{x_{1i}-\mu}{\sqrt{2}\sigma}\right]  \;.
\end{align*}

The minimum $\chi^{2}=0$ of $\chi^{2}$ is obtained with the settings $c=1$,
$\mu=0$, $\sigma=1\,$, $n_{i}=t_{i}$. The uncertainties of the parameters
$\mu$, $\sigma$, correspond to the boundary $\chi^{2}=1$. Of interest is the
dependence of the errors $\delta_{\mu}$ and $\delta_{\sigma}$ on the bin
width. To be independent of the number of events, we consider the ratios
$\delta_{\mu}/\delta_{\mu\infty}$ and $\delta_{\sigma}/\delta_{\sigma\infty}$
where $\infty$ stands for the limit of an infinite number of bins of zero
width. In Fig. \ref{resolbin} the ratios are plotted as a function of the bin
width in units of $\sigma_{0}$. For a bin width equal to $\sigma_{0}$ the
resolution of $\sigma$ is reduced by about $9.3\,\%$ with respect to the
optimal resolution and for $\mu$ the degradation is about $4.1\,\%$. For the
Nyquist estimate $w=\sigma/\sqrt{2\text{ }}$ the numbers are $4.8\,\%$ for
$\sigma$ and $2.1\,\%$ for $\mu$ respectively .

Remark: This estimate does not apply to the non-parametric case of probability
density estimation (PDE). There the quality of the agreement of the estimated
distribution $\hat{f}(x)$ which in our case is a histogram, is usually
measured with the mean integrated square error $MISE$:%
\[
MISE=\int_{-\infty}^{\infty}\left[  \hat{f}(x)-f(x)\right]  ^{2}dx\;.
\]

The formula for the$MISE$ assumes a \emph{uniform distribution} inside the
bin. Here $\hat{f}(x)$ corresponds to the lowest order spline approximation,
while in physics applications the bin content is an estimate of the
\emph{integral of the distribution} and the quality of the agreement is mostly
based on the goodness-of fit statistics $\chi^{2}$.

\section*{Appendix 5: Simplex convergence}

\label{appendixsimplex}Simplex is a wide spread robust method to find the
parameters that maximizes of a non-linear function. Contrary to many other
minimization methods it does not require to determine derivatives of the
function that has to be maximized. It has been shown that Simplex converges in
deterministic problems but not necessarily in problems where the data are
affected by noise. The problem increases with the number of parameters.

A description of the simplex algorithm can be found in many textbooks and need
not be repeated here. There are many variants (see refs. in \cite{tomik95} of
the original version of Nelder and Mead (\cite{nelder-mead}). Standard Simplex
\cite{nelder-mead} has been used in all fits of this manuscript. With $N$
parameters, we initialize the first of the $N+1$ parameters vectors (points in
the $N$-dimensional parameter space) to the true parameter values known in our
Monte Carlo studies. The remaining vectors differ each in one parameter value
by the expected uncertainty from the true value. If the number of parameters
is large, and especially if the parameters are correlated, Simplex fits have
the tendency to stop without having reached the function minimum
\cite{tomik95}. Simplex may choose shrinkage while a reflection of the worst
parameter point could be the optimal choice. Finally, all points have almost
equal parameter coordinates such that the convergence criterion is fulfilled.
Further improvement steps are so small that reducing the convergence parameter
does not change the result. The convergence problem is studied in great detail
in \cite{tomik95} and a solution which introduces stochastic elements in the
stepping process is proposed.

In this manuscript a different approach is followed. After Simplex signals
convergence, the fit is repeated where the best point so far obtained is kept
and the remaining points are initialized in the same way are before.
Alternatively these points are chosen randomly centered at the best value.

Both ways have been applied. The following parameters were used: The maximum
number of steps in Simplex was set to a rather low value of $5000$. The
starting values of the parameters were set to the true values which are known
for the Monte Carlo studies. The convergence parameter was $10^{-12}$. If the
log-likelihood values of two subsequent steps differ by less than this value,
the fit is stopped. The fit is repeated up to $200$ times each time keeping
the best parameter point but modifying for each of the remaining $N$ points
one parameter. Then a maximum of $1000$ or sometimes $10000$ additional fits
with random initialization of the $N$ points are added. If $20$ subsequent
fits do not change the double precision function value within the precision of
the computer, the fitting is terminated. The fit with curvature penalty is
especially problematic. This is due to the strong correlation of the
neighboring bin contents. With entropy and norm penalties usually $50$
repetitions are sufficient.

\section*{Appendix 6: Expectation Maximization Algorithm}

The EM method finds iteratively the MLE in situations where the statistical
model depends on latent variables. The method goes back to the sixties, has
been invented several times and has been made popular by Dempster, Laird and
Rubin \cite{em}. A very comprehensive introduction to the expectation
maximization (EM) algorithm is given in the Wikipedia article\ Ref.
\cite{wikiEM}. The EM algorithm exists in many different variants. We will
restrict our discussion to its application to classification problems.

To get an idea of the method, we consider a simple standard example. Let us
assume that we have a sample of observations $x_{1},...,x_{N}$ each drawn from
one of $M$ overlapping normal distribution $f_{m}(x|\mu_{m})\thicksim
\mathcal{N}(\mu_{m},s)$ with unknown mean values $\mu_{1}$,..., $\mu_{M}$ and
given standard deviation $s$. The log-likelihood for the parameters is
\[
\ln L(\mu_{1},...,\mu_{N})=\sum_{m=1}^{M}\frac{\Sigma_{i=1}^{N}z_{mi}x_{i}%
}{\Sigma_{i=1}^{N}z_{mi}}-\mu_{m})
\]
where the classification variable $z_{mi}=1$ if $x_{i}$ belongs to the normal
distribution $m$ and $z_{mi}=0$ otherwise. If we know the classification
variables, we get the MLE of the parameter $\mu_{m}$:%
\[
\hat{\mu}_{m}=\frac{\Sigma_{i=1}^{N}z_{mi}x_{i}}{\Sigma_{i=1}^{N}z_{mi}}\;.
\]
If this is not the case, we can estimate $z_{mi}$ from the observed
distribution. In the EM formalism $z_{mi}$ is called \emph{missing} or
\emph{latent} \emph{variable}.\ We can solve our problem iteratively with two
altenating steps, an expectation and a maximization step. We start with a
first guess \ $\mu_{m}^{(1)}$ of the parameters of interest and estimate the
missing data. In the \emph{expectation step} $k$ we compute the probability
$g_{mi}^{(k)}$that $x_{i}$ belongs to subdistribution $m$. It is proportional
to the value of the distribution $f_{m}(x_{i}|\mu_{m})$ at $x_{i}$:%
\[
g_{mi}^{(k)}=\frac{f_{m}(x_{i}|\hat{\mu}_{m}^{(k)})}{\Sigma_{j=1}^{M}%
f_{j}(x_{i}|\hat{\mu}_{m}^{(k)})}\;.
\]
The probability $g_{mi}$ is the expected value of the latent variable $z_{mi}%
$.The \emph{expected} log-likelihood is%
\[
Q(\vec{\mu},\widehat{\vec{\mu}}^{(k)})=\sum_{m=1}^{M}\left(
{\displaystyle\sum\limits_{i=1}^{N}}
g_{mi}^{(k)}x_{i}-\mu_{m}\right)  \;.
\]
In the \emph{maximization} step we obtain the MLEs%
\[
\hat{\mu}_{m}^{(k+1)}=\Sigma_{i=1}^{N}g_{mi}^{(k)}x_{i}\;.
\]
which are used in the following expectation step. The iteration converges to
the overall MLE.

Let us generalize this procedure. Given be a probability distribution
$p(\vec{x},\vec{z}|\vec{\theta})=g(\vec{z}|\vec{x}$ ,$\vec{\theta})p_{1}%
(\vec{x}|\vec{\theta})$ depending on a parameter vector $\vec{\theta}$ and a
sample of observations $x_{1},...,x_{N}$. The distribution $g$ of the latent
variables $\vec{z}$ is a function of $\vec{\theta}$ and $\vec{x}$.

\begin{itemize}
\item \emph{Expectation step}: For the parameter vector $\vec{\theta}^{(k)}$
we compute the distribution $g$ of the hidden variables $\vec{z}$. We form the
log-likelihood function $\ln L(\theta|\vec{x},\vec{z})$ which is a random
variable as it depends on the random $z$. Averaging over $z$, we obtain the
expected value $Q(\vec{\theta},\widehat{\vec{\theta}}^{(k)})$ of the
log-likelihood:
\[
Q(\vec{\theta},\widehat{\vec{\theta}}^{(k)})=\boldsymbol{E}_{z|x,\theta^{(k)}%
}\left[  \ln L(\vec{\theta}|\vec{x},\vec{z})\right]  \;.
\]
The conditional expectation means that we average over $\vec{z}$ given the
distribution of $g(\vec{z})$ obtained for fixed values $\vec{x}$ and
$\widehat{\vec{\theta}}^{(k)}$:%
\[
Q(\vec{\theta},\widehat{\vec{\theta}}^{(k)})=\int_{\mathcal{Z}}\ln
L(\vec{\theta}|\vec{x},\vec{z})g\left(  \vec{z}|\vec{x},\widehat{\vec{\theta}%
}^{(k)}\right)  dz\;.
\]
If the values of the vector components $\vec{z}$ are discrete, the integral is
replaced by a sum over all $J$ possible values:%
\[
Q(\vec{\theta},\widehat{\vec{\theta}}^{(k)})=%
{\displaystyle\sum\limits_{j=1}^{J}}
\ln L(\vec{\theta}|\vec{x},\vec{z}_{j})g\left(  \vec{z}_{j}|\vec{x}%
,\widehat{\vec{\theta}}^{(k)}\right)  \;.
\]
Alternatively, somewhat less efficient, we can insert the expected values of
the latent variables:%
\[
Q(\vec{\theta},\widehat{\vec{\theta}}^{(k)})=\ln L\left(  \vec{\theta}|\vec
{x},\boldsymbol{E}(\vec{z})\right)  \;.
\]

\item \emph{Maximization step}: The MLE $\vec{\theta}^{(k+1)}$ is computed:%
\[
\vec{\theta}^{(k+1)}=\arg\max_{\theta}Q(\theta|\theta^{(k)})\;.
\]

\end{itemize}

The procedure is started with a first $\vec{\theta}^{(1)}$ guess of the
parameters and iterated. It converges to a minimum of the log-likelihood. To
avoid that the iteration is caught by a local minimum, different starting
values can be selected. It is especially useful in classification problems in
connection with p.d.f.s of the exponential family\footnote{To the exponential
family belong among others the normal, Poisson, exponential, gamma,
chi-squared distrribution.} where the maximization step is relatively simple.

\begin{myexample}
Unfolding a histogram

Experimental data are collected in form of a histogram with $N$ bins. The
number of events in bin $i$ be $d_{i}$. The experiment suffers from an
imperfect resolution and from acceptance losses which we have to correct for.
The "true" histogram with $M$ bins contains $\theta_{j}$ events in bin $j$.
Knowing the measurement device we can simulate the experimental effects and
determine the matrix $A$ which relates $\vec{\theta}$ with the expected values
of the numbers $\vec{d}$: $\boldsymbol{E}(d_{i})=\Sigma_{j=1}^{M}A_{ij}%
\theta_{j}$. The element $A_{ij}$ is the probability to observe an event in
bin $i$ which belongs to the bin $j$ in the true histogram. The missing
information is the number of events $d_{ij}$ in an observed bin $i$ that
belong to the true bin $j$. Hence there are $M$ missing variables per bin. The
number $d_{ij}$ is Poisson distribution with mean $A_{ij}\theta_{j}$. The
likelihood depends only on the hidden variables:
\[
\ln L(\vec{\theta}|d_{11},...,d_{NM})=%
{\displaystyle\sum\limits_{j=1}^{M}}
{\displaystyle\sum\limits_{i=1}^{N}}
[-A_{ij}\theta_{j}+d_{ij}\ln A_{ij}\theta_{j}]\;.
\]
The following alternating steps are repeated:

\begin{itemize}
\item Expectation step: We have%
\begin{align*}
Q(\vec{\theta},\widehat{\vec{\theta}}_{j}^{(k)})  &  =\boldsymbol{E}_{d_{ik}%
}\ln L\\
&  =%
{\displaystyle\sum\limits_{j=1}^{M}}
{\displaystyle\sum\limits_{i=1}^{N}}
[-A_{ij}\theta_{j}+\boldsymbol{E}(d_{ij}^{(k)})\ln A_{ij}\theta_{j}]\;.
\end{align*}
The expected value $\boldsymbol{E}(d_{ij}^{(k)})$ conditioned on $d_{i}$ and
$\widehat{\vec{\theta}}^{(k)}$ is given by $d_{i}$ times the probability that
an event of bin $i$ belongs to true bin $j$:
\[
\boldsymbol{E}(d_{ij}^{(k)})=d_{i}\frac{A_{ij}\hat{\theta}_{j}^{(k)}}{%
{\displaystyle\sum\limits_{j=1}^{M}}
A_{ij}\hat{\theta}_{j}^{(k)}}\;.
\]
We get%
\[
Q(\vec{\theta},\widehat{\vec{\theta}}_{j}^{(k)})=%
{\displaystyle\sum\limits_{j=1}^{M}}
{\displaystyle\sum\limits_{i=1}^{N}}
[-A_{ij}\theta_{j}+d_{i}\frac{A_{ij}\hat{\theta}_{j}^{(k)}}{%
{\displaystyle\sum\limits_{m=1}^{M}}
A_{im}\hat{\theta}_{m}^{(k)}}\ln A_{ij}\theta_{j}]\;.
\]

\item Maximization step:
\end{itemize}

The computation of the maximum of $Q$ is easy, because the components of the
parameter vector $\vec{\theta}$ appear in independent summands.
\begin{align*}
\frac{\partial Q}{\partial\theta_{j}}  &  =%
{\displaystyle\sum\limits_{i=1}^{N}}
\left[  -A_{ij}+d_{i}\frac{A_{ij}\hat{\theta}_{j}^{(k)}}{%
{\displaystyle\sum\limits_{j=1}^{M}}
A_{ij}\hat{\theta}_{j}^{(k)}}\frac{1}{\theta_{j}}\right]  =0\;,\\
\hat{\theta}_{j}^{(k+1)}%
{\displaystyle\sum\limits_{i=1}^{N}}
A_{ij}  &  =%
{\displaystyle\sum\limits_{i=1}^{N}}
d_{i}\frac{A_{ij}\hat{\theta}_{j}^{(k)}}{%
{\displaystyle\sum\limits_{j=1}^{M}}
A_{ij}\hat{\theta}_{j}^{(k)}}\;,\\
\theta_{j}^{(k+1)}  &  =%
{\displaystyle\sum\limits_{i=1}^{N}}
d_{i}\frac{A_{ij}\theta_{j}^{(k)}}{%
{\displaystyle\sum\limits_{j=1}^{M}}
A_{ij}\theta_{j}^{(k)}}\left/  \alpha_{j}\right.  \;.
\end{align*}

$\Sigma_{i=1}^{N}A_{ij}=\alpha_{j}$ is the average acceptance of the events in
the true bin $j$.
\end{myexample}

\end{document}